%% file: cernrepexa.tex
\documentclass{cernrep}
\RequirePackage[dvipsnames]{xcolor}
\usepackage{hyperref}
\usepackage{xspace}
\usepackage{relsize}
\usepackage{xcolor}
\usepackage{soul}

\begin{document}
\title{Heavy quarks and jets as probes of the QGP}

\author{Liliana Apolinário\textsuperscript{1,2}\thanks{liliana@lip.pt}, Yen-Jie Lee\textsuperscript{3}\thanks{yen-jie.lee@cern.ch}, Michael Winn\textsuperscript{4}\thanks{michael.winn@cea.fr}}
\institute{\textsuperscript{1}LIP, Av. Prof. Gama Pinto, 2, 1649-003 Lisboa, Portugal \\
\textsuperscript{2} Instituto Superior Técnico (IST), Universidade de Lisboa, Avenida Rovisco Pais 1, 1049-001 Lisbon, Portugal\\
\textsuperscript{3}Massachusetts Institute of Technology (MIT), Cambridge, Massachusetts, USA\\
\textsuperscript{4}Département de Physique Nucléaire (DPhN), Institut de Recherche sur les lois Fondamentales de l'Univers (IRFU)
CEA - Saclay, F-91191 Gif-sur-Yvette Cedex, France}

\begin{abstract}
Quark-Gluon Plasma (QGP), a QCD state of matter created in ultra-relativistic heavy-ion collisions, has remarkable properties including, for example, a low shear viscosity over entropy ratio. By detecting the collection of low-momentum particles that arise from the collision, it is possible to gain quantitative insight into the created matter. However, its fast evolution and thermalization properties remain elusive. Only the usage of high momentum objects as probes of QGP can unveil its constituents at different wavelengths. In this review, we attempt to provide a comprehensive picture of what was, so far, possible to infer about QGP given our current theoretical understanding of jets, heavy-flavor, and quarkonia. We will bridge the resulting qualitative picture to the experimental observations done at both the LHC and RHIC. We will focus on the phenomenological description of experimental observations, provide a brief analytical summary of the description of hard probes, and an outlook towards the main difficulties we will need to surpass in the following years. To benchmark QGP-related effects, we will also address nuclear modifications to the initial state and hadronization effects. 
\end{abstract}


\maketitle

\clearpage
\tableofcontents
\clearpage
\input{introduction.tex}
\input{motivation.tex}

\input{initialstate.tex}
\input{acceleration2.tex}
\input{QGPDensity3.tex}
\input{QGPcontent.tex}
\input{time.tex}

\input{outlook.tex}

\input{acknowledgements.tex}

\newpage

\bibliographystyle{utphys}   

\bibliography{bibliography}

\end{document}

%% file: introduction.tex
\section{Introduction}

The strong interaction is one of the four known forces in physics. In the experimentally accessible energy regime, it is described by a quantum field theory: quantum chromodynamics (QCD). Albeit we believe that we know the underlying equations, the theory's degrees of freedom, quarks and gluons (force carriers) are not directly observed in nature. They are confined in complex objects, the hadrons, such as protons and neutrons, of the size of $10^{-15}$~m. QCD is a non-abelian gauge theory where gluons also carry strong interaction 'color' charges allowing them to interact with each other. As such, in contrast to electromagnetic and weak interactions, calculations for low momentum exchanges are not amenable to perturbation theory. Interestingly, the building blocks of nuclei, hadrons such as protons and neutrons, make up most of the known ordinary energy in the universe, and the gluon field inside them dominates their mass.

The hadron's fundamental constituents only reveal themselves at a large-momentum exchange or, equivalently, short distances. An example is the production of high momentum particles or jets, which serve as a proxy for the outgoing quark or gluon resulting from such large energy exchange interaction. One may wonder, then, how can we 'see' these quarks and gluons move freely over larger distances than the typical hadron size? The answer is "we can do it indirectly". We heat a macroscopic volume of these hadrons in the laboratory to release them and form a new QCD matter state named quark-gluon plasma (QGP).

\subsection{Why do we study quark-gluon plasma?}

Quark-gluon plasma~(QGP) is a high-temperature matter phase built from up-to-now fundamental QCD particles. This thermodynamic system can be characterized by the temperature and the chemical potentials associated with quantum numbers conserved under strong interaction, such as the baryon number. At asymptotically large temperatures and low chemical potentials, the strong interaction becomes weakly coupled due to asymptotic freedom. The thermodynamic properties can be well described by weak-coupling perturbation theory~\footnote{For a recent review see~\cite{Ghiglieri:2020dpq} and for the complete expression of the free energy~\cite{PhysRevD.53.3421}}: a free plasma of quarks and gluons appears at zeroth order of the coupling constant. But can we trust the extrapolation of our knowledge of single-particle scattering processes and present theory calculations from asymptotically high energies and temperatures to lower scales?

At temperatures around $T = 150~\rm{MeV}$, we approach the hadronization transition. However, the thermodynamic system becomes non-trivial before being characterized as a gas of hadrons. Its conceptual and quantitative description is currently an open quest. For zero baryochemical potential, we know from lattice gauge theory calculations of the equation of state that the transition between hadrons and QGP is a rapid cross-over. Beyond this theoretical insight, we do not understand the degrees of freedom of this matter when probed at different energy scales. Studying strongly interacting matter in extreme conditions tests our understanding of parton confinement inside hadrons and mass generation via the strong interaction, the two properties of QCD defining matter as we know it. The concepts applied to QGP range from solving the gauge theory on the lattice, stretching weak coupling techniques to their limits and employing the AdS/CFT correspondence to explore the strongly coupled regime of gauge theory~\cite{Casalderrey-Solana:1639593}. 

Experimental observations allowed us to establish the applicability of close-to-ideal hydrodynamics when modeling the produced matter. Hence, they have set the paradigm of the strongly coupled QGP. The shear viscosity over entropy ratio appears to be close to AdS/CFT calculations indicating the smallest observed value in nature. This makes these strongly coupled systems resemble ultra-cold fermi gases that can be produced in modern atomic physics set-ups in the laboratory (see for a comparison of the different systems in~\cite{Schafer:2009dj} and for a recent extraction of $\eta/s$ in~\cite{PhysRevLett.119.065302}).

The interest in QGP is much broader than studying fundamental QCD interactions at the different phase-space regions. In a sense, the questions addressed in heavy-ion collisions or the tools used to study the created matter are shared with nearby physics fields research as condensed matter physics, high-energy physics, hadron structure, and astrophysics.

To understand QGP characteristics, it is crucial to understand the effects originated by the incoming nuclei, the initial state of the collision, from the subsequent evolution of the created matter (QGP itself). As it turns out, the colliding nuclei in the initial state are not a simple superposition of nucleons even at the high collision energies. Most of the particle production is driven by a densely packed system dominated by gluons, and it is still unclear if this system shows signs of saturation\footnote{This assumed property of hadronic wave functions at very large energies results from non-linear recombination phenomena that limit the continuous growth of the partons occupancy level inside hadrons (see section~\ref{sec:initial} for more details).}. Its observation is still one of the open questions in hadron structure studies. 

As for tools used to study QGP, high-transverse momentum jets analyzed in heavy-ion collisions are one of the building blocks of high-energy physics searches in the hunt for new physics. The tools and object definition strategies naturally overlap between the two fields, benefiting both communities. 

The application of thermodynamics and hydrodynamics implying very fast 'hydrodynamization' of the system~\footnote{The word hydrodynamization refers to situations where the fluid cells are far from local equilibrium, but the system can already be described by hydrodynamics.}, brings the field of QGP physics in contact with more general questions related to universal features of out-of-equilibrium physics and their approach to thermal equilibrium. One can ask: When does a hydrodynamic description become suitable~\cite{Nagle:2018nvi}, and how do we get to it~\cite{Schlichting:2019abc}?  

Beyond the understanding of strong interactions and the approach to equilibrium in a gauge theory, the early universe was in this state of matter during its lifetime about microseconds after its creation. The core of neutron stars may be composed of quark matter relevant to the physics of neutron star mergers~\cite{Annala:2019puf}. Quark matter physics is therefore connected to cosmology and astrophysics.  

Quantifying QGP and its transition to ordinary hadrons is essential for understanding the strong interaction and connecting to other physics fields. In contrast to particle physics, QGP physics, as a field of many-body physics, does not follow a reductionist path to understanding nature but studies emergent phenomena at the smallest length scales available.

\subsection{Accessing QGP in heavy-ion collisions: matter production and 'external' probes}

To experimentally study QGP, ultra-relativistic heavy-ions are brought to collisions at accelerator facilities at Brookhaven National Laboratory (USA) and CERN (Europe). We can create a deconfined matter at close to zero net-baryon number at midrapidity by reaching high enough energy densities. The lifetime of the created medium is, however, very short, around $10$ fm/$c$. Therefore, the properties of this state of matter have to be extracted indirectly from detecting particles emitted from the collision point at much later time scales. Thermodynamic and hydrodynamic concepts are applied to soft particle production, namely, to the descriptions of light-flavor hadron production and correlations at low transverse momentum scales. They carry the vast majority of the deposited energy in the collision zone. These small transverse momenta and, hence, long-wavelength modes can be used to characterize the system within a hydrodynamic modeling set-up. Namely, its space-time evolution, the matter velocity fields, thermodynamic and transport properties for energy-momentum or quantum numbers conservation. However, these modes decouple late from the system's time evolution and allow only to directly access the time evolution of macroscopic properties and details of the decoupling\footnote{Soft electromagnetic radiation allows inferring the information with a probe that decouples at the moment of the photon or virtual photon emission to a very good approximation.}. To study the microscopic properties at early times, we need to test the system with higher resolution using 'external' probes that can be identified from the soft hadronic bulk of the collision.

Fortunately, particles from high-momentum exchanges are produced at the nuclei's impact and can be handled as external probes that are not fully thermalized. Given a sufficiently large momentum exchange, their production rate can be calculated within perturbative QCD (pQCD). These so-called 'hard' probes traverse the created matter. By using the different high-momentum objects produced at the collision, we can resolve different length scales and access information not available from the bulk particle emission. Furthermore, because their production is amenable to pQCD, they can be used to infer information on the initial state of the heavy-ion collision via system measurements where QGP modifications are small. This review focuses on the two most prominent hard probes of this type: jets and heavy flavor particles. 

This document attempts to provide an overview of heavy quarks and jets as tools to investigate the properties of QGP produced in heavy-ion collisions. The following chapters are built following the physics-oriented topics that these objects can probe. In chapter~\ref{sec:motivation}, we attempt to motivate the use of high-momentum objects as unique probes of QGP. Any attempt to obtain information regarding the created medium needs precise knowledge regarding the initial state. As such, in chapter~\ref{sec:initial}, we will focus on the current uncertainties on nuclear Parton Distribution Functions (PDFs), their improvement based on jets and heavy-flavor measurements and how they are currently affecting the interpretation of the present experimental results. It follows chapter~\ref{sec:acceleration} where we provide a general picture of the current experimental results on jets and heavy flavor. This chapter also provides a brief state-of-the-art theoretical description of QGP-induced modifications on these objects. While much of our current understanding is due to qualitative features observed in data, we can provide quantitative estimates for QGP transport coefficients. The procedure to infer these quantities from hard probes and the current landscape when compared with the soft sector are presented in chapter~\ref{sec:QGPDensity}. Despite its high-momentum scale, jets and heavy flavor are subject to non-perturbative effects such as hadronization. In chapter \ref{sec:hadronization}, we discuss the current uncertainties on hadronization mechanisms when extrapolating from $e^+e^-$ to $pp$ and $AA$ collisions. Lighter systems are currently an open problem as it is still unclear whether there are the necessary conditions for developing a QGP droplet. We will motivate the use of hard probes as an additional handle to solve this puzzle in chapter \ref{sec:smallsystem}. With the current leverage on hard probes and estimates of average QGP quantities, the next natural step would be to exploit hard probes as tomographic probes of the created medium. Our perspective on the future use of jets and heavy flavor particles to unveil QGP time evolution will be addressed in chapter \ref{sec:timescale}. Finally, some general conclusions and outlook follow in chapter \ref{sec:outlook}.

%% file: motivation.tex
\section{Why do we study jets and heavy flavor particles?}
\label{sec:motivation}

After the two Lorentz-contracted ion disks collide, numerous interactions occur between quarks and gluons (also referred to as partons). Those mainly involve small transverse momentum transfers that will contribute to QGP formation. By measuring angular distributions and correlation functions of the low transverse momentum (soft) particles, it has been found that the obtained results are consistent with the expectation from a strongly interacting QGP~\cite{Busza:2018rrf}. Namely, the magnitude of the azimuthal anisotropy of the emitted final-state particles is compatible with models built from relativistic hydrodynamics with minimal shear viscosity. These low transverse momentum particles (known as 'soft probes') have provided several insights into QGP. However, there are still many unresolved questions that could not yet be answered directly with the studies of soft particles:

\begin{enumerate}
    \item How does a strongly coupled QGP emerge from an asymptotic free gauge theory?
    \item What is the substructure of QGP when probed at various length scales?
    \item What are the QGP transport properties?
\end{enumerate}

To answer the first question, ideally, we would like to trace a colored probe with known four-momentum that is initially far from thermalization and study how this object evolves in the medium as a function of time. To probe the QGP substructure, the most desirable experiment would be a scattering experiment, such as the one performed by Ernest Rutherford for the studies of the internal structure of an atom~\cite{Akiba:2015jwa,DEramo:2018eoy}. One could then learn about the transport properties of the medium by checking how the hard probes get scattered when plowing through the created medium. In addition, those hard probes could leave wakes beyond as they propagate through the plasma. The propagation of the wake trailing the probe could provide a further understanding of QGP transport properties, such as the speed of sound that Lattice QCD-based models can calculate. As such, hard probes allow us to study QCD in a high color density environment over an extensive range of energy scales involving not only $\Lambda_{\rm QCD}$ but also other scales related to the medium, e.g., the medium temperature and the Debye screening length. The theoretical developments associated with modifications of hard probes will push the limits of applicability of perturbative QCD and provide important tests on the scales where the perturbative approach breaks down. 

Although it is not yet feasible to probe QGP with an external probe due to the short lifetime of the medium (at the order of 10 yoctoseconds), large momentum transfer scatterings between partons from the nuclei happen occasionally. Those scatterings originate quarks and gluons with large transverse momentum that will develop concurrently with the remaining collision. These fast-moving and non-thermalized partons propagate through the produced medium and will interact with it losing energy. Among those, heavy quarks are particularly interesting. Because of the relatively small abundance in the QGP medium, the heavy-quark flavor (charm and beauty) rarely changes in the medium through elastic scatterings via strong interactions. It can only change via weak interaction with decay times of the order of picoseconds. Due to their large masses, they are expected to have a much longer relaxation time than light flavor partons (light quarks and gluons). This enables us to trace a slow-moving colored probe which opens a window to extract the in-medium color force. Moreover, heavy quarks are also expected to lose a smaller amount of energy through radiation due to the dead-cone effect\footnote{Suppression of small-angle gluon radiation induced by the lower heavy quark velocity at the same kinetic energy compared to light quarks. See section~\ref{sec:acceleration} for more details.} that could provide us an opportunity to test energy loss models.

Theoretically, the production rate of hard probes is calculable with perturbative QCD due to the large transverse momentum transfer involved in those processes. Those calculations have been tested extensively in elementary collision systems such as $e^+e^-$ and proton-(anti-)proton collisions. Moreover, those probes are expected to be produced early, so they experience the whole medium evolution. This means that the modification of hard probes is accumulated throughout their journey while traversing the rapidly cooling QGP medium.

\subsection{Different probes of QGP evolution}

\subsubsection{Jets for the studies of high momentum partons}

Experimentally, due to color confinement, one cannot directly detect high momentum quarks and gluons. They evolve and radiate into a beam of partons. At the end of the shower, these partons eventually hadronize, and final state hadrons are produced. The resulting hadron transverse momentum spectrum carries information about the modification of the parent parton produced in the initial hard scattering. However, on average, individual hadrons have a small fraction of the initial parton momentum. Due to the steeply falling momentum spectra characteristic of a hard scattering process and the shape of the fragmentation function of the outgoing partons, studies of transverse momentum hadron spectrum are also more sensitive to hadronization effects. 

To get a complete picture, jets, objects based on clusters of final-state particles originating from a quark or a gluon, are currently the closest experiment of this kind that could capture a more significant fraction of the mother parton momentum compared to individual hadrons. Jets are defined by jet clustering algorithms and can be used as a good proxy for the initiating quark or gluon. The most commonly used jet finding algorithm, in both proton-proton and heavy-ion collisions, is the anti-$k_{T}$. With the distance parameter set to $R$, this algorithm essentially behaves like a cone algorithm with radius $R$. In hadronic colliders, $R$ is usually defined as $R=\sqrt{\Delta y^2 + \Delta \phi^2}$, where $\Delta y$ is the rapidity difference between two particles and $\Delta\phi$ is the azimuthal angle difference. In the presence of a QGP, medium-induced gluon radiation and elastic energy loss could transfer the parton energy to distances larger than the jet area. By comparing jet momentum spectra in proton-proton and heavy-ion collisions, it is possible to extract the amount of the initiating parton energy that the jet algorithm cannot cluster into the main jet. Theoretically, the amount of energy loss is closely related to the medium's color density, stopping and elastic scattering power, which will be discussed in Sections~\ref{sec:acceleration} and ~\ref{sec:QGPDensity}. Moreover, jets are intrinsically multi-scale probes. Due to the probabilistic nature of the parton fragmentation process, partons of the same initial energy could form different parton showers. Classifying jets according to their substructure gives access to the parton shower history. One has thus more differential information about QGP structure. As we will discuss in Section~\ref{sec:timescale}, jets and heavy flavor have the potential to gain more insights into the QGP time evolution.

Attenuation of the jets, often referred to as jet quenching, is essential experimental evidence for the parton energy loss in the presence of a QCD matter. However, in addition to the medium-induced modification, jet observables such as jet momentum spectra~\cite{CMS:2016uxf,ATLAS:2018gwx} and dijet angular distributions~\cite{ATLAS:2010isq,CMS:2011iwn} are sensitive to the initial parton distribution functions of the heavy nuclei. Studies of nuclear modification of the parton distribution functions using jets in smaller systems such as proton-ion collisions, which will be discussed in Section~\ref{sec:initial}, are of great interest for interpreting the results in heavy-ion collisions.

\subsubsection{Heavy flavor hadrons for the studies of heavy quarks}

We face a similar issue in detecting heavy quarks, and the experimental observables are heavy-flavor hadrons or heavy-flavor-tagged jets. 
Heavy flavor particles are versatile probes because of their large mass compared to $\Lambda_{QCD}$ and the medium temperature. Therefore, their production cross-sections are calculable by perturbative QCD. They do not change identity via strong interaction inside the medium since QGP rarely contains heavy quarks with masses larger than its temperature. They are also produced early due to the large momentum transfer involved, and most of them will carry low momenta. With low momentum heavy-flavor hadrons, we hope to extract the diffusion coefficients of QGP and reveal the in-medium color force based on theoretical input (see section~\ref{sec:QGPDensity}). Measurements of intermediate to high transverse momentum heavy-flavor hadron~\cite{ALICE:2015vxz,CMS:2017uoy} and jets~\cite{CMS:2013qak} provide tests on jet quenching models. Mainly, those measurements are more sensitive to the elastic scattering power of QGP when compared to the light flavor, as will be discussed in section~\ref{sec:acceleration}. 

The organization process of quarks and gluons into color-neutral hadrons, the hadronization process, is a fundamental question in QCD. Due to its very low momentum transfer processes, the description of the hadronization is a challenging problem, even in vacuum. Studies of heavy quarks in heavy-ion collisions provide a unique opportunity to study the hadronization process in environments with different surrounding parton densities. In a high parton density environment, novel mechanisms such as those realized in \textit{parton recombination}, where hadrons are formed via a combination of nearby quarks, could change the heavy flavor baryons-to-meson multiplicity ratios. Moreover, the relative abundance of heavy-flavor hadrons with and without strangeness could reveal the flavor role of QGP's hadronization (see Section~\ref{sec:hadronization} for more information).

\subsubsection{Quarkonia with various size, $B_c$ and exotica}

More than 95\% of the produced heavy-quarks at the RHIC and LHC hadronized to heavy-flavor baryons and mesons. A small fraction of them forms Quarkonium states ($Q\bar{Q}$) and $B_c$ mesons. Those states are particularly interesting probes because of two heavy quarks in the hadron. Quarkonia with different hadron sizes and binding strengths are probes of the Debye screening lengths in the medium and the medium-induced decays since these finite temperature modifications produce different surviving probabilities inside QGP. Moreover, ab-initio calculations of equilibrium properties on the lattice are available, allowing for a data-model comparison.

Quarkonia spectra are also sensitive to the parton recombination involving two heavy quarks. The recombination may happen at the phase boundary or even in the hadron phase, which may be studied using exotic hadrons such as X(3872), a tetraquark (or hadron molecule) candidate. Together with heavy-flavor hadrons, one can draw an interesting connection between the in-medium color force, deconfinement, and hadronization of  QGP.

%% file: initialstate.tex
\section{The initial state of heavy-ion collisions with jets and heavy quarks}
\label{sec:initial}

The initial state of heavy-ion collisions at high energies is defined by the partons in the colliding nuclei. A nucleus is characterized by an average partonic density that determines the production rate of the released partons after the initial impact. In addition, there are event-by-event fluctuations induced by the relative position of the nucleons inside the nucleus and sub-nucleonic fluctuations. These will affect the energy deposition in the plane transverse to the colliding beams, thus being relevant for the azimuthal anisotropies of particle production in heavy-ion collisions.

In this chapter, we will introduce the concept of nuclear parton densities in section~\ref{subsec:partdens}. It follows a discussion of its influence on QGP physics in nucleus-nucleus collisions in section~\ref{subsec:impact}. We then address jet and heavy-flavor measurements in proton-nucleus collisions and their effects on nuclear PDF estimations in~\ref{subsec:partdens}. In this subsection, we also briefly address current limitations and possible caveats and discuss the impact parameter dependence of nPDFs. We conclude with a discussion of future directions in~\ref{subsec:new}.

\subsection{Parton distribution functions and their modification in nuclear collisions}
\label{subsec:partdens}

The production cross-section of jets and heavy quarks in hadronic collisions can be calculated in pQCD assuming factorization. We neglect, for the moment, the modifications of the final state kinematics induced after the initial hard scatterings in heavy-ion collisions (jet quenching, see chapter~\ref{sec:acceleration}). In this perspective, the production cross-section of a specific flavor parton is expressed as a convolution of the parton densities in the initial state and a hard matrix element calculated in pQCD. Therefore, the parton densities determine the abundance of QGP hard probes produced early in the collision. 

In collinear factorization, the total cross-section for the process $O$ with a single hard scale $Q$ in the collision of hadron $h1$ and hadron $h2$ can be expressed as:
\begin{align}
\label{eq:factorization}
    \sigma_O (s, Q^2) =&\\ \nonumber \mathlarger{\mathlarger{\sum}_{n=0}^{\infty}} \alpha_S^n(\mu_R^2) \cdot \mathlarger{\mathlarger{\sum}}_{i,j} &  \int \text{d}x_1 \text{d}x_2 f_{i/h1} (x_i, \mu_F^2) f_{j/h2} (x_j, \mu_F^2) \times \hat{\sigma}_{i,j \to O +X} (Q(x_i,x_j,s), \mu^2_R, \mu_F^2) \, , 
\end{align}
where the outer sum runs over powers of the coupling constant $\alpha_S$ and the inner sum runs over the $i$ and $j$ indices representing different parton species. Eq.~\ref{eq:factorization} considers only the leading powers in $1/Q$. The perturbative calculation for a specific parton pair $i,j$ is denoted by $\hat{\sigma}_{i,j \to O + X}$ and it is controlled by the hard scale of the process, $Q$, the renormalization scale, $\mu_R$, and the factorization scale, $\mu_F$. The non-perturbative functions $f_{i/h1},f_{j/h2}$ are the parton distribution functions~(PDFs) that depend on the fractional momentum carried by the parton $x_i$,$x_j$ and factorization scale, $\mu_F$.  A detailed introduction to perturbative calculations in collider physics can be found in Ref.~\cite{Campbell:2017hsr}.  

The presence of $\mu_F$ is a consequence of the factorization between the hard process and the non-perturbative PDF that absorbs the collinear infrared divergences of parton splittings below this scale. Thus, the PDF definition depends on the factorization scheme's choice and must be evaluated consistently. Nevertheless, a common choice is to set the factorization scale to the external hard scale. One often refers, for simplicity, to the 'scale' dependence of PDFs.    
The scale dependence of the non-perturbative PDF is given by the perturbative Dokshitzer-Gribov-Lipatov-Altarelli-Parisi equations~(DGLAP). At leading order in $\alpha_S$, it reads as:
\begin{align}
    \frac{\text{d}f_i(x,\mu_F)}{\text{d} \log \mu_F^2} &= \mathlarger{\mathlarger{\sum}}_j \int_x^1 \frac{dz}{z} \frac{\alpha_s}{2 \pi} P_{i\longleftarrow j }(z) f_j (\frac{x}{z}, \mu_F) = \frac{\alpha_S}{2 \pi} \mathlarger{\mathlarger{\sum}}_j (P_{i \longleftarrow j} \otimes f_j ) (x, \mu_F),
    \label{eq:DGLAP}
\end{align}
where $P_{i\leftarrow j }(z)$ is the splitting kernel from the parton type $j$ to parton of type $i$, and $z$ is the fractional longitudinal momentum carried by parton $j$ relative to parton $i$.  

In the case of single hadron production, where the hadron is not definable on the partonic level, the cross-section computation contains an additional convolution with a fragmentation function. It provides the transition from the perturbatively calculated object $O$ to the hadron in the final state. These functions will also depend on an additional factorization scale that marks the transition between the perturbative and non-perturbative physics with respect to the final state (for a review on fragmentation functions see~\cite{Metz:2016swz}~\footnote{ Under the assumption of factorization concerning the final state, fragmentation functions in a nucleus-nucleus collision can be used to encode the QGP-induced modifications, see section~\ref{sec:acceleration}.}).

Altogether, at finite order in perturbation theory, the result of the perturbative QCD calculation depends on the choice of the factorization and renormalization scales via logarithms of the type $\log(\mu/Q)$. It is common to set $Q= \mu_F = \mu_R$ and to vary this choice to assess the impact of higher-order corrections not included in the perturbative calculation. A discussion of scale dependence can be found in~\cite{Plehn:2009nd}. 

The PDFs are determined by fits to experimental data, and we use the DGLAP evolution equation to connect constraints from processes at different $Q$, but at the same $x$~\footnote{Constraints from lattice QCD become available, see for a discussion in~\cite{Lin:2017snn}.}. The amount of experimental data and the systematic experimental uncertainties will naturally dictate the constraining power of the PDF fitting method. A larger statistic and smaller systematic experimental uncertainties will reflect minor PDF uncertainties. An experimental observable can be used to further constrain the PDF, provided that its experimental uncertainty is smaller than the current PDF uncertainty. In addition, the observable must be calculable within collinear perturbative QCD with theoretical uncertainties smaller than the size of the PDF uncertainties
~\footnote{A discussion and implementation of theoretical uncertainties in PDF determinations is given in reference~\cite{NNPDF:2019ubu}.}. 

\begin{figure}
    \centering
    \includegraphics[width=0.2\textwidth]{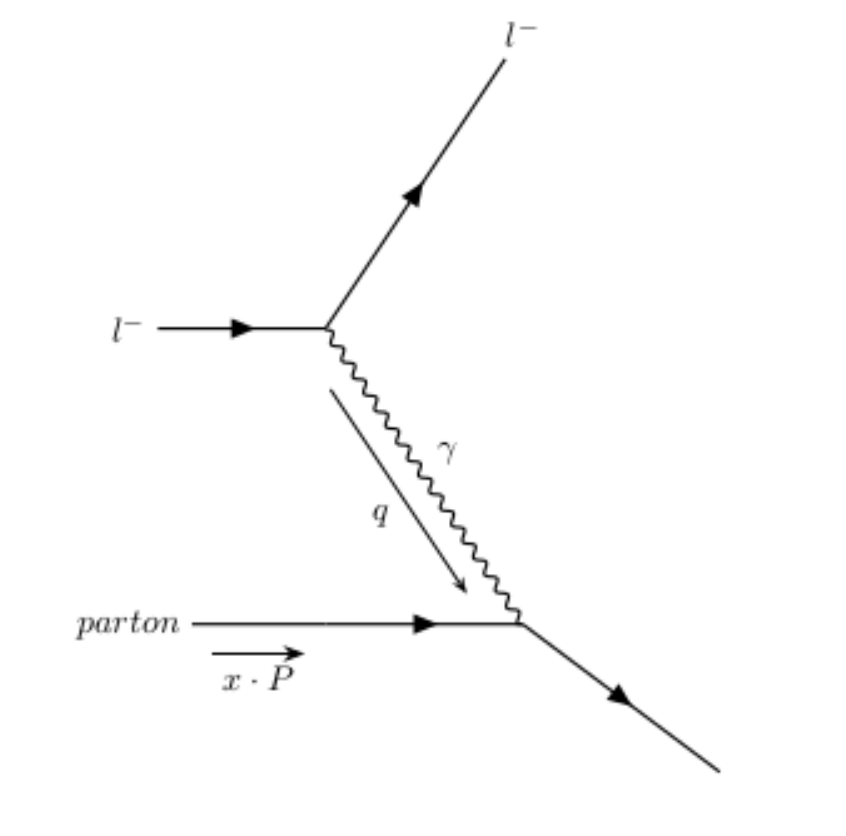} 
    \includegraphics[width=0.39\textwidth]{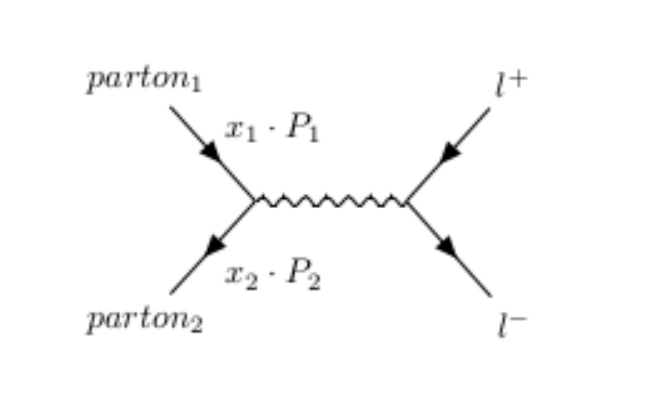} 
    \includegraphics[width=0.39\textwidth]{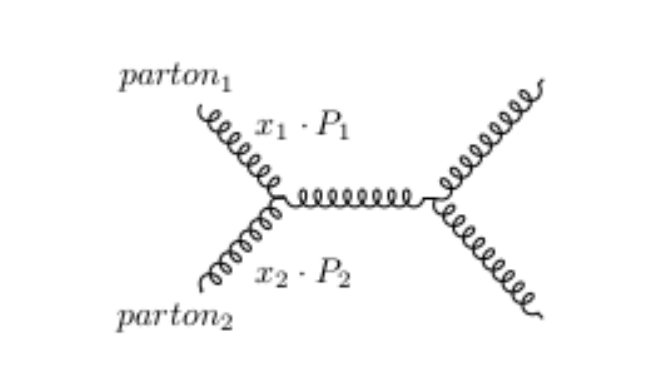} 
    \caption{Parton scattering Feynman diagrams contributing at leading order to deep-inelastic scattering in electron-hadron collisions (left), neutral Drell-Yan production in hadron-hadron collisions (middle), dijet production in hadron-hadron collisions (right).}
    \label{fig:var}
\end{figure}

The bulk of experimental data constraining PDFs stems from deep inelastic scattering~(DIS) of a lepton with a hadron. In DIS, the four-momentum transfer $Q^2$ is calculated from the incoming and outgoing lepton four momenta $q = p_{\rm out}-p_{\rm in}$: $Q^2 = -q^2$. The kinematic variable Bjorken-$x$ is calculated as $Q^2/(2\cdot Pq)$, where $P$ is the nucleon's four-momentum. In the $Q^2>>m_{proton}^2$ limit, the Bjorken-$x$ can be identified with the longitudinal momentum fraction of the probed parton in the nucleon. Fig.~\ref{fig:var}, on the left, shows the scattering of the lepton with a quark via a virtual photon in DIS. In addition, the Drell-Yan process in hadronic collisions, the production of virtual photons, W and Z bosons, are used to put constraints on PDFs. The lepton-pair invariant mass or transverse momentum provides the hard scale while the Bjorken-$x$ is correlated with the lepton kinematics in the final state. This is illustrated in Fig.~\ref{fig:var} (middle) for the case of lepton-antilepton production from a virtual photon in a hadron-hadron collision. Based on the involved hard perturbative scale and factorization, the production of jets and heavy quarks can also be used to constrain PDFs. Prototypically, a Feynman diagram contributing to dijet production is shown in Fig.~\ref{fig:var} (right).

The parton distributions of the proton are known with reasonable precision over a broad range of Bjorken-$x$ down to about $10^{-4}$ based on lepton-proton collider data from HERA and, more recently, W, Z and dijet measurements from the LHC~\cite{Ethier:2020way}. 

\begin{figure}
    \centering
    \includegraphics[width=0.5\textwidth]{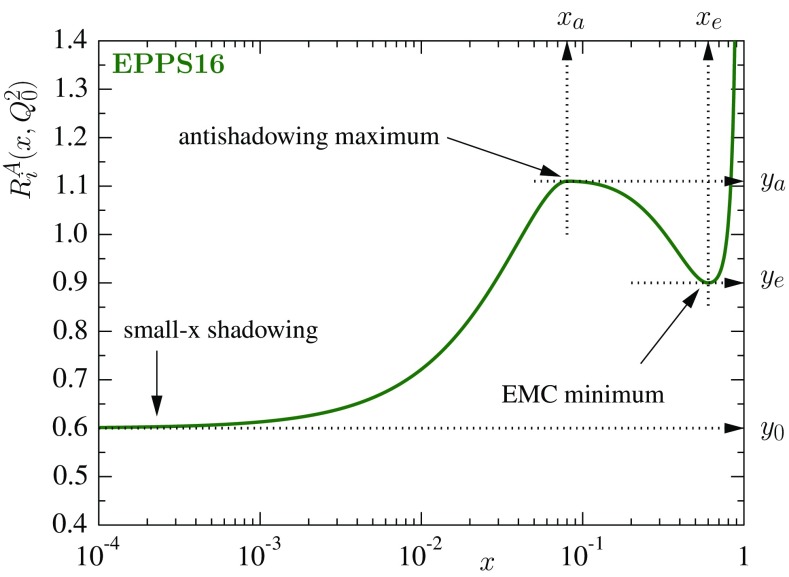} 
    \caption{Different regions of nuclear modification in the nuclear PDF parameterization of EPPS16 taken from Ref.~\cite{Eskola:2016oht}.} 
    \label{fig:shadowingregions}
\end{figure}

A nuclear PDF (nPDF) must be defined in hadronic collisions involving nuclei rather than protons. In this case, one can define the nPDF in the following way:
\begin{align}
    f_{i,(A,Z)} (x,\mu_F) = \frac{Z}{A} f_{i,p(A,Z)} (x,\mu_F) +\frac{A-Z}{A} f_{i,n(A,Z)} (x, \mu_F), 
\end{align}
generalizing Eq.~\ref{eq:factorization} for nucleon-nucleon collisions and taking into account the nuclear mass number $A$ and charge number $Z$ of the nucleus. The average protons and the neutrons PDFs contained inside the nucleus are denoted as $f_{i,p(A,Z)}$ and $f_{i,n(A,Z)}$ respectively. In the simplest scenario, the nuclear PDF $f_{i,(A,Z)}$ is a superposition of the free protons' and neutrons' PDF. This assumption leads to a scaling of the cross-section with the nuclear mass number $A$ (neglecting the difference between the neutrons' and protons' valence quarks) when substituting the interacting nucleon for a nucleus of the same mass number $A$. However, deviations from this simple expectation were reported from the SLAC experiment in 1972 when measuring the total hadronic photoabsorption cross-section for carbon, copper, and lead nuclei~\cite{PhysRevD.7.1362}. In the 1980s, the European Muon Collaboration~(EMC) saw strong nuclear modifications of DIS cross-sections (see, e.g., in~\cite{Aubert:1983xm}). Since then, this phenomenon has been measured with increasing precision. 

There is no quantitative theoretical description for the observed nuclear effects until today. They are associated to multiple scattering phenomena at low and intermediate $x$~\cite{Armesto:2006ph} or consequences of gluon saturation at low $x<10^{-2}$~\cite{Gelis:2010nm}. Antishadowing is commonly discussed in terms of momentum and baryon number sum rules balancing the modifications at lower $x$. The EMC-effect is often associated with short-range correlations at high-$x$~\cite{Schmookler:2019nvf} and/or with Fermi-motion at high-$x$~\cite{PhysRevLett.125.262002}. The resulting nuclear modification of PDFs~(ratio between nPDF and proton PDF) is also found experimentally to depend on the nuclear mass number of the nucleus. Accordingly, one separates phenomenologically four approximate regions of the nuclear modification of parton distribution functions as depicted in Fig.~\ref{fig:shadowingregions}:
\begin{itemize}
\item at the lowest Bjorken-$x$, the shadowing region  
\item the antishadowing region around $x=10^{-1}$ 
\item the EMC-effect region around $x=5\cdot 10^{-1}$ 
\item the Fermi-motion at $x \approx 1$.
\end{itemize}

As for the $Q^2$-dependence of nuclear modifications, it has been argued to follow the DGLAP evolution as the nucleon's PDF itself~\cite{PhysRevLett.65.1725,Eskola:1992zb}. This assumption has been adapted to absorb the nuclear modifications into process-independent nuclear PDFs extracted from a fit to nuclear DIS and Drell-Yan data. Given the conceptual simplicity and the phenomenological success of this approach within the current experimental precision, see for recent extractions at NLO~\cite{Kovarik:2015cma,Khanpour:2016pph,Eskola:2016oht,Walt:2019slu,AbdulKhalek:2019mzd,AbdulKhalek:2020yuc,Khanpour:2020zyu,Eskola:2021nhw,Khalek:2022zqe}, this assumption can be considered as an agnostic framework to discuss nuclear modifications of parton densities. We adapt it for this manuscript and discuss different phenomena potentially relevant to the context of the observables considered in this chapter afterward.

\subsection{Impact on the extraction of properties of QGP}
\label{subsec:impact}

A natural quantification of QGP-induced modifications for hard scale QCD processes are comparisons between $pp$ and nucleus-nucleus collisions, where the $pp$~measurement is treated as a \textit{vacuum baseline}. This approach allows canceling, approximately, uncertainties in analytical calculations\footnote{As an example, the estimated uncertainties in a pQCD calculation from scale variations.} and in the experiment. However, modifications induced by the initial state of heavy-ion collisions (the presence of nuclear PDFs) are still present. These must be taken into account for the extraction of QGP properties. 

\begin{figure}
    \centering
    \includegraphics[width=0.8\textwidth]{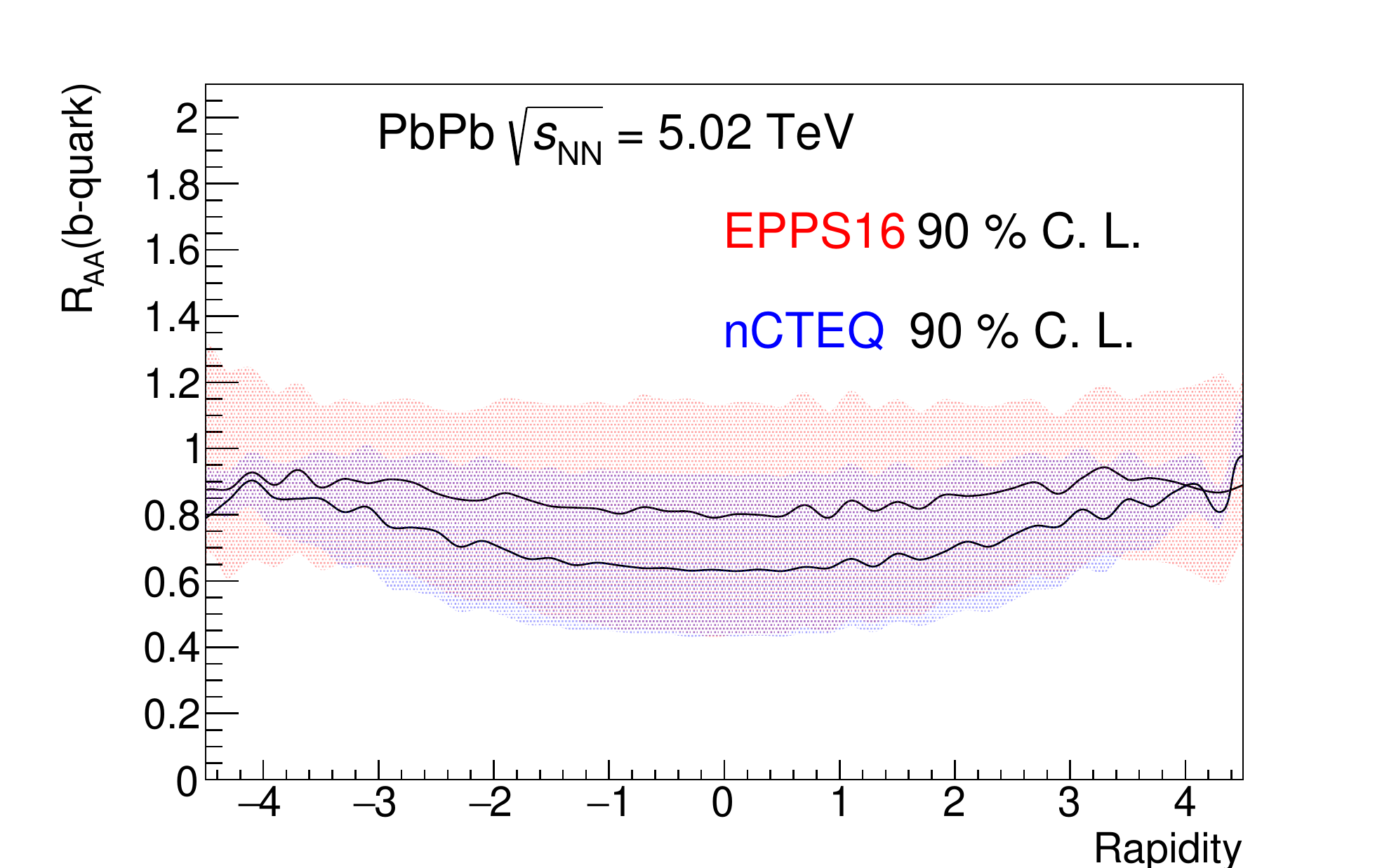}
    \caption{Single beauty quark R$_{AA}$ as function of rapidity with EPPS16 and nCTEQ15 in PbPb collisions at $\sqrt{s_{NN}}=$5.02~TeV only displaying the uncertainty from nuclear PDFs. The calculation is performed at fixed next-to-leading order with a scale choice of $\mu_F = \mu_R = m_{beauty}$ calculated with MADGRAPH5$_A$MC@NLO ~\cite{Alwall:2014hca}.  }
    \label{fig:JPSIRAA}
\end{figure}

To illustrate the impact of our limited knowledge of the initial state, we can take the nuclear modification factor $R_{AA}$ as an example. In the nuclear modification factor, yields or cross-sections measured in nuclear collisions are normalized by the proton-proton equivalents:
\begin{align}
    R_{AA} = \frac{Y^X_{AA}}{\langle T_{{AA}}\rangle \cdot \sigma^X_{pp}} \, ,
    \label{eq:sec3_RAA}
\end{align}
where $Y^X_{AA}$ is the yield of the signal $N^X$ normalised to the number of inelastic nucleus-nucleus collisions $N^{evt}_{cent}$ in a given centrality class $cent$. The centrality class is defined as the quantile of an experimentally measured quantity's distribution that is strongly correlated with the impact parameter of the nucleus-nucleus collision. Typically, the number of particles or energy in a given acceptance. The nuclear overlap function $\langle T_{AA}\rangle$ is defined for a given centrality class and calculated within a Glauber model~\cite{Miller:2007ri}. If the nuclear collision is a simple superposition of proton-proton collisions, the nuclear modification factor is equal to one~\footnote{Due to the presence of neutrons inside the nucleus, this implies that even in the absence of any nuclear modifications of parton distribution functions of bound nucleons w.r.t. free nucleons, the nuclear modification factor is not expected to be exactly one.}. For the centrality integrated case, the nuclear modification factor can be expressed as the cross-section ratio in the nucleus-nucleus collision divided by $A^2$ times the cross-section in pp~collisions. Thus, the nuclear modification factor has a direct sensitivity to nuclear PDFs. As we will see below, the uncertainties induced by the missing knowledge on nuclear PDFs can be sizeable. Most notably for gluons (see section~\ref{subsec:constraints}). 

We can construct the centrality integrated nuclear modification factor as the ratio of cross-sections only considering the nuclear PDFs as the origin of the modification for PbPb collisions at the LHC at $\sqrt{s_{NN}}=5$~TeV. In this case, $R_{AA}$ contains the nuclear PDFs in the numerator and the proton PDFs in the denominator through the cross-section (eq.~\eqref{eq:factorization}). 

Taking beauty production as an example (see Fig.~\ref{fig:JPSIRAA}), we consider the $p_T$-integrated inclusive heavy-quark cross-section. It is dominated by the kinematic domain of low transverse momentum that directly measures heavy-quark diffusion, thermalization, and heavy-quark pair interaction, as will be discussed later. The hard scale of the process is given by the mass scale of the beauty quark. At LHC collision energy, these comparatively low $Q$ scales are, to a large extent, produced by midrapidity partons at low Bjorken-$x$, down to $10^{-3}$ (ratio of the hard scale and the center-of-mass energy). Gluons in the initial state largely dominate the production in this kinematic regime. There are no clean constraints for nuclei from DIS and DY fixed-target data for this low-$x$. Consequently, the nuclear PDF in this regime is largely unconstrained, and the uncertainties are largely based on nPDF parameterization extrapolations. 

The rapidity distribution of beauty nuclear modification factors is shown in Fig.~\ref{fig:JPSIRAA}, taking two nuclear parton distributions often used in the community for illustration purposes~\footnote{These nuclear parton distributions do not take into account the constraints based on the most recent dijet measurements and based on heavy-flavor hadron production discussed in sec.~\ref{subsec:constraints}.}. As discussed in chapter~\ref{sec:acceleration_HFQQ}, the measurement of the rapidity dependence is essential for investigating QGP physics. The nuclear modification factor amounts to 0.65 for nCTEQ and 0.8 for EPPS16 at midrapidity but with uncertainties that stretch the full dynamic range of plausible values. In this case, the interpretation of the nuclear beauty production at low transverse momentum is dominated by the uncertainties of the initial production. The situation is worse for charm production probing smaller Bjorken-$x$ as it requires even larger extrapolation towards low Bjorken-$x$. This is shown in~\cite{ALICE:2021rxa} where the $90\%$ confidence level of EPPS16 for the $p_{\rm T}$ integrated nuclear modification factor of D$^0$ at midrapidity stretches from 0.2 to 1.05. Therefore, any additional knowledge of nuclear modification of parton distribution functions is crucial for interpreting these observables in nucleus-nucleus collisions.

The inclusive production of hadrons at high transverse momentum and jet production is also affected by nuclear parton distribution functions. The possible effects of vacuum nuclear modification factor on high transverse momentum hadrons and jets at the LHC are discussed in~\cite{Huss:2020dwe}. They show that the vacuum nuclear modification factor from EPPS16 is consistent with unity within present uncertainties. However, the central value of the nuclear modification factor at rapidity $y=0$ and integrated over centrality deviates up to 10~\% for inclusive hadron production above 10~GeV/$c$ in transverse momentum. The maximal deviation is observed around $p_T = $50~GeV/$c$ and reaches a deviation above 5~\% at the highest momenta. While the results are consistent, precise knowledge of nuclear parton distributions will become necessary for studies of parton energy loss in nucleus-nucleus collisions aiming at the sub-10~$\%$~level precision. In addition, the modifications can become more critical for studies exploring small collision systems that we cover in chapter~\ref{sec:smallsystem}.  

Self-normalized quantities are typically less sensitive to nuclear parton distribution functions, but they are not entirely removed. As shown in~\cite{Huss:2020dwe}, jet and Z-boson coincidence measurements show even more significant uncertainties from nuclear PDFs than single object inclusive measurements.

Moreover, differences between nPDF effects for different parton species can also induce dissimilarities between proton-proton and nucleus-nucleus collisions. For instance, quark compared to gluon jet fragmentation variations can cause, in principle, nuclear modifications of jet fragmentation functions. However, within the current precision, the fragmentation function in proton-proton collisions and proton-nucleus collisions at the same collision energy is consistent with unity~\cite{Aaboud:2017tke}.

\subsection{Constraints on nuclear parton distributions from jets and heavy-flavor in pPb collisions}
\label{subsec:constraints}

In the following, we discuss the usage of jet and heavy-flavor production in proton/deuteron-nucleus collisions to constrain gluon nPDF. Due to the large medium-induced modifications~(see chapter~\ref{sec:acceleration}), nucleus-nucleus collision observables with hadronic final states are not well suited to constrain nPDFs. The situation differs in proton or deuteron-nucleus reactions, where standard collinear perturbative QCD calculations analog to pp~collisions are compared with experimental data.

First, inclusive single hadron production in proton-nucleus or deuteron-nucleus collisions has been used in the nPDF EPS09~\cite{Eskola:2009uj} to constrain the gluon nPDF that is otherwise primarily unconstrained at low-$x$. This nPDF set relied on PHENIX data from pion production in d-Au collisions~\cite{PHENIX:2006mhb} at RHIC energies. The underlying assumption is that the transition from the hard process to the final-state hadron is not modified in the d-Au environment.
However, the usage of hadronic final states in proton/deuteron-nucleus collisions has been criticized since the observed nuclear modifications in these collisions may be caused by different phenomena than the initial nuclear state. The DSSZ group assigned the nuclear modification of pion production in d-Au collisions at RHIC to the nuclear modification of the fragmentation function~\cite{deFlorian:2011fp} that they extracted from HERMES and RHIC data in~\cite{Sassot:2009sh}~\footnote{One of two most recent global nPDF fits, the EPPS fit~\cite{Eskola:2021nhw}, the successor of the EPS group fits, includes this data set. In contrast, the other recent nPDF fit nNNDPF3.0 does not consider this data set~\cite{Khalek:2022zqe}. However, due to the D-meson and dijet data, the role of the deuteron-gold inclusive pion production is no longer as prominent as it used to be in the EPS09 analysis.}. These two view points persist in the literature. 

In addition to this possible ambiguity, the observation of long-range angular correlations in p(d)-nucleus collisions discussed in chapter~\ref{small:azim} might have an impact on the interpretation of nuclear modifications of hadronic observables in p(d)-nucleus collisions in terms of nuclear PDFs. The explanations of these phenomena may compromise the applicability of standard collinear perturbative QCD calculations used in current PDF determinations.

As for LHC energies, two types of hadronic final-state data sets have been used as constraints for nuclear parton distribution functions: dijet and heavy-flavor hadron measurements in proton-lead collisions. They are interesting for nuclear PDFs constraints due to:
\begin{itemize}
    \item a large sensitivity to gluon densities induced by the dominance of gluon-induced production in most of the phase space~\footnote{For the heavy-flavor hadron production, the production is mostly driven by gluons. We refer the reader to ref.~\cite{Helenius:2018uul} for more information on scheme-dependent subtleties at NLO in the context of pPb LHC charm production data. For jet production, at least one of the two incoming partons is a gluon, see in ref.~\cite{Rojo:2014kta} for a more detailed discussion as function of the jet kinematics.},
    \item the access to a large kinematic range relevant for heavy-ion nucleus-nucleus collision measurements,
    \item the available NLO and higher-order collinear pQCD calculations for dijet and heavy-flavor production.
\end{itemize}
In addition, jet production is particularly attractive since using infrared and collinear safe jet algorithms minimizes the impact of non-perturbative corrections~\footnote{In chapter~\ref{sec:motivation}, the commonly used jet algorithms at hadron colliders are introduced.}. This choice allows comparing experiment and theory without needing non-perturbative fragmentation functions. 

Heavy-flavor production also has a particular interest. Its relatively small mass scale, but still well above $\Lambda_{QCD}$, probe the nuclear PDF at Bjorken-$x$ values below $\approx 10^{-3}$, where nuclear modifications are expected to be significant. This region is unexplored so far. However, the currently available data of single heavy-flavor hadrons necessitates the usage of fragmentation functions to compare the experiment with theory calculations. In addition, nuclear effects unrelated to nPDFs may play a significant role. 

To get an approximate indication of the Bjorken-$x$ and $Q^2$ phase space probed at the LHC, we need to define auxiliary variables $x_2$. For the dijet: 
\begin{align}
  x_2 = \frac{ p_{\rm T,jet1}+ p_{\rm T,jet2}}{\sqrt{s_{\rm NN}}} \cdot e^{-\eta_{\rm dijet}} \, ,
  \label{eq:nPDFdijet}
\end{align} 
where $p_{\rm T,jet1}+p_{\rm T,jet2}$ is the sum of the two jet transverse momenta and $\eta_{dijet}$ the pseudorapidity of the dijet system in the nucleon-nucleon-center-of-mass frame. In this coordinate convention, the incoming proton momentum direction defines the $z$-axis. For the heavy-flavor case, we define:
\begin{align}
    x_2 = \frac{m_T}{\sqrt{s_{\rm NN}}} \cdot e^{{-y}}
    \label{eq:nPDFhf}
\end{align},
with $m_T=\sqrt{(2 m_c)^2+ p_{\rm T,h}^2 }$ and with $y$ as rapidity of the hadron in the nucleon-nucleon center-of-mass frame. These experimentally defined quantities correlate with Bjorken-$x$ of the Pb-nucleus and the $Q^2$ of the process. These kinematic relationships are based on leading-order QCD and are washed out by initial and final state radiation, which we comment on later. 
 
\begin{figure}
    \centering
    \includegraphics[width=0.45\textwidth]{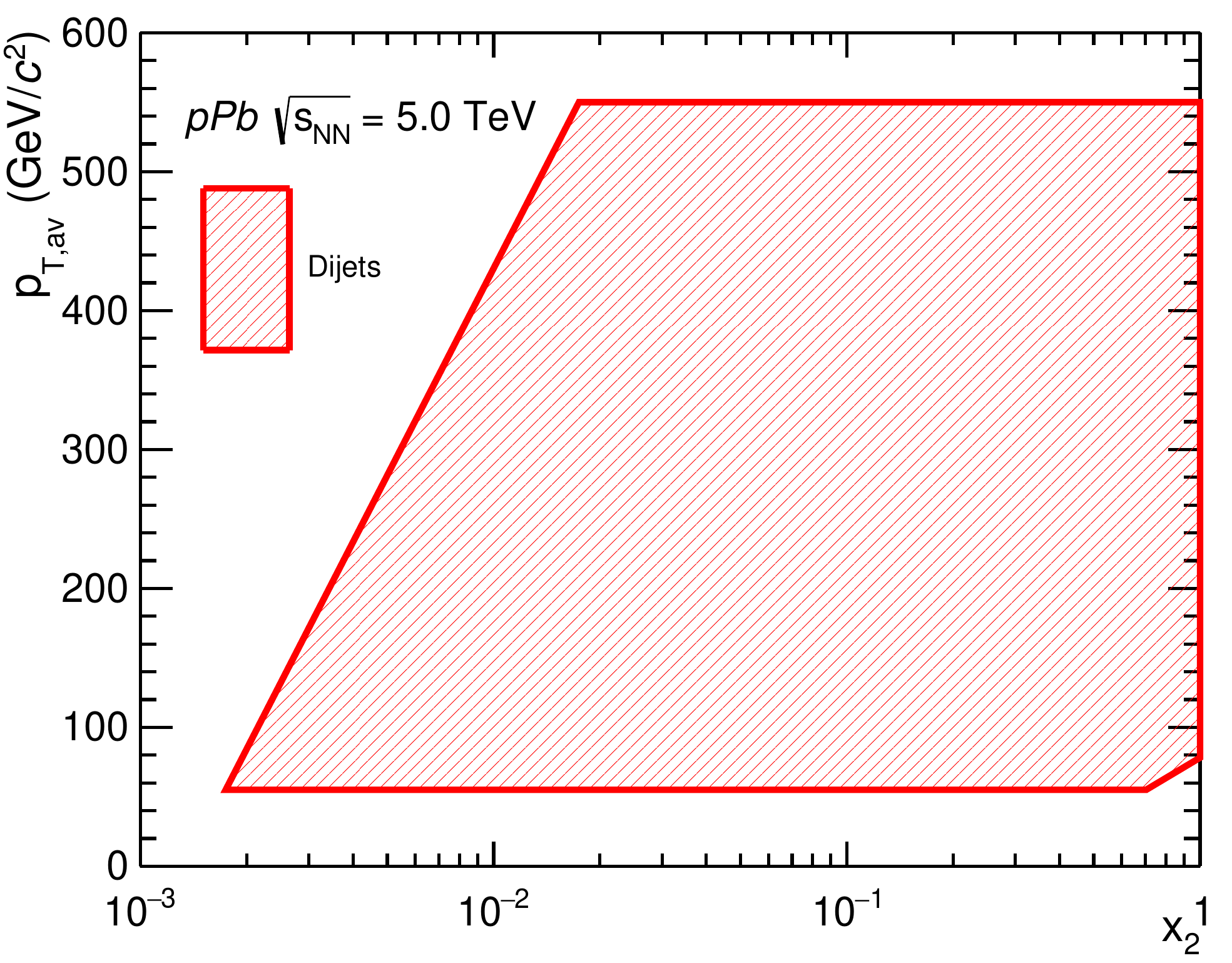}
    \includegraphics[width=0.45\textwidth]{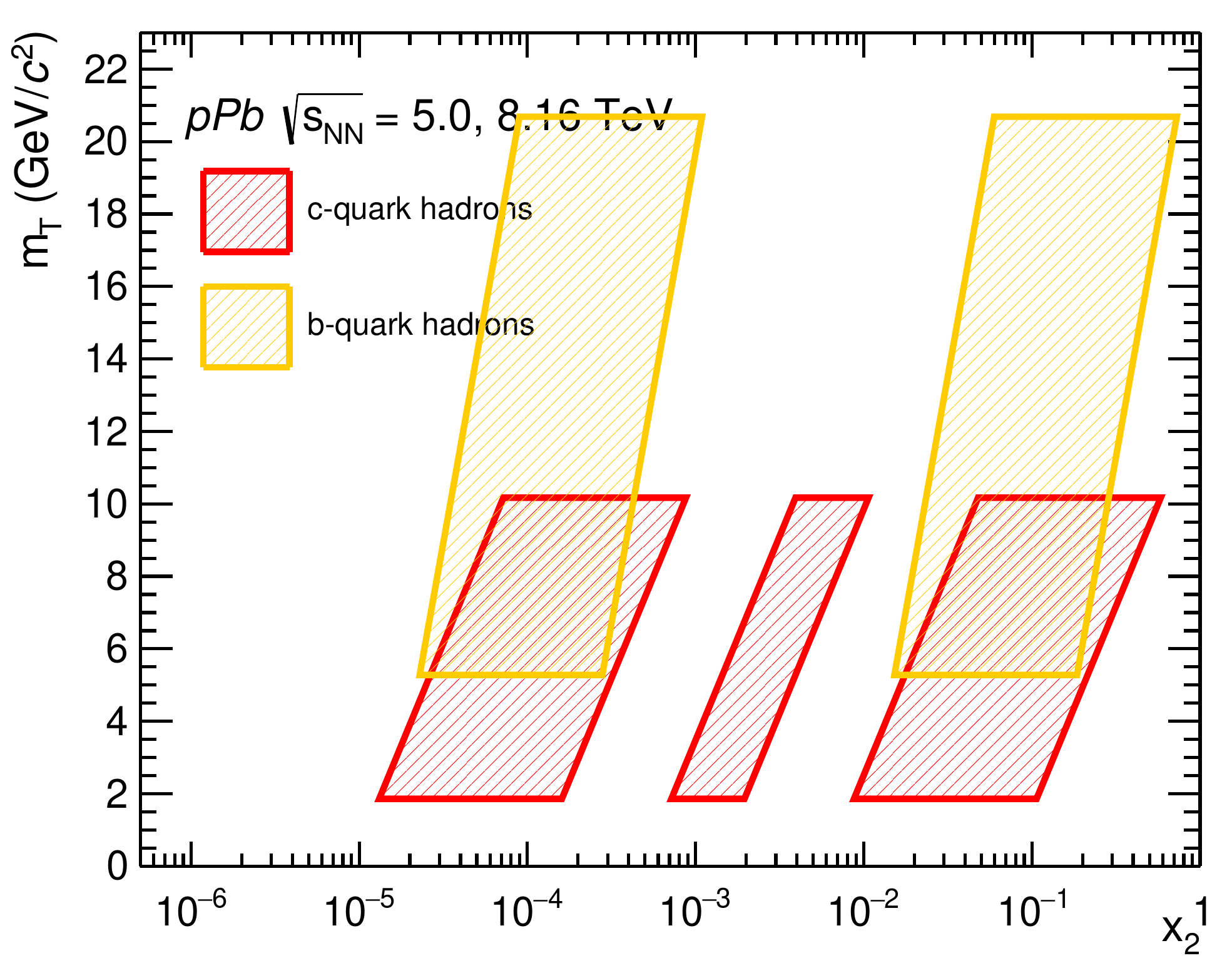}
    \caption{Simple estimates of measurement sensitivity to probe nuclear PDFs with pPb collisions: left dijet measurement case by CMS~\cite{Sirunyan:2018qel}, right open heavy-flavor measurements selection with lowest transverse momentum coverage by LHCb~\cite{Aaij:2019lkm,Aaij:2017gcy} and ALICE \cite{Acharya:2019mno}. }
    \label{fig:kinematicplane}
\end{figure}

The $x_2$ and $2*p_T$ or $m_T$ phase space is illustrated in Fig.~\ref{fig:kinematicplane} showing the broad kinematics of available measurements at the LHC. 
 
\subsubsection{Dijet measurements}

The first CMS dijet measurement~\cite{Chatrchyan:2014hqa} in pPb collisions has been already used in EPPS16~\cite{Eskola:2016oht}. Recent CMS dijet measurement includes more $\langle p_{\rm T} \rangle$ bins with improved precision and the corresponding measurement in pp collisions for normalization~\cite{Sirunyan:2018qel}. It also uses angular variables less affected by the dominating uncertainties due to the jet energy scale on the experimental side and scale variation uncertainties on the theoretical side. In addition, the pseudorapidity distribution of the dijet $\eta_{\rm dijet}$ is exploited, profiting from the dominance of $2 \to 2$ kinematics in the dijet population. This generic feature of jet production allows relating the final state dijet kinematics to the initial parton kinematics despite some smearing due to initial and final state radiation. A thorough discussion of theoretical uncertainties can be found in~\cite{Eskola:2013aya}. The ratio between pPb and pp data is defined as:
\begin{align}
    R^{\text{norm}}_{\text{pPb}} = \frac{\frac{1}{\text{d}\sigma^{\text{pPb}}/\text{d} p_{\text{T}}^{\text{ave}}} \text{d}^2 \sigma^{\text{pPb}}/\text{d} p_{\text{T}}^{\text{ave}} d \eta_{\text{dijet}} }{\frac{1}{\text{d}\sigma^{\text{pp}}/\text{d} p_{\text{T}}^{\text{ave}}} \text{d}^2 \sigma^{\text{pp}}/\text{d} p_{\text{T}}^{\text{ave}} \text{d} \eta_{\text{dijet}}},
    \label{eq:dijetobs}
\end{align}
where $p_T^{ave}$ is the average of the transverse momenta of the two jets. The variable $\eta_{\rm dijet}$ is related via the $2 \to 2$ kinematics in the nucleon-nucleon center-of-mass frame to the Bjorken-$x$ of the proton and the lead nucleus by the relation: $\eta_{\rm dijet} = \frac{1}{2} \ln{x_{p}/x_{Pb}}$, where the momentum direction of the proton is defined as the positive beam direction.
 
The large range of dijet pseudorapidities $|\eta_{\text{dijet}}|<3.0$ in the nucleon-nucleon center-of-mass frame allows to probe three different regions of nuclear modification for $x> 3 \cdot 10^{-3}$: the shadowing region ($\eta_{\text{dijet}}<1.5$), the antishadowing region ($-0.5<\eta_{\text{dijet}}<1.5$), and EMC effects ($\eta_{\text{dijet}}< -0.5$). The five kinematic bins in transverse momentum, average between the two jets in the interval 55-400~GeV/$c$, probe a wider Bjorken-$x$ range, via the variation of the hard scale of the process at fixed collision energy and rapidity, and adds redundancy to the measurement. 
 
\begin{figure}
    \centering
    \includegraphics[width=0.45\textwidth]{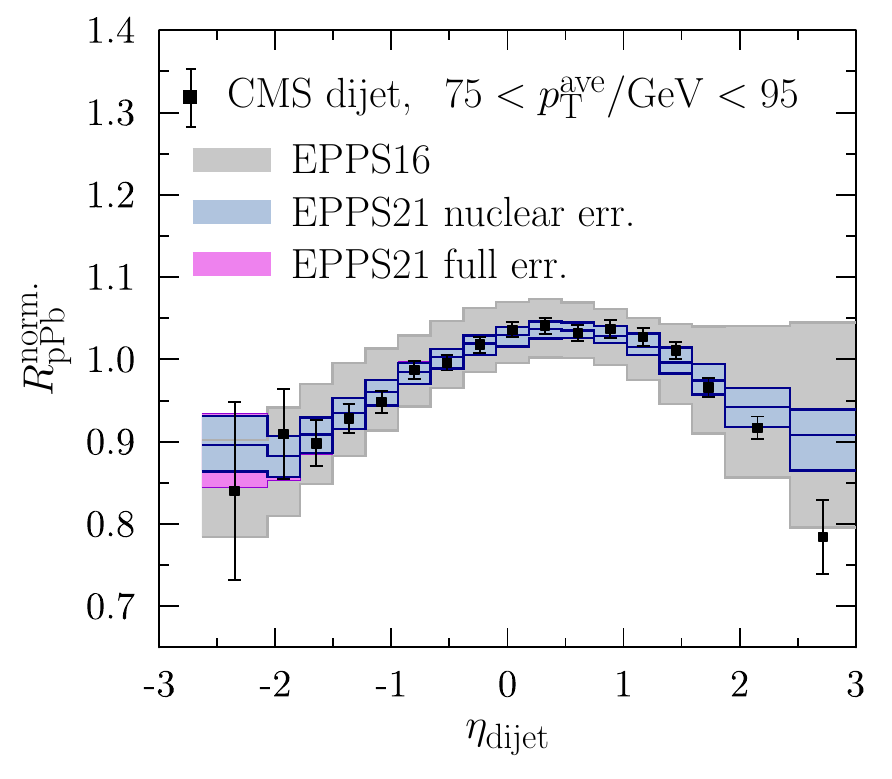}
    \includegraphics[width=0.52\textwidth]{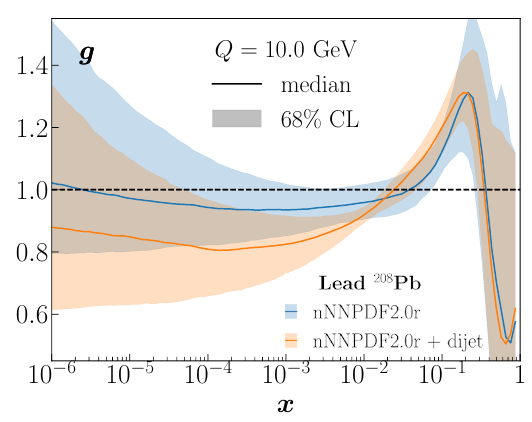}
    \caption{Left: Comparison of dijet distribution double ratio with EPPS16~(gray) and EPPS21~(blue) with 90~\% confidence level uncertainties, adapted from~\cite{Eskola:2021nhw}. Right: The ratio of the lead gluon PDF divided by the proton gluon PDF for nNNPDF2.0r~(blue) and nNNPDF3.0(no D-mesons) as a function of Bjorken-$x$, adapted from~\cite{Khalek:2022zqe}.}
    \label{fig:jetcons}
\end{figure}

This data set~\cite{Sirunyan:2018qel} has been first used in a nPDF reweight of EPPS16~\cite{Eskola:2019dui} showing a strong impact on the resulting precision. A PDF reweight based on an existing global PDF fit estimates the compatibility of a new data set with the global fit, the modification of the PDF central values, and the PDF uncertainties based on the new data with reduced computational cost. However, it comes with the price of relying fully on the assumptions of the previous fit introducing potential biases~\footnote{PDF reweighting have been introduced in~\cite{Giele:1998gw}. Examples for Bayesian methods in the context of nuclear PDFs can be found in~\cite{Armesto:2013kqa,Armesto:2015lrg,Kusina:2016fxy}. An example of a Hessian method can be found in~\cite{Eskola:2019dui}.}. This data set has now been recently introduced in the global PDF fits EPPS21~\cite{Eskola:2021nhw} and nNNPDF3.0~\cite{Khalek:2022zqe}. Fig.~\ref{fig:jetcons} (left) shows the CMS $\eta_{dijet}$-ratio (eq.~\ref{eq:dijetobs}) in one $p_T^{ave}$ bin compared to EPPS16~(in gray) and the new EPPS21~(in blue) nuclear PDF~\footnote{EPPS21 contains also the constraints from other observables in particular D-meson production discussed below. The blue uncertainty band indicates the nuclear part of the EPPS21 uncertainty, whereas the pink error band indicates the total uncertainty, including the propagated proton~PDF uncertainty.}. While the nPDF EPPS16 uncertainties were larger than the experimental data, the new EPPS21 data set continues to be compatible with the experimental results but with significantly smaller uncertainties. 
The nNNDPF group~\cite{Khalek:2022zqe} also shows that the inclusion of the dijet data can help to reduce the uncertainty on the gluon nPDF in the range $10^{-3}<x<0.4$~\footnote{The study also shows a moderate impact on quark nuclear modification.}. Fig.~\ref{fig:jetcons} (right) shows the ratio of the gluon nPDF and proton gluon PDF for the nuclear PDF nNNPDF2.0r without the dijet data~(blue) and the nNNPDF2.0r with the dijet data~(red) provided in ref.~\cite{Khalek:2022zqe}. The latter continues to show a strong reduction of the uncertainties in the Bjorken-$x$ range of $10^{-3}\lesssim x \lesssim 0.4$. 

In summary, the reweighting study~\cite{Eskola:2019dui} and the two recent nPDF fits~\cite{Eskola:2021nhw,Khalek:2022zqe} show a considerable improvement in the knowledge of gluons in nuclei based on the CMS dijet data. They indicate evidence for small-$x$ shadowing and strong gluon antishadowing peaking at $x \approx 0.1-0.2$ in the lead nucleus. 

\subsubsection{Heavy-flavor measurements}

As the second type of measurement, we discuss constraints on the gluon content of nuclei by heavy-flavor quark production in proton-nucleus collisions. Forward rapidity heavy-flavor production~\footnote{Forward rapidity refers to a positive rapidity value based on the convention for eq.~\ref{eq:nPDFdijet} and eq~\ref{eq:nPDFhf}.} at the LHC is particularly interesting since it explores the low-$x$ regime that is currently not directly accessible via dijets~\footnote{Measurements at backward (negative) rapidity probe the transition between shadowing and antishadowing at the LHC. However, in this regime, the dijet measurements and other more indirect constraints provide better restrictions due to smaller scale variation uncertainties in the cross-section calculations~\cite{Eskola:2021nhw,Khalek:2022zqe}.}. Collinear factorization may break down in this kinematic region, and non-linear evolution equations become relevant due to gluon saturation~\cite{Gelis:2010nm}. Therefore, the test of collinear factorization in this regime is essential for the field of hadron structure and non-linear QCD. In this context, studies have been conducted to constrain the proton gluon content at low-$x$ by LHCb through heavy-flavor measurements in pp~collisions~\cite{Gauld:2015yia,Gauld:2015kvh,Garzelli:2016xmx,Zenaiev:2019ktw,Khalek:2022zqe}. 

In heavy-flavor hadron production, the Bjorken-$x$ of the nucleus is correlated with the $x_2$ as given in Eq.~\ref{eq:nPDFhf}. This relation is only approximate since fragmentation and higher-scale processes dilute the correlation between parton and final-state hadron kinematics. However, the sensitivity to Bjorken-$x$ values down to $10^{-5}$ for forward rapidity charm production in the nucleon-nucleon center-of-mass frame at $ 3< y < 3.5$ has been shown to be preserved in NLO pQCD calculations~\cite{Helenius:2018uul,Eskola:2019bgf,Eskola:2021nhw,Khalek:2022zqe}. Therefore, the rapidity and transverse momentum differential cross-section measurements can constrain the nuclear parton distribution function down to very low Bjorken-$x$. The measurement of charm and beauty allows for adding redundancy with slightly different $Q^2$-values.

In all nPDF studies using heavy-flavor observables, the nuclear modification factor $R_{pPb} = \sigma_{pPb}/(A \sigma_{pp})$ is used as default observable. This choice minimizes the impact of theoretical uncertainties related to missing higher order corrections estimated by the variation of the renormalization and factorization scales~\cite{Lansberg:2016deg,Kusina:2017gkz,Eskola:2019bgf,Kusina:2020dki,Eskola:2021nhw,Khalek:2022zqe}.    

Both open heavy-flavor hadron production and heavy quarkonium production measured in hadronic collisions can constrain nPDFs. However, additional complications may affect quarkonium-bound state formation in hadronic collisions, as recently discussed in~\cite{Lansberg:2019adr}. Despite prompt heavy quarkonium not being considered for nPDFs by most groups~\cite{Chapon:2020heu}, the reweighting of nuclear PDFs suggests that the constraints of prompt ground state quarkonium production are consistent with constraints from open heavy-flavor~\cite{Lansberg:2016deg,Kusina:2017gkz,Kusina:2020dki}. In agreement with recent practice~\cite{Eskola:2021nhw, Khalek:2022zqe}, we neglect quarkonium data, and focus on the discussion of open-heavy-flavor production at forward rapidity. The most precise and differential data is the D$^0$-meson production measurement by LHCb~\cite{LHCb:2017yua}. 

\begin{figure}
    \centering
    \includegraphics[width=0.48\textwidth]{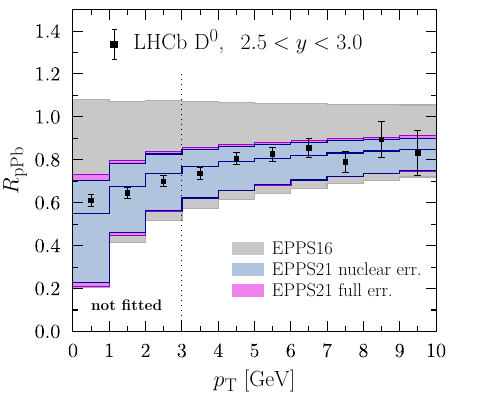}
    \includegraphics[width=0.50\textwidth]{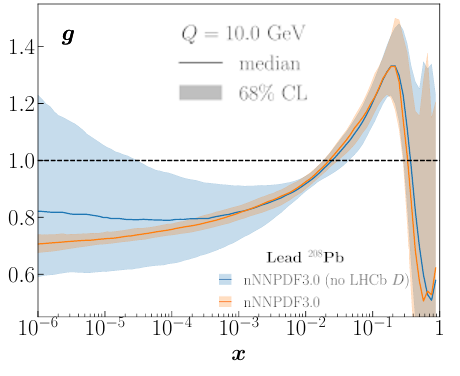}
    \caption{Left: Comparison of nuclear modification factor with EPPS16~(gray) and EPPS21~(blue) with 90~\% confidence level uncertainties, adapted from~\cite{Eskola:2021nhw}. Right: The ratio of the lead gluon PDF divided by the proton gluon PDF for nNNPDF3.0 without D-mesons constraint~(blue) and nNNPDF3.0 with D-mesons constraint as a function of Bjorken-$x$, adapted from~\cite{Khalek:2022zqe}.}
    \label{fig:hfcons}
\end{figure}

Heavy flavor hadron production has been used in reweights of existing global nPDF fits~\cite{Lansberg:2016deg,Kusina:2017gkz,Eskola:2019bgf,Kusina:2020dki}.The two most recent global nPDF fits, EPPS21~\cite{Eskola:2021nhw} and nNNPDF3.0~\cite{Khalek:2022zqe}, include the LHCb D-meson data to constrain the low-x gluons~\footnote{In case of nNNPDF3.0~\cite{Khalek:2022zqe}, in the form of a reweight.}. Fig.~\ref{fig:hfcons} (left) shows the $D^0$-meson nuclear modification factor as measured by LHCb (black points) for one of the rapidity bins. For comparison, it is included the EPPS16 fit prior to the inclusion of the data~(in gray) and the EPPS21 nPDF fit~(in blue)~\footnote{The blue uncertainty band indicates the nuclear part of the EPPS21 uncertainty, whereas the pink error band indicates the total uncertainty including the propagated proton PDF uncertainty.} that takes into account this data set applying a transverse momentum cut of 3~GeV/$c$. It is visible that the data is well compatible with both nPDF sets and that the new EPPS21 fit shows strongly improved uncertainties. Fig.~\ref{fig:hfcons} (right) shows the ratio of gluon PDF in the nNNPDF3.0 fit with and without the constraint from the D-meson data. A dramatic improvement of the derived uncertainty on the gluon shadowing is observed down to the lowest probed Bjorken-$x$ values. 

Both global PDF fits~\cite{Eskola:2021nhw,Khalek:2022zqe} as well as first nPDF reweightings considering forward heavy-flavour production at the LHC~\cite{Lansberg:2016deg,Kusina:2017gkz,Eskola:2019bgf,Kusina:2020dki}, support evidence for strong shadowing at low Bjorken-$x$ down to $x \approx 10^{-5}$. The constraints are also compatible with the ones obtained from dijet measurements~\cite{Eskola:2021nhw,Khalek:2022zqe}. The forward rapidity data is also compatible with calculations based on the effective field theory in the low-$x$ regime of QCD, the color glass condensate~\cite{Ducloue:2015gfa}~\footnote{Charmonium production has also been shown to be described by color glass condensate calculations in this kinematic domain~\cite{Ma:2015sia,Ducloue:2015gfa}. }. 

\begin{figure}
    \centering
\includegraphics{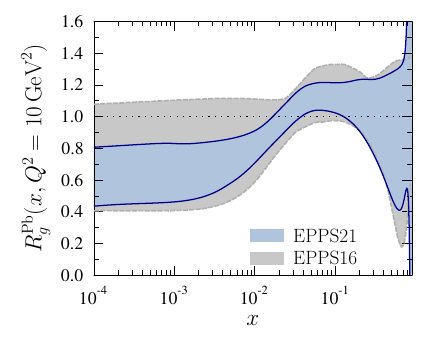}
    \caption{The ratio of the lead gluon PDF divided by the proton gluon PDF for EPPS16 without D-meson and dijet constraints~(gray) and EPPS21 with D-mesons and dijet constraint as a function of Bjorken-$x$ with 90~\% confidence level uncertainties, adapted from~\cite{Eskola:2021nhw}. }
    \label{fig:EPPS}
\end{figure}

Despite the current limitations and caveats discussed below, the dijet and the heavy-flavor data show a consistent picture supporting strong gluon shadowing at low Bjorken-$x$ and a sizeable antishadowing~\cite{Eskola:2020yfa,Eskola:2021nhw,Khalek:2022zqe} for the nuclear PDF of the lead nucleus. The overall improvement in the gluon PDF between the EPPS16 fit without the two data sets previously discussed and the new EPPS21 fit is shown in Fig~\ref{fig:EPPS}. 

\subsubsection{Limiting factors and caveats}
\label{subsubinitiallim}
Concerning dijet production, we remark that the normalized pseudorapidity distribution in pp or pPb~collisions provided in the CMS dijet data~\cite{Sirunyan:2018qel} are not well described by theoretical calculations at NLO~\cite{Eskola:2019dui,Khalek:2022zqe}. The precise origin of this discrepancy is, at the moment, unknown. It has been proposed that missing higher-order pQCD corrections of NLO calculations are a possible origin~\cite{Khalek:2022zqe}. Studies at NNLO for other jet observables~\cite{AbdulKhalek:2020jut} may lead to an improved description. Initially, there were concerns about using dijet data in nPDF fits due to the possibility of jet quenching in pPb collisions. The experimental investigations in~\cite{Chatrchyan:2014hqa,Aaboud:2017tke} have not found any evidence of jet quenching, confirming the dominance of nPDF effects in the observed jet kinematics comparison between pp and pPb collisions. 

A limiting factor for constraints on nPDFs based on charm production at the LHC is the scale uncertainties for charm production at low transverse momentum~\cite{Lansberg:2016deg,Kusina:2017gkz,Eskola:2019bgf,Kusina:2020dki,Khalek:2022zqe}. These uncertainties, related to the relatively low hard scale of the process,
are smaller for beauty production due to its larger mass scale. However, the statistical precision in experimental data at the forward rapidity is not yet at the same level of charm production based on currently available luminosities, see e.g., in~\cite{LHCb:2017ygo,Aaij:2019lkm}. Upcoming data takings with larger luminosity can improve the constraining power without additional theory work~\cite{Citron:2018lsq}. In addition, the first publication of beauty production at NNLO shows a reduced scale dependence than the previous NLO calculation, potentially improving theory-related uncertainties~\cite{Catani:2020kkl}.

Since the constraints from heavy-flavor hadron production are based on measurements at relatively low $Q^2$ and in a dense hadronic environment, several caveats have been presented in the literature that we discuss in the following. 

A possible complication is hadronization, which we will discuss in more detail in chapter~\ref{sec:hadronization}. The hadronization fractions of heavy-flavor are found to be modified in proton-proton and proton-lead collisions compared to the hadronization fractions in $e^+e^-$ and electron-proton collisions at low transverse momentum~\cite{Aaij:2011jp,Aaij:2014jyk,Acharya:2017kfy,Aaij:2018iyy,Aaij:2019lkm,ALICE:2020wla,ALICE:2020wfu}. In particular, the fraction of heavy-quarks hadronizing to baryons is larger at the LHC than in $e^+e^-$ collisions. This modification in the hadronization can impact nuclear parton distribution constraints from proton-nucleus data. However, the experimentally measured $\Lambda_c/D$~ratio~\cite{Acharya:2017kfy,ALICE:2020wla,ALICE:2020wfu} and the $\Lambda_b/B$~ratio~\cite{Aaij:2019lkm} are either consistent within uncertainties between pp and pPb~collisions or very close to each other. Furthermore, the limitations of nuclear PDF constraints imposed by hadronization modifications could be overcome in the future by measuring the ratio of inclusive charm or beauty production as a function of rapidity. This is experimentally challenging, but it minimizes the dependence on the modelization of the heavy-quark production and heavy-quark hadron kinematics relation. In proton-proton collisions, a measurement of the inclusive charm production at midrapidity, including several baryons, has been achieved recently by ALICE~\cite{ALICE:2021dhb}.

An additional concern is the modification of the hadron kinematics via further interactions not taken into account in pQCD calculations. Measurements on the transverse momentum spectra of D-mesons in proton-nucleus collisions were reported by ALICE~\cite{Acharya:2019mno}. Using an experimental estimate of impact parameter, the results show a modification of transverse momentum spectra for the most central events. While one of the proposed explanations was attributed to the radial flow induced by an expanding medium (as in nucleus-nucleus collisions~\cite{Acharya:2019mno}), the transverse momentum distributions can also be affected by multiple scattering in the nucleus inducing $k_T$ broadening~\cite{Arleo:2020rbm}. The precise understanding of these observations remains open at the current stage. 
 
An alternative scenario as the dominant mechanism for the strong nuclear modifications observed for heavy flavor is cold energy loss in nuclear matter~\cite{Arleo:2012rs}. It has been first discussed in the context of quarkonium production and has been recently extended to heavy-flavor mesons~\cite{Arleo:2021bpv}. The mechanism leads to a rapidity shift in the distribution of the produced heavy-flavor hadrons. The effect is also expected in dijet production, although the impact is considered small at high $Q^2$, since the lost energy is proportional to $1/Q$. The data-model comparisons in~\cite{Arleo:2021bpv} show that about half of the suppression of heavy-flavor mesons at forward rapidity can be accounted for with this mechanism within the presented model. It might be necessary to account for the cold nuclear energy loss in a global nPDF fit as suggested in~\cite{Arleo:2021bpv}.

Nonetheless, independently of specific interpretations, the EPPS21 nPDF fit is based on a transverse momentum selection~($p_T>3$~GeV/$c$), and no tension is observed with the lower transverse momentum bins~\cite{Eskola:2019bgf}.  

We discussed so far measurement in proton-lead collisions to constrain the lead nPDFs. However, the body of experimental data at lower collision energies is not given for the lead nucleus. This is also true for any other nucleus. Therefore, the nuclear PDF fits rely on modeling nuclear modifications as a function of the mass number. The inherent assumptions of these parameterizations are one of the major limitations of current nuclear PDFs, see e.g. in~\cite{Eskola:2021nhw,Khalek:2022zqe}.

\subsubsection{Spatial dependence of nuclear PDFs}

The nPDFs constraints from proton/deuteron-nucleus collisions discussed so far are based on inclusive particle production. The extracted nuclear modifications represent averages over the entire spatial extent of the nucleus. In principle, one expects that the nuclear PDFs show a spatial dependence if this can be properly defined\footnote{For a discussion of conditions required to be fulfilled in quantum field theory, see~\cite{Wu:2021ril}}. This effect has been considered early on in nucleus-nucleus collisions~\cite{PhysRevC.61.044904}. Provided that an impact parameter can be experimentally defined in deuteron-nucleus or proton-nucleus collisions, it is possible to constrain the spatial dependence of nPDFs in these collisions. Including this dependence can remove a systematic bias when comparing yield calculations and experimental results in centrality intervals in nucleus-nucleus collisions. However, experimentally, the determination of the collision impact parameter via final state particle multiplicities or forward energy deposition in deuteron-nucleus or proton-nucleus collisions has a resolution of similar size to the full range of possible track multiplicities and energies~\cite{Adam:2014qja}. In addition, it is difficult to eliminate systematic biases in a controlled way, see for a discussion in~\cite{Adam:2014qja}. As an example, soft particles are correlated with both the impact parameter and the hard scale used for the nPDF constraint. This renders direct experimental constraints difficult. Nevertheless, a measurement of the spatial dependence of nuclear shadowing has been proposed based on J/$\psi$~production in deuteron-nucleus collisions~\cite{PhysRevLett.91.142301}. For the J/$\psi$ nuclear modification factor, a measurement in classes of centrality in deuteron-gold collisions has been performed by PHENIX~\cite{PHENIX:2010hmo}. The measurement is fitted with nuclear absorption and nPDF parameterizations considering different spatial dependencies of nPDFs in~\cite{McGlinchey:2012bp}. Independent of direct experimental inputs, an attempt to parameterize the spatially dependent nPDFs can be found in~\cite{Helenius:2012wd}. This work exploits the nPDFs as function of nuclear mass number to derive a spatial dependence. The comparison in ref.~\cite{McGlinchey:2012bp} shows that the derived nPDF parameterization by~\cite{McGlinchey:2012bp} features a much stronger radial dependence than the results in ref.~\cite{Helenius:2012wd}.

\subsection{New directions}
\label{subsec:new} 

We have discussed inclusive heavy-flavor and jet production measurements at the LHC in proton-nucleus collisions as the only hadronic final state measurements considered in nuclear PDF determinations with LHC data. The precision of the data offers an excellent opportunity, but the inclusion in global PDF fits will profit from independent inputs and more theory work to build a coherent picture. We briefly discuss further options with heavy-flavor and jets and possibilities to test the corresponding findings with improved and independent measurements.

The upcoming data takings of LHC will provide increased luminosities in pPb collisions by about a factor of 5-10~\cite{Citron:2018lsq} compared to previous data takings. The luminosity and instrumentation improvements should allow for stretching the kinematic reach in rapidity and energy~\cite{Citron:2018lsq}. For dijets, larger statistics, as anticipated by the projection of CMS~\cite{CMS-PAS-FTR-18-027}, extend the measurement by one bin down to lower dijet pseudorapidity bins and, hence, to higher Bjorken-$x$. The systematic uncertainty over the full kinematic range is assumed to be reducible by a factor of 2. This is achieved by the larger statistics and, hence, smaller data-driven uncertainties on the jet energy scale. Given the higher anticipated collision energy for future data takings at $\sqrt{s_{NN}}=8.8$~TeV instead of $\sqrt{s_{NN}}=$~5.02~TeV, the higher kinematic reach at negative dijet $\eta$ will allow keeping the reach towards high Bjorken-$x$ approximately the same whereas the low-$x$ reach will be extended by about 40~\%. In addition, the larger luminosities will also provide similar precision for beauty hadron measurements~\cite{LHCb-CONF-2018-005} as already achieved for charm hadrons. In the charm sector, the full reconstruction of two charm hadrons, as first measured in~\cite{Aaij:2020smi}, opens the possibility of validating theoretical calculations through a more differential comparison with data. In addition, it will enable the potential selection of specific kinematic domains that may reduce theory uncertainties or improve the sensitivity to specific Bjorken-$x$ values. 

Measurements at the relativistic heavy-ion collider~(RHIC) did not achieve yet the precision of the LHC measurements for fully reconstructed open heavy-flavor hadrons. RHIC can further provide an additional handle, particularly with the new sPHENIX detector~\cite{PHENIX:2015siv}. It has a foreseen start of data taking in 2023. 

Very large Bjorken-$x$ gluons ($x>0.5$) are not well constrained for the nuclear case (see Figs.~\ref{fig:jetcons},\ref{fig:hfcons} and \ref{fig:EPPS}). At the LHC, top production has been proposed as a handle for high gluon~$x$~\cite{dEnterria:2015mgr} although achievable luminosities will provide only limited constraining power~\cite{Citron:2018lsq}. A collision energy variation allows access to different areas within the Bjorken-$x$-$Q^2$ plane and to test the collinear factorization assumption via DGLAP evolution. A future high-luminosity fixed-target program at the LHC can be used for nPDF constraints of gluons at high Bjoerken-$x$ via heavy-flavor measurements if energy loss effects are not dominant and if pQCD theory uncertainties are sufficiently small~\cite{Hadjidakis:2018ifr}. The first measurement of charm production in fixed-target mode at the LHC has been published by LHCb~\cite{Aaij:2018ogq}. This measurement based on proton-helium and proton-neon collisions is not yet at a similar level of statistical precision and availability of kinematic bins as the proton-lead results. In addition, in the available kinematic regime relatively close to the target rapidity, hadronization modifications or other effects not accounted for in standard collinear pQCD calculations may play a role.  

The best environment to measure nuclear PDFs are DIS facilities. The electron-ion collider in the USA has started, and it is planned to take data starting at the beginning of the 2030ies~\cite{Accardi:2012qut} with center-of-mass energy up to 140~GeV. It will provide precision constraints down to Bjorken-$x$ of 10$^{-3}$~\cite{Aschenauer:2017oxs} via inclusive DIS measurements. 

However, the collision energy of the beam species limits the Bjorken-$x$ reach of the EIC detector. To probe the very low Bjorken-$x$ regime, the LHC beams will be required to achieve large enough collision energies to provide sensitive measurements. This is relevant for defining the initial state at the LHC and searching for saturation at low-$x$. Beyond the possibilities outlined so far to exploit with precision the partonic structure, a deep inelastic facility with the LHC beams, the LHeC, would be ideal for measuring the nuclear PDFs~\cite{AbelleiraFernandez:2012cc,AbdulKhalek:2021gbh}.
 
To solidify the shadowing found with hadronic probes for nuclear modifications of partons at low Bjorken-$x$ at the LHC, measurements with electromagnetic final or initial state are desirable. LHCb proposed the measurement of Drell-Yan~\cite{LHCb-CONF-2018-005} production at the forward rapidity that is sensitive to the gluons in the lead nucleus~\cite{Citron:2018lsq} down to similar Bjorken-$x$-values as heavy-flavour production~\footnote{Although Drell-Yan production proceeds via coupling to quarks, the strong dominance of gluons at low-$x$ allows to constrain them via Drell-Yan~\cite{Citron:2018lsq}.}. The ALICE collaboration will construct a highly granular calorimeter to measure photons at the forward rapidity~\cite{ALICECollaboration:2020rog} to probe the gluons at Bjorken-$x$ down to 10$^{-6}$. 

In addition, the ATLAS collaboration demonstrated the feasibility of dijet photoproduction measurements in nucleus-nucleus collisions at the LHC in a preliminary measurement~\cite{ATLAS:2017kwa}. This process is accessible at the LHC in ultraperipheral collisions (UPC)~\cite{Baltz:2007kq}. In this data set, long-range azimuthal correlations have also been measured by ATLAS~\cite{ATLAS:2021jhn}. A further understanding of these phenomena is desirable to judge implications for the applicability of collinear pQCD calculations used in nPDF fits. Preliminary results on exclusive dijet production in PbPb collisions have been reported by the CMS collaboration~\cite{CMS:2020ekd}. Prospects for these observables are indicated in~\cite{Citron:2018lsq}. 

Exclusive quarkonium production in UPC has been advertised as a probe of shadowing~\cite{Guzey:2013xba}. In the limit of the momentum transfer $t \to 0$ between the incoming photon and the incoming hadron, the cross-section is at leading order proportional to the square of the gluon distribution~\cite{Ryskin:1992ui,Brodsky:1994kf}. The already available experimental measurements~\cite{ALICE:2012yye,ALICE:2013wjo,CMS:2016itn,ALICE:2021gpt,LHCb:2021bfl} are precise and they constrain nuclear shadowing within given model assumptions~\cite{Guzey:2013xba,Guzey:2020ntc}. Their precision will improve with the luminosity increase of about 1-2 orders of magnitude compared to available publications~\cite{Citron:2018lsq}. As in the case of inclusive heavy-quark production at low transverse momentum in proton-nucleus collisions, there are large-scale uncertainties for NLO fixed-order calculations~\cite{Ivanov:2004vd}. 
Generally, the non-perturbative object probed in exclusive quarkonium production is the generalized parton distribution function~(GPD)~\cite{Diehl:2003ny}. First attempts to translate these data into PDF constraints have been made for the proton~\cite{Flett:2019pux} based on the relations presented in~\cite{Shuvaev:1999fm,Shuvaev:1999ce}. Hopefully, further theoretical work will allow the inclusion of exclusive quarkonium data in nPDF fits. 

%% file: acceleration2.tex
\section{QGP-induced modifications on the hard probes: the short wave-length behavior}
\label{sec:acceleration}

QGP characteristics can be indirectly accessed from modifications imprinted on high momentum particles propagating through an extended medium that spreads after the collision for about 5-10 \rm{fm/c}. These high energy objects, \textit{hard probes}, are particularly sensitive to the short wave-length behavior of the medium as the transverse resolution of a radiated quanta can be of the order of $\lambda \sim 1/Q \ll 1~\rm{fm}$ (e.g., $Q = 0.1~\rm{TeV} \Rightarrow \lambda = 0.05~\rm{fm}$). Its momentum scale is well above the QCD scale at these high energies, $\Lambda_{QCD} \simeq 200~\rm{MeV}$~\cite{Zyla:2020zbs}, making them optimal candidates to be described by perturbative methods. Results from proton-proton collisions (\textit{vacuum physics}) can provide a reliable reference for experimental results and theoretical descriptions. While we refer the reader to~\cite{Campbell:2017hsr} for a review on the topic, we briefly recall the main physical ingredients that provide the baseline to accommodate medium-induced modifications: vacuum physics (see section~\ref{sec:acceleration_vacuum}). We will then provide a model-independent overview and general picture concerning the experimental observations on jets and heavy-flavor in section \ref{sec:acceleration_model_independent}. A brief review of the analytical description of particle propagation through a hot and dense QCD medium that supports the current picture follows in section ~\ref{sec:acceleration_energy_loss}.

\subsection{From \textit{vacuum} physics to heavy-ions}
\label{sec:acceleration_vacuum}

Large momentum transfer processes can produce heavy particles or jets with high transverse momenta during a hadronic collision. Cross-sections for such scattering process can be computed, in collinear factorization, through a separation of initial state effects (PDFs, see section~\ref{sec:initial}), parton-level cross-section and the integration over the corresponding phase space. At leading order (tree-level), these describe the momenta of outgoing jets or final-state particles, see Eq.~\eqref{eq:factorization}. Nonetheless, higher orders (emission of additional gluons) must be considered if one wants to access the jets' internal structure and the accompanying particles' distribution. These effects are often simulated by resorting to parton showers, a conventional approach in which a multi-particle system is simulated based on a succession of emissions from the incoming parton or colored quark - anti-quark dipole. Both techniques yield equivalent results, and the reader can find more details specific to each approach and their current implementation on widely used Monte Carlo event generators in~\cite{Buckley:2011ms}. Nonetheless, the common building block is based on QCD bremsstrahlung. 

Vacuum gluon emission is dominated by soft and collinear radiation. The probability of emitting a gluon from an incoming quark, $dS$, in the limit of soft and collinear radiation, reads:
\begin{equation}
    dS = \frac{2 \alpha_s C_F}{\pi} \frac{dE}{E} \frac{d\theta}{\theta} \frac{d\phi}{2\pi}\, ,
\label{eq:sec4_gluonrad}
\end{equation}
where $E$ is the energy of the outgoing gluon and $(\theta,\phi)$ the gluon polar and azimuthal angles with respect to the quark. Accounting for finite energy corrections, one obtains:
\begin{equation}
    dS = \frac{\alpha_s}{\pi} P_{g \leftarrow q}(z) \frac{dE}{E} \frac{d\theta}{\theta} \frac{d\phi}{2\pi}\, ,
\end{equation}
where $P_{g \leftarrow q}(z) = C_F (1 + (1-z)^2)/z$ is one of the four (unregularized) Altarelli-Parisi splitting functions, the probability of emitting a gluon from a quark with fraction of energy $z$. The remaining three splitting functions, each corresponding to an elementary QCD vertex, are:
\begin{eqnarray}
P_{q\leftarrow q} (z) & = & C_F \left( \frac{1+z^2}{1-z} \right) \\
P_{q\leftarrow g} (z) & = & T_R (z^2 + (1-z)^2) \\
P_{g\leftarrow g} (z) & = & 2 C_A \left( \frac{z}{1-z} + \frac{1-z}{z} +z(1-z) \right) \, ,
\end{eqnarray}
where the Casimir color factors sum to $C_F = (N^2 -1)/(2N)$, $T_R = 1/2$ and $C_A = N$ with the number of colors $N$.
The procedure to handle the soft and collinear divergences present in the emission probability can vary. However, the most general approach is introducing a cut-off in an arbitrary scale (e.g., transverse momentum) to account for resolvable emissions effects on the final observable. The contribution below this cut-off is then added to the hard process's loop contribution, yielding a finite result. The probability of not producing resolvable branchings up to that cut-off is then identified as the Sudakov form factor. Its evolution with the cut-off scale is given by the QCD DGLAP evolution equations~\cite{Gribov:1972ri,Altarelli:1977zs,Dokshitzer:1977sg}. In fact, these correspond to the \textit{renormalization equation} for the PDFs (when \textit{evolving} the PDFs from one perturbative resolution scale to another - see section~\ref{sec:initial}), but the evolution kernel to describe parton showers is the same. This same prescription allows building a sequence of gluon emission processes based on a Markov chain approach. In particular, vacuum parton showers are based on the probabilistic interpretation that subsequent radiation is constrained solely by the previous emission kinematics. This picture can be easily derived when considering a QCD antenna's emission pattern in the soft limit. The result of this exercise, for a QCD antenna of opening angle $\theta_{12}$, yields:
\begin{equation}
    dS = \frac{\alpha_s C_F}{\pi} \frac{dE}{E} \frac{d\cos(\theta)}{1-\cos(\theta)} \Theta(\theta_{12} - \theta)\,
\end{equation}
thus showing that jet evolution, when built as a sequence of independent parton splittings, has to consider an angular ordering condition~\cite{Mueller:1981ex,Ermolaev:1981cm}, i.e., a monotonic decrease of successive opening angles along the cascade.

While the above perturbative prescription suits particles whose mass is below $\Lambda_{QCD}$, heavy quarks (like charm, beauty and top) require a description in all phase space, including regions where $Q^2 \sim m_{quark}$, being $m_{quark}$ the quark mass. At higher energies, the same QCD shower description is relevant to the development of the heavy-quark showers, as long as modifications induced by the presence of mass terms are included. Those will change the kinematics and introduce a constraint on radiation phase space suppressing the QCD radiation (dead-cone effect~\cite{Dokshitzer:1991fd}). Namely, the radiation from a particle of mass $m$ and energy $E$ is suppressed within an angular size of $m/E$. In proton-proton collisions, the measurement of the dead-cone was, until recently, experimentally challenging as the effect is too small or the expected gap in the phase space was populated by the decay products of the incoming heavy particle. Dead-cone impact has been indirectly observed in several measurements, such as heavy-flavor meson production~\cite{DELPHI:2000edu}. However, the direct observation of collinear radiation suppression was done in $D^0$-meson tagged jets by ALICE~\cite{ALICE:2021aqk}.
For low energies (compared to $m_{quark}$), the relevant physics may be relatively different from light quark masses. Pure relativistic perturbative methods fail, and non-relativistic or lattice calculations are employed instead. The former requires an effective Lagrangian to describe the low energy interaction, whose expansion is done in orders of $\alpha_s$ (straightforward application of QCD perturbation theory) and $1/m_{quark}$ (non-perturbative contributions). This Lagrangian can be obtained from the usual QCD Lagrangian:
\begin{equation}
    \mathcal{L} = \bar{\Psi} i \slashed{D} \Psi - m \bar{\Psi} \Psi \, ,
    \label{eq:sec4_lagrangian}
\end{equation}
where $\Psi$ represents the the heavy-quark field, with mass $m$, and the covariant derivative $D_\mu = \partial_\mu - i g T^a A_\mu^a$. To start, one needs to consider that a heavy-quark bound inside a hadron will move approximately at the same velocity, $v$. Thus, one can single out this momentum $m v$ from the heavy-quark momentum $p$:
\begin{equation}
    p = mv + k \, ,
\end{equation}
where $k = \mathcal{O}(\Lambda_{QCD}) \ll m$ is the residual momentum. The quark field is then decomposed into lower and upper components:
\begin{equation}
    H_v(x) = e^{i m v \dot x} \frac{1-\slashed{v}}{2} \Psi(x) \ \ \ , \ \ \ h_v(x) = e^{i m v \dot x} \frac{1+\slashed{v}}{2} \Psi(x) \, ,
\end{equation}
such that:
\begin{equation}
    \Psi(x) = e^{-imv\dot x} (H_v(x) + h_v(x)) \, .
\end{equation}
To finally reach the heavy-quark effective Lagrangian\footnote{One can use the same parameterization to describe a heavy anti-quark by simply reversing the signs in $v$.}, some algebra is required. One needs to multiply the equation of motion obtained from the Lagrangian \eqref{eq:sec4_lagrangian} by each of the projectors above, $(1\pm \slashed{v})/2$ to work out the following expression:
\begin{equation}
    \mathcal{L} = \bar{h}_v i v \dot D h_v + h_v + \bar{h}_v i \slashed{D}_\perp \frac{1}{i v \cdot D + 2 m} i \slashed{D}_\perp h_v \, 
\end{equation}
where $D_\perp^{\mu} = D^\mu . v^\mu v\cdot D$ is orthogonal to the heavy quark velocity.
The second term can now be expanded in powers of $\Lambda_{QCD}/m$. Keeping only the leading-power correction, one finally gets:
\begin{equation}
    \mathcal{L}_{eff} = \bar{h}_v i v \cdot D h_v + \frac{1}{2m} \bar{h}_v (i D_\perp)^2 h_v + \frac{g}{4m} \bar{h}_v \sigma^{\mu \nu} G_{\mu \nu} h_v + \mathcal{O}\left( \frac{1}{m^2} \right)\, ,
\end{equation}
where $G_{\mu \nu} = (i g)^{-1} \left[ D^\mu, D^\nu \right]$.

So far, the discussion was limited to the QCD degrees of freedom (quarks and gluons). However, to translate the theoretical concept of partons into final-state particles, an additional prescription is required: hadronization. Jets allow creating observables theoretically and experimentally robust to these non-perturbative effects. Nonetheless, small-radius jets, jet substructure or single-particle measurements, including heavy-flavor, require a further hadronization scheme to be applied to the final multi-particle system. These will add an additional dependence of the final observable on non-perturbative effects, as discussed in section~\ref{sec:hadronization}. Nevertheless, once all these physical ingredients are put together, we have all the building blocks to compare our analytical and phenomenological tools to proton-proton\footnote{Proton-proton collisions have been the standard baseline to assess jet quenching effects. However, in light of the experimental results compatible with collectivity in $pA$ or $pp$ high multiplicity, systems like $ep$~\cite{ZEUS:2019jya} or $e^+e^-$~\cite{Badea:2019vey,Belle:2020mdh} can provide an additional handle as the underlying event contribution is minimal to nonexistent.} experimental data. 

Heavy-ion collisions will require additional physics input to account for QGP production. While the previous formalism is suitable to describe the result from high-momentum transfer processes in proton-proton collisions, we expect modifications induced by multiple interactions with the medium in the presence of hot and dense QCD matter. In order to describe these modifications, it is necessary to start with an assumption about the produced medium. In particular, how to characterize its degrees of freedom. Numerous experimental measurements indicate a liquid-like behavior for the produced \textit{soup} of quarks and gluons. In particular, there is strong evidence that ultra-relativistic heavy-ion collisions form a strongly coupled hydrodynamic fluid, with a minimal shear viscosity to entropy ratio, $\eta/s$. How such collective behavior arises from the elementary building blocks of a QCD theory is among the most prominent questions in the field. In strong-coupling approaches, the dissipation of the shear stress is compensated by the value of the coupling itself. The transfer of energy-momentum across nearby fluid elements takes place with almost no change in the velocity gradient, thus yielding a small $\eta/s$. In these conditions, the mean free path is minimal, and one cannot apply the description of elementary \textit{quasi-particles}. Despite the solid evidence for the formation of an extended strongly-coupled liquid, it is expected that at the shortest length scales, as possibly resolved by the interactions of high-energy objects such as hard probes, QGP constituents can be resolved as quasi-particles. These quarks and gluons with effective masses are regarded as elementary excitations of the thermodynamical system. This class of weak-coupling approaches yields, nonetheless, a large $\eta/s$ as they require the mean free path to be larger than the distance between the medium constituents. The transition between these two regimes will depend on the resolution scale dictated by the interaction between probe and medium. 

The multitude of hard probes is given by:
\begin{itemize}
    \item multi-scale objects as jets;
    \item heavy-quarks with different masses;
    \item quarkonium states, $B_c$ mesons and exotica whose survival and generation probability will vary according to their binding potentials.
\end{itemize} 
Altogether, they provide a range of probing resolutions that can help identify the medium constituents and how it evolves as a function of the probing resolution energy. 

With such distinct approaches for treating the medium constituents, current QGP studies rely on a systematic comparison with experimental data to validate the underlying assumptions of in-medium modifications and, ultimately, withdraw (average) QGP characteristics (see section \ref{sec:QGPDensity}\footnote{A further note on the time-dependence of QGP parameters will be discussed in section \ref{sec:timescale}}). Despite their differences, in-medium descriptions of parton propagation agree that interaction with the medium will degrade the fast-moving particle's energy. While differing in detail, this consequence is independent of the scattering centers' realized nature or interaction scale. It has been put forward by Bjorken as one of the smoking gun signals of a QGP-like medium~\cite{Bjorken:1982tu}, together with strangeness enhancement~\cite{PhysRevLett.48.1066}\footnote{This can be understood as thermal production from statistical hadronization with strangeness in equilibrium. Therefore, strange hadron production in nucleus-nucleus collisions is described by thermal particle production as described in chapter~\ref{sec:hadronization}.}, chiral symmetry restoration~\cite{Pisarski:1983ms} and quarkonium suppression~\cite{Matsui:1986dk}.  
Since the pioneer works of Bjorken~\cite{Bjorken:1982tu}, several effects were thought to be realized by the presence of a hot and dense QCD state of matter:
\begin{itemize}
    \item quantum interference between successive scatterings, caused by the LPM effect in a QCD matter would lead to the suppression of the radiation spectrum concerning a scenario of independent emissions\footnote{This reduction of the bremsstrahlung and pair production cross-sections was initially derived in the context of QED \cite{Landau:1953um,Migdal:1956tc}, and only later for QCD.};
    \item momentum deflection of the incoming particle induced by the interaction with the medium scattering centers (\textit{diffusion});
    \item suppression of interference effects between successive emissions changing the vacuum color coherence phenomena\cite{Mehtar-Tani:2010ebp,Casalderrey-Solana:2011ule,Mehtar-Tani:2011hma,Casalderrey-Solana:2012evi};
    \item presence of additional color fields whose effects can vary between (i) screening the real part of the interaction potential of colored bounded states (mechanism of the original \textit{quarkonia melting} idea)~\cite{Matsui:1986dk} (ii) the imaginary part of the interaction potential~\cite{Laine:2006ns,PhysRevD.78.014017}~\footnote{In the context of kinetic rate equations, the in-medium breakup can be understood represents such an imaginary part of the potential, see for a detailed discussion in~\cite{Rapp:2008tf}.  } and (iii) regeneration of quarkonium states from unbound quarks~\cite{Braun-Munzinger:2000csl,Thews:2000rj}. The interplay between these competing effects will result in an overall suppression or enhancement, depending on QGP properties and the abundance of heavy-flavor quarks thermalized within QGP. 
    \item screening radiation effects off massive quarks even in the presence of a medium (\textit{dead-cone} effect);
\end{itemize}

The following section will discuss these effects to a more extent by identifying the overall picture that has now become more transparent from experimental results. The constant data-model comparison has been crucial to understanding the QCD dynamics in hot and dense conditions and unraveling average QGP properties. In fact, from this process, it was realized that there is an additional contribution of energy and momentum originated from medium constituents excited by the propagation and interaction of the incoming parton, \textit{medium response}~\cite{CMS:2015hkr,CMS:2021otx}\footnote{See for instance, calculations in Refs.~\cite{Sadofyev:2021ohn,Barata:2022krd}}. Simultaneously, novel experimental methods were developed to determine a particular phenomenon's validation/falsification unambiguously~\cite{Apolinario:2017qay}. The next challenge will be to accommodate the fast evolution of the medium on current theoretical descriptions. The time evolution of the medium parameters requires input from hydrodynamical models, whose uncertainty on the initial conditions~(see section ~\ref{sec:initial}) as well as thermalization speed~\cite{Schlichting:2019abc} is still considerable. Pioneering systematic investigations of the several components, including a wide variety of space-time evolution models for heavy-flavor~\cite{Rapp:2018qla,Cao:2018ews} have already allowed scrutinizing the available phase space of each modeling assumption. However, since current observations are always the result of the integrated propagation on a fast-expanding medium, an accurate determination of the interaction mechanisms has been elusive. Nonetheless, hard probes accommodate diverse scales whose potential to provide time differential measurements has just started. These are further explored in sections \ref{sec:timescale} and \ref{sec:outlook}. But before covering methods to analyze QGP evolution with hard probes, let us first discuss which average quantitative and qualitative features are, up to now, firmly established.

\subsection{Data-driven interpretation of in-medium particle propagation}
\label{sec:acceleration_model_independent}

The plethora of heavy-ions experimental results has brought a unique insight into QCD dynamics at large densities and a glimpse of QGP evolution. While some initial expectations have been confirmed, novel qualitative features challenged some initial assumptions underlying the theoretical description of in-medium propagation. We will briefly discuss these in the following by addressing single-particle measurements, jets, heavy-flavor, and quarkonia. 

Single-particle measurements are extremely useful to withdraw QGP properties. They travel through the expanding medium, thus carrying information on the interaction processes. For very high transverse momentum, $p_T \gtrsim 10~\rm{GeV}/c$, the hadronization timescale is expected to be larger than the medium lifetime. This seems reasonably consistent with the experimental observations so far~\cite{CMS:2014jjt}, thus providing an extra theoretical handle to identify unambiguously QGP-induced effects. Nonetheless, the description of high $p_T$ particles is bounded by the uncertainty of non-perturbative effects, such as hadronization. Jets, sprays of particles, are, by construction, less sensitive to these effects. These objects, whose clustering history can mimic QCD-inspired interference patterns, are yet another tool to scrutinize QGP interactions experienced by a multi-parton system. As we will see in the following, jet substructure opens a multitude of possibilities to infer the resolution properties of the produced medium and its transparency to the passage of high-energy particles. Jets' exploitation as probes of QGP has not yet matured to the point of extracting quantitative properties of the created matter. But the qualitative new phenomena brought by jet observables are substantial. 

Finally, we will address heavy quarks and quarkonia as unique probes of QGP. Proxies to heavy quarks, both objects interact differently with QGP matter compared to the light flavor. Due to their large mass, these objects can be kinematically selected to be non-ultra-relativistic. They are thus strongly affected by changes in the strong interaction potential. In addition, its interaction with QGP constituents will lead to a combination of radiative and collisional energy loss, thus being very sensitive probes of the medium transport properties in phase space regions not accessible by light-flavor single-particle or jet measurements. A complete survey of QGP characteristics and QCD dynamics can only be accomplished by simultaneously analyzing these three different hard probes.

\subsubsection{Nuclear Modification Factor}

\begin{figure}
    \centering
    \includegraphics[width=0.45\textwidth]{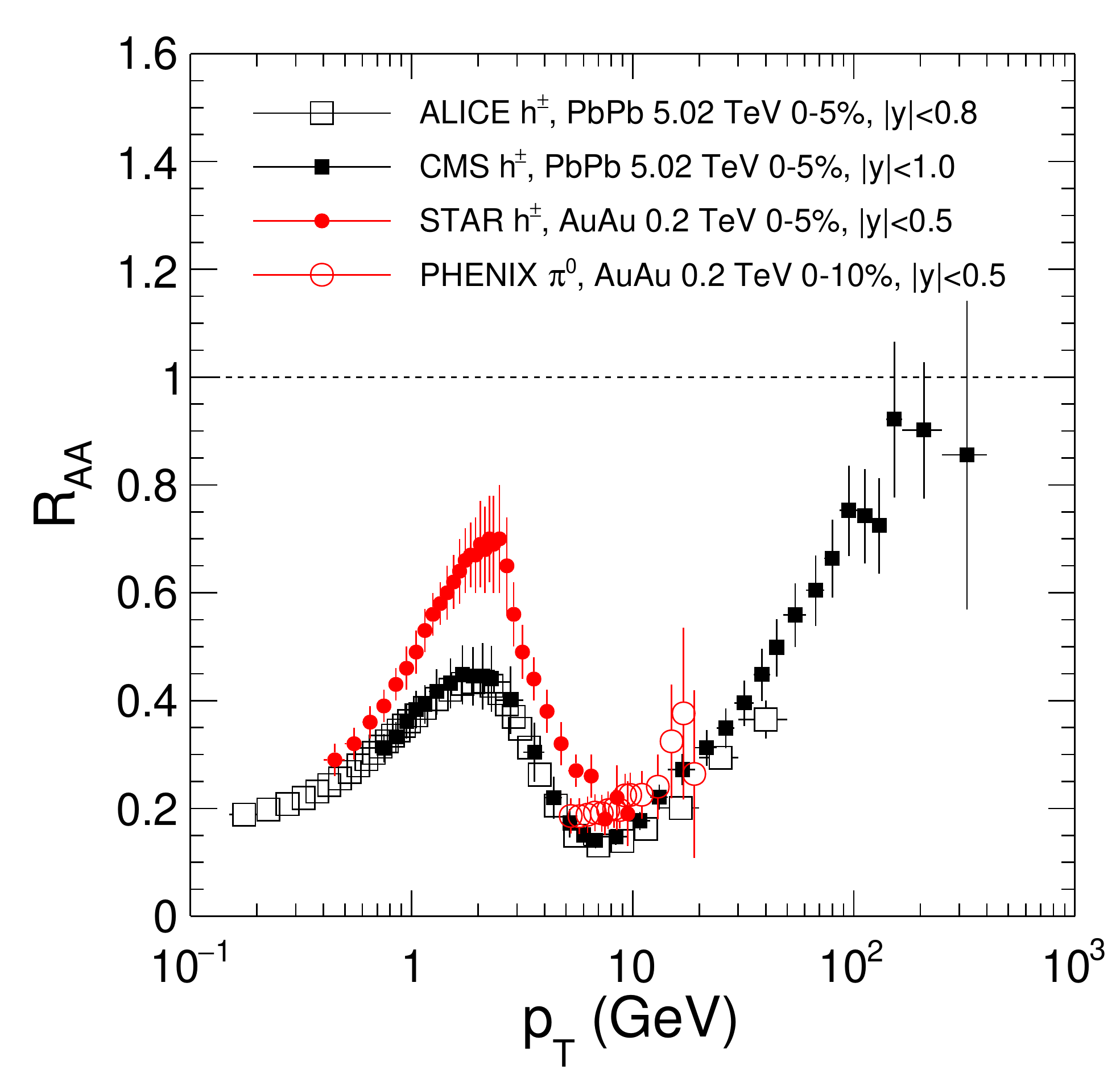}
    \includegraphics[width=0.45\textwidth]{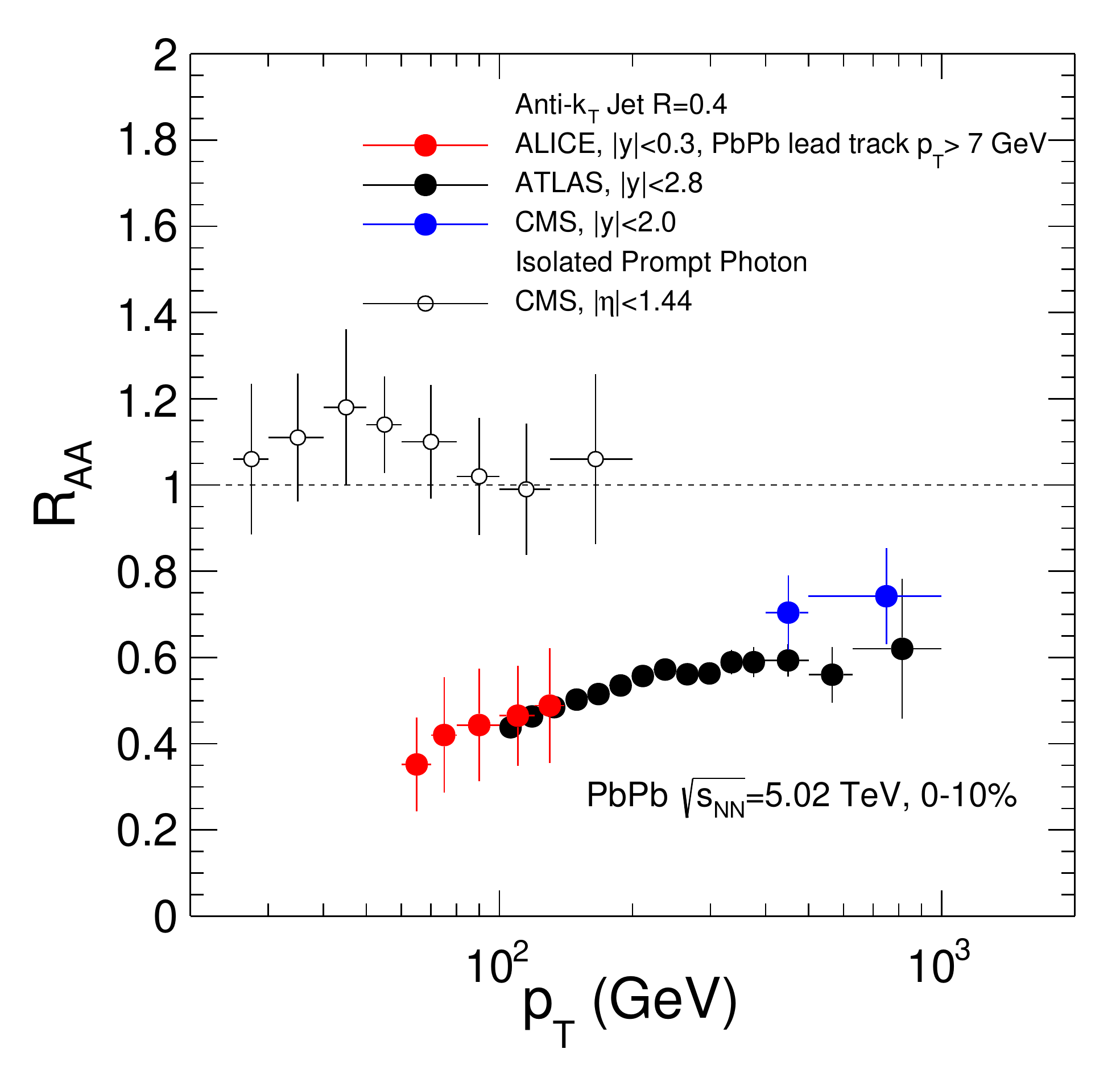}
    \caption{Summary of the nuclear modification factors $R_{AA}$ from single hadrons~\cite{CMS:2016xef,ALICE:2018vuu,STAR:2003fka,PHENIX:2012jha} (left), isolated prompt photons~\cite{CMS:2020oen} and jets~\cite{ALICE:2019qyj,ATLAS:2018gwx,CMS:2021vui} (right) in central heavy-ion collisions at RHIC and the LHC.}
    \label{fig:sec4_energyloss}
\end{figure}

In-medium energy loss was observed across a multitude of different probes. One of the first observables in $AuAu$ collisions at RHIC, followed by $PbPb$ collisions at the LHC, was the relative yield of objects measured in such dense systems to the equivalent of $pp$ collisions that could be superimposed in the nuclear overlap: the nuclear modification factor, $R_{AA}$ (eq. \eqref{eq:sec3_RAA}). This observable has been measured for hadrons, inclusive jets, heavy-flavor, quarkonia, b-jets, among others. Some of them are summarised in Fig.\ref{fig:sec4_energyloss} left (single light particle) and right (jets and prompt photon). 

As expected from a QCD medium, neutral color probes, such as $W^{\pm}$, $Z^0$ and prompt $\gamma$ bosons produced at a hard scattering\footnote{Not to be confused by the thermal photon production that results into an excess of the photon yield at low-$p_T$}, do not experience any medium-induced modification~\cite{CMS:2012oiv,ATLAS:2012qdj,CMS:2011zfr,CMS:2012fgk}. The resulting isolated prompt photon yield (open white circles in Fig.~\ref{fig:sec4_energyloss}, right panel) is consistent with a picture in which $PbPb$ collisions can be modeled as a superposition of $pp$ collisions, aside from nuclear PDFs effects. Colorful probes, such as single-particle measurements and jets, are suppressed when compared to a simple superposition model. This feature is observed up to the largest transverse momentum, where nPDFs results have a limited influence (see ~\ref{sec:initial}).

The results on single-particle $R_{AA}$ for RHIC ($AuAu$ collisions at $\sqrt{s} = 0.2~\rm{TeV}$) and LHC ($PbPb$ collisions at $\sqrt{s} = 5.02~\rm{TeV}$), shown in Fig.~\ref{fig:sec4_energyloss} (left panel), refer to central pseudo-rapidity. Despite the absolute value differences, there is a suppression in the intermediate $p_T$ part ($p_T \sim 10~\rm{GeV}$) of the spectrum with a slow increase towards unity when going to higher transverse momentum. The low part of the spectrum contains both final- and initial-state effects from nuclear PDFs, making it more difficult to draw quantitative conclusions on medium-induced radiation properties. However, from $p_T > 10~\rm{GeV}$, the evolution becomes monotonic, and modifications to the initial $pp$ spectrum carry information about QGP produced in these collisions. RHIC and LHC results seem to be overly consistent within the current experimental uncertainties in this region. This evolution with $p_T$ for light-flavor particles shows that energy loss becomes negligible compared to the particle's initial energy. As it will be discussed in section \ref{sec:acceleration_energy_loss}, the probability of a parton losing some energy is mostly governed by the medium parameters (such as the medium length). As such, the total energy loss will be, proportionally, smaller for high-$p_T$ particles, contributing to an $R_{AA} \sim 1$. Current uncertainties on CMS measurements do not allow claiming if the spectra coincide with the $pp$ reference at high-$p_T$, but future measurements will unlock higher precision on this tail of the spectrum. 

Another observation from Fig.~\ref{fig:sec4_energyloss} is the consistency of the light flavor particles $R_{AA}$ between RHIC and LHC energies. These two colliders produce different QGP temperatures and spatial extents, and, naively, one would expect differences in the nuclear modification factor (in particular, $R_{AA}^{LHC} < R_{AA}^{RHIC}$). Even though the $R_{AA}$ yields the same value, the amount of energy lost in these two systems is, in fact, different. The $p_T$ spectrum of particle production falls very quickly with the particle's transverse momentum. As such, the resulting $R_{AA}$ at a given $p_T$ bin is mainly driven by the migration of particle population towards smaller $p_T$ bins due to energy loss effects. Conversely, positive contributions from higher $p_T$ bins are, proportionally, negligible. While this observation also holds for LHC energies, at RHIC, the $p_T$ spectrum is even steeper. Consequently, even with a smaller energy loss, RHIC results can yield the same $R_{AA}$ as LHC. Even though the $R_{AA}$ is indeed a proxy for energy loss, it is necessary to account for additional kinematic selections to withdraw quantitative conclusions.

In addition to single-particle measurements, jets are also widely used in heavy-ion collisions, as discussed above. To reconstruct jets, a jet clustering algorithm with the corresponding radius parameter, $R$, needs to be specified. As defined in the Snowmass Accord of 1990~\cite{Huth:1990mi}, a jet definition needs to "yield a finite cross-section at any order of perturbation theory", i.e., it needs to be infrared and collinear (IRC) safe so that non-perturbative effects are suppressed by powers of $\Lambda_{QCD}/p_T$ and each order of the perturbative expansion results into a smaller contribution with respect to the previous one. The usually employed algorithm to identify jets out of the final-state event - the anti-$k_T$ algorithm - belongs to the generalized-$k_T$ family, which meets the IRC safe criteria. This particular algorithm uses the information from tracking and/or calorimeter towers, and it involves a distance measure between pairs of tracks/cells, $i,j$, such that:
\begin{equation}
    d_{ij} = d_{ji}= \min (p_{T,i}^{2p}, p_{T,j}^{2p}) \frac{\Delta R_{ij}^{2p}}{R^{2p}} \, .
\label{eq:sec4_dij}
\end{equation}
where $p_{T,i(j)}$ is the transverse momentum of the input object and $\Delta R_{ij} = \sqrt{ (y_i - y_j)^2 + (\phi_{i}-\phi_{j})^2}$ the distance in the (rapidity, azimuthal angle) plane between them. The $p$ parameter is arbitrary, but the most widely used exponents refer to the anti-$k_T$ ($p = -1$)~\cite{Cacciari:2008gp}, $k_T$ ($p = 1$) ~\cite{Catani:1993hr,Ellis:1993tq} and Cambridge/Aachen~\cite{Dokshitzer:1997in,Wobisch:1998wt} ($p = 0$) algorithms. Results on anti-$k_T$ $R_{AA}$ jets with $R = 0.4$ are shown on the right panel of Fig.~\ref{fig:sec4_energyloss}. As opposed to single-particle measurements, the $R_{AA}$ is slowly increasing with the jet transverse momentum up to a plateau of $R_{AA} \sim 0.7$. Being multi-particle systems, jets can contain more than one in-medium radiator, thus losing energy depending on its fragmentation pattern. For the medium to resolve all the jet constituents, the transverse medium resolution needs to be smaller than the transverse distance between two particles. As such, even though high-$p_T$ jets are usually correlated with a higher multiplicity, and so an increasing number of hypothetical in-medium emitters, particles originating by the incoming parton will be highly boosted. In this case, instead of resolving all jet constituents, it is possible to recover vacuum-like coherence effects~\cite{Mehtar-Tani:2010ebp,Casalderrey-Solana:2011ule,Mehtar-Tani:2011hma,Casalderrey-Solana:2012evi}, thus effectively suppressing in-medium radiation (see 
~\ref{sec:acceleration_energy_loss} for more details). Additionally, the jet $R_{AA}$ is also quite sensitive to the shape of the $p_T$ spectrum, as discussed for the particle $R_{AA}$. Negative contributions towards lower transverse momentum will mostly drive the contribution for each $p_T$ bin. The jet population that is less modified by the medium corresponds to a fragmentation pattern that yields a smaller number of particles/emitters than that in $pp$ collisions, even at high $p_T$. This overall picture is usually attributed to the flatness of the $R_{AA}$. Recent analytical calculations that include energy loss and coherence effects~\cite{Mehtar-Tani:2021fud} yield a significant jet suppression up to the $~\rm{TeV}$ scale, thus confirming that the $R_{AA}$ suppression is mostly driven by jets whose structure was resolved by the medium. Future measurements will confirm these findings, thus helping to put constraints on QGP resolvable transverse structure.
\begin{figure}
    \centering
    \includegraphics[width=0.45\textwidth]{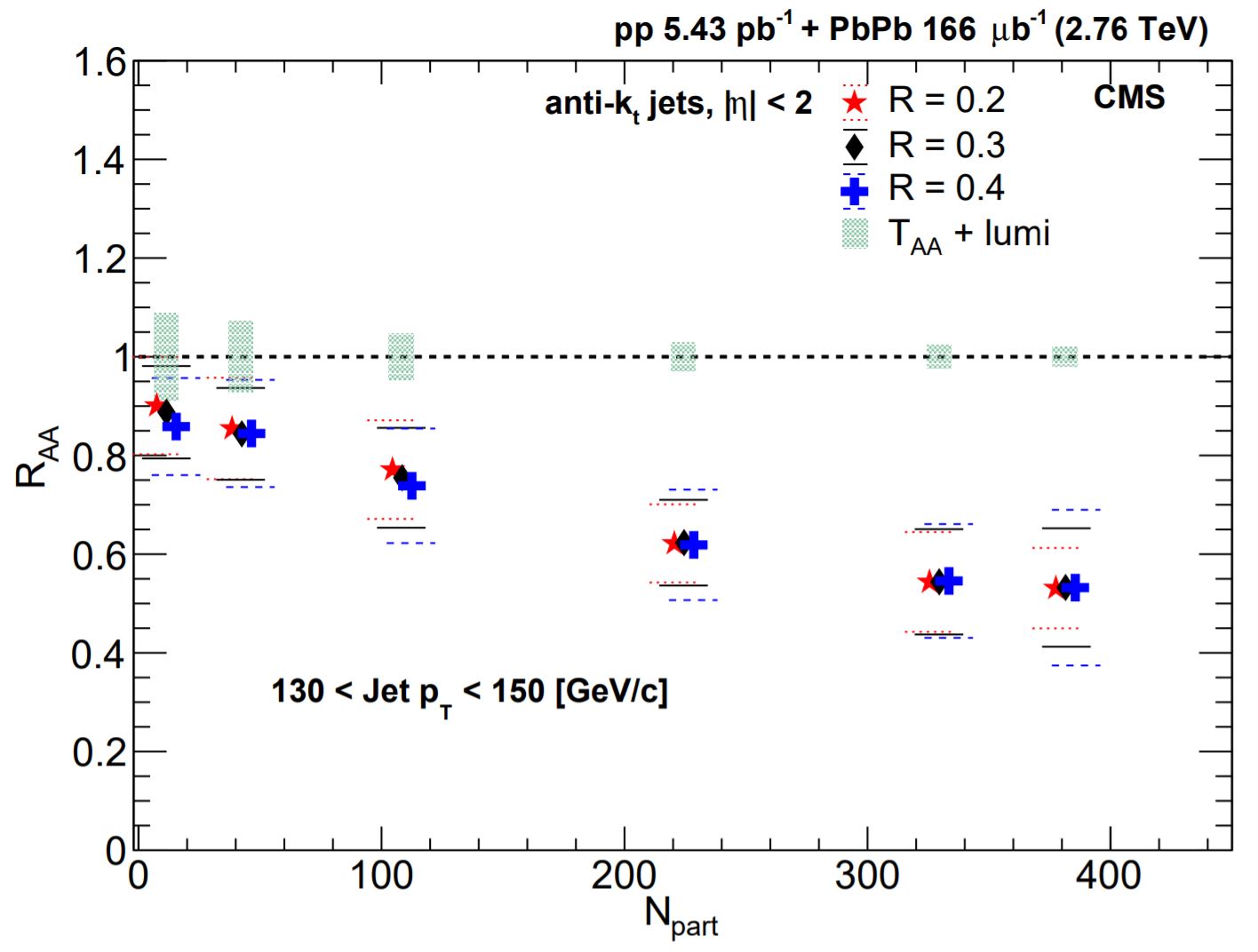}    \includegraphics[width=0.45\textwidth]{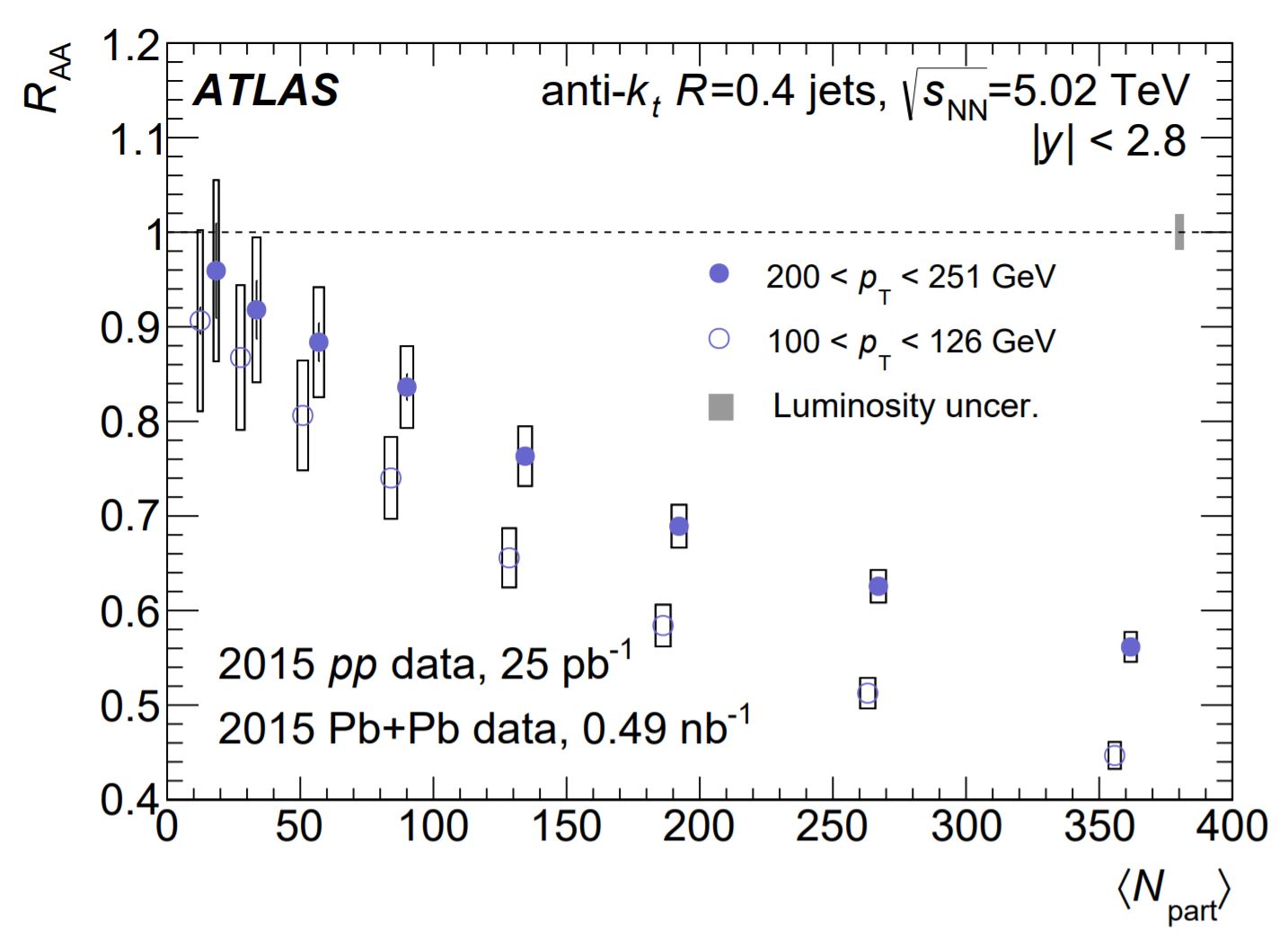}
    \caption{Jet $R_{AA}$ as a function of $N_{\rm part}$ in PbPb collisions at 2.76~\cite{CMS:2016uxf} (left) and 5.02 TeV~\cite{ATLAS:2018gwx} (right).}
    \label{fig:sec4_energylossVsCentrality}
\end{figure}

While jet $R_{AA}$ does not depend significantly on the jet $p_T$, energy loss does have a clear path-length dependence. The jet $R{AA}$ for two jet $p_T$ bins as a function of the average number of participants in the collision, $N_{\rm{part}}$, is shown in the left panel of Fig.~\ref{fig:sec4_energylossVsCentrality}. These are the number of wounded nucleons that participated in the collision, directly contributing to QGP production. From a geometric model, as the Glauber model~\cite{Miller:2007ri}, the average number of participants is proportional to the overall region between the two ions, i.e., the impact parameter. The more central the collision (smaller the impact parameter), the larger the number of $N_{\rm{part}}$ and the extent of the medium that is created. One thus expects energy loss effects to deplete for a small number of participants. As ATLAS and CMS results show, the jet $R_{AA}$ does evolve towards unity for small $N_{\rm{npart}}$, only having a mild dependence with the jet transverse momentum. The offset between the two jet $R_{AA}$ is almost constant for a larger number of participants, slowly reducing for smaller QGP lengths. On the right panel, the jet $R_{AA}$ is shown for a fixed jet $p_T$ bin but different jet radius, from $R = 0.2$ to $R = 0.4$. Interestingly, the jet $R_{AA}$ does not depend very strongly on the jet radius, meaning that no further energy is recovered when going from $R = 0.2$ to $R = 0.4$. One would expect to recover the jet's initial energy with increasing jet radius. Understanding where and how this radiated energy is recovered can help us connect to the produced medium's thermalization and transport properties.

\begin{figure}
    \centering
    \includegraphics[width=0.45\textwidth]{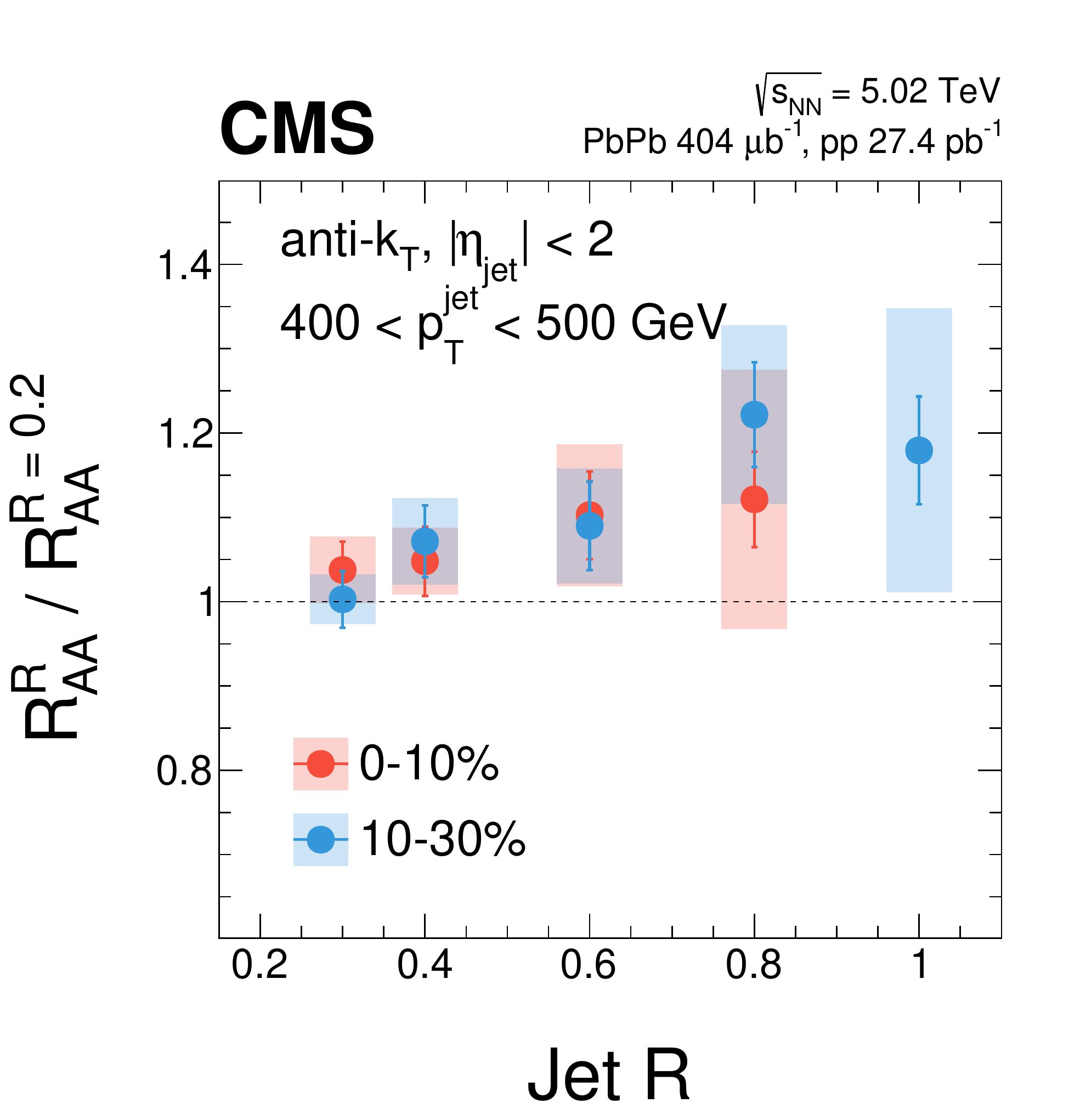}
    \includegraphics[width=0.45\textwidth]{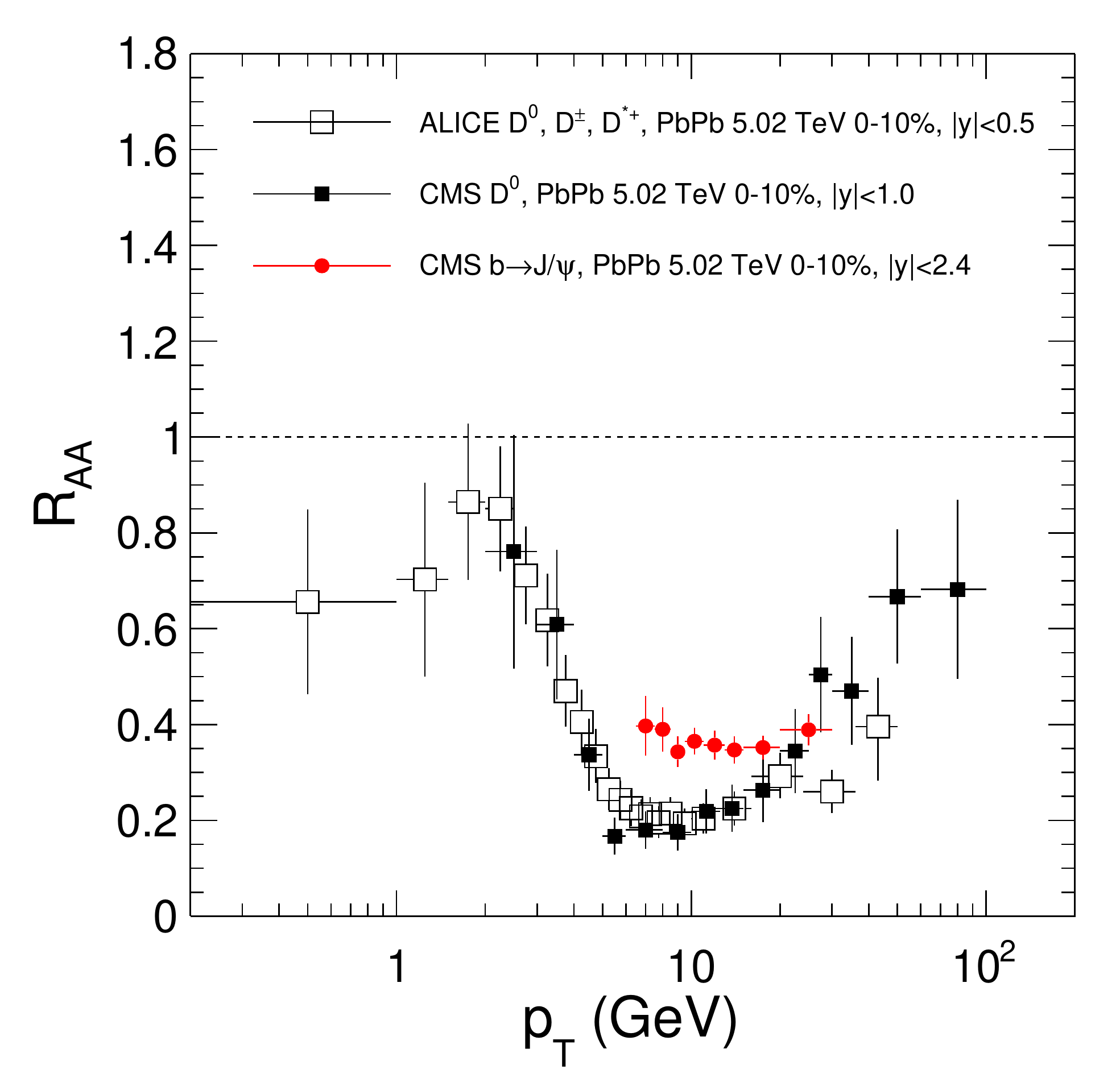}
    \caption{(Left) anti-$k_{T}$ jet nuclear modification factors measured as a function of resolution parameter $R$ in central PbPb collisions at 5.02 TeV~\cite{CMS:2021vui} (Right) Summary of non-prompt $J/\psi$ from b decay~\cite{CMS:2017uuv} and D meson~\cite{CMS:2017qjw,ALICE:2021rxa} nuclear modification factors in central PbPb collisions at 5.02 TeV.}
    \label{fig:sec4_HF_jet}
\end{figure}

Large jet radius measurements in a heavy-ion environment are experimentally challenging. The calorimeter coverage is often limited, and for $R = 1$, only CMS and ATLAS are technically equipped for such analyses. Even so, they must deal with the large background contamination from the collision bulk naturally captured in such large extended jet areas. CMS published results on the double ratio of the nuclear modification factor, taking $R = 0.2$ as reference. These are shown in figure~\ref{fig:sec4_HF_jet}, left panel, for a fix jet $p_{T} \in [500; 1000]~\rm{GeV}$, as a function of the jet radius, $R$. The observed evolution confirms that the jet $R_{AA}$ has a very mild dependence with the jet radius, up to $R = 1.0$, for the most central $PbPb$ collisions. This feature cannot be accommodated by many of the current jet quenching Monte Carlo or theoretical approaches tuned to describe jet spectra and jet substructure reconstructed with smaller jet radius~\cite{Sirunyan:2021pcp}. However, this is just the first example of the benefits of using larger jet radii. Despite the inherited difficulties\footnote{To handle large radii jets, better experimental control is needed on the underlying event, energy correction uncertainties, and also larger statistics.}, they are, by construction, less sensitive to the known jet selection bias discussed above. This selection is induced by the inherent fluctuations from the parton shower. For small $R$ jets, the resulting particles captured inside these jets will be driven mainly by originally more collimated patterns and/or that fluctuated less towards wider opening angles. With a large radius, we also allow jets with different fragmentation patterns to be included in the same $p_T$ bin, improving the comparison between $pp$ and $AA$ at the same transverse momentum. Future studies to understand the role of energy loss fluctuations in jet development can be finally accessible.

In addition to this selection bias, parton flavor will also vary when comparing $pp$ to $AA$ at a given $p_T$. Because of the larger color charge, gluons typically radiate more than quarks. They will likely occupy the low end of the $p_T$ spectrum as they will be more quenched. As for the quark sector, heavy-flavors are expected to experience smaller energy loss compared to light-flavor due to the dead-cone effect. In the presence of a medium, this effect is likely to survive. A strong energy loss hierarchy, namely, $\Delta E_{b} < \Delta E_{c} < \Delta E_q < \Delta E_{g}$ would then be reflected in the corresponding $R_{AA}$\footnote{See, for instance, the calculation of heavy quark jet in Ref.~\cite{Blok:2019uny}}. In particular, we would expect the same trend provided that the $p_T$ spectra of each species would have the same slope. Even though collisional energy loss, induced by elastic scatterings with the medium, contributes in the opposite direction, recent results (right panel of Fig.~\ref{fig:sec4_HF_jet}) at the LHC ($\sqrt{s} = 5.02~\rm{TeV}$) show that there is a hierarchy in the energy loss experienced by heavy-flavor particles in the region of $p_T \simeq 10$~\rm{GeV}. As provided by ALICE and CMS, beauty decay products have a larger $R_{AA}$ than charm (prompt D-mesons). Current models attempt to draw the magnitudes of diffusion and drag induced by the medium interactions. However, we refer the reader to chapter~\ref{sec:QGPDensity} for this discussion. At very high energy, the $R_{AA}$ is consistent with light flavor results as mass-induced effects become negligible. 

\subsubsection{Jet Substructure and Coincidence measurements}

\begin{figure}
    \centering
    \includegraphics[width=0.45\textwidth]{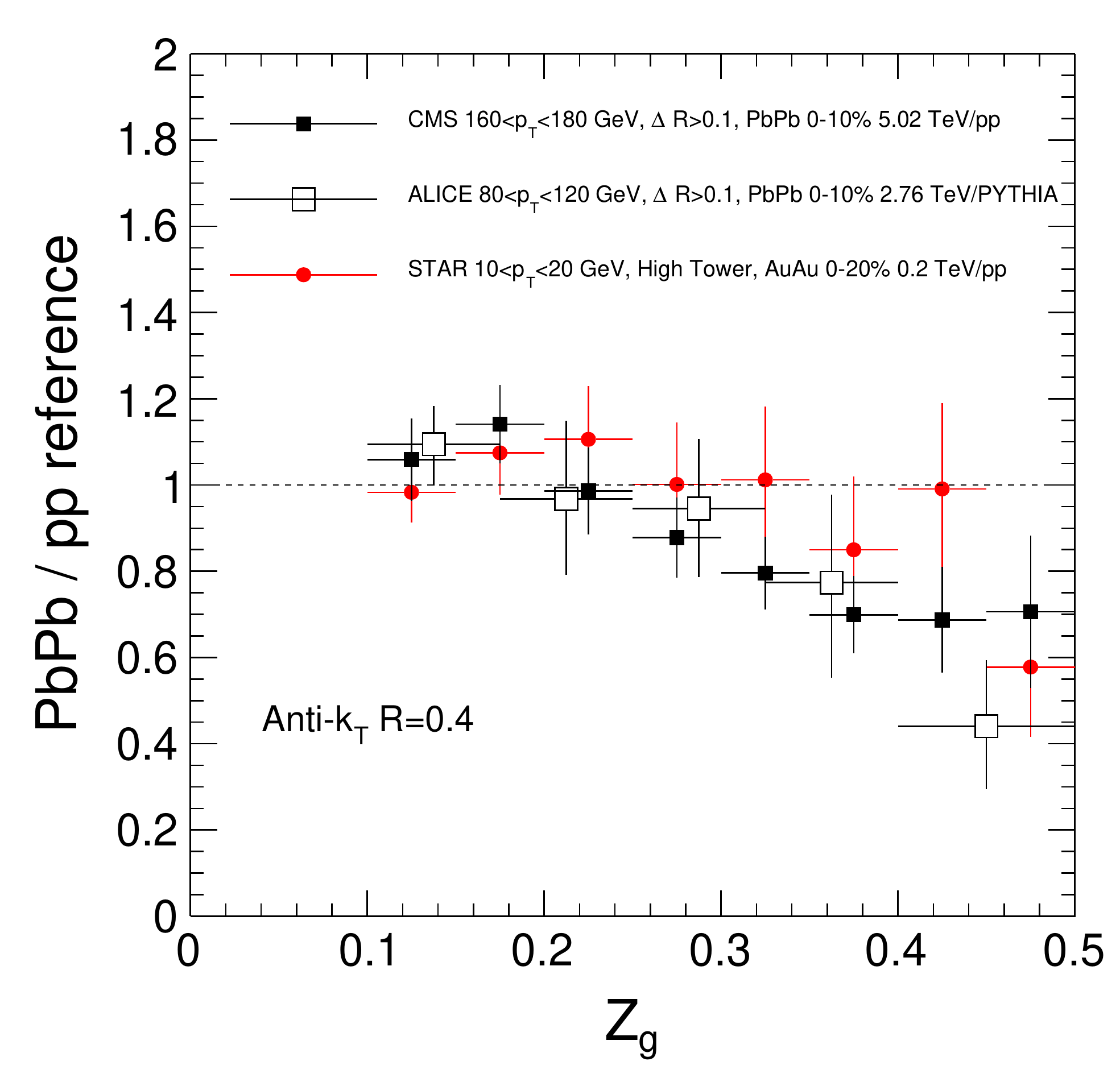}
    \includegraphics[width=0.45\textwidth]{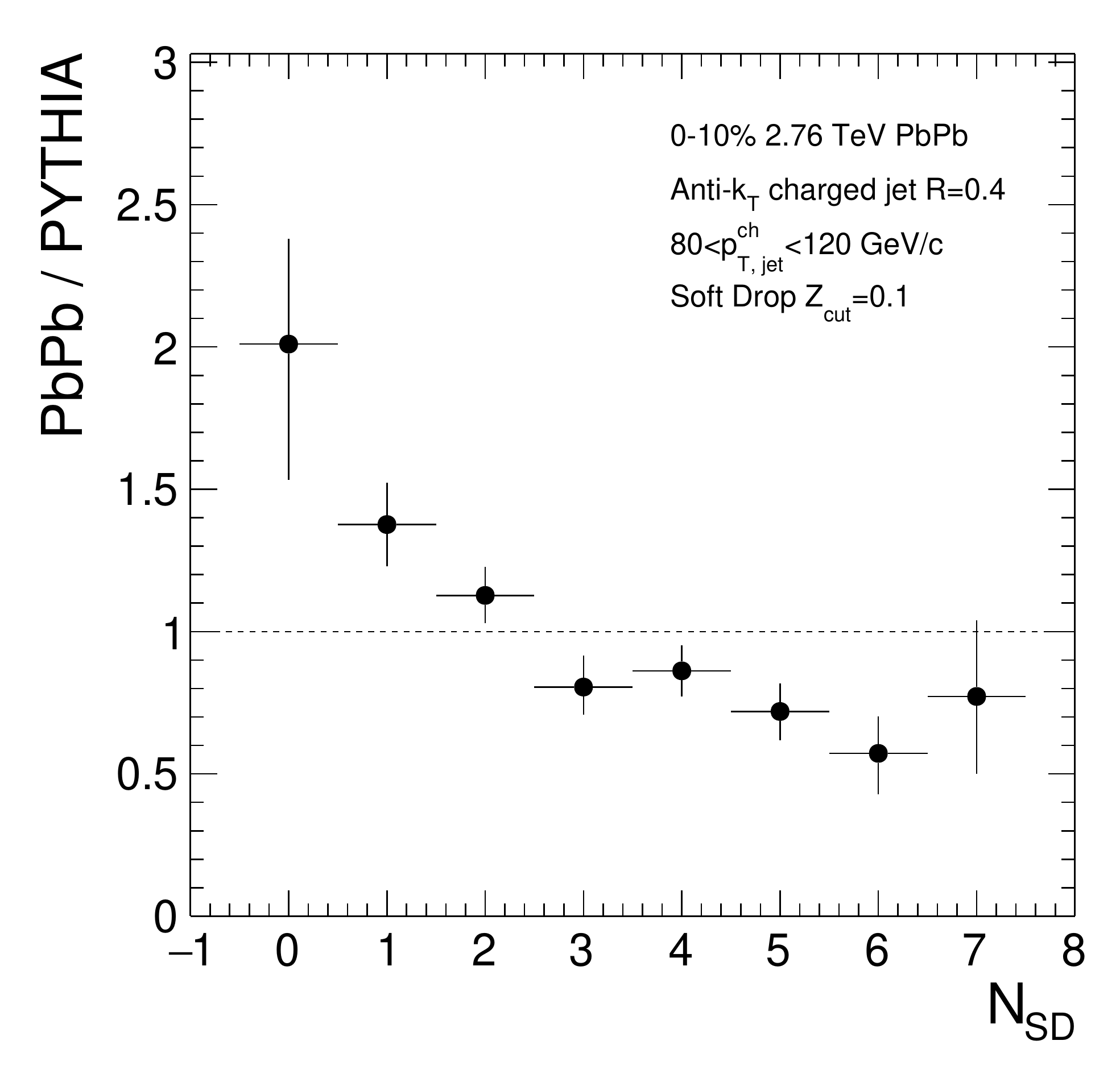}
    \caption{ (Left) Ratio of normalized momentum sharing spectra of two-prong soft-drop substructure, $z_g$, in central PbPb and pp reference published in  ALICE~\cite{Acharya:2019djg}, CMS\cite{CMS:2017qlm}. Those results are compared to preliminary STAR measurement in central AuAu collisions at 0.2 TeV. (Right) Ratio of $N_{SD}$ distributions from 0-10\% PbPb data at 2.76 TeV and pp reference predicted by PYTHIA anti-$k_{T}$ R=0.4 charged jets~\cite{ALICE:2019ykw}.}
    \label{fig:sec4_zg}
\end{figure}

While $R_{AA}$ is generally reliable in providing a reference to tune jet quenching models, extracting further details on energy loss phenomena requires more exclusive observables. So far, we have discussed how medium-induced radiation will induce energy loss compared to $pp$ expectations. This phenomenon has been demonstrated experimentally in isolated photon+jet and Z boson+jet coincidence measurements, showing that the associated jets are quenched and carry less transverse momentum compared to the electroweak boson~\cite{CMS:2012ytf,CMS:2017ehl,ATLAS:2018dgb,CMS:2017eqd}. Nonetheless, not all the energy flows out of the jet area. A fraction of those particles resulting from medium-induced radiation will be captured inside the reconstructed jet, affecting the jet substructure. The relatively recent application of jet reclustering tools to the heavy-ion field has unlocked multiple ways to identify this extra radiation's kinematics. Among several observables, the first of these tools, initially designed to jet substructure studies in $pp$ collisions, is the momentum sharing of two-prong substructure exposed via grooming, $z_g$~\cite{Larkoski:2015lea}, and the number of soft-drop emissions, $N_{SD}$. The procedure to obtain these quantities goes as follow: once a jet of radius $R$ has been identified (usually through the anti-$k_T$ algorithm, $p = -1$ in eq.~\eqref{eq:sec4_dij}), the jet constituents are reclustered with Cambridge/Aachen (C/A) algorithm (p = 0). The obtained sequence is then unclustered, step-by-step. At each unclustering step, the transverse momentum of the resulting objects (sub-jets) is identified ($p_{T,1}$ and $p_{T,2}$) as well as their relative distance in the $(y,\phi)$ plane ($\Delta R_{12}$). If the condition:
\begin{equation}
    \frac{\rm{min} (p_{T,1},p_{T,2})}{p_{T,1}+p_{T,2}} > z_{cut} \left( \frac{\Delta R_{12}}{R} \right)
    \label{eq:sec4_SD}
\end{equation}
is satisfied, then:
\begin{equation}
    z_g = \frac{\rm{min} (p_{T,1},p_{T,2})}{p_{T,1}+p_{T,2}}
    \label{eq:sec4_zg} \, .
\end{equation}
Otherwise, the emission is discarded, and the process continues along the hardest branch (largest $p_T$) until the Soft-Drop condition given by eq.~\eqref{eq:sec4_SD} is fulfilled. The $N_{SD}$ represents the number of emissions along the primary (hardest) branch that meet the Soft-Drop criteria, or, equivalently, the number of hard QCD splittings within the captured parton shower. The parameters $z_{cut}$ and $\beta$ are generally, arbitrary, but to expose the \textit{jet splitting function} ($z_g$), these are set to $z_{cut} = 0.1$ and $\beta = 0$. Both RHIC and the LHC experiments have measured $z_g$, normalized to the number of groomed jets (jets that contain at least one Soft-Drop emission), and the results are shown in Fig.~\ref{fig:sec4_zg} (left), as the ratio of PbPb to the pp reference. 

The results by CMS~\cite{CMS:2017qlm} (black squares) and ALICE~\cite{ALICE:2019ykw} (open squares) refer to jets with different transverse momentum and different center-of-mass energies. Nonetheless, both indicate that the number of asymmetric splittings (lower $z_g$) is enhanced compared to more symmetric configurations. The modifications could be attributed to the dense medium created at LHC energies that can resolve the two emitters. Jet constituents will scatter with the medium particles independently and originate medium-induced radiation whose kinematics ($z_g$) differ from vacuum physics. In particular, one would expect additional low-$z_g$ emissions, provided that the medium-induced jet splitting would survive the soft-drop condition. The per groomed jets normalization adopted in Fig.~\ref{fig:sec4_zg}, chosen to allow a comparison between the different collaborations, does not allow us to conclude the absolute difference with respect to proton-proton collisions. ALICE's original results~\cite{ALICE:2019ykw} were done with a different normalization, chosen per jet instead. In this case, the effect is driven by a suppression of the most symmetric (large $z_g$) configurations. This also hints at the previously discussed jet selection bias: configurations that are narrower (and therefore with a typically smaller energy-momentum sharing between the prongs) dominate the selected PbPb population. Nonetheless, an accurate description of the $z_g$ generally requires more than a jet selection effect. As discussed in~\cite{Milhano:2017nzm}, the introduction of medium recoil effects seems to provide a reasonable description of the results. In addition, CMS has an intrinsic cut of $\Delta R > 0.1$. A softening of these constraints could help pin down the medium-recoil effects on the jet splitting function, as the medium response will typically stay near the hard fragments' direction of flight. Conversely, if the medium resolves both emitters, the resulting $\Delta R$ distribution would be depleted. 

STAR results from RHIC are also shown in red. However, the current statistical uncertainties do not allow claiming if the $z_g$ distribution, at $0.2~\rm{TeV}$ is modified or, conversely, if the jet splitting function does not show any apparent modifications, despite the clear signatures of energy loss as provided by the $R_{AA}$ (see Fig.~\ref{fig:sec4_energyloss}). Future measurements at $\sqrt{s} = 0.2\rm{TeV}$ will allow us to understand if the less extended and diluted medium produced at RHIC energies can be a decisive factor in the differences in the intra-jet modification. 

On the right panel of Fig.~\ref{fig:sec4_zg} it is shown the $N_{SD}$ distribution measured by ALICE~\cite{ALICE:2019ykw} and normalized to the $pp$ description as given by PYTHIA. These correspond to the same events shown in Fig.~\ref{fig:sec4_zg} (left panel) but now considering the normalization with the total number of jets. The number of jets with a higher number of \textit{hard emissions} (larger $N_{SD}$) is indeed suppressed with respect to fragmentation patterns will less activity (that produce less multiplicity). This observation demonstrates the jet selection bias introduced when comparing jets with the same final transverse momentum, $p_T$, in $pp$ and $PbPb$. This bias constantly undermines any interpretation regarding medium-induced effects and QGP characteristics, making the $PbPb$ to $pp$ comparison less obvious. Recent works have proposed alternative methods to bin jet samples to avoid this bias. In~\cite{Brewer:2018dfs}, it was suggested to use the cross-section distribution, similarly to how a collision centrality is defined. In the latter, centrality classes are specified by the event activity, which is highly correlated with the impact parameter. For instance, one can select the $[0-10]\%$ most central events. In the same way, one can sort events by the integrated cross-section and compare the quantiles, $Q_{AA}$ of this distribution between $pp$ and $PbPb$ events. While free of jet selection biases, this method can only be safely applied to extract an average energy loss. Since jet energy loss fluctuations are a key feature in interpreting current observations, this observable is not yet used in an experimental environment. Future tests in boson+jet could help to validate the effectiveness of the proposed $Q_{AA}$ for jet quenching studies. 

\begin{figure}
    \centering
    \includegraphics[width=0.35\textwidth]{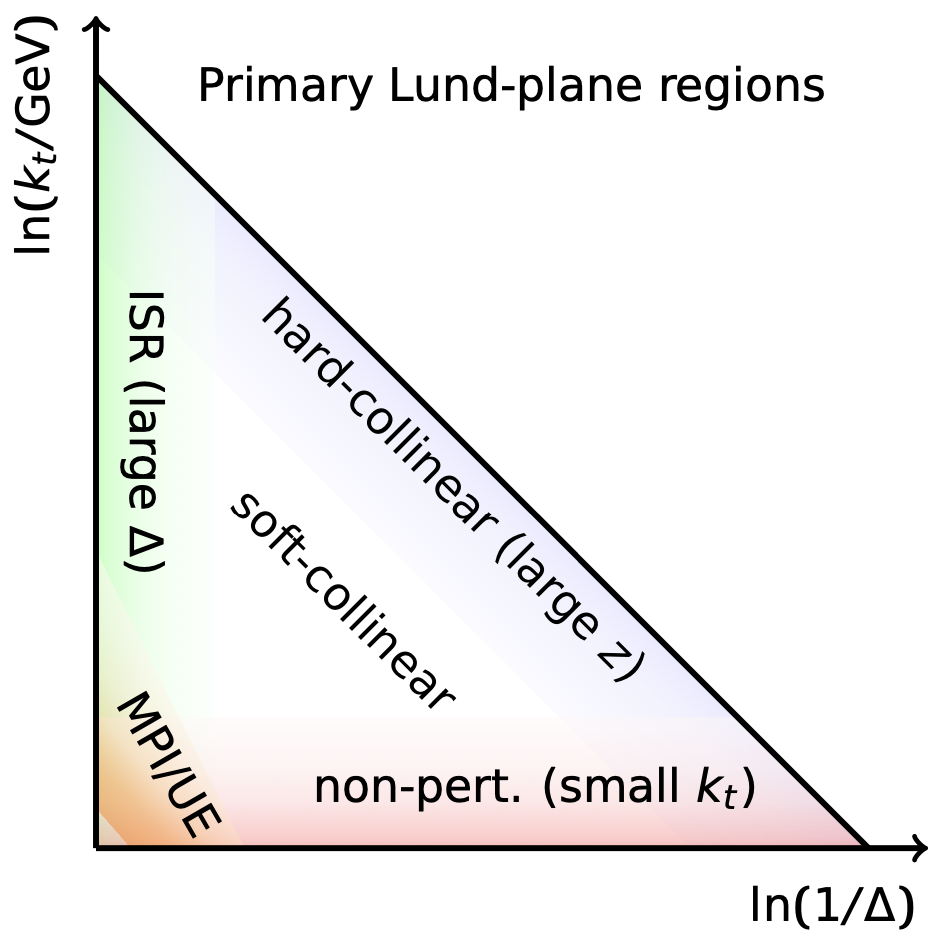}
    \includegraphics[width=0.55\textwidth]{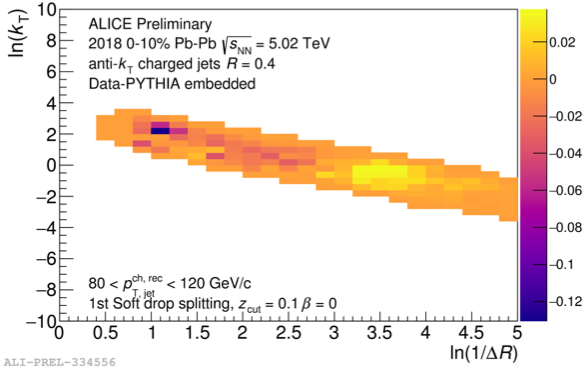}
    \caption{Illustration of the different regions of the jet Lund plane in (left) vacuum (Fig. taken from~\cite{Dreyer:2018nbf}) and (right) Preliminary PbPb results by ALICE (Fig. taken from~\cite{Havener:2709654})}
    \label{fig:sec4_lundplane}
\end{figure}

Jet substructure measurements are not restricted to the jet splitting function, $z_g$ and $N_{SD}$. In fact, with the same Soft-Drop criteria, it is possible to classify the subsequent radiation by populating the kinematics from the unclustering tree into a 2D plane: the jet Lund plane. Lund planes were initially proposed to illustrate the QCD radiation pattern from an incoming high-energy parton~\cite{Andersson:1988gp}. This two-dimensional representation could be in $(\log(z), \log \theta)$ space, as inspired by eq.~\eqref{eq:sec4_gluonrad} (after integrating over the azimuthal angle $\phi$), or any other kinematic equivalent variable, such as $(\log(k_T), \log \theta)$, where $k_T \simeq z E \theta$ is the transverse momentum of the radiated parton with respect to the parent. More recently, the work~\cite{Dreyer:2018nbf} showed that these representations could also be applied at the jet level, provided that the jet clustering history was reconstructed with an algorithm closely resembling QCD angular ordering pattern. The typical choice is the C/A since the clustering increases the pair distances in the $(y,\phi)$ plane. From this history tree, it is possible to identify the primary jet Lund-plane corresponding to the unclustering steps from the leading branch. In contrast, radiation from subsequent branches can be mapped into a succession of orthogonal planes. For better visualization, the primary jet Lund-plane is usually employed. An illustration for a quark-initiated jet is shown in Fig.~\ref{fig:sec4_lundplane} (left), in the variables $\Delta_{ab} = \sqrt{ (y_a - y_b)^2 + (\phi_a-\phi_b)^2}, k_t = p_{t,b} \Delta_{ab}$ for the (sub)leading sub-jets as identified by $(b)a$. The regions corresponding to FSR, ISR, hadronization effects of contamination from the underlying event (UE) or multi-parton interactions (MPI) are explicit in the figure. They would correspond to very different regions of angle and momentum.

Jet Lund planes have numerous applications in $pp$ collisions, ranging from top-tagging, quark-gluon discrimination, and now, more recently, as clear visualization of the dead-cone effect in $pp$ collisions~\cite{Cunqueiro:2018jbh}. In heavy-ion collisions, its application is again manifold. For example, one can use them to identify the kinematics of medium-induced radiation, potentially isolating those from typical vacuum emissions. In the soft and collinear emissions limit, the jet Lund plane would appear uniformly distributed (see eq.~\eqref{eq:sec4_gluonrad}). In the presence of a medium, additional radiation would populate the plane depending on their kinematics. In particular, due to this radiation's nature, an enhancement of soft particles would be expected. A more careful analysis of the medium scales and the different regions that in-medium radiation would populate Fig.~\ref{fig:sec4_lundplane} (right) is done in the following section. Preliminary results from ALICE~\cite{Havener:2709654} hint to a an enhancement of radiation at soft ($\log(k_T) \sim 0$) and small angles $\log(1/R) \sim 3.5$. Comparisons between these results and analytical or Monte Carlo results are still unavailable. They seem to be in qualitative agreement with theoretical expectations (enhancement of soft fragments), but details are still heavily under debate. In addition, the collinear enhancement is also compatible with the bias toward selecting narrower jets, which again blurs a possible interpretation of the medium scales that govern medium-induced radiation. 

\begin{figure}
    \centering
    \includegraphics[width=0.75\textwidth]{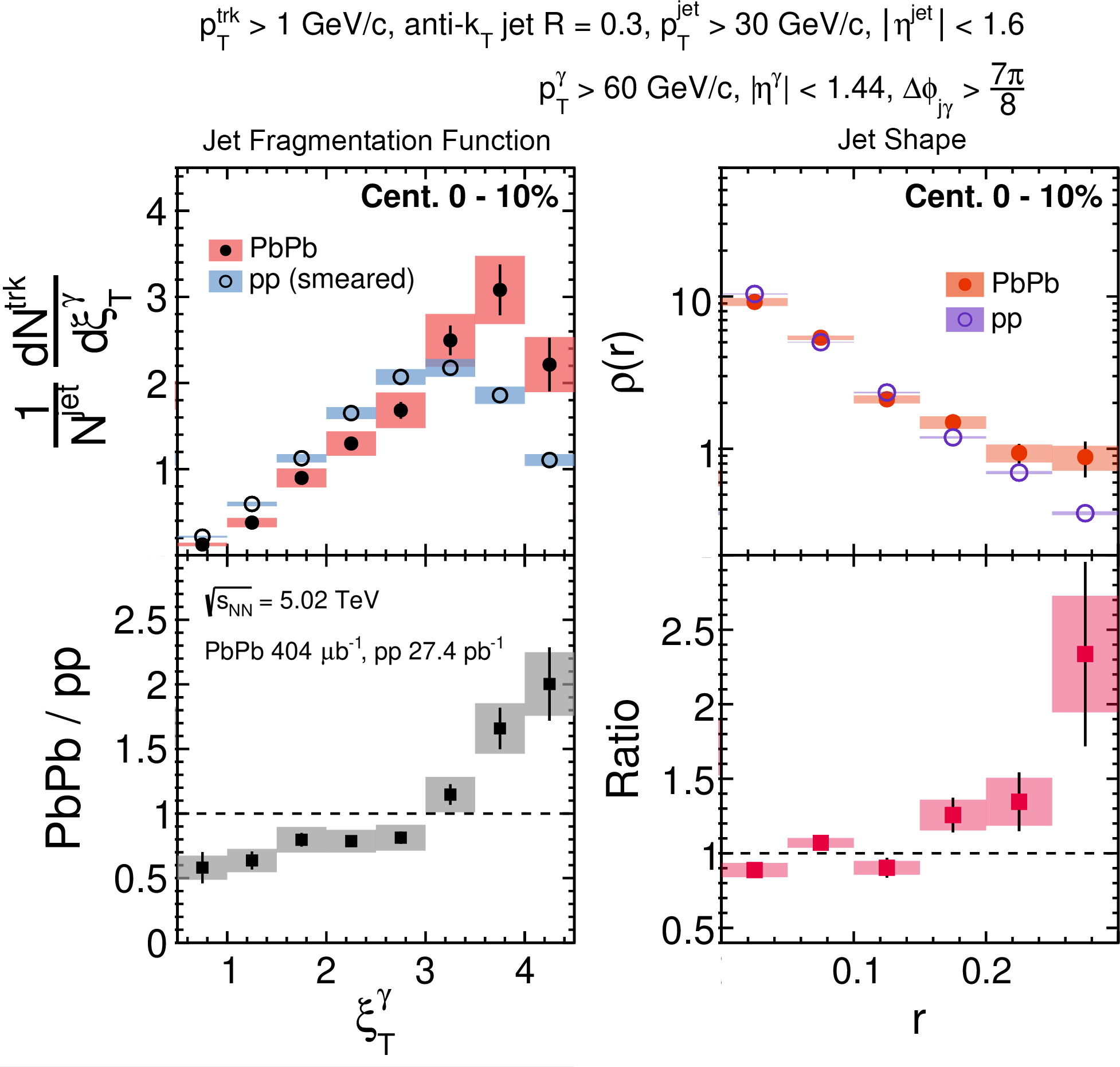}
    \caption{Summary of photon-tagged jet shape and jet fragmentation function in pp and 0-10\% PbPb collisions at 5.02 TeV~\cite{CMS:2018mqn,CMS:2018jco} }
    \label{fig:sec4_jetshape}
\end{figure}

How fast high energetic particles deposit energy in the medium allows inferring the plasma's thermalization properties. The mechanisms by which this thermalization occurs are still largely unknown. However, as discussed later in section~\ref{sec:QGPDensity}, within a kinetic theory approach, the interaction processes responsible for the thermalization at the early stage of the bulk production are related to the thermalization of hard momenta deposition inside of the plasma. Their determination from both sides of the spectrum of the available probes is necessary to constrain current modeling assumptions on this QGP characteristic. Soft fragments originating from in-medium interactions will travel along the incoming parton direction and are detected as positive energy-momentum fluctuations concerning the remaining event. Several observables help build a picture where these fragments' energy is soft and spread at relatively large angles (or early enough to undergo further rescattering), thus originating an enhancement of soft constituents. For example, one can look at the combination of jet fragmentation functions and jet transverse momentum profile. These quantities are defined as a distribution of
\begin{equation}
    \xi_T = \log \left( \frac{ |\vec{p}|^2 }{ (\vec{p}^{trk} \cdot \vec{p}^{jet}) } \right) \, 
\label{eq:sec4_jetFF}
\end{equation} \, 
and
\begin{equation}
    \rho(r) = \frac{1}{\delta r} \frac{1}{N_{particles}} \sum_{particles} \frac{p_T (r - \delta r/2, r + \delta_r/2)}{p_T(0,R)} \, ,
\label{eq:sec4_jetshape}
\end{equation}
respectively. The 3-momenta of the jet is denoted by $\vec{p}$ while the 3-momenta of charged particles by $\vec{p}^{trk}$. The accumulated momentum within $p_T (a, b)$ refers to the momentum accumulated within the radial distribution $\Delta R = \sqrt{ (y_a-y_b)^2 + (\phi_a - \phi_b)^2 }$. Results for photon-tagged jet events by CMS are shown in Fig.~\ref{fig:sec4_jetshape}. These coincidence measurements require a photon with $p_T^\gamma > 60~\rm{GeV/c}$ and a jet with $p_T^{jet} > 30~\rm{GeV/c}$ that is located at $\Delta \phi_{\gamma J} = |\phi_{gamma} - \phi_{jet}| > 7 \pi /8$. They have the advantage of having the reconstructed boson $p_T^{\gamma}$ as a proxy for the initial jet $p_T$, alleviating possible jet selection biases. They still require a selection on the $p_T^{jet}$, but it is possible to withdraw a clearer physical picture from such calibrated channels. The jet fragmentation functions, as a function of $\xi^{\gamma}$ are shown in the left column of Fig.~\ref{fig:sec4_jetshape}. In this definition, one uses the $p_T^\gamma$ instead of $p_{T}^{jet}$ in eq.~\eqref{eq:sec4_jetFF}, together with a negative sign so that the observable is positively defined. Smeared $pp$ data, to account for the heavy-ion environment, is shown in blue (top panel), while the $PbPb$ for $[0-10]\%$ central events are in red (top panel). The $PbPb/pp$ ratio appears in the bottom panel. The constituents with the more significant momentum fraction (small $\xi_T^\gamma$) are slightly suppressed with respect to $pp$ collisions. Generally attributed to the effect of medium-induced radiation, these high-p$_T$ particles originate afterward a soft bulk (large $\xi_T^\gamma$) that propagates towards large distances concerning the jet direction. The resulting jet shape is shown in the right column of Fig.~\ref{fig:sec4_jetshape}. While the core of the jet is almost unmodified with respect to $pp$, there is an excess of fragments (with a large $\xi_T^\gamma$) located at radial distances $R = \sqrt{ (y-y_i)^2 + (\phi - \phi_i)^2 } > 0.25$ away from the jet axis, but still within the jet area. While these features are in qualitative agreement with the characteristics of medium-induced radiation, the magnitude of particles accumulated at large distances motivated an additional source of soft fragments that could hint directly towards QGP thermalization properties: the medium response to the jet propagation. Due to the low momentum scale associated with this phenomenon, this component is the most model-dependent description of current jet quenching models. Its typical signature yields an increase of fragments towards large radial distances, but the relative magnitude concerning a baseline is still under study. 

\begin{figure}
    \centering
    \includegraphics[width=0.65\textwidth]{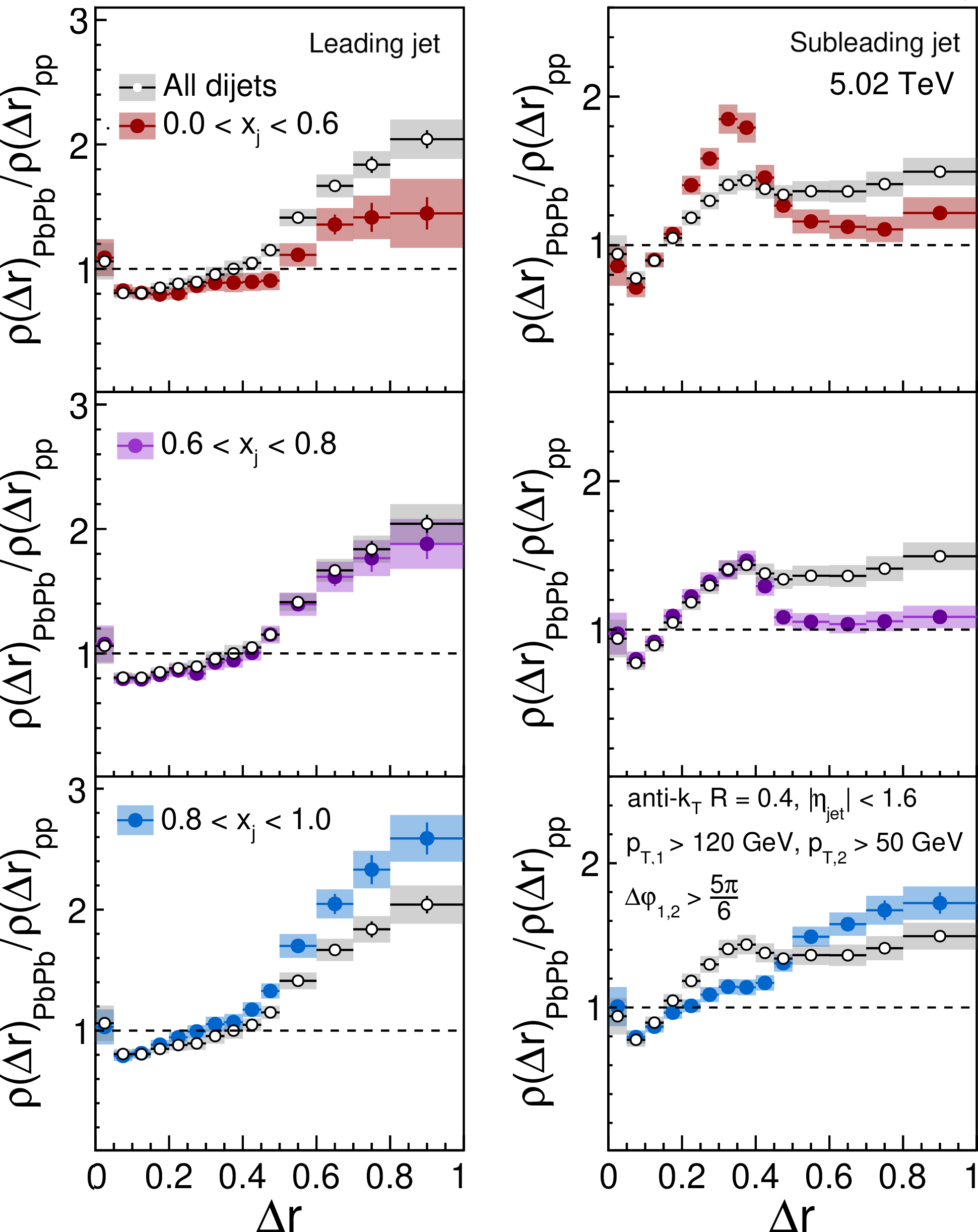}
    \caption{Ratios of leading (left) and subleading (right) jet shapes in 0-10\% PbPb and pp collisions measured as a function of dijet transverse momentum ratio $x_{j}$ by CMS~\cite{CMS:2021nhn}}
    \label{fig:sec4_jetshape_dijet}
\end{figure}

Jet transverse momentum profile was also analyzed in dijet systems reconstructed with $R = 0.4$, extending the particle constituents range to $R = 1.0$. Results by CMS on the ratios of leading (left) and subleading (right) jets in $[0-10]\%$ $PbPb$ events to $pp$ at $\sqrt{s} = 5.02$~TeV are shown in Fig.~\ref{fig:sec4_jetshape_dijet}. On each panel, it is shown a different dijet transverse momentum ratio, $x_j$, from imbalance systems (small $x_j$, top panel) to more balance configurations (large $x_j$, bottom panel), simultaneously with the distribution from all dijets (no selection in $x_j$, in gray). As with $\gamma$+jet, the jet transverse momentum radial profile increases towards large distances. In the subleading jet, the excess of particles to $pp$ starts at $\Delta r \simeq 0.2$ and stays relatively constant up to $R = 1$, consistent with the observed $R_{AA}$ dependence with the jet radius. As for the leading jet, the excess of particles appears only from $\Delta r \simeq 0.5$, as, by construction, the jet with a narrower fragmentation lost less energy.

Nonetheless, the leading and subleading jets have distinct features depending on the dijet asymmetry. For asymmetric systems, the leading jet becomes very narrow (due to the selection bias towards collimated jets), while the subleading, by definition, is broader. However, for the subleading jet to survive the transverse momentum threshold, it needs to contain still part of the energy inside the jet clustering radius employed in this analysis ($R = 0.4$). The population of jets likely to survive this selection will appear with a peak structure near the reconstructed radius. As we move towards more balanced dijets, the less constrained the jet selection criteria are for the jets to survive the cuts. As such, we see that progressively, both leading and subleading jet radial profiles become alike.

\begin{figure}
    \centering
    \includegraphics[width=0.55\textwidth]{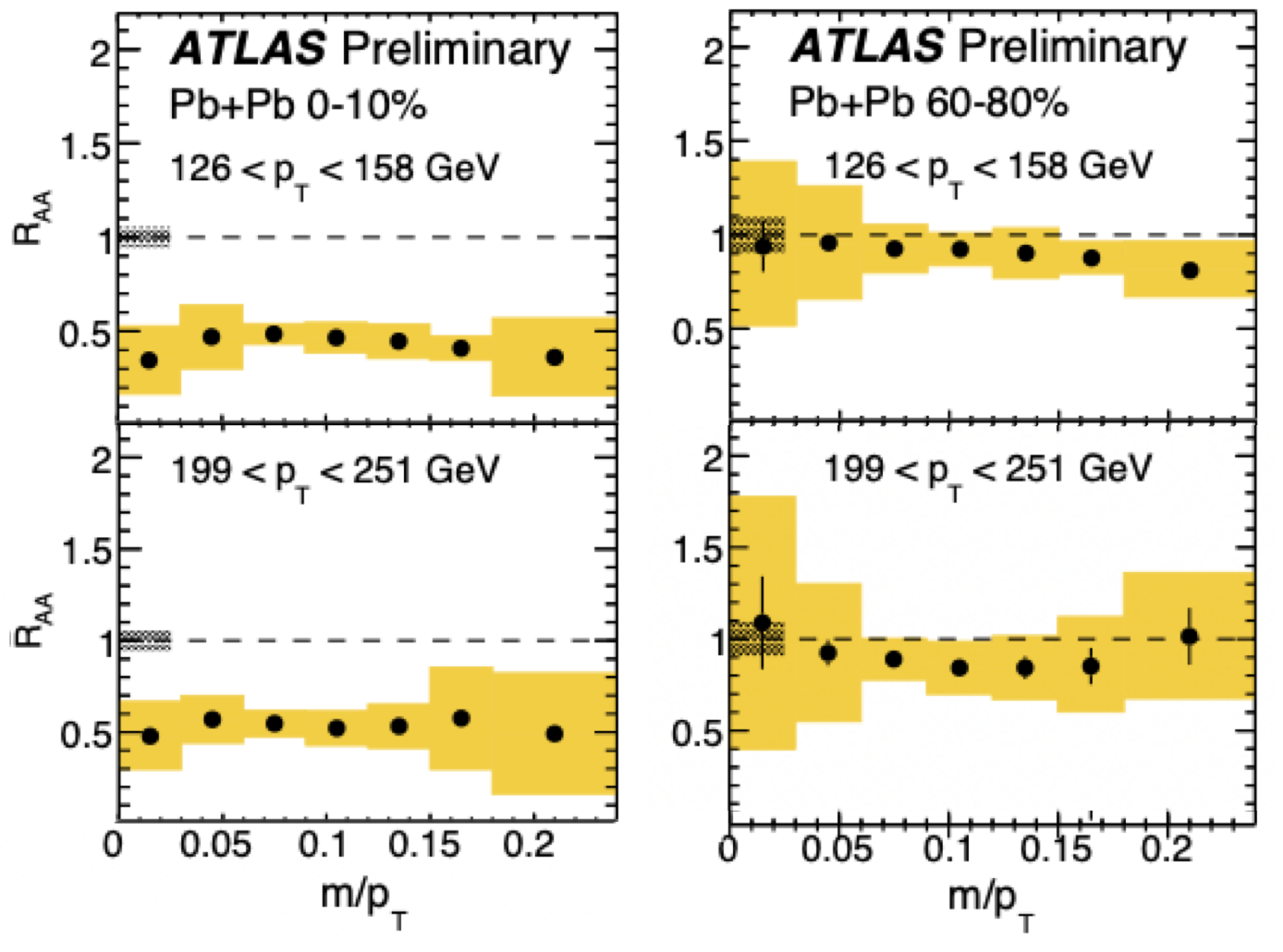}
    \includegraphics[width=0.35\textwidth]{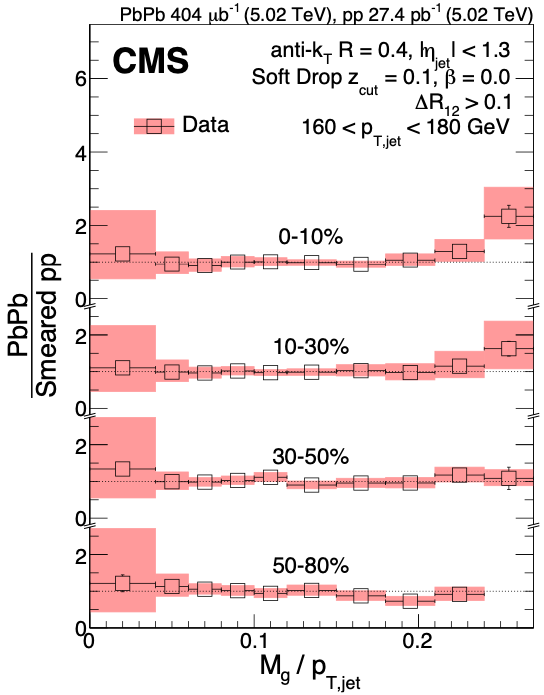}
    \caption{Ratios of ungroomed~\cite{ATLAS-CONF-2018-014} (left) and groomed~\cite{CMS:2018fof} (right) of the jet mass over jet $p_T$.}
    \label{fig:sec4_jetmass}
\end{figure}

The reconstructed jet mass depends on the momentum and angle distribution of particles inside the jet. It is, by definition, susceptible to both in-medium radiation and the presence of medium response inside the jet. Results on the jet mass over the jet transverse momentum, $m/p_{T}$ for jets with $R = 0.4$ have been reported by both ATLAS (ungroomed) and CMS (groomed) for different jet transverse momentum and different centrality classes~\cite{ATLAS-CONF-2018-014,CMS:2018fof}. CMS uses Soft-Drop grooming before determining the jet quantities and takes the smeared $pp$ result as baseline (see the first panel of Fig.~\ref{fig:sec4_jetmass}). There is a hint of a slight enhancement of the jet population with a large mass. The nuclear modification factor, $R_{AA}$, as a function of the jet mass to momentum ratio, $m/p_{T}$, as measured by ATLAS is shown in two-right panels of Fig.~\ref{fig:sec4_jetmass}. Contrasting CMS, these results do not show a significant dependence on $m/p_{T}$.  

The combination of jet radial profile, jet fragmentation functions and jet mass results defy our current understanding of jet quenching. A careful enhancement of soft particles at large angles with a depletion of the energy near the center must be obtained to maintain the resulting jet mass. While in agreement with the previous experimental results, most jet quenching models typically predict a considerable enhancement of the jet mass induced by additional medium-induced radiation and resulting soft particles from jet-induced medium interactions. So far, a consistent description of these three jet substructure observables has not been yet achieved. Other Soft-Drop grooming settings that provide sensitivity to different parts of the phase space might be helpful to constrain particular modeling phenomena of the current models.

Jet charge, the momentum-weighted ($p_{T,i}$) sum of the electric charges ($q_i$) of particles ($i$) inside a jet:
\begin{equation}
    Q\kappa = \frac{1}{(p_{T}^{jet})^\kappa} \sum_{i \in jet} q_i p_{T,i}^\kappa \, ,
\end{equation}
has been measured recently by CMS in $PbPb$ collisions~\cite{CMS:2020plq}. This observable is, by construction, sensitive to the charge of the initiating jet particle, being initially used to measure the electric charge of the parton initiated jet in DIS experiments~\cite{FIELD19781,Fermilab-Serpukhov-Moscow-Michigan:1979zgc}. Most recently, it has been measured in $pp$ collisions to characterize the contributions of quark and gluon fragmentation to jet production~\cite{CMS:2017yer}. Exploring these fractions in PbPb can also provide additional information on in-medium energy loss processes since quarks and gluons lose energy with $C_F = 4/3$ and $C_A = 3$, respectively. Therefore, one would expect the resulting population of quark to be larger than the gluon fraction for high-$p_T$ jets. In addition, this measurement was referenced to separate the isospin effects~\cite{Li:2019dre}, thus gaining novel insight into the in-medium dependence of non-perturbative fragmentation functions. However, the results show that the extracted fractions of gluon-like and quark-like do not change, as all $PbPb$ centrality classes are compatible with $pp$.

While jet substructure observables are useful for understanding the medium transport properties, complementary information, such as the characterization of QGP constituents, can help build a more consistent picture. To this aim, jet accoplanarity was proposed as a possible avenue to understand if, at the energy scales provided by the high momentum particles, QGP could be characterized as a collection of quasi-particles~\cite{DEramo:2012uzl,Kurkela:2014tla,DEramo:2018eoy}. These suggestions were based on a Rutherford-like experiment in which partons plowing through QGP would scatter off a medium constituent. Although rare, the high momentum transfer process would induce a large-angle deflection, referred to as \textit{Molière scattering}, that could be detected experimentally~\footnote{See for instance, calculations with heavy quark jet in Ref.~\cite{Blok:2020kkn,Blok:2020jgo}}. The jet accoplanarity is assessed in boson (Z/$\gamma$)+jet events, or on a trigger hadron + recoiling jet, by measuring the distribution of the azimuthal angle of particles regarding the trigger object direction, $\Delta \phi = |\phi_{boson/hadron} - \phi_{part}|$. In $pp$ collisions, we expect this distribution to be highly peaked at $\pi$ as a natural consequence of the QCD vacuum shower. In contrast, in $PbPb$ collisions, effects of the medium response, medium-induced momentum broadening and/or Molière scattering would show up as deviations from this baseline. Nonetheless, while the first two contributions dominate the $Delta \phi \sim \pi$ region, large-angle deflections would induce a yield enhancement in the tail of this distribution. 

\begin{figure}
    \centering
    \includegraphics[width=0.29\textwidth]{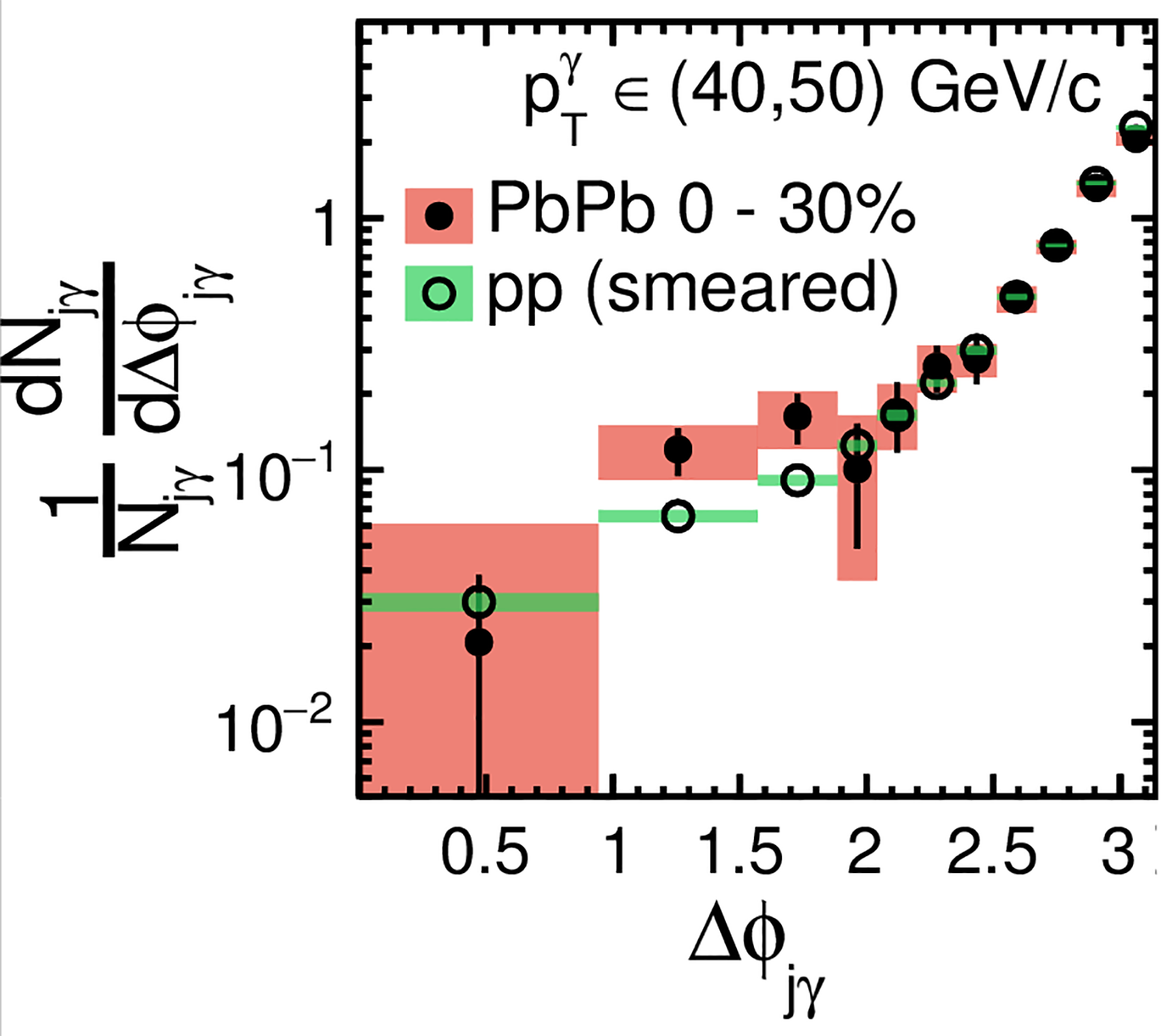}
    \includegraphics[width=0.31\textwidth]{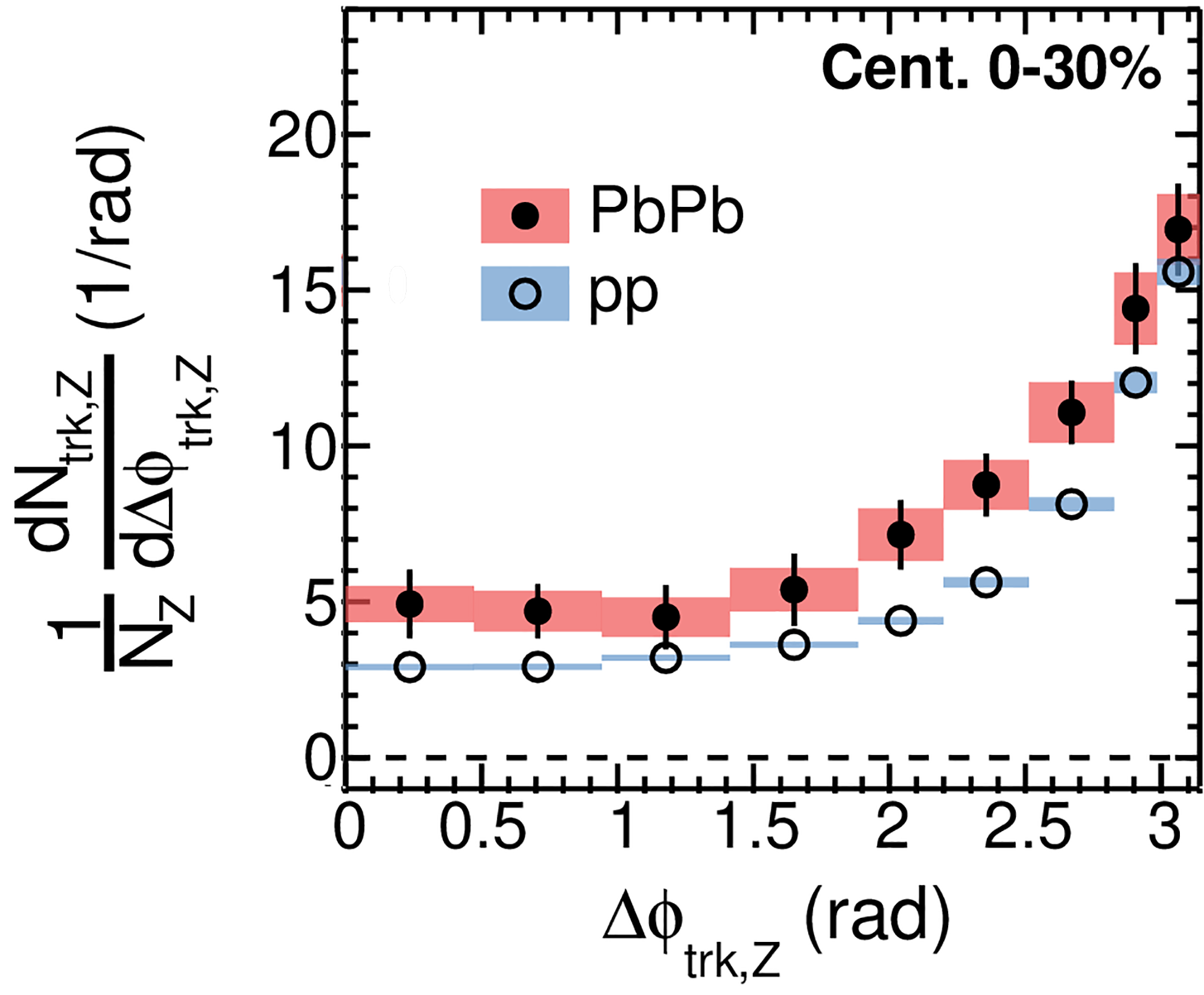}
    \includegraphics[width=0.335\textwidth]{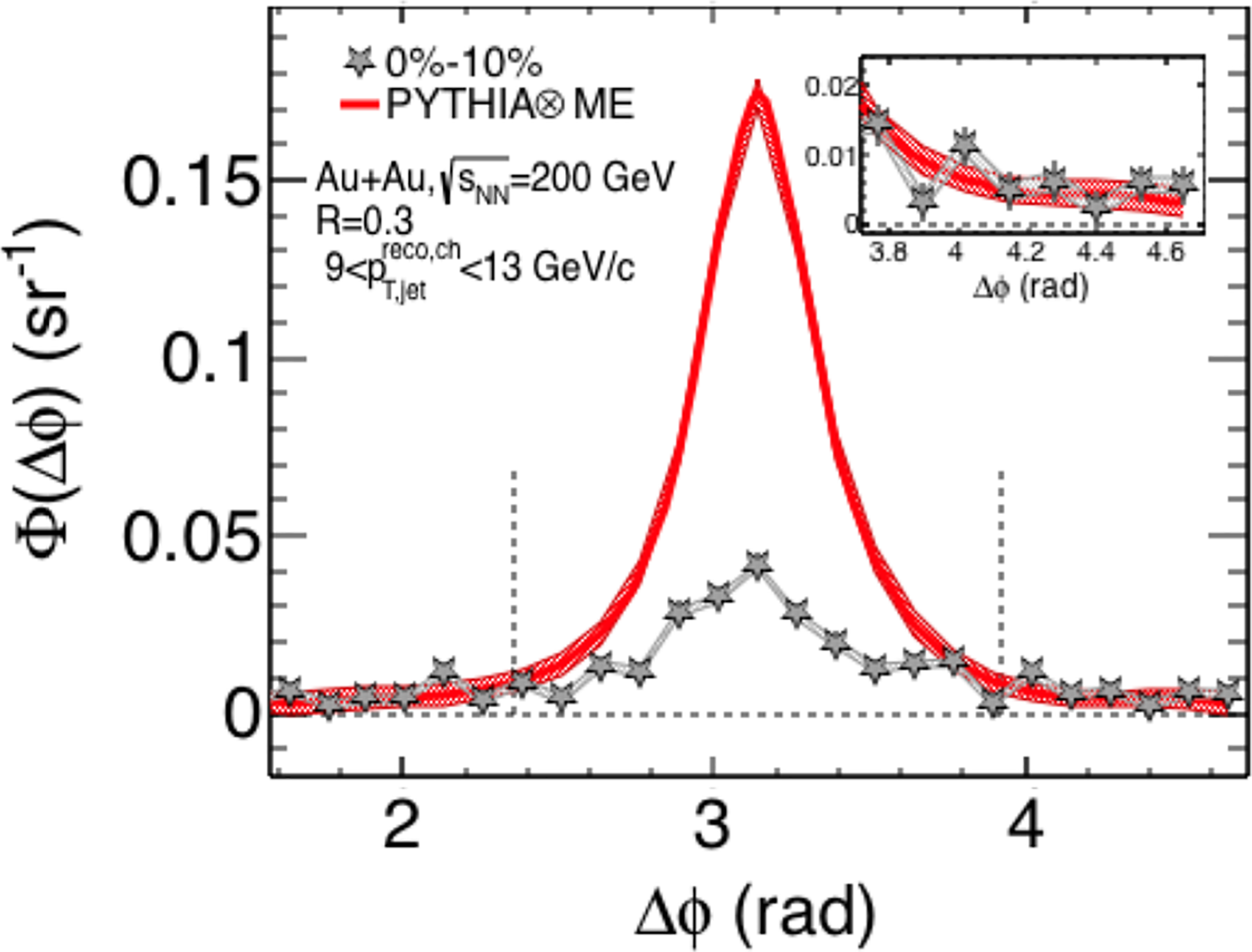}
    \caption{(Left) Azimuthal angle difference distributions between isolated photon and jets in 0-30\% PbPb and pp collisions at 5.02 TeV~\cite{CMS:2017ehl} (Middle) Background subtracted azimuthal angle difference distributions between photons and jets in 0-30\% PbPb and pp collisions at 5.02 TeV~\cite{CMS:2021otx} (Right) Azimuthal angle difference between hadron and charged jets in AuAu collisions at 0.2 TeV~\cite{STAR:2017hhs} }
    \label{fig:sec4_broadening}
\end{figure}

Results by STAR, ALICE, and CMS (see Fig.\ref{fig:sec4_broadening}) do not show yet any sign of azimuthal decorrelation with respect to $pp$ expectations. The first panel compares the azimuthal angle difference distributions for $\gamma$+jet between $PbPb$ and smeared $pp$ events. The second panel is the associated yield to the $Z$ boson, where the $PbPb$ results include background subtraction. There is a uniform enhancement in the number of particles associated with a $Z$ over $\Delta \phi_{trk, Z}$. Recent studies~\cite{Yang:2021qtl} show that medium-modification of initial multiple parton interactions (MPI) originates a soft component that distributes uniformly in azimuthal angle. As such, the observed enhancement is typically driven by quenching of MPI for $\Delta \phi \simeq 0$ while medium-effects dominate the jet direction. The third panel shows STAR results on the azimuthal angle difference between a trigger hadron and recoiling charged jets at $\sqrt{s} = 0.2~\rm{TeV}$. While these findings are compatible with broadening from medium-induced radiation, current observations seem to favor most jet quenching models that include the response of the medium. With the continuous efforts to improve the analytical description of perturbative effects, medium response description will also be more constrained, allowing to withdraw more solid conclusions from the current experimental data. 

\subsubsection{Heavy-Flavor and Quarkonia}
\label{sec:acceleration_HFQQ}

In addition to investigating energy loss with different flavor partons, measurements of hadrons containing heavy quarks in heavy-ion collisions are appealing since they allow experimentalists to tag perturbatively produced heavy quarks down to low transverse momenta~\footnote{In this context, 'perturbative production' of heavy quarks may not only happen in the initial hard scatterings calculable via collinear factorization but also during the parton shower or the first stages of the collision at very high energy density.}. To quantify QGP properties, we discuss two key aspects: (i) the diffusion of heavy quarks and their kinetic thermalization at low momentum and (ii) the formation of quarkonium-bound states. 

The production of open heavy-flavor hadrons dominates the total charm and beauty cross-sections. The main interest of these hadrons in heavy-ion collisions can be summarized as follows:
\begin{itemize}
    \item constrain heavy-quark in-medium transport, including the pertinent spatial diffusion coefficient $D_s$ of heavy-quarks inside a QGP
    \item test the degree of thermalization at small transverse momenta,
    \item investigate energy loss of heavy-flavor quarks,
    \item provides a reference for quarkonium measurements as a heavy quark reservoir.
\end{itemize}
In addition, the perturbative production of the quarks themselves can be used as an additional anchor point for hadronization studies (see section~\ref{sec:hadronization}). 

Open heavy-flavor hadrons have been indirectly measured using their decay products, both leptonic (around $10~\%$) and hadronic. The former, via single lepton and dilepton channels, have been carried out at RHIC since early on (see~\cite{Andronic:2015wma} and references therein). Exclusive hadronic decay channels of charm and beauty hadrons at RHIC and LHC, pioneered by ALICE~\cite{ALICE:2012ab} and CMS~\cite{CMS:2017uoy} were unlocked with more modern silicon vertex detectors. These measurements with hadronic decay channels offer direct access to the hadron kinematics. This section aims to present the most precise experimental data and the current global qualitative understanding. We focus on charm since the measurements of beauty at the LHC are not yet at the same level of precision at low transverse momentum. We note, however, that they corroborate a qualitative picture within the current experimental precision~\cite{ALICE:2016uid,CMS:2016mah,CMS:2017uuv,CMS:2017uoy,CMS:2018bwt,ALICE:2020hdw,ATLAS:2020yxw,ALICE:2020hdw,ATLAS:2021xtw}. 

\begin{figure}
    \centering
    \includegraphics[width=0.5\textwidth]{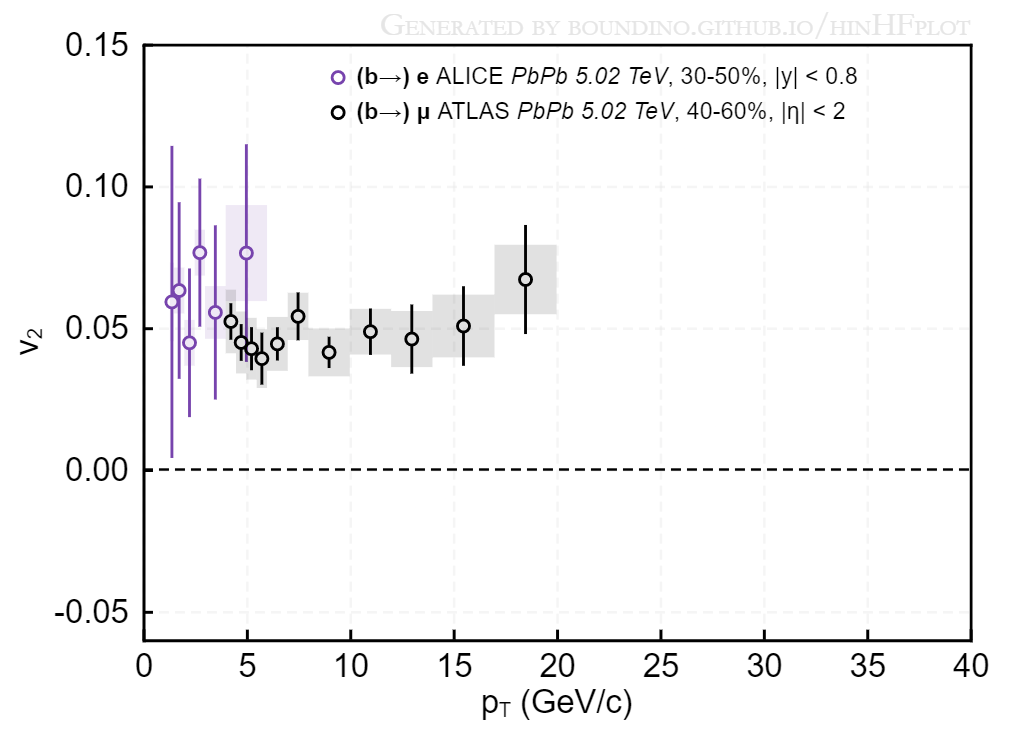}
    \caption{Elliptic flow $v_2$ of leptons from weak decays of beauty hadrons measured by ALICE~\cite{ALICE:2020hdw} and ATLAS~\cite{ATLAS:2021xtw} in PbPb collisions at $\sqrt{s_{NN}}=5.02$~TeV. }
    \label{fig:HFbeautyflow}
\end{figure}

\begin{figure}
    \centering
    \includegraphics[width=0.45\textwidth]{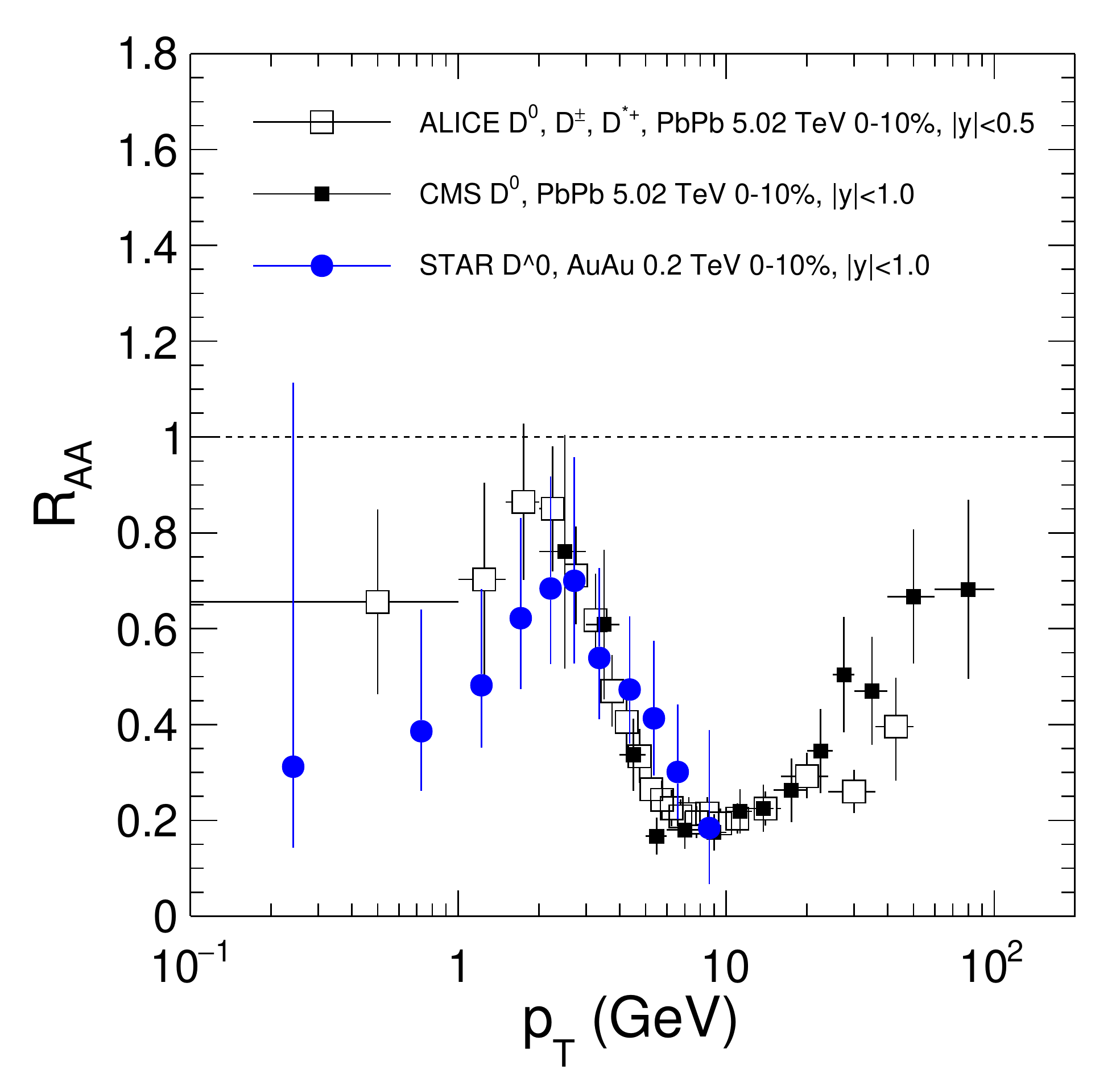}
    \includegraphics[width=0.45\textwidth]{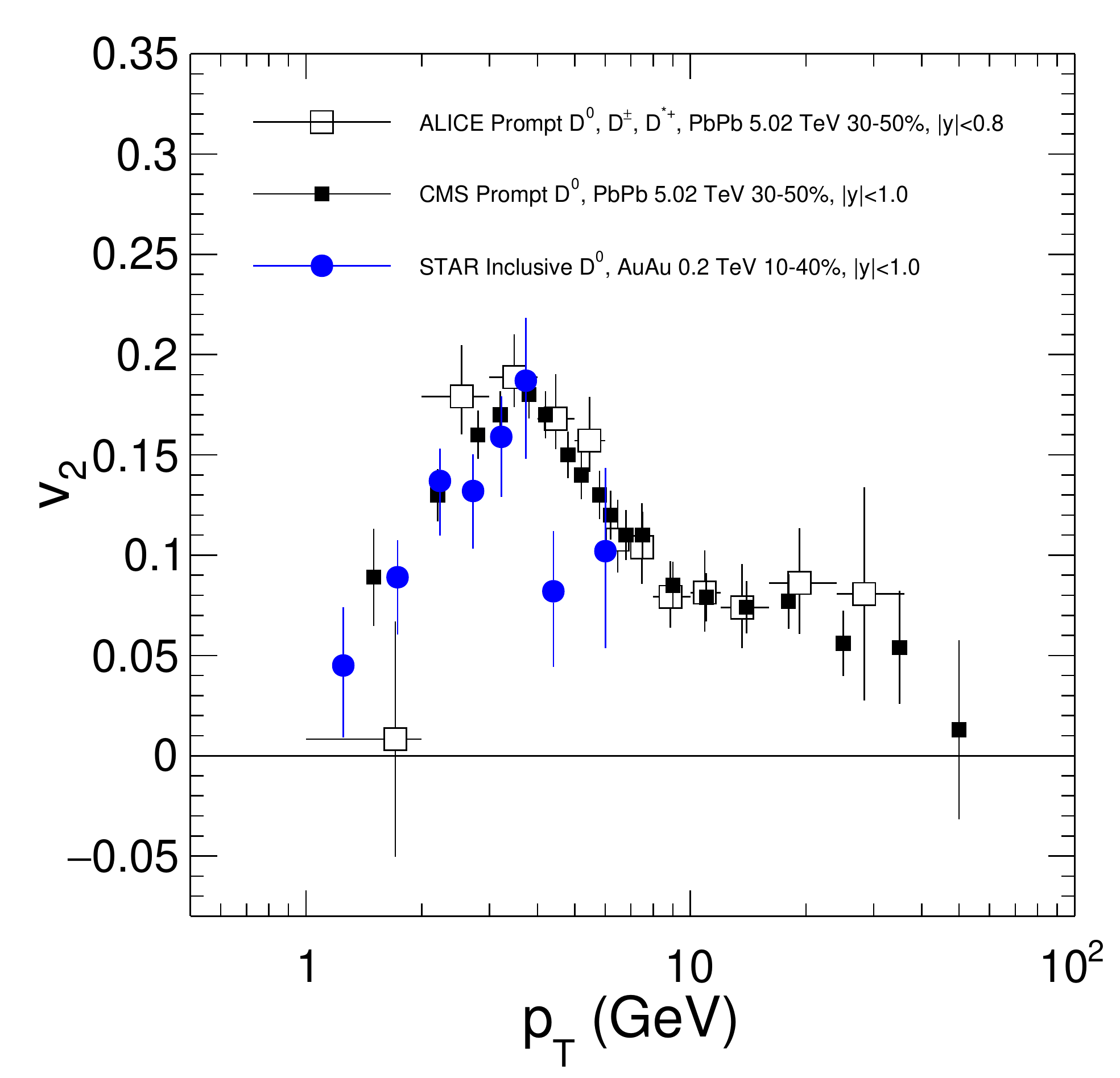}
    \caption{(Left) $D$ meson nuclear modification factors as a function of $D$ $p_{T}$ in 0-10\% PbPb collisions at 5.02 TeV~\cite{CMS:2017qjw,ALICE:2021rxa} and AuAu at 0.2 TeV (Right) $D$ meson elliptic flow as a function of $D$ $p_{T}$ in in 30-50\% PbPb collisions at 5.02 TeV~\cite{CMS:2017vhp,ALICE:2020iug} and 10-40\% AuAu at 0.2 TeV~\cite{STAR:2017kkh}. }
    \label{fig:sec4_Dmesons}
\end{figure}

We consider the D-meson measurements containing u- or d-type quark as a second valence quark. The measurements of $D_s$~production~\cite{STAR:2021tte,ALICE:2015dry,ALICE:2018lyv}, a meson with a net-charm and net-strangeness content, and charm baryon production~\cite{Acharya:2018ckj,Sirunyan:2019fnc,Adam:2019hpq} are discussed in the hadronization chapter~\ref{sec:hadronization}. We focus on the nuclear modification factor $R_{AA}$~\footnote{We discuss the results in terms of the nuclear modification factor to compactify the qualitative discussion. However, the particle spectra contain additional information and should be used with increasing precision comparisons.} and the elliptic flow coefficient $v_2$. The $v_2$ is the second coefficient of the Fourier decomposition of azimuthal distributions of two or more particles in a heavy-ion collision. A finite $v_2$ coefficient was introduced as a signature of collective expansion~\cite{Ollitrault:1992bk}, and it has been regarded as one of the key hallmarks of a strongly-coupled QGP produced at RHIC~\cite{Ludlam:2005cfp}. The picture becomes transparent for non-central collisions between identical spherical nuclei, where the spatial nuclear overlap is usually asymmetric (resembling an almond shape). In this case, the pressure gradient along the smaller axis is larger, and so is the transverse expansion. The result is an enhancement of the particles emitted in this direction. For this reason, a positive elliptic flow (anisotropies on the momentum space of the produced particles) is identified as the transmutation of the initial-state spatial asymmetries of the energy deposition in the overlap region. Naturally, because the $v_2$ is a consequence of the bulk of the produced particles, it is expected to appear mostly for low transverse momenta. In a central collision, the reaction plane is more symmetric, and initial state fluctuations cause the remaining elliptic flow coefficient (see for a pedagogic introduction in~\cite{Ollitrault:2007du} and for a review of hydrodynamic modeling in~\cite{Gale:2013da}). 

Recent result on $R_{AA}$ in central collisions and v$_2$ in semi-central collisions of non-strange D-mesons measured at RHIC and at the LHC are shown in Fig.~\ref{fig:sec4_Dmesons}. The results by  CMS~\cite{CMS:2017qjw,CMS:2017vhp} and ALICE~\cite{ALICE:2020iug,ALICE:2021rxa} indicate a good agreement among LHC experiments. The measurements by STAR at RHIC~\cite{STAR:2014wif,STAR:2017kkh} and the LHC results show the same qualitative features in all transverse momentum regions.  

The strong nuclear suppression at high transverse momenta and the observed finite azimuthal anisotropies above 6-10~GeV/$c$ are interpreted as a signature of parton energy loss, analog to light flavor hadron observables. They can be described within different model calculations~\cite{Xu:2014tda,Xu:2015bbz,Shi:2018izg,Zigic:2018ovr,Zigic:2018smz,Stojku:2020tuk,Kang:2016ofv} as shown in~\cite{ALICE:2021rxa}. In this momentum region, the elliptic flow is usually associated with path-length energy loss dependence. Its magnitude also depends on effects other than collision geometry, such as heavy-quark-medium interaction and hadronization effects.

The nuclear modification factor at lower transverse momentum is characterized by a bump structure in the STAR data and a hint of a similar structure in the ALICE data in most central collisions. 
The second Fourier coefficient $v_2$ is measured for single heavy-flavored quark hadrons by ALICE~\cite{ALICE:2013olq ,ALICE:2014qvj,ALICE:2017pbx,ALICE:2018gif,ALICE:2020iug}, CMS~\cite{CMS:2017vhp} and STAR~\cite{STAR:2017kkh}. The elliptic flow coefficient in Fig.~\ref{fig:sec4_Dmesons}~(right) shows a large, significant value both at RHIC and the LHC at around 3 GeV/$c$. The measured coefficients are slightly smaller than the azimuthal anisotropies observed for light-flavor hadrons, see e.g. in ~\cite{ALICE:2020iug} for a comparative compilation. The structures in $R_{AA}$ and $v_2$ can be understood in models that embed the heavy-flavor quark interaction in a hydrodynamically expanding medium~\cite{Nahrgang:2013xaa,Beraudo:2014boa,Song:2015sfa,Cao:2016gvr,Cao:2017hhk,Beraudo:2017gxw,Scardina:2017ipo,Ke:2018jem,Katz:2019fkc,He:2019vgs}. In addition, there is experimental evidence for a significant elliptic flow $v_2$ for leptons from weak decays of beauty hadrons~\cite{ALICE:2020hdw,ATLAS:2021xtw} as shown in Fig.~\ref{fig:HFbeautyflow} down to relatively low transverse momentum. This observation indicates a strong interaction of the charm and beauty quarks with the created QGP. 

The azimuthal anisotropy for charmed mesons can also be partially associated with the bulk matter's collective motion via the hadron formation mechanism, hadronization. In this context, the observation of sizeable but yet smaller J/$\psi$~\footnote{A bound state of charm and anti-charm quark.} $v_2$~\cite{ALICE:2013xna,CMS:2016mah,ALICE:2017quq,ALICE:2018bdo,ALICE:2020pvw} is interesting. Assuming that part of the charm-anticharm quark pairs building the J/$\psi$ is not originating from the same hard scattering, the sizeable J/$]psi$~$v_2$ points towards the collective motion of the charm quarks. It appears that, although heavy-flavor partons have a longer relaxation time in the QGP compared to light-flavor partons by a factor $M/T$, the heavy quarks are strongly interacting and participate in the collective motion of the medium. This is also corroborated with the observation of sizeable $v_3$ for non-strange D-mesons~\cite{CMS:2017vhp,ALICE:2021rxa}. 
 
In addition to the medium properties and the heavy-quark medium interaction, the nuclear modification factor and the azimuthal anisotropies at low and intermediate transverse momentum are sensitive to the initial state and the hadronization mechanism. For instance, the transverse momentum integrated nuclear modification factor depends on nuclear PDFs. The recently measured $p_T$-integrated D$^0$ nuclear modification factor~\cite{ALICE:2021rxa} amounts to $0.689\pm 0.054 $ $  \rm{(stat.)}^{+0.104}_{-0.106} $ $ \rm{(syst.)}$ in $0-10~\%$. This value is within the expectations of 0.3 and 0.6 at $90\%$ C.L. extracted from reweight of the nCTEQ15 nPDF~\cite{Lansberg:2016deg,Kusina:2017gkz,Kusina:2020dki} based on the pPb heavy-flavor measurements. We refer the reader to chapter~\ref{sec:initial} concerning the initial state and to chapter~\ref{sec:hadronization} for hadronization effects.

A small fraction of the perturbatively produced heavy-flavor quarks is observed as decay products of heavy-quarkonium states.  
Experimentally, heavy-quarkonium are states with masses below the threshold for open heavy meson pair production. Their vacuum lifetimes are larger than the lifetime of the hydrodynamic description of relativistic heavy-ion collisions. In particular, the vector states ($J^{PC}=1^{--}$) with a significant probability of decaying via a virtual photon to electron-positron or muon-antimuon pairs can be measured with good precision in nucleus-nucleus collisions. It naturally raised the question of whether heavy quarkonium production is changed in nucleus-nucleus collisions due to the presence of QGP and what information this modification carries on the produced matter. While the heavy quarks are typically formed early~\footnote{See Chapter~\ref{sec:timescale} for a discussion of quarkonium from jet fragmentation.}, the formation of a bound state takes a finite time and has to be put in relation to the medium time scales and its interactions with it. Historically, the starting point for investigating quarkonium in heavy-ion collisions was introduced in~\cite{Matsui:1986dk}. The proposed signature of deconfinement was the suppression of the lightest $c\bar{c}$ vector state (J/$\psi$) via color screening of the potential at finite temperature.

For the description of quarkonium observables, it is important to note that several states are below the open heavy-quark meson pair production threshold. The heavier states are characterized by smaller binding energy with respect to the open heavy-quark meson pair production threshold. At finite temperatures, the color screening leads to a disappearance of these bound states at lower temperatures when compared to the lightest state. Assuming we implant thermally equilibrated quarkonium hadrons into a QGP, these states would disappear sequentially via color screening with increasing QGP temperature. This behavior is often called sequential 'melting'. The yield of observed lightest quarkonium vector state hadrons contains a sizeable fraction of strong or electromagnetic decays of these heavier quarkonium states, so-called feed-down contributions. Therefore, sequential melting or any other effects affecting differently the various states impacts the lightest observable state via feed-down. 

We focus on the prominent experimental findings and describe their dominant interpretation, the open questions and current limitations. In addition, we also point out the connection to open heavy-flavor and jet quenching at high transverse momentum.   

\begin{figure}
    \centering
    \includegraphics[width=0.45\textwidth]{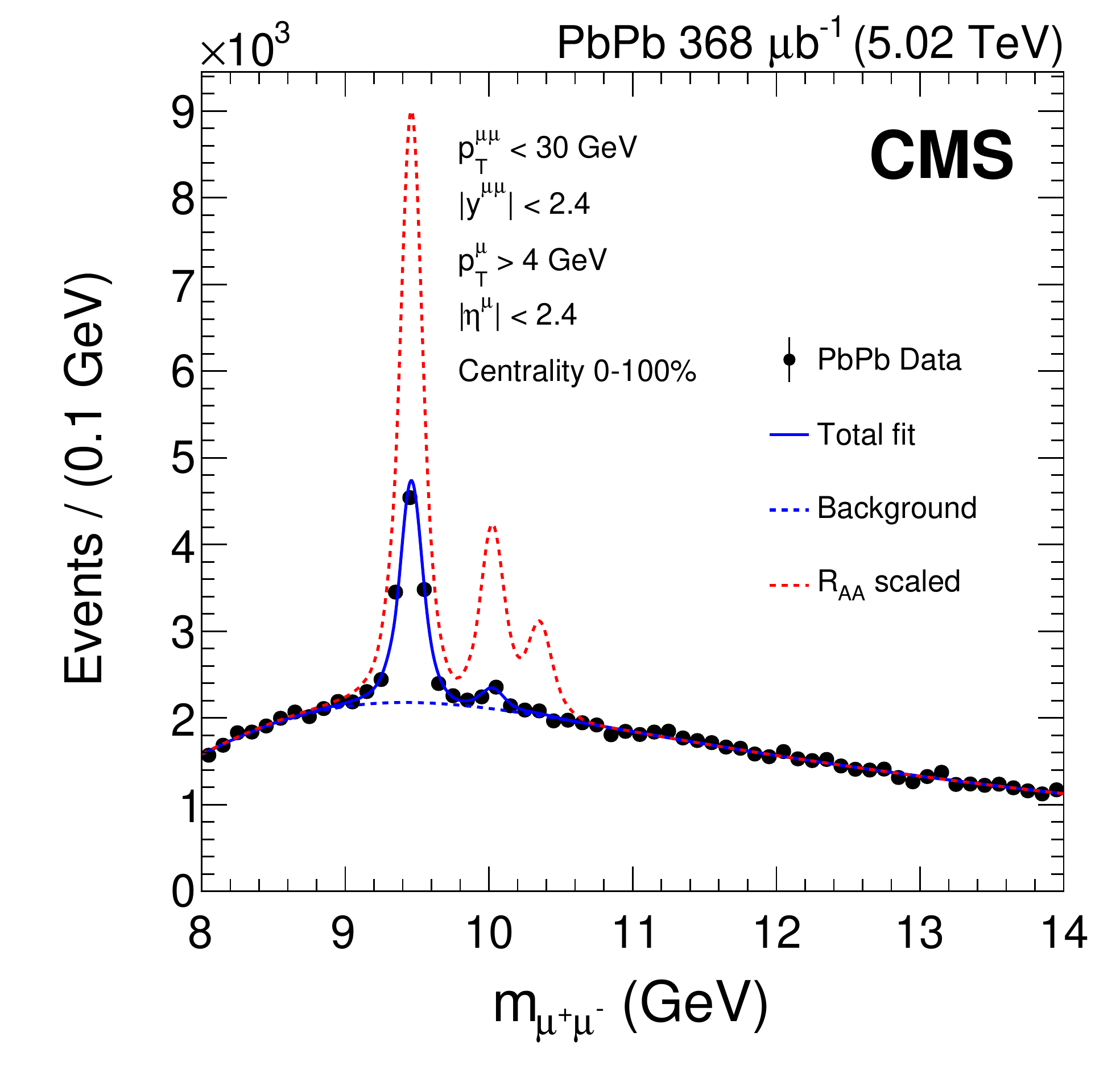}
    \includegraphics[width=0.45\textwidth]{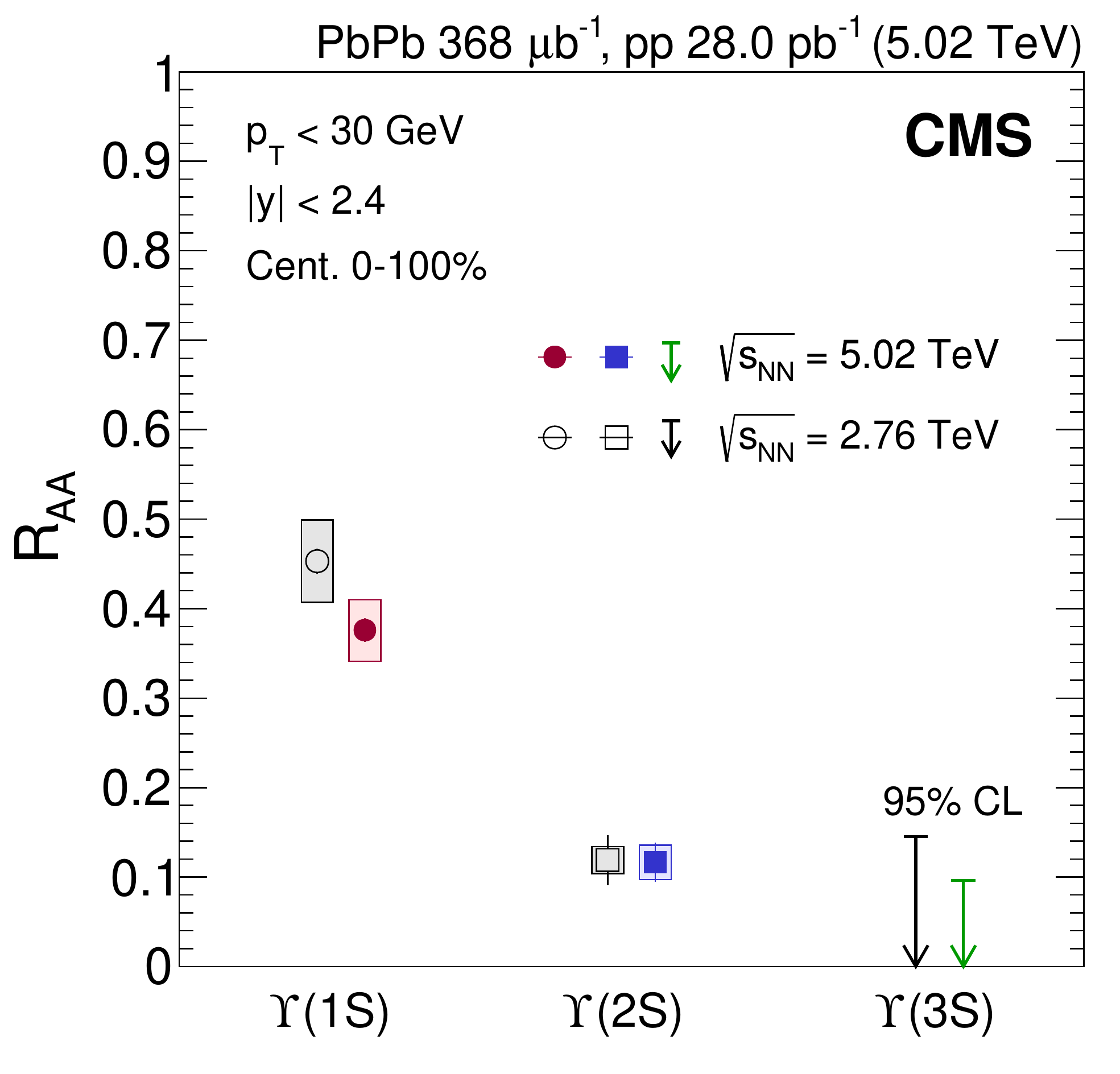}
    \caption{Most recent measurement of sequential suppression pattern of $\Upsilon$-states observed by CMS~\cite{CMS:2018zza}. The invariant mass spectrum in PbPb collisions together with a expectation of the signal assuming a nuclear modification of one for all three $\Upsilon$-states~(left) and the corresponding nuclear modification factors are shown. }
    \label{fig:sec4_quarkonia}
\end{figure}

The $\Upsilon$(1S,2S,3S) states are part of the bottomonium family, bound states of a beauty and an antibeauty quark. Since the typical time scale of the beauty-antibeauty pair production is small (O(1/(2mb) $=0.02$~fm/$c$)), the pairs produced in hard initial scatterings witness the time evolution of the collision system. The production of $\Upsilon$-meson states has been measured in nucleus-nucleus collisions by CMS~\cite{CMS:2011all,CMS:2012bms,CMS:2012gvv,CMS:2016rpc,CMS:2017ycw,CMS:2018zza}, by ALICE~\cite{ALICE:2014wnc,ALICE:2019pox,ALICE:2020wwx} and ATLAS~\cite{ATLAS-CONF-2019-054} at the LHC as well as STAR~\cite{STAR:2013kwk,STAR:2016pof} and PHENIX~\cite{PHENIX:2014tbe} at RHIC.
For the interpretation of the data, the assumptions about the feed-down contributions from excited states in the absence of medium effects are taken from measurements in pp or $p\bar{p}$~collisions at collider energies. P-wave states are usually measured via their electromagnetic decays to a vector state quarkonium and a real or virtual photon. However, the measurements of these final states are limited to relatively large transverse momenta. A good overview of experimental measurements and feed-down implications can be found in~\cite{Lansberg:2019adr}. 

Strong nuclear modifications of the vector state abundance with respect to pp~collisions have been observed and are illustrated by the most recent CMS result in Fig.~\ref{fig:sec4_quarkonia}. There is a clear suppression of the ground state in the nuclear modification factor concerning pp~collision expectation (about 0.4) and an even stronger suppression of the excited states. The stronger suppression of the excited 2S state (factor 3 smaller) and the, so far, non-observation of the 3S-state (0.096 at 95\% confidence level~\cite{CMS:2018zza}) have been advertised early on by CMS as a signature of a QGP-modified production modification~\cite{CMS:2012gvv}. This argumentation is based on the fact that the dominant non-final-state nuclear modification at the LHC, such as parton density modifications and energy loss scenarios discussed in Chapter~\ref{subsubinitiallim}, impact all three $\Upsilon$ states similarly. The nuclear absorption of the bound state while traversing the nucleus that is assumed to be sizeable at collision energies up to RHIC~\cite{Rapp:2008tf}, should become negligible at the LHC since the crossing time becomes small relative to the quarkonium formation time (see~\cite{Andronic:2015wma} for a discussion and references therein). The results by ALICE observe a similar suppression at forward rapidity for the $\Upsilon$(1S) and the $\Upsilon$(2S). At RHIC, STAR measured nuclear modification factors around 0.5 in gold-gold and uranium-uranium. ALICE~\cite{ALICE:2019pox} and CMS have measured the elliptic flow of $\Upsilon$(1S)~\cite{CMS:2020efs}, consistent with no observed signal within sizeable uncertainties. The latter observation is in line with model expectations that include QGP~induced effects~\cite{ALICE:2019pox,CMS:2020efs}.  We discuss the current modeling of bottomonium in detail and connections to theory in~\ref{sec:quarkoniumtheo}.
 
Charmonium production has been measured from SPS energies up to LHC energies. In this review, we focus on the findings at RHIC and at the LHC, where these studies are embedded in the 'standard model' of heavy-ion collisions with a time scale separation between the equilibration phase, the hydrodynamic phase and the transition to hadrons. A number of largely consistent experimental measurements at RHIC~\cite{PHENIX:2006gsi,PHENIX:2008jgc,PHENIX:2011img,STAR:2013eve,STAR:2016utm} and at the LHC~\cite{ALICE:2012jsl,CMS:2012bms,ALICE:2013osk,ALICE:2015nvt,ALICE:2015jrl,ALICE:2016flj,CMS:2016mah,ALICE:2017quq,CMS:2017uuv,ATLAS:2018hqe,ATLAS:2018xms,ALICE:2018bdo,ALICE:2019lga} allow us to draw a qualitative picture of charmonium at these energies.

In 2000, two phenomenological publications~\cite{Braun-Munzinger:2000csl,Thews:2000rj} introduced the picture of quarkonium regenerated at a late stage of the collision history, either during the deconfined state or at the phase boundary between QGP and hadrons. In the transport approach~\cite{Thews:2000rj}, the fraction of charm quark pairs finding themselves in quarkonium bound states is controlled by a transport equation with a dissociation rate destroying and generating charmonium bound states during the medium lifetime. Alternatively, the initially produced charm quarks are distributed between different charmed hadrons only at the freeze-out according to a common charm fugacity factor per charm quark and their thermal weight~\cite{Braun-Munzinger:2000csl}. The latter picture assumes a complete kinetic equilibration of charm quarks at the transition to hadrons. It is in contrast to the original concepts where charmonium in nucleus-nucleus collisions is inhibited from being formed or destroyed by the in-medium modification of quark-antiquark potential or by hadronic interactions. In the regeneration scenario, the charmonium production is directly coupled to the total charm quark production  since the charmonium bound state is formed from charm-anti-charm pairs from the same hard scattering or from pairs where each charm quark comes from a different hard scattering. In both approaches, transport and thermal model, the regenerated component is approximately proportional to the square of the total charm production. 

Experimentally, in addition to the feed-down contributions from heavier charmonium states, depending on collision energy and kinematics, there is a sizeable contribution from weak decays of b-hadrons as well. These weak decays with typical lifetimes of $c\tau \sim 500 \mu $m can be separated with modern vertex detectors from the prompt production, exploiting the finite displacements at finite momentum. For the most measured lightest vector state, the J/$\psi$, the contribution from b-decays is well measured at the LHC in pp~collisions and amounts to $5-15\%$ at $sqrt{s}=$~5~TeV~\cite{LHCb:2021pyk,ALICE:2021edd} at low transverse momentum and rises up to $60\%$  at 30~GeV/$c$ at $\sqrt{s}=$5~TeV~\cite{CMS:2017exb,ATLAS:2018hqe}, whereas the fraction is smaller at RHIC~\footnote{A discussion for RHIC is given in~\cite{Lansberg:2019adr}.}. This contribution is quantified in pp, pPb and PbPb~collisions. Although the available measurements are limited in precision at low transverse momentum~\cite{ALICE:2015nvt}, the impact of the non-prompt part remains limited in heavy-ion collisions, see the discussion in~\cite{ALICE:2015jrl}. Consequently, the available inclusive J/$\psi$ production measurements at low transverse momentum can be mainly discussed via modifying the prompt component. The prompt feed-down chains of charmonium yielding to J/$\psi$ are better known than for the $\Upsilon$-states. At low transverse momentum, direct J/$\psi$ production amounts to around 80~\% of prompt production, 10-15~\% from P-wave $\chi_c$-states $\chi_{c1}$ and $\chi_{c2}$ and the remaining fraction from $\psi$(2S), the radially excited vector state~\cite{Lansberg:2019adr}.

\begin{figure}
    \centering
     \includegraphics[width=0.45\textwidth]{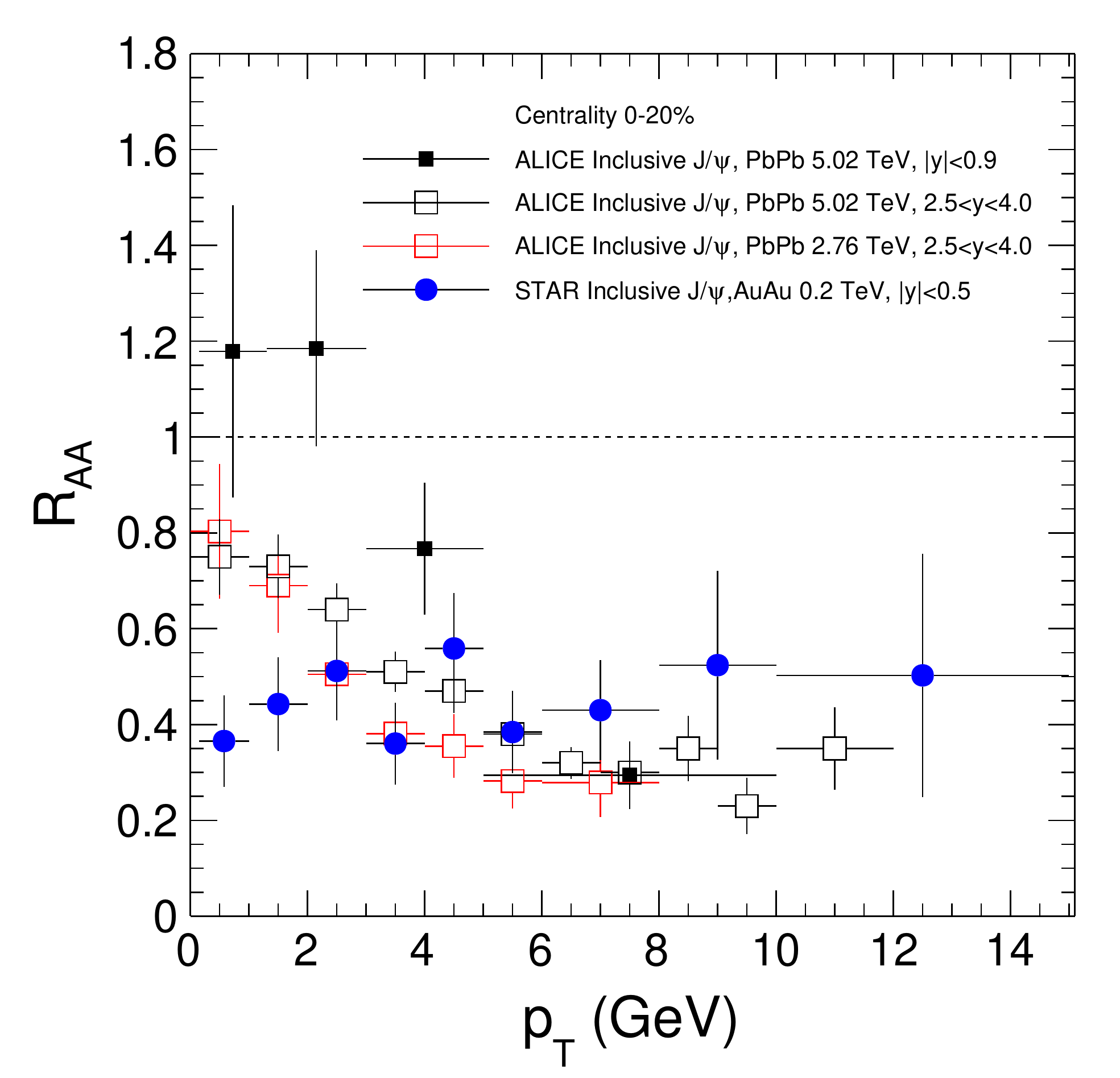}
    \includegraphics[width=0.45\textwidth]{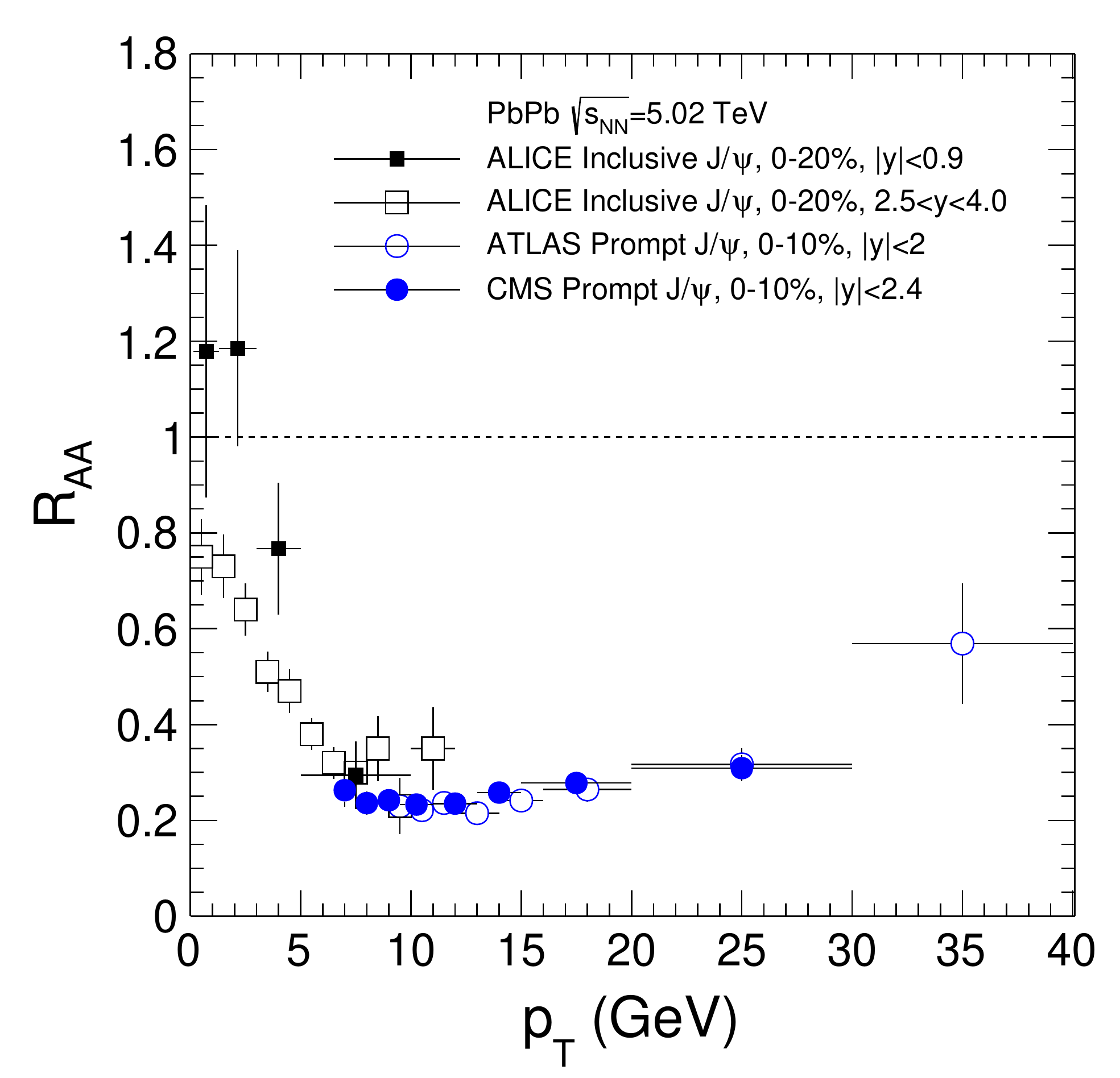}
    \caption{Left: The nuclear modification factor at low transverse momentum measured by ALICE at forward rapidity~\cite{ALICE:2015jrl,ALICE:2019lga} at $\sqrt{s_{NN}} = 2.76$~TeV and  $5.02$~TeV and at midrapidity~\cite{ALICE:2019nrq} at  $\sqrt{s_{NN}} = 5.02$~TeV compared to the result of STAR~\cite{STAR:2013eve} at midrapidity and $\sqrt{s_{NN}}= 0.2$~TeV. The nuclear modification factor measurements by ALICE~\cite{ALICE:2019lga,ALICE:2019nrq}, ATLAS~\cite{ATLAS:2018hqe} and CMS~\cite{CMS:2017uuv} extending to high transverse momentum at $\sqrt{s_{NN}} = 5.02$~TeV.  }
    \label{fig:sec4_charmonium}
\end{figure}

The nuclear modification factor of inclusive J/$\psi$ is larger at the LHC than at RHIC~\ref{fig:sec4_charmonium} for the bulk of the production at low transverse momentum as expected in the regeneration scenario in central collisions. This can be seen in Fig.~\ref{fig:sec4_charmonium} on the left-hand side. The nuclear modification factor measured by PHENIX and confirmed by the measurement by STAR depicted in the figure is around 0.4 and rather flat as a function of transverse momentum. At the LHC, the nuclear modification factor measured at forward rapidity rises towards the lowest transverse and reaches 0.8. On the right-hand side of Fig.~\ref{fig:sec4_charmonium}, it is shown a compilation of the most recent J/$\psi$~nuclear modification factor up to the highest transverse momentum and for two different rapidity ranges at low transverse. At midrapidity, the nuclear modification factor is larger than the one measured at forward rapidity and exceeds even unity, although the uncertainties prevent strong conclusions. Indeed, a smaller nuclear modification factor is expected in a regeneration scenario due to the approximate dependence on the square of the charm cross-section and the decrease of charm production at forward rapidity. 

In addition, a large sizeable elliptic flow of J/$\psi$ has been observed at the LHC by ALICE and CMS~\cite{ALICE:2013xna,CMS:2016mah,ALICE:2017quq,ALICE:2018bdo,ALICE:2020pvw}. This large elliptic flow at low transverse momentum below 4~GeV/$c$~\cite{ALICE:2017quq,ALICE:2018bdo,ALICE:2020pvw}, reaching $ v_2 \sim 0.1$, is associated, in the available models~\cite{Du:2015wha,Zhou:2014kka,He:2021zej}, with the regenerated J/$\psi$~component. The observation is therefore consistent with a picture where charm quarks participate strongly in the collective motion of the expanding fluid and form this bound state. The measurement of finite $v_3$ with 5$\sigma$ significance in~\cite{ALICE:2020pvw} in $2<p_T<5$~GeV/$c$ is also in line with this hypothesis. 

The experimental data as a function of transverse momentum, rapidity and collision energy at low transverse momentum show that regeneration is a dominant J/$\psi$ production mechanism at LHC energies. This indicates a deconfined medium where charm and anti-charm quarks meet to form quarkonium. The hadronic cross-section estimates do not allow generating specifically the small spatial extent bound states J/$\psi$ in such abundance, see for instance in~\cite{Braun-Munzinger:2000uqj}. In~\ref{sec:quarkoniumtheo}, we discuss the requirements for a picture of the precise mechanism, deciding whether the generation of charmonium is dominantly from the phase boundary or during the deconfined stage.


\subsection{Energy loss, diffusion and bound states}
\label{sec:acceleration_energy_loss}

When a particle propagates through a QCD-medium, it can scatter elastically with the medium constituents. The effect will be a change in its momentum, resulting in the particle diffusion concerning its incoming direction. Another possible outcome of this interaction process is medium-induced radiation that will induce energy loss. In the following, we will address each of those phenomena separately, showing the several main prescriptions to analytically describe each of them. Our focus will be on comparing them, emphasizing the differences and common limitations. We will show how this motivates additional modeling in Monte Carlo event generators and, subsequently, the current status of the experimental extraction of medium parameters. In addition, the formation of a bound state can inform us about the properties of the QCD-medium. As such, we will introduce the connections between the observed phenomena and theoretical developments in this probe.

\subsubsection{Radiative Energy Loss}
\label{sec:accleration_radiative_energy_loss}

\begin{figure}
    \centering
    \includegraphics[width=0.45\textwidth]{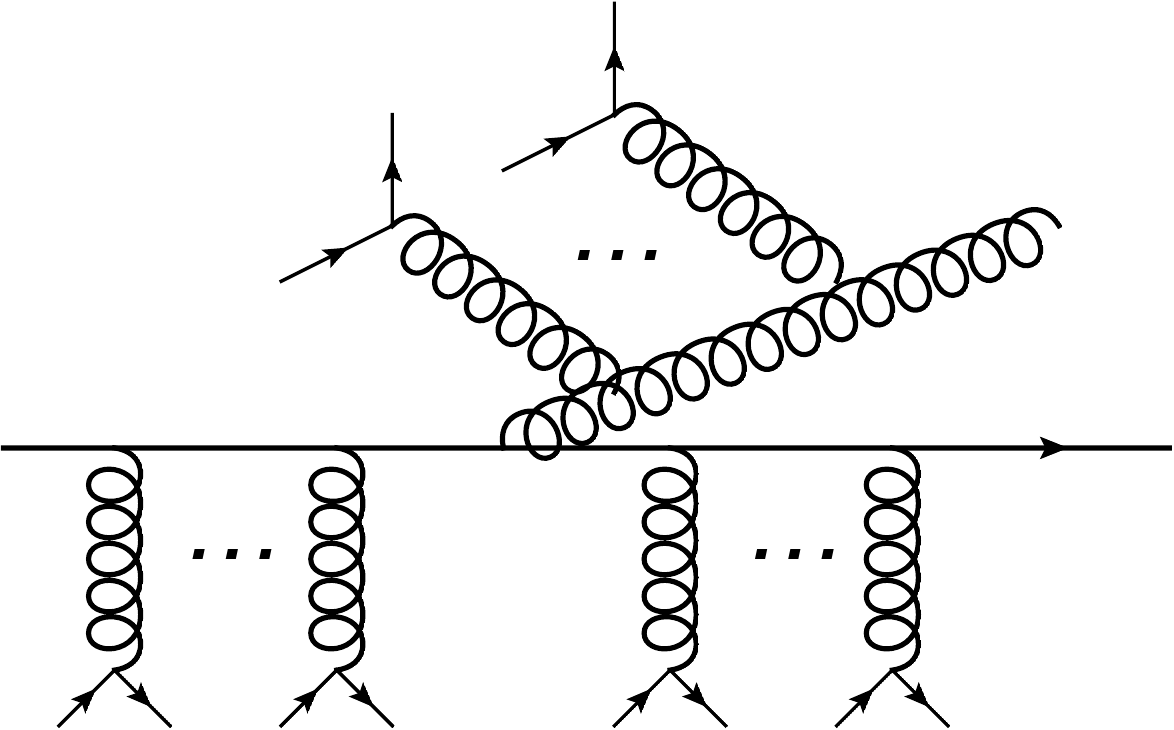}
    \caption{Illustration of in-medium single gluon emission within a perturbative QCD model. The incoming quark and the emitted gluon undergo multiple elastic scatterings with the medium constituents.}
    \label{fig:sec4_gluon}
\end{figure}

From a perturbative perspective, the building block of medium-induced energy loss is single gluon radiation from an incoming parton that undergoes elastic scatterings within the medium (see Fig.~\ref{fig:sec4_gluon}). In a first approximation, the latter can be modeled as a collection of independent scattering centers with a Debye screened potential and a thermal mass.
The number of scatterings a particle undergoes by each emission is arbitrary, as well as the resulting kinematics. There are different formalisms to analytically describe the resulting radiation. The two that were first applied to phenomenological studies were: multiple scatterings based on a path-integral formalism (BDMPS-Z~\cite{Baier:1996sk,Baier:1996kr,Zakharov:1996fv}) or an opacity expansion in terms of the number of scattering centers (GLV~\cite{Gyulassy:2000fs,Gyulassy:2000er,Wiedemann:2000za}). Both assume that the energy of the radiated gluon, $\omega$, is small with respect to the original parton energy $E$ ($\omega \ll E$). Until recently, these approaches were limited to consider two limiting cases: multiple soft scatterings approximation (done within BDMPS-Z formalism, the ASW quenching weights~\cite{Salgado:2003gb}) or single hard scattering limit (leading term of the GLV formalism). 

The medium-induced gluon radiation spectrum is given by:
\begin{equation}
    \omega \frac{dI}{d\omega d^2 \mathbf{k}} = \frac{2 \alpha_s C_R}{(2\pi)^2 \omega^2} \text{Re} \int_{0}^{\infty} dt^\prime \int_{0}^{t^\prime} dt \int \frac{d^2 \mathbf{p}}{(2\pi)^2} \frac{d^2 \mathbf{q}}{(2\pi)^2} \, \mathbf{p} \cdot \mathbf{q} \, \Tilde{\mathcal{K}}(t^\prime, \mathbf{q}; t, \mathbf{p}) \mathcal{P} (\infty, \mathbf{k}; t^\prime, \mathbf{q}) \, ,
\end{equation}
for a gluon of energy $\omega$ and two-dimensional transverse momentum $\mathbf{k}$. The incoming particle propagates along the longitudinal coordinate along the medium and emits a gluon at time $t$ ($t^\prime$) in the (conjugate) amplitude. The emission kernel and momentum broadening, both in momentum space, are denoted by $\Tilde{\mathcal{K}}(t^\prime, \mathbf{q}; t, \mathbf{p})$ and $\mathcal{P} (\infty, \mathbf{k}; t^\prime, \mathbf{q})$ respectively. The emission kernel, in coordinate space, can be written explicitly as:
\begin{equation}
    \mathcal{K} (t^\prime, \mathbf{z}; t, \mathbf{y}) = \int_{r(t) = y}^{r(t\prime) = z} \mathcal{D} \mathbf{r} \exp \left[ \int_{t}^{t^\prime} d\xi \left( \frac{i\omega}{2} \dot{\mathbf{r}}^2 - \frac{1}{2} n(\xi) \sigma (\mathbf{r}) \right) \right] \, 
    \label{eq:sec4_kernel}
\end{equation}
while the momentum broadening factor takes the form:
\begin{equation}
    \mathcal{P} (\infty, \mathbf{k}; t^\prime, \mathbf{q}) = \int d^2\mathbf{z} \, e^{-i (\mathbf{k} - \mathbf{q}) \cdot \mathbf{z}} \exp \left[ -\frac{1}{2} \int_{t^\prime}^{\infty} d\xi n(\xi) \sigma (\mathbf{z}) \right] \, .
\end{equation}
The medium characteristics are encoded in the linear medium density, $n(\xi)$, and in the strength of a single scattering, i.e, the dipole cross-section:
\begin{equation}
    \sigma (\mathbf{r}) = \int \frac{d^2 \mathbf{q}}{(2\pi)^2} V(\mathbf{q}) ( 1 - e^{i \mathbf{q} \cdot \mathbf{r}} ) \, .
\end{equation}
All the details regarding the parton-medium interaction are given through the elastic collision rate $V(\mathbf{q})$. This can be characterized by a screened Coulomb potential with momentum transfer $q$: 
\begin{equation}
    V(\mathbf{q}) = \frac{8 \pi \mu^2}{(\mathbf{q}^2 + \mu^2)^2} \, ,
    \label{eq:sec4_yukawa}
\end{equation}
where, in a thermal medium, the screening mass $\mu$ is related to the Debye mass, $\mu = \mu_D$. Other functional forms might also be used, but in general, they will require a form that goes as $V(q) \sim q^{-4}$ to correctly describe the UV physics (point-like interactions at short distances). The GLV approach consists of expanding the exponent of the path integral (eq.~\eqref{eq:sec4_kernel}) in a series of $n(\xi) \sigma(r)$ (\textit{opacity}) for a particular form of the potential. Although this prescription is exact order by order, generalizations to account for many interactions are analytically challenging. In the BDMPS-Z approach, suitable when the number of scattering centers is very large, the procedure is to approximate the dipole cross-section to the lowest order (leading logarithm) and take the small distance component to be:
\begin{equation}
    n(\xi) \sigma (\mathbf{r}) \approx \frac{1}{2} \hat{q} (\xi) \mathbf{r}^2 + \mathcal{O}(\mathbf{r}^2 \ln \mathbf{r}^2 ) \, .
\end{equation}
In this case, the medium is fully characterized by the jet transport coefficient, $\hat{q}$, that translates the mean transverse momentum acquired by the incoming particle per mean-free path:
\begin{equation}
    \hat{q} = \left\langle k_{T}^2 \right\rangle/\lambda \, .
    \label{eq:sec4_qhat}
\end{equation}
The transport coefficient, $\hat{q}$ can also be retrieved from the elastic interaction cross-section, as long as a prescription to relate the cross-section distribution to the medium parameters~\cite{Arnold:2009mr,Mehtar-Tani:2019tvy} is provided. In general, $\hat{q}$ is defined as the first moment of the Fourier transform of the elastic cross-section to momentum space:
\begin{equation}
    \hat{q} = \rho \int d^2 \mathbf{q} \, \mathbf{q}^2 \, \frac{d\sigma}{d^2 \mathbf{q}} \, ,
    \label{eq:sec4_qhat2}
\end{equation}
where $\rho$ is the density of scattering centers within the medium. However, the momentum integration is logarithmic divergent and needs to be regulated by a cut-off, thus introducing some arbitrariness. 

A different approach to energy loss formalism, initially proposed in~\cite{Guo:2000nz,Wang:2001ifa}, consists of making a collinear expansion of the scattering amplitude for the incoming particle to factorize out the effect of the nuclear PDFs. This higher-twist (HT) formalism was initially developed for single scattering per emission, but there are generalizations to account for multiple scatterings~\cite{Majumder:2009ge}. Nonetheless, only the leading moment of the exchanged transverse momentum distribution is considered (i.e., the limit of hard collinear gluon radiation).  

Alternatively, QGP characterization can also be regarded as a weakly-coupled medium in thermal equilibrium, as initially developed by Arnold-Moore-Yaffe (AMY)~\cite{Arnold:2001ba,Arnold:2002ja}. In this case, the medium is characterized by its temperature, $T$. At leading order in the coupling, $\alpha_s = g_s^2/(4\pi)$, the medium temperature can be related to the interaction potential~\cite{Aurenche:2002pd}, by denoting the Debye mass as
\begin{equation}
    \mu^2 (T) = 8 \pi^2 \alpha_s T^3 \left( 1 + \frac{N_f}{6} \right) \, 
\end{equation}
with $N_f$ the number of active quark flavors at temperature $T$. The corresponding hard-thermal-loop~(HTL) potential can be used as a refinement of the Yukawa potential above (eq.~\eqref{eq:sec4_yukawa}) as it should provide a more accurate description of the thermal interactions. Because the medium is no longer modeled by static scattering centers, both collisional and radiative energy loss can be formally considered within the same approach. 

These four main formalisms to address in-medium radiative energy loss have as common limitations a strong ordering in energy and momentum scales between the medium constituents, radiated gluons and the incoming propagating particle. In all of them, the incoming parton energy, $E$, and radiated gluon energy, $\omega$, are larger than the exchanged momentum, $E,\omega \ll q$, so that propagating particles travel along eikonal trajectories. Several improvements towards the inclusion of finite energy corrections, and generalizations beyond-eikonal effects, among others, have been put forward in the last years. Some examples targeting each of the described formalisms above include:
\begin{itemize}
    \item BDMPS-Z/ASW \& GLV: Recent numerical approaches could generalize these calculations by providing a full resummation in the number of scattering centers~\cite{Andres:2020vxs,Feal:2019xfl}. Other analytical improvements consider an expansion around the harmonic oscillator approximation~\cite{Barata:2020sav,Barata:2020rdn}. However, the radiated gluon's energy is still limited to be relatively soft with respect to the parent parton. 
    \item AMY: New analytical developments~\cite{Park:2016jap} now consider finite gluon radiation effects formerly limited to instantaneous gluon radiation. In the previously considered rate equations, partons propagate independently and radiate only after their phase-space separation satisfies the uncertainty principle.
    \item Higher-Twist: Several analytical developments focused on extending the modified DGLAP evolution equations to include induced gluon radiation within higher-twist contribution to energy loss~\cite{Deng:2009ncl}. However, the most recent developments address Monte Carlo implementation that couples both radiative energy loss to a dynamic evolution of the bulk medium in real time~\cite{Chen:2020tbl}.
\end{itemize}
However, all of them are still derived in restricted kinematical limits and address single gluon emission.

Generalizations from single to multiple gluon emission must be considered if one wishes to perform phenomenological studies targetting jets at colliders. A complete calculation of a multi-parton state requires the analytical description of in-medium interferences between different emitters, as done for a QCD vacuum shower (see~\ref{sec:acceleration_vacuum}). Medium-induced effects on the antenna setup have been addressed within the BDMPS-Z/ASW prescription: a collinear antenna that radiates a soft gluon~\cite{Mehtar-Tani:2010ebp,Casalderrey-Solana:2011ule,Mehtar-Tani:2011hma,Casalderrey-Solana:2012evi}. Medium interactions will change the interferences between the two antenna particles. In particular, within a multiple soft scattering approximation, in the limit of very soft gluon emissions, $\omega \rightarrow 0$, the integrated azimuthal spectrum of radiated gluon from a quark - anti-quark antenna with opening angle $\theta_{12}$ yields:
\begin{equation}
    \left. dN \right|_{soft} = \frac{\alpha_s C_R}{\pi} \frac{d\omega}{\omega} \frac{ d\theta}{\theta} \left[ \Theta(\theta_{12} - \theta) - \Delta_{med} \Theta(\theta - \theta_{12}) \right] \, ,
    \label{eq:sec4_antenna}
\end{equation}
with $C_R$ being the Casimir color factor of the parent parton and $\theta$ the emission angle of the radiated gluon with respect to the parent parton. The decoherence parameter, $\Delta_{med}$, describes the probability of the pair losing its initial color correlation after propagating a time $t$ through the medium~\cite{Mehtar-Tani:2011hma,Casalderrey-Solana:2012evi}:
\begin{equation}
    \Delta_{med} (t) = 1 - \exp\left\{ -\frac{1}{12} r^2 (t) Q_s^2(t) \right\}\, ,
\end{equation}
where $r(t) = \theta_{12}t$ reflects the transverse resolution of the quark - anti-quark antenna and $Q_s^2(t) = \hat{q} t$ characterizes the transverse resolution of the medium. The interpretation of this result is simplified if the two extreme limits are considered. When $\Delta_{med} \xrightarrow{r \ll Q_s^{-1}} 0$, i.e., the medium cannot probe both emitters independently since the antenna is very collimated, the vacuum result is recovered, and coherence between multiple emitters is restored. In this case, the pair remains correlated after $t$ and will continue to emit as a single color charge. For $\Delta_{med} \xrightarrow{Q_s^{-1} \ll r} 1$, each of the antenna partons can be resolved independently by the medium. Interferences between the different emitters are suppressed, and the two particles can emit independently, as given by the second term of eq.~\eqref{eq:sec4_antenna}. This anti-angular ordering additional term is weighted by $\Delta_{med}$ and, in the limit $\Delta_{med} \rightarrow 1$, the resulting spectrum is a superposition of two independent emissions. Generalizations from this result to multiple parton emissions are being explored in~\cite{Caucal:2018dla,Caucal:2020uic,Barata:2021byj}. In \cite{Caucal:2018dla}, the authors show that in a double-logarithmic approximation, vacuum-like and medium-like emissions are factorized between an initial parton cascade (vacuum-like) and medium-induced emissions, triggered by the subsequent collisions with the plasma constituents. The latter is described by BDMPS-Z formalism, which considers coherence effects during multiple in-medium scatterings while a parton is emitted. As for the multiple in-medium branchings, a rate equation that describes the instantaneous formation of in-medium gluons during a finite size medium is usually employed (similar to what is employed in the AMY formalism). This, and other similar approaches, assume that to go from single to multi-gluon emission, the number of emitted gluons along a trajectory path follows a Poisson distribution~\cite{Salgado:2003gb}. Its characteristics (the mean) are provided by the integral of each formalism's single gluon emission probability. The resulting parton energy loss can afterward be calculated, but without taking the degradation of the primary parton energy or the medium's dynamic evolution into account. The first limitation can yield a total energy loss larger than the initial energy of the incoming particle. However, in practice, it can simply represent the probability that the medium fully absorbs the particle. For the dynamic evolution, it was noted that, within a multiple soft scattering approximation~\cite{Salgado:2002cd,Adhya:2019qse}, a scaling law could, to a large extent, relate the resulting medium effects from an arbitrary dynamical expansion to an equivalent static scenario. Nonetheless, the resulting values of the quenching parameters are significantly different between the two cases~\cite{Adhya:2019qse}, indicating the importance of considering medium expansion effects on the extracted jet transport coefficients (see below) and its time-dependence instead of the average quantities (see section ~\ref{sec:timescale}). Alternatively, transport approaches can also be used to study elastic energy loss and in-medium parton propagation. For example, the AMY model includes transport equations for the quark and gluon energy and momentum distributions as a function of the momentum loss rate of a given parton. This prescription naturally considers the scenario where the incoming particle loses all its initial energy. It also has the advantage of keeping track of all particles (as opposed to the Poisson convolution), which allows accounting more naturally for elastic processes and medium-recoils. 

Nonetheless, all these prescriptions neglect possible coherence effects between subsequent emitters. Most recently, in~\cite{Barata:2021byj}, it was proposed a modified evolution equation that also includes interferences between different emitters. The setup follows a two-gluon configuration with no overlapping formation times and constitutes a first correction to the previous picture where all emitters radiate independently. The formation time associated with medium-induced radiation is usually small, and therefore, effects regarding finite formation time are discarded. However, they are known to suppress radiation, and in the near future, it will be necessary to incorporate them for future precision studies.

\begin{figure}
    \centering
    \includegraphics[width=0.45\textwidth]{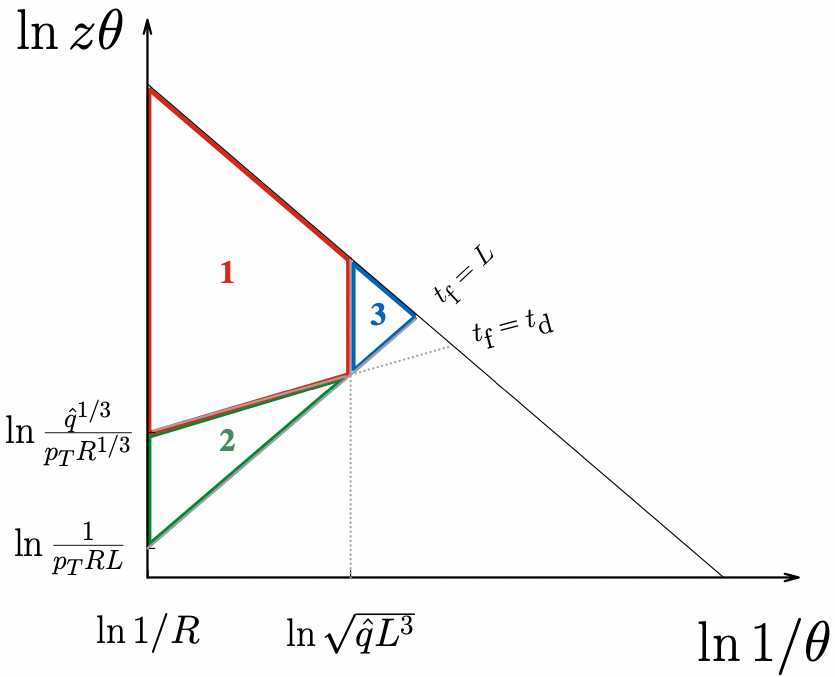}
    \caption{Illustration of a jet Lund plane with the medium scales depicted. The different regions are discussed in the text. Figure taken from~\cite{Andrews:2018jcm}}
    \label{fig:sec4_lund}
\end{figure}

The impact of considering finite formation time effects has been explored in~\cite{Dominguez:2019ges}, by using the jet Lund Plane to illustrate the modifications of color decoherence to the radiation pattern compared to the vacuum baseline. A Lund diagram written in $(\log(1/\theta), \log (z\theta))$, for the first for in-medium radiation (see Fig.~\ref{fig:sec4_lund}), allows identifying the different medium scales. This representation is equivalent to the $(\log(1/\theta), \log (k_T))$, since $k_T \simeq z E \theta$ where $E$ is the energy of the parent parton (equivalent to the jet transverse momentum, $p_T$). The medium scales are the transport coefficient $\hat{q}$ (diffusion in transverse space), the medium length $L$ and the decoherence time $t_d$ (when two in-medium radiators start to emit independently). Medium-induced radiation will be located at $t \leq L$. It is thus useful to include the characteristics scales of constant $t$. Let us take a system formed by an off-shell quark of energy $E$ emitting a gluon with a fraction of its energy $z$ emitted at an angle $\theta$ with respect to the quark direction. The time it would take for the gluon to become an independent source (the gluon formation time) can be estimated as:
\begin{equation}
    t_{form} \simeq \frac{E}{M^2} \simeq  \frac{2}{z (1-z) E \theta^2},
\end{equation}
where $M$ is the off-shell mass of the incoming quark. 

The characteristics equation of constant $t_{form}$ in the Lund plane depicted in Fig.~\ref{fig:sec4_lund} then follows $y = x + \ln \frac{2}{p_{T} t_{form}}$, where $y \equiv \ln z \theta$ and $x \equiv \ln(1/\theta)$. Medium-induced radiation can only be found in the region above the line that marks $t_{form} = L$. The radiation that will fall inside the jet will be constrained to the region limited by this line, and the vertical line of $\theta = R$, being $R$ the jet radius. We can now plug in the medium characteristic scales to better understand the physical picture one could expect from medium-induced radiation. Taking the transport coefficient $\hat{q}$, the characteristic equation for in-medium gluons, that acquire momentum broadening from in-medium interactions during its formation time ($\left\langle k_T^2 \right\rangle = \hat{q} t_{form}$), then follows $y = 1/3 \, x + \ln (2 \hat{q})^{1/3}/p_T$. The intersection between the two lines allows identifying the timescale $t_{dec} \simeq (\hat{q} \theta^2)^{-1/3}$ and, correspondingly, the angle $\theta_c \sim 1/\sqrt{\hat{q}L^3}$ (neglecting factors of 2). These new scales are now related to QGP characteristics and set the time (and angle) from which medium-induced gluons behave as independent in-medium sources. Taking some parametric estimates, such as $\hat{q} \sim 2~\rm{GeV^2/fm}$, $R = 1$, $p_T = 100~\rm{GeV}$ and $L \sim 5\rm{fm}$, one arrives to the Lund plane from the right panel of Fig.~\ref{fig:sec4_lundplane}, where $\theta_c \lesssim 0.1$. As such, we can now identify three regions:
\begin{enumerate}
    \item $t < t_d < L$ and $\theta_c < \theta < R$;
    \item $t_d < t < L$.
    \item $t < t_d < L$ and $\theta < \theta_c < R$;
\end{enumerate}
Regions (1) and (3) correspond to emissions with a short formation time that is not resolved by the medium ($t_f < t_d$). They will behave as vacuum-like emissions, and their emission probability will be uniform in this plane. Region (2) corresponds to emissions that will acquire momentum broadening during its formation time. Thus, its spectrum will not follow vacuum expectations, and this region will no longer be uniform. This is the region where medium-induced effects will be more pronounced. Comparison with experimental results (Fig.~\ref{fig:sec4_lundplane}) is, so far, not possible as most likely the visible signatures in the collinear and soft region are due to the jet selection bias towards narrower structures. Future results on jet Lund planes with other self-calibrated channels, e.g.: $Z+$jet, will establish if this qualitative picture is realized in experimental data.

\subsubsection{Elastic Energy Loss and Diffusion}
\label{sec:acceleration_elastic_energy_loss}

In addition to radiative energy loss, particles undergo elastic scatterings with the medium constituents. At leading order, in a perturbative field theory at finite temperature, the longitudinal momentum loss and diffusion resulting from momentum transferred between the incoming particle and medium constituents can be characterized through a $2\rightarrow 2$ elastic process. The resulting elastic energy loss rate (drag) will be quantified by:
\begin{equation}
    \hat{e} = \frac{dE}{dt} \, ,
\end{equation}
and the elastic diffusion by:
\begin{equation}
    D_s = \frac{d(\Delta E)^2}{dt} \, ,
\end{equation}
similarly to the transverse momentum diffusion of the propagating hard parton, $\hat{q} = dk_T^2/dt$). As discussed in section ~\ref{sec:acceleration_model_independent}, the resulting phenomena will be particularly significant for heavy-flavor at intermediate to low energies. For instance, the lower momentum part of the nuclear modification factor and the $v_2$ coefficient (see section ~\ref{sec:acceleration_HFQQ}) can be described by transport approaches with a diffusion treatment. For this reason, Gyulassy-Levai-Vitev (GLV) and Higher-Twist (HT) formalisms have been extended to incorporate dynamical scattering centers (DGLV~\cite{Djordjevic:2008iz,Djordjevic:2006tw}), or by adding the corresponding longitudinal drag and diffusion as an additional component (HT~\cite{Qin:2009gw,Majumder:2007hx,Qin:2012fua}).

Other approaches, such as jet transport models, can naturally include diffusion and keep track of all propagating particles. Examples are
MARTINI~\cite{Schenke:2009gb,Park:2018acg}, that makes use of rate equation and the Linear Boltzmann Transport (LBT) model~\cite{Li:2010ts,Wang:2013cia,He:2015pra,Cao:2016gvr,Cao:2017hhk,Ke:2018jem}, that describes jet parton evolution by linear Boltzmann equations. Some of these approaches contain radiative energy loss~\cite{Schenke:2009gb,Park:2018acg,Wang:2013cia,He:2015pra,Cao:2016gvr,Cao:2017hhk,Ke:2018jem}, via AMY or HT approach, further improved via the inclusion of finite formation time effects and running coupling. Others are only focused on elastic processes~\cite{Li:2010ts} since they play a more significant role at low transverse momentum. In fact, dedicated heavy-quark transport model implementations with, or without radiative energy loss, include the TAMU~\cite{Ravagli:2007xx,Riek:2010fk,He:2011qa,He:2019vgs}, the POWLANG~\cite{Beraudo:2009pe,Beraudo:2014boa,Beraudo:2017gxw}, the Nantes~\cite{Gossiaux:2009mk,Aichelin:2013mra,Nahrgang:2013saa,Nahrgang:2013xaa},  the Duke~\cite{Cao:2011et,Cao:2013ita}, URQMD~\cite{Lang:2012nqy}, the Catania~\cite{Das:2013kea,Das:2015ana,Scardina:2017ipo,Plumari:2019hzp}, the PHSD~\cite{Song:2015sfa,Song:2015ykw}, the LIDO~\cite{Ke:2018jem}, the LGR~\cite{Li:2019lex} and the DAB-MOD model~\cite{Katz:2019fkc}. In most of them, the heavy-quark propagation is governed by a Langevin equation. While a linearized-Boltzmann equation is suitable for a quasi-particle description where heavy quarks can interact via perturbative scatterings, the Langevin (diffusion) equation assumes frequent and soft interactions without the need for a quasi-particle description. While more suitable for parameterizing non-perturbative effects, the interaction terms need external input for the drag and diffusion coefficients usually derived or constrained via pQCD calculations or lattice QCD simulations. Within a pQCD picture, the elastic term can be computed via the $t-$channel amplitude in the high energy limit (incoming particle energy larger than the medium temperature) where the collinear divergence is regularized by the Debye screening mass $\mu_D$.

These approaches also include QGP medium evolution and need to specify a bulk medium. For this same reason, the ones addressing light quarks and gluon-initiated jets are optimal laboratories to investigate the medium response's role to the passage of a high-energy parton (see \ref{sec:QGPDensity}). Conversely, even though these approaches are more suitable for treating parton-medium interactions due to the lower virtuality scale of an already developed QCD parton shower, these prescriptions are still dominated by the uncertainties in defining perturbative interactions at low momentum scales. Consequently, the magnitude of medium response provided by each model is radically different, counterbalanced by radiative energy losses to provide a good description of experimental results(see, for instance~\cite{Sirunyan:2021pcp}). Progress has been made recently to identify differences in the models in well-controlled benchmark scenarios and to define paths to improve the different model assumptions related to the probe-QGP interaction, hadronization and the initial state~\cite{Rapp:2018qla,Cao:2018ews}. Recently, the experimental collaborations STAR and ALICE tried to extract transport coefficient based on data-model agreement~\cite{STAR:2017kkh,ALICE:2021rxa}. We will compare model-based extractions with theory and with other QGP properties in the following chapter~\ref{sec:QGPDensity}. As we will see, the qualitative findings on the heavy sector agree with the strongly coupled QGP picture derived from the soft particle sector. They indicate a relaxation time of charm quarks of the same order of magnitude as the QGP lifetime. Even more, the full kinetic and chemical thermalization of charm quarks in QGP is a potential scenario for central heavy-ion collisions at hadron colliders at low transverse momentum. This idea was recently detailed in Ref.~\cite{Andronic:2021erx}. In case of full thermalization, the momentum spectra of charmed hadrons would follow the velocity profile of modern hydrodynamic modeling, and the particle abundance of different bound states would follow the predictions of thermal models as the one in~\cite{Andronic:2021erx}. At the current stage, the experimental precision at the lowest transverse momenta of the ground state mesons and the missing knowledge on the initial state, abundance of baryons and their feed-down from excited states do not allow excluding this extreme possibility. In this case, charm quarks at the lowest transverse momentum would not be able to provide direct information on the heavy-quark QGP interaction as expressed in transport coefficients. The direct sensitivity would remain from measurements at intermediate transverse momentum. However, the charm sector at low transverse momentum would be an additional test of the freeze-out concept and deconfinement since the charm quarks would recombine statistically as we detail in the following section on quarkonium.

Elastic energy loss can also be described by taking a strongly coupled gauge theory. Even though the concept of a \textit{jet} in this theory is not directly compared to QCD, all features between the two objects are identical~\cite{Chesler:2014jva}. This holographic approach to jet quenching considers the drag experienced by each parton in the jet during its propagation. The energy loss experienced a parton, with incoming energy $E_{inc}$, at a given distance $x$, will be~\cite{Chesler:2014jva,Chesler:2008uy,Gubser:2008as}:
\begin{equation}
\frac{1}{E_{in}} \frac{dE}{dx} = - \frac{4}{\pi} \frac{x^2}{x^2_{stop}} \frac{1}{\sqrt{x_{stop}^2 - x^2}} \ \ \ , \ \ \ x_{stop} = \frac{1}{2 \kappa_{sc}} \frac{E_{in}^{1/3}}{T^{4/3}} \, ,    
\end{equation}
where $\kappa_{sc}$ is a model parameter encoding the difference between Super Yang-Mills and QCD (degrees of freedom difference) and $x_{stop}$, the stopping distance. Implementations of this model in Monte Carlo event generators~\cite{Casalderrey-Solana:2014bpa,Casalderrey-Solana:2016jvj,Hulcher:2017cpt} also allow tracing back medium-recoiling particles resulting from the passage of a jet. We will discuss the different Monte Carlo implementations and corresponding limitations more carefully in section~\ref{sec:acceleration_transport}.

\subsubsection{Quarkonium: bound state probes of QGP}
\label{sec:quarkoniumtheo}

QGP acts on the propagation of the heavy quark-antiquark pair taking into account several effects. First, there is the modification of the attractive potential of the quark-antiquark pair forming a color singlet state. Furthermore, transitions from color singlet to color octet states and transitions between different quarkonium states must be considered. This picture can be related to QCD in different limits of coupling strength and quarkonium size in semi-classical transport approaches via:
\begin{itemize}
\item Boltzmann~\cite{Yao:2018nmy,Yao:2020xzw} or Langevin equations allowing the treatment of multiple heavy-quark pairs~\cite{Blaizot:2017ypk};
\item Schrödinger equation approaches~\cite{Akamatsu:2011se,Miura:2019ssi,Islam:2020bnp}
\item and, in open quantum system calculations, using the Lindblad equation~\cite{Akamatsu:2014qsa,Brambilla:2019tpt,Brambilla:2021wkt}~\footnote{The open-quantum approaches separate the quarkonium degrees of freedom as a small interacting system from the environment, a QGP heat bath.}.
\end{itemize}
A detailed review of the theory's progress in the field can be found in~\cite{Rothkopf:2019ipj}.

Different models based on these developments are compared to experimental data on bottomonium. For instance, recent direct comparisons with data are done within transport model approaches~\cite{Du:2017qkv, Yao:2020xzw}, real-time Schrödinger equation approaches~\cite{Islam:2020bnp} and open-quantum system approaches~\cite{Brambilla:2021wkt}. Overall, they are able to describe the trends observed in the experimental data. 

Although the latter approaches are quite different, future progress from lattice QCD, effective field theory and real-time modeling may permit us to confirm and specify better the physical picture. In the long run, we may be able to over-constrain the experimental data to extract information on heavy-quark-medium information from experimental data. 

For instance, a sensitivity study in a semi-classical approach~\cite{Du:2019tjf} showed that the experimental nuclear suppression data on $\Upsilon$(1S) is directly sensitive to the inter-quark force and can be constrained within the model. Furthermore, a connection to open-heavy flavor can be established assuming a dominance of the coulombic part of the potential, a reasonable assumption for $\Upsilon$(1S), and for $\pi T >> E_{binding}$. In this case, the heavy-quark momentum diffusion coefficient is proportional to the in-medium modified thermal width of the $\Upsilon$(1S), see e.g. in~\cite{Brambilla:2016wgg,Brambilla:2019tpt}. This coefficient is one of the two non-perturbative parameters governing the in-medium propagation with a Lindblad equation in this specific limit. It can be related to spatial diffusion coefficient and the heavy-quark relaxation, see e.g.~\cite{Caron-Huot:2009ncn}. Hence, there is a direct theoretical connection between the non-perturbative parameters relevant for heavy-quarkonium physics and single heavy-quark hadron physics in QGP.

Therefore, $\Upsilon$~production has the potential to give us insight into QGP properties related to heavy-quark movement and the mechanism of deconfinement as a static quark-pair proxy. However, the published measurements on $\Upsilon$~production data can be described by a simple effective model that implements the scattering of the quarkonium with so-called comover particles~\cite{Ferreiro:2018wbd}~(partonic and hadronic). This model shows the need for experimental precision and dialogue on how to falsify specific hypotheses and define a set of minimal ingredients to capture the complete set of observations. In particular, there is a need for better experimental knowledge on parton densities and potentially other physical effects unrelated to QGP physics that may be relevant as cold energy loss and hadronic rescatterings. For instance, the rapidity dependence of the nuclear modification factor measured by CMS and ALICE, shown in~\cite{ALICE:2020wwx}, is in contrast with expectations from models~\cite{Yao:2018nmy,Krouppa:2016jcl}, in which the rapidity dependence is driven by the rapidity dependence of the initial conditions of the medium~\cite{Krouppa:2016jcl} or the effect of nuclear PDFs on bottomonium production~\cite{Yao:2018nmy}. Nevertheless, although not quantitatively, it follows the trend expected from coherent energy loss in the nucleus implementing a rapidity shift between the production in pp and nucleus-nucleus collisions~\cite{Arleo:2014oha}. Furthermore, the improvement and extension of existing feed-down measurements in pp~collisions down to lower transverse momentum would improve the control of the model setup. The goal of confirming the physical picture behind $\Upsilon$ production in nucleus-nucleus collisions and using it to extract information on the heavy-quark QGP interaction will rely on future experimental efforts. 

In the case of charmonium production, the experimental measurements indicate a significant contribution of regenerated J/$\psi$ in nucleus-nucleus collisions.
However, at the moment, the conceptually different model calculations, transport-type models~\cite{Zhao:2007hh,Ferreiro:2012rq,Zhou:2014kka}~\footnote{We list here also the comover interaction model since its rate equation also includes a regeneration component~\cite{Ferreiro:2012rq}.}, and the statistical hadronization model~\cite{Andronic:2019wva}, are consistent with the experimental data. Due to the coupling of the regenerated component and the overall charm quark production, the total charm production is the largest source of common uncertainty among the models. However, the charm quark production yields in the models deviate strongly from each other. For example, in ref.~\cite{ALICE:2019nrq}, the difference amounts to 50~\% for the midrapidity measurement. A precise measurement of the total charm cross-section in both rapidity ranges in nucleus-nucleus collisions with low transverse momentum quarkonium measurements is, therefore, one of the most critical goals in narrowing down the possible model scenarios. In this context, determining the nuclear parton densities via measurements in proton-nucleus and photon-nucleus collisions could also be used to calculate the nuclear modification of the total charm cross-section in the nucleus-nucleus collisions based on the charm production in pp~collisions as starting point. Comparing the direct charm production measurement in nucleus-nucleus collisions and the estimation via pp~collision measurements and nPDF determinations would also allow checking the applicability of collinear factorization to nucleus-nucleus collisions~\footnote{For instance, the thermal production of charm is expected to be small. However, the modeling uncertainties of the very first instances in preequilibrium are large~\cite{Schlichting:2019abc}. In principle, there could be a small but non-negligible contribution. A discussion of different calculations given the future circular collider at higher collision energies is given in~\cite{Mangano:2017tke}.}.

In addition, the ratio of prompt $\psi$(2S) over J$/\psi$ production is predicted to be strongly modified independently of the total charm cross-section. Thus, it can be used as a validation of the regeneration and, potentially, as an additional test for different scenarios in future measurements~\cite{ALICE:2012dtf,Citron:2018lsq}.   

The relation of quarkonium with respect to jets can allow the development of a better picture of the fragmentation and the contribution of isolated J/$\psi$~production and in jets to inclusive production. A first measurement has been published by LHCb in pp~collisions~\cite{LHCb:2017llq}. CMS has recently published the first measurement of J/$\psi$ in PbPb collisions~\cite{CMS:2021puf} that we discuss in Section~\ref{sec:timescale} given the time structure of the parton shower.

\subsubsection{Medium-modified parton showers and jet transport coefficients}
\label{sec:acceleration_transport}

Analytical description of in-medium QCD processes is challenging. The diversity of energy scales from hard propagating particles to the interactions with a thermal medium defies a consistent treatment within the same formalism. Given all the assumptions and limiting kinematical validity regions of each approach, a data-theory comparison is still impossible for the majority of experimental observables. Monte Carlo event generators can help make this bridge, but they require further modeling to fully incorporate jet quenching effects. Extensions beyond the validity regime of employed approximations, time-dependence of medium parameters, implementations of medium-modified QCD parton showers, spatial and time distribution of shower and medium particles, and generalizations from 1D to 3D expansion, among others, require further phenomenological modeling. Altogether, they contribute to the multiplication of the available tools to be applied to heavy-ion studies. 

These Monte Carlo event generators are not fully based on first-principles calculations. In addition to the phenomenological models that address non-perturbative QCD effects, analytical results on medium-induced modifications are usually restricted to kinematic approximations, lacking a full coverage of the full phase-space for additional radiation and medium interactions. There is still a need to include further modeling based on free parameters that yield additional freedom to fit observations. Nonetheless, these tools help to thoroughly compare different analytical approaches. An example are the jet transport coefficients as implemented in each model ($\hat{q}$, $\hat{e}$, $D_s$). As mentioned in section \ref{sec:accleration_radiative_energy_loss}, $\hat{q}$ is related to the elastic potential and, consequently, to the Debye mass. However, there is some arbitrariness on how to properly set this relation. Even so, given that this regime is the dominant one for a dense and extended medium (like the one produced in collisions at the LHC), much attention has been given to extracting this parameter from experimental data. For a weakly-coupled scenario and high temperatures, this parameter was shown to be related to the shear viscosity, $\eta$~\cite{Majumder:2007zh}. In particular, it was conjectured a similar relation between $\hat{q}/T^3$ and the shear viscosity to entropy ratio, $\eta\,/s$, in a weakly coupled plasma, while $T^3/\hat{q} \ll \eta/s$ for a strongly-coupled scenario. Thus, its quantitative measurement would allow establishing the interaction strength of the plasma and its transition from strong to weak coupling. The longitudinal diffusion $\hat{e}_2$, was also shown to be related to the shear viscosity of the medium. Together with longitudinal energy loss resulting from the diffusion process, $\hat{e}$, these two parameters also carry valuable information on the interaction properties and strength of the QGP coupling.

Given the direct relationship between these parameters and the medium characteristics, this exercise was pursued early on to: compare the expectations provided by the different formalisms in ideal medium conditions (a "QGP brick"~\cite{Armesto:2011ht}); identify differences in the models on heavy-flavor in well-controlled benchmark scenarios and to define paths to improve the different model assumptions related to the probe-QGP interaction, hadronization and the initial state~\cite{Rapp:2018qla,Cao:2018ews}; and extract quantitative values from both RHIC and LHC heavy-ion experimental results~\cite{Burke:2013yra,Feal:2019xfl,STAR:2017kkh,ALICE:2021rxa}. Even though the transport coefficients might be model-dependent, these exercises resulted in a quantitative assessment of the obtained average transport coefficients. We will present this effort more thoroughly in the following section (see section~\ref{sec:QGPDensity}).

%% file: QGPDensity3.tex
\section{Medium properties}
\label{sec:QGPDensity}

Results on soft probes, such as inclusive particle production and angular correlation analysis of low momentum hadrons, have provided insights into the produced medium. Theoretical calculations, which include a high energy density, low specific shear viscosity, and thermalized medium, can describe the data from soft probes. Medium properties have been extracted from Bayesian analyses using relativistic hydrodynamics-based models that depend on the underlying assumptions in various stages in the evolution. The hard sector of a collision can provide additional constraints. In the previous chapter, we described the medium-induced acceleration and deceleration of the hard probes. We will focus on the current medium properties extracted using jets, heavy quarks and quarkonia with the discussed theory setup.

Hard probes differ from the medium particles in their transverse momentum or mass. These high momentum objects can extend the study of the created medium to different phase spaces because they are produced early at higher temperatures and smaller length scales. While soft particles lose all the memory of their initial momentum, hard probes are sensitive to the earlier stages of QGP because they have a longer relaxation time. In the following, we will address the constraints from jets and heavy quarks on the density, stopping power (\ref{subsec:QGPdensitySopping}), and shear viscosity (\ref{subsec:QGPdensityShear}).

\subsection{Density and Stopping power}
\label{subsec:QGPdensitySopping}

The maximum energy density occurs at the instant when the two ions overlap and collide. This density amounts to more than 12 GeV/$\rm{fm}^3$ based on the charged hadron $dN/d\eta$, $\langle p_T\rangle$ and transverse energy distribution $dE_t/d\eta$ measurements~\cite{ALICE:2010mlf,CMS:2012krf,CMS:2011aqh,ALICE:2016igk}. The energy density then decreases as a function of time as the system expands and cools down. Note that the energy density is different from the color charge density. For instance, the energy density in a high-energy electron-positron annihilation is large, but the produced color charge density is low. Based on measurements of total energy using inclusive particle spectra, we know that the initial energy density of the system is considerable. On the other hand, the analysis of the jet quenching imprints indicates a strongly interacting system characterized by a very high color density. These jet quenching analyses typically rely on several theoretical models to extract the medium properties. We will start by showing how the theory described in chapter \ref{sec:acceleration_energy_loss} can be related to the medium properties, in section \ref{subsec:qhatToMediumProperties}, followed then by their application to jets (\ref{subsec:qhatJets}), heavy-flavor (\ref{subsec:qhatHF}) and quarkonia (\ref{subsec:qhatQuarkonia}).

\subsubsection{From jet transport coefficient to medium properties}
\label{subsec:qhatToMediumProperties}

A crucial part of the experimental programs of RHIC and the LHC is to determine the nature of the matter produced in ultra-relativistic heavy-ion collisions. There are several experimental evidences~\cite{STAR:2005gfr, PHOBOS:2004zne,PHENIX:2004vcz,BRAHMS:2004adc} that the quark-gluon plasma produced at both RHIC and LHC is a strongly coupled fluid. Nonetheless, within a quasi-particle framework, a class of weak coupling approaches is applied to describe the interaction of high-momentum particles with the short-wavelength partonic excitations of QGP. In this picture, the transport coefficients introduced above within inelastic and elastic energy loss scenarios can be related to the (non-perturbative) bulk properties of QGP, such as the shear-viscosity, $\eta$. Depending on the formalism one uses to address the determination of a jet transport coefficient, it is also possible to relate these to other bulk properties, such as energy density, $\epsilon$, or entropy, $s$. This section will focus on kinetic-transport theory as it can describe low- and high-$p_T$ particle production within the same physics dynamics. This theory provides an interpolation between free streaming (in small systems) and almost ideal fluid dynamics (in large systems). In fact, since the works in Ref.~\cite{Arnold:2002zm} with the introduction of an effective kinetic theory (EKT) to describe the transport QCD matter, EKT was applied to jet quenching phenomenology, as well as flow phenomena~\cite{Kurkela:2015qoa}. While we refer the reader to Ref.~\cite{Muller:2021wri} for details on how to derive a relation between the jet transport coefficient, $\hat{q}$ and QGP bulk properties, we present a heuristic derivation to illustrate how these two can be related. 

As discussed in section~\ref{sec:acceleration_energy_loss}, the jet transport coefficient is related to the differential interaction cross-section between the incoming parton and a single scattering center, $d\sigma/ dq_\perp^2$:
\begin{equation}
    \hat{q} = \rho \int d^2 \mathbf{q} \, \mathbf{q}^2 \, \frac{d\sigma}{d^2 \mathbf{q}} \, \, .
    \label{eq:sec5_qhat}
\end{equation}
It translates the average transverse momentum distribution exchanged between incoming parton and medium scattering center, $\left\langle k_\perp^2 \right\rangle$ per mean free path, $\lambda$. Within the framework of kinetic theory, the shear viscosity scales with the mean-free path of a constituent's particle of momentum $p$ in the medium, as well as its average thermal momentum, $\left\langle p \right\rangle$: 
\begin{equation}
    \eta \sim \frac{1}{3} \left\langle p \right\rangle \lambda \rho \, , 
\end{equation}
where $\rho$ is the density of scattering centers within the medium. Regarding the medium as a collection of quasi-particles, $\lambda = (\rho \sigma_{tr})^{-1}$, where $\sigma_{tr}$ is the total transport cross-section. When soft scattering dominates, this can be related to the differential cross-section as: 
\begin{equation}
    \sigma_{tr} = \int d\Omega \frac{d\sigma}{d\Omega} \approx \frac{4}{\hat{s}} \int d q_\perp^2 q_\perp^2 \frac{d\sigma}{dq_\perp^2} = \frac{4 \hat{q}}{\hat{s} \rho} \Rightarrow \lambda = \left(\frac{\hat{s}}{4 \hat{q}} \right)\, ,
\end{equation}
where $\sqrt{\hat{s}}$ is the center-of-mass energy and $\Omega$ the solid angle. Taking the medium as a gas of massless thermal particles, one can use the corresponding equation of state $\left\langle p \right\rangle \approx 3T$, and the relations $\left\langle \hat{s} \right\rangle \approx 18 T^2$ and $s \approx 3.6 \rho$, to finally derive: 
\begin{equation}
\frac{\eta}{s} \approx 1.25 \frac{T^3}{\hat{q}} \, .
\end{equation}
Despite the additional QGP thermodynamic ingredients, our purpose here is to illustrate that measuring transport coefficients related to the high momentum objects, such as $\hat{q}$, can elucidate the nature of the coupling strength within the medium. This link is one of the main reasons for the multiple attempts to obtain a numerical value for this phenomenological parameter. In particular, a larger value of $\hat{q}$ will imply that the medium is strongly coupled. In contrast, a smaller $\hat{q}$ could indicate the presence of a marginally weak coupling medium~\cite{Muller:2021wri}. 

QGP equation of state is one of the needed inputs to relate $\hat{q}$ with the medium properties (like medium energy-density $\epsilon$, entropy $s$, or temperature $T$). Above, we mentioned the ideal gas scenario, but lattice results can also be used as input (see, e.g., \cite{HotQCD:2014kol}). Nonetheless, such choices do not hold for the entire medium evolution, given the multitude of phases that the result of the collision will go through, from the pre-equilibration up to the hadronic re-scattering phase. Energy losses during the pre-equilibration phase are still heavily under debate. Most works indicate that an absence of quenching during this initialization time seems to describe the high-$p_T$ flow harmonics present in less central collisions better~\cite{Noronha-Hostler:2016eow,Andres:2019eus}. On the other hand, it was suggested in~\cite{Noronha-Hostler:2016eow} that event-by-event fluctuations could be the main factor for the high-$p_T$ flow harmonics. 

As for the hadronic phase, one can consider a hadron resonance gas (HRG) description, thus using a hybrid equation of state using lattice and HRG inputs. The introduction of an evolving medium and subsequent time dependence on the jet transport coefficients also contributes to the significant theoretical uncertainties obtained on the $\hat{q}$ parameterization and when trying to retrieve an average $\hat{q}$ out of experimental results.

The jet transport coefficient translates the amount of transverse momentum broadening acquired by the incoming particle. While defined as the first moment of the elastic scattering cross-section\footnote{There is an additional ambiguity with respect to the logarithmic divergence (see chapter ~\ref{sec:acceleration_energy_loss})}, the $\hat{q}$ can be further related to the energy loss. However, such a link will now depend on the model under consideration. Different approaches will naturally yield different results on $\hat{q}$ and its dependencies on QGP parameters. However, once this is inferred, the value of $\hat{q}$ can be compared across the different models. For example, one can look to \cite{JET:2013cls}, where five different approaches were used to retrieve the corresponding $\hat{q}$ value using high-$p_T$ single hadron measurements from both RHIC and LHC. Other similar attempts include refs.~\cite{Andres:2016iys,Feal:2019xfl,Xie:2020zdb,JETSCAPE:2021ehl}, where the use of di-hadron and $\gamma$-hadron correlations~\cite{Xie:2022ght} and jet inclusive spectra~\cite{Ke:2020clc} is also being explored. The models used in~\cite{JET:2013cls} were Higher-Twist (BW and M)~\cite{Chen:2011vt,Majumder:2011uk}, MARTINI~\cite{Schenke:2009gb}, and McGill-AMY~\cite{Qin:2007rn}. Each of them has a different $\hat{q}$ dependence on QGP parameters. In the Higher-Twist-BW, the definition of $\hat{q}$ is assumed to be proportional to the local partonic QGP density, $\rho_{QGP}$, modeled as an ideal gas at a temperature $T$. Further considerations of a fast-evolving medium yield the dynamical parameterization on the space, $r$ and time $\tau$ yield:
\begin{equation}
\hat{q} (\tau, r) = \left[\hat{q}_0 \frac{\rho_{QGP} (\tau, r)}{\rho_{QGP} (\tau_0, 0)} (1-f(\tau,r)) + \hat{q}_h (\tau, r) f (\tau,r) \right] \frac{p \cdot u}{p_0} \, ,
\end{equation}
where $\hat{q}_0$ denotes the transport coefficient for a parton located at the center of the QGP bulk ($r = 0$) and initial time ($\tau_0$), $p^\mu$ is the 4-velocity of the jet and $u^\mu$ the four flow velocity in the collision frame. The $f$ denotes the fraction of the hadronic phase. The $\hat{q}_h$ is approximated by the meson and baryon densities, $\rho_M$ and $\rho_B$m at a given temperature $T$:
\begin{equation}
\hat{q}_h = \frac{\hat{q}_N}{\rho_N} \left[ \frac{2}{3} \sum_M \rho_M (T) + \sum_B \rho_B(T) \right] \,. 
\end{equation}
The parameter $\hat{q}_N$ refers to the jet transport coefficient from cold nuclear matter effects, extracted from DIS experiments, and $\rho_N$ is the nucleon density in the center of a large nucleus. By contrast, in MARTINI, the interaction potential is derived from a hard-thermal loop, where, at leading order, $\hat{q} = \hat{q} (E, T, \mu_D)$, with $E$ the energy of the incoming particle, $T$ the medium temperature and $\mu_D$ the Debye mass from the elastic potential, and an additional free parameter $C$:
\begin{equation}
    \hat{q} \approx C_R \frac{42 \zeta(3)}{\pi}\alpha_s ^2 T^3 \ln \left( \frac{2 C E T}{4 \mu_D^2} \right) \, .
\end{equation}
The other models will yet use different QGP dependencies, and we refer the reader to the original manuscripts. These examples emphasize the uncertainties of obtaining a $\hat{q}$, which may involve additional free parameters.

While these exercises help build a qualitative picture of QGP's inner workings, an accurate determination of its properties is still bounded by the current theoretical uncertainties to describe simultaneously perturbative and non-perturbative interactions. With the continuous progress to achieve a unified description of a coupled jet-medium evolution, more systematic studies will allow extracting, with higher accuracy, fundamental properties of QGP using hard probes.  

\begin{figure}
    \centering
    \includegraphics[width=1\textwidth]{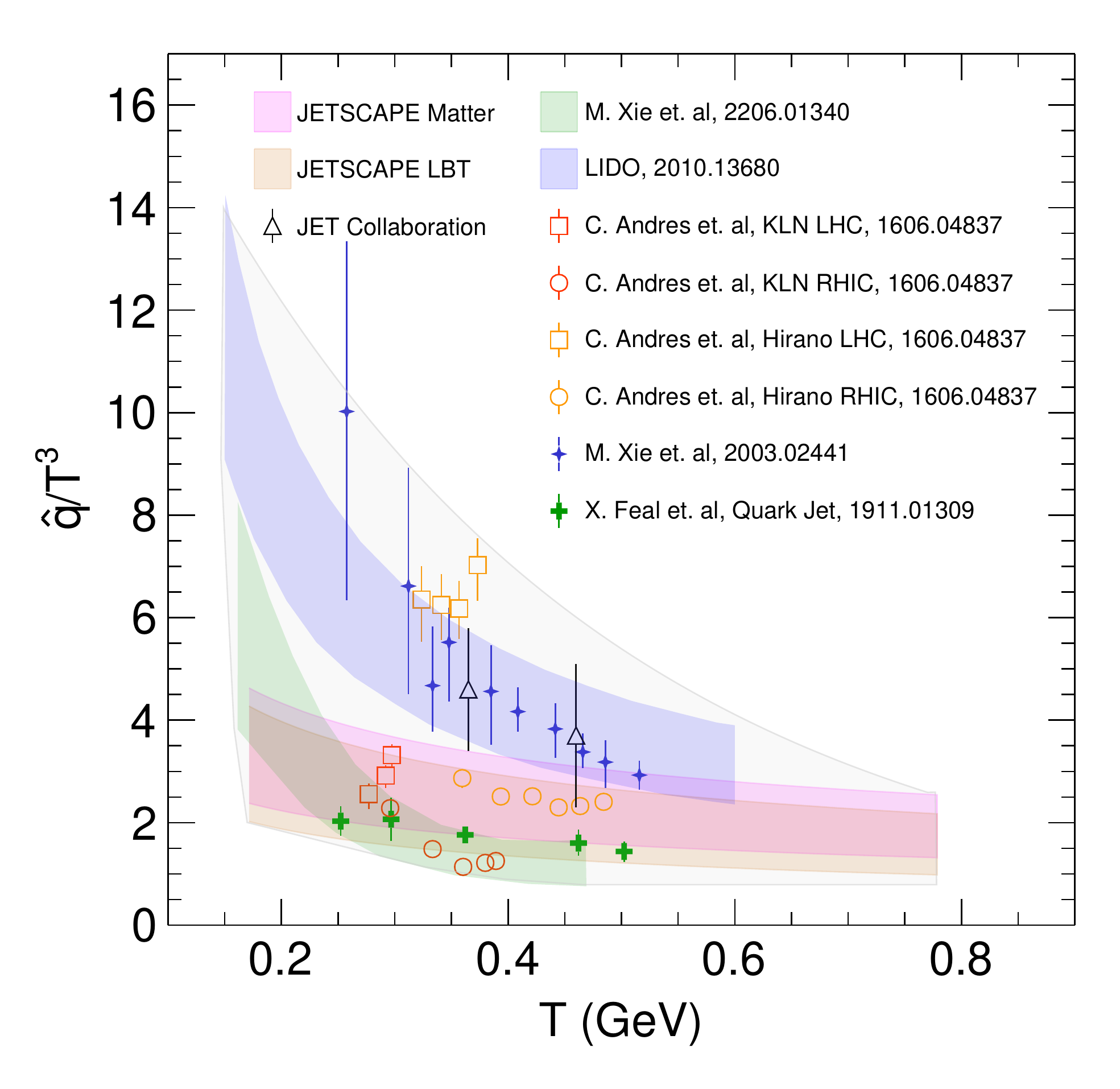}  
\caption{A snapshot of $\hat{q}$ extracted from charged hadron spectra~\cite{JETSCAPE:2021ehl,Andres:2016iys,JET:2013cls,
Feal:2019xfl,Xie:2020zdb,Xie:2022ght,Ke:2020clc}
di-hadron, $\gamma$-hadron correlations~\cite{Xie:2022ght} and jet inclusive spectra~\cite{Ke:2020clc} as a function of temperature. A gray area is also shown to cover the central values of the extracted $\hat{q}$ from different models for demonstration purposes.}
    \label{fig:CompilationQhat}
\end{figure}

\subsubsection{Input from light-hadron and jet measurements}
\label{subsec:qhatJets}

The nuclear modification factors of high transverse momentum charged particles are significantly lower than one, showing that the fast-moving hadrons are suppressed. Moreover, with the photon-tagged and Z-tagged jet measurements, a significant modification of the jet $p_{T}$ spectra is observed. The mean ratio of the jet and electroweak boson transverse momenta is shifted to a lower value, providing model-independent evidence that the quarks and gluons lose energy inside the medium. Since the quarks and gluons mainly interact with the medium through strong interaction, the color density of the produced state is incompatible with a hadron gas. Nuclear modification factors of jets are also studied in nucleus-nucleus collisions in different centrality classes. The suppression of the jet is strongest in head-on collisions and the weakest in peripheral events. It is not yet conclusive when the jet quenching effect has become negligible in AA collisions. On the other hand, many searches for jet quenching effects have been carried out in smaller collision systems such as XeXe and pA collisions. In XeXe collisions, the results are consistent with the expectation from theoretical models, replacing the Pb ion with a smaller Xe ion. The situation in pA and dA collisions is not yet clear. While a significant flow-like signal is observed in pA collisions, there was no convincing evidence that jet quenching is observed in those smaller collision systems. Hard and soft probes are consistent with a picture of a high color density system in PbPb and gold-gold (AuAu) collisions. However, as discussed in section~\ref{sec:smallsystem}, it is not yet conclusive whether this continues to hold in smaller and more dilute systems.

As discussed in the previous section, to quantify the stopping power of the medium, the jet quenching parameter $\hat{q}$ has been extracted using the nuclear modification factors of the high transverse momentum particles. Examples of such model-dependent extractions are summarized in Figure~\ref{fig:CompilationQhat}, which is currently extracted only from charged particle spectra or dihadron correlation functions. In those models, many different assumptions are used in the extraction of $\hat{q}$:
\begin{itemize}
    \item The initial state of the heavy-ions right before the collision and its factorization with respect to final state effects;
    \item The evolution of the collision bulk, namely the temperature and energy-density time-evolution profiles;
    \item The jet-medium interaction before the formation of a QGP;
    \item The jet quenching mechanism, including start and end time of the quenching effects during the bulk evolution (energy loss can occur during the partonic phase or in both partonic and hadronic phases);
    \item The hadronization of the medium and hard probes;
    \item The considered medium degrees of freedom  (gluon gas, $N_f = 3$ quark flavors,...) and the equation of state;
\end{itemize}
In addition, $\hat{q}$ is also linked to the propagating particle since quarks lose energy with $C_F = 4/3$, while gluons lose energy with $C_A = 3$. In the current literature, the transport coefficient is often reported for a propagating quark and can be denoted by $\hat{q}_F$. Nevertheless, X. Feal et al. (\cite{Feal:2019xfl}) report their results for a propagating gluon, while others present an average $\hat{q}/T^3$ that will contain a mixture of the two contributions. In Fig.\ref{fig:CompilationQhat}, we re-scaled the results from~\cite{Feal:2019xfl} by $C_F/C_A = 4/9 \simeq 0.45$ (green markers) to include only the equivalent of the quark for comparison with the remaining models. 

The results for the extracted $\hat{q}/T^3$ of quark-initiated jets illustrated in Fig.~\ref{fig:CompilationQhat} do not agree between the different approaches as the remaining ingredients used in each model differ substantially. The assumptions regarding the QGP initialization time usually range from $\tau_0 = 0.6$~fm/$c$ up to $\tau_0 = 1.0$~fm/$c$. All but~\cite{Andres:2016iys} (C. Andrés et al.) neglect energy loss effects before this initialization time. The effect of solely changing the $\tau_0$ assumption is illustrated in this reference~\cite{Andres:2016iys}. In Fig.~\ref{fig:CompilationQhat}, we decided to include only the free-streaming option with $\tau_0 = 0.6$~fm/$c$. As for the model adopted for the medium, in \cite{Feal:2019xfl, JET:2013cls}, it is assumed a Stefan-Boltzmann gas of quarks and gluons, while in \cite{Andres:2016iys,JETSCAPE:2021ehl,Xie:2020zdb,Xie:2022ght,Ke:2020clc} a hydrodynamical model is employed instead. In this case, the initial conditions must be specified. In~\cite{Andres:2016iys}, it was explicitly considered in their calculations two examples: a factorized Kharzeev-Levin-Nardi (KLN) model~\cite{Drescher:2006pi} (open red markers) and the Hirano model~\cite{Hirano:2001eu} (open orange markers). We kept RHIC and LHC results separated as open circles and squares. To illustrate the sensitivity of $\hat{q}$ to these effects, the corresponding results are also shown in Figure~\ref{fig:CompilationQhat}. The jet-medium interaction model is also different in all considered approaches. For instance, in~\cite{JETSCAPE:2021ehl} (Jetscape), it was reported a $\hat{q}$ when using MATTER to describe energy loss processes (pink shaded region) or a Linear Boltzmann Transport Model (LBT, orange shaded region). In Xie et. al (2003.02441, blue markers), it was used a higher-twist approach to energy loss, while the initial conditions were defined through a Monte Carlo Glauber model and the space-time evolution of the QGP given by a viscous hidrodynamical model. 

Most of the data sets used in these approaches are constrained to single charged particles to avoid reconstruction biases. However, to overcome this effect, in~\cite{Xie:2019oxg} (not shown in Fig.~\ref{fig:CompilationQhat}), a global $\chi^2$ analysis of the nuclear modification factor for single and double hadron production at RHIC and LHC energies is adopted. While at RHIC, the results from using one or both data sets are consistent, there are differences in the used data set at LHC energies. The extracted $\hat{q}$, while within the range of values presented in Fig.~\ref{fig:CompilationQhat}, would be larger for double hadron production, given that the observed suppression in this observable is also proportionally larger. Following works (~\cite{Xie:2022ght,Ke:2020clc}) addressed the use of different data sets to extract $\hat{q}$. To use different sets of data, it is customary to apply Bayesian statistical inference methods (as used in JETSCAPE). However, these methods rely on explicit parameterizations, including the temperature dependence of $\hat{q}$, whose phase-space can extend into regions not probed by the experimental data. The result can therefore include fictitious constraints in the physics parameters. In M. Xie et al., (2206.011340)\cite{Xie:2022ght}, they developed an Information Field approach to reduce these effects and apply it to single-inclusive hadron, di-hadron and $\gamma$-hadron spectra from both RHIC and LHC energies. The resulting $\hat{q}$ exhibits a much stronger temperature dependence (green shaded region) concerning the remaining approaches. Jet suppression also encodes information on $\hat{q}$, although an additional challenge is the presence of the medium back-reaction that can blur an accurate evaluation of the transport coefficients. The work~\cite{Ke:2020clc} (blue shaded region) applied the transport model LIDO to describe jet-medium interactions and used single inclusive jet and single hadron suppression data from central nuclear collisions at both RHIC and LHC. They reported a larger $\hat{q}$ compared to JETSCAPE, which also includes transport equations. According to the authors, the LIDO approach has the jet-medium interactions restricted to the transport equation (final stage of the shower), thus overestimating the magnitude of $\hat{q}$ to compensate for the absence of medium effects during the early stages of the shower.

While an accurate determination of the extracted $\hat{q}/T^3$ is not yet possible (the central value can vary by a factor of up to 6), approximate temperature dependence of this parameter can already be withdrawn. First attempts indicated an additional dependence with $\sqrt{s}$ as results from RHIC and LHC data were inconsistent for the same medium temperature (see Andrés et al. in Fig.~\ref{fig:CompilationQhat}). With the most recent efforts to include more accurate descriptions of jet-medium interaction, it is clear that $\hat{q}/T^3$ does not depend on the collision system but rather only on the medium characteristics. With the continuous progress in the analytical description of jet quenching phenomena, we expect novel qualitative features from the hard sector to contribute to narrowing the current uncertainties. 

Future use of additional data sets to extract $\hat{q}$ will also help to bring a more unified description. Inclusive charged hadron spectra focus only on the leading fragment of the event. It was recently shown that jet substructure observables such as jet splitting functions, an almost entirely orthogonal measurement to jet and hadron spectra, are sensitive to the magnitude of $\hat{q}$~\cite{Mehtar-Tani:2016aco,Chien:2016led,Milhano:2017nzm,Chang:2017gkt}. Determination of jet transport coefficients from jet spectra, jet substructure, or both could be an important test in the future and complementary to the studies done with inclusive hadron spectra and dihadron correlation functions.

In the angular correlation measurements of two hard observables at large angular differences, such as photon-jet~\cite{Chatrchyan:2012gt,CMS:2017ehl,ATLAS:2018dgb}, dijet~\cite{ATLAS:2010isq,CMS:2011iwn}, Z-jet~\cite{CMS:2017eqd} and hadron-jet~\cite{ALICE:2015mdb,STAR:2017hhs}, one hopes to learn about the medium structure through Rutherford-experiment-like measurements. Currently, all such measurements in AA collisions are consistent with pp references, and there is no indication of back-scattering in the experimental data. This shows that the produced medium is very smooth using the currently accessible high momentum particles, consistent with the physics pictures used in the soft sector modeling. More detailed studies with high statistics data at higher and higher particle transverse momentum could help resolve the QGP structure at shorter scales~\cite{CMS:2017dec,Citron:2018lsq,Busza:2018rrf}.

\subsubsection{From heavy quark diffusion to medium properties}
\label{subsec:DsEtaOverS}

Heavy quarks have a thermalization time comparable to the QGP lifetime. Moreover, the heavy quark mass is larger than typical medium temperatures. Therefore the propagation of low-momentum heavy quarks is like a Brownian motion with many small momentum transfer elastic collisions with the medium. In the transport framework of the Fokker-Planck equation, the heavy quark distribution function $f_Q$ can be written as
\begin{eqnarray}
\frac{\partial}{\partial t} f_Q(t,p) = \frac{\partial}{\partial p} p A(p) f_Q(t,p)+ \frac{\partial^2}{\partial^2\vec{p}} B(p) f_Q(t,p) \ .
\label{fp}
\end{eqnarray}
where the medium properties are encoded in the transport coefficients $A$ and $B$, which depend on medium temperature and heavy quark momentum. In this framework, $A$ and $B$ represent the heavy quark's thermal relaxation rate and momentum diffusion, respectively. The transport coefficient $D_s$ is defined by:
\begin{equation}
{D}_s = \frac{T}{m_Q A(p=0)} \ , 
\label{Ds}
\end{equation}
and it characterizes the long-wavelength limit of the heavy-quark transport framework. One can form a dimensionless scaled heavy quark diffusion coefficient, ${D}_s(2\pi T)$, and the inverse of this coefficient characterizes the coupling strength of QGP. Moreover, the heavy quark mass is divided out in the equation above, removing its dependence on charm and beauty quark masses. It has also been suggested a proportionality of the dimensionless medium quantities ${D}_s (2\pi T) \sim \eta/s \sim \sigma_{\rm EM} / T$ in Ref~\cite{Rapp:2009my}, and the ratio of these quantities is expected to depend on the nature of the medium. For instance, the ratio ${D}_s (2\pi T) / (\eta/4\pi s)$ is around 2.5 in a weakly coupled QGP, while in the strong coupling limit, this ratio is found to be 1~\cite{Policastro:2002se}.  

Based on the connections described above, we use the following relation in the following discussion:
\begin{eqnarray}
\frac{\eta}{s}=\frac{Ds(2\pi T)}{ 4\pi k} \, ,
\end{eqnarray}
for the translation between specific shear viscosity and the pertinent spatial diffusion coefficient, where the scale factor $k$ ranges between 1 and 2.5 to cover the possible difference between pQCD and AdS/CFT-based medium models~\cite{Dong:2019byy}. 

\subsubsection{input from heavy flavor hadrons}
\label{subsec:qhatHF}

In chapter~\ref{sec:acceleration}, we reviewed a collection of precise measurements from the LHC experiments and the RHIC upgrades that have brought heavy-flavor hadron measurements to a new level of precision. This was accomplished by experimentally allowing the reconstruction of weak decays with lifetimes between $50-500$ $\mu$m in exclusive decay chains of specific hadrons via modern silicon vertex detectors. The measurements of heavy flavor hadrons are well in line with the expectations from a strongly coupled QGP. Despite the expected longer relaxation time, there is a strong nuclear modification of production yields and participation in the collective motion visible via the large elliptic and triangular flows. A large number of models can describe the open-heavy flavor data by implementing different types of transport equations. However, the model constructions and inputs vary significantly and need to be brought on a common assumption as outlined in Ref.~\cite{Rapp:2018qla} to allow for more quantitative conclusions. Even full thermalization is a scenario not excluded by experimental data~\cite{Andronic:2021erx}. Furthermore, to scope the complex problem, today's modeling makes a schematic idealization of the initial state, mostly semi-classical propagation descriptions in the presence of a QGP and hadronization prescriptions. However, it deals, nevertheless, with a large assumption and parameter space. This picture has served us well in identifying qualitative interpretations with quite different approaches. With increasing experimental precision and the desire to extract matter properties, one needs to test the following aspects of the model systematically:
\begin{itemize}
\item the assumption of factorization in the description of the initial state
\item the role of pre-equilibrium dynamics and escape mechanism
\item the approximations required to justify a specific transport equation suited as a starting point
\end{itemize}
 Independent of conceptual model limitations and ambiguities, it is mandatory to improve the knowledge of the initial state and hadronization with the current understanding and within all available 'frameworks'. The associated uncertainties largely dominate standard observables like nuclear modification factor and elliptic flow, as we illustrate in chapter~\ref{sec:initial} and~\ref{sec:hadronization}. The latter inputs on the experimental side will soon improve significantly with increasing statistics from the experiments. 

\begin{figure}
    \centering
    \includegraphics[width=1\textwidth]{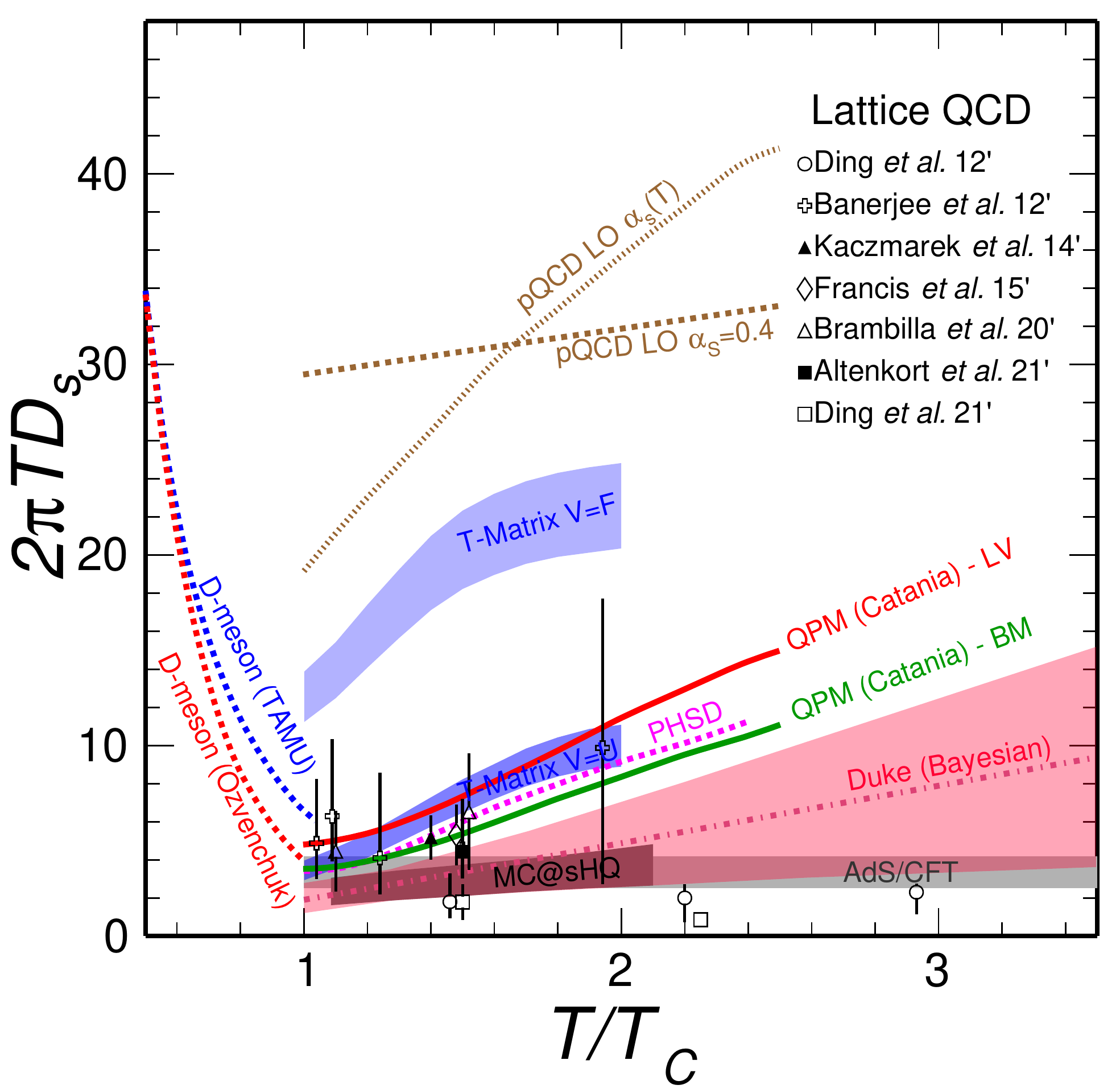} 
\caption{A compilation of the charm-quark spatial diffusion coefficient, $2\pi T{\cal
D}_s$, presented over a wide range of reduced temperature $T/T_{\rm C}$. Below the pseudo-critical temperature $T_{\rm C}$, the results for $D$ meson diffusion using effective hadronic interactions~\cite{He:2011yi,Tolos:2013kva} are shown. Above $T_{\rm C}$, results from quenched lattice QCD \cite{Banerjee:2011ra,Ding:2012sp,Kaczmarek:2014jga,Brambilla:2020siz,Ding:2021ise,Francis:2015daa,Altenkort:2020fgs} are shown as the data points with vertical error bars and compared to anti-de Sitter/conformal field theory (AdS/CFT)-based calculations~\cite{Horowitz:2015dta}. In addition, various model calculations based on LO pQCD~\cite{Moore:2004tg,vanHees:2004gq}, quasi-particle model (QPM) calculations utilizing Boltzmann-based extractions (BM) or Langevin-based extractions (LV)~\cite{Das:2015ana}, a dynamical QPM model in Parton-Hadron-String Dynamics (PHSD)~\cite{Song:2015sfa}, $T$-matrix based calculations with free energy ($F$) or internal energy ($U$) as potential~\cite{Riek:2010fk} and the Monte Carlo at Heavy Quarks (MC@sHQ) perturbative approach with running coupling~\cite{Andronic:2015wma} are also shown. Finally, a data-driven extraction based on a Bayesian analysis is shown as a red band, presenting the 90\% confidence interval in the Duke hydro/transport model. }
    \label{fig:CompilationDs}
\end{figure}

Heavy-flavor observables can be used to constrain the heavy-quark diffusion coefficient. As in the light-flavor sector, one needs to consider a specific model ansatz and fix the input assumptions. Bayesian fits have demonstrated the coefficient sensitivity to the experimental results in several approaches. As an example, the attempt to extract the spatial quark diffusion coefficient $D_{s}$ of a heavy quark is similar to the extraction of matter properties in the soft sector~\cite{Xu:2017obm}. A compilation of the results from the theoretical calculations~\cite{Rapp:2018qla,Cao:2018ews} together with the mentioned data-constrained extraction~\cite{Xu:2017obm} are shown in Figure~\ref{fig:CompilationDs}.

The extracted $D_s (2\pi T)$, between 1 and 3 close to the critical temperature, is significantly below the perturbative QCD calculations. Moreover, the results are also close to predictions from quenched lattice QCD and consistent with the expectation from an AdS/CFT-based calculation.

\subsubsection{Input from Quarkonia}
\label{subsec:qhatQuarkonia}

The investigation of quarkonia in heavy-ion collisions at RHIC and at the LHC brings two qualitatively new insights: the sizeable regeneration of J/$\psi$~\cite{ALICE:2012jsl,ALICE:2013osk,ALICE:2015nvt,ALICE:2015jrl,ALICE:2016flj,ALICE:2017quq,ALICE:2018bdo,ALICE:2019lga} and the sequential suppression of the $\Upsilon$-states~\cite{CMS:2011all,CMS:2012bms,CMS:2012gvv,STAR:2013kwk,PHENIX:2014tbe,ALICE:2014wnc,CMS:2016rpc,STAR:2016pof,CMS:2017ycw,CMS:2018zza,ALICE:2019pox,ATLAS-CONF-2019-054,ALICE:2020wwx}. Both observations are interpreted as most directly possible signatures of deconfinement. The observation of their modification and its interpretation connects them with open-heavy flavor observables. First, there is a connection via the allowed 'coupling' of the single heavy quark and the quarkonium via regeneration. In addition, on the theoretical ground, when QGP 'resolves' the quarkonium constituents, it is possible to link the interactions of heavy quarkonium with those of single heavy quarks. 
We just started to go beyond qualitative findings. The degree and moment of regeneration cannot be decided for charmonium states based on current experimental results. Here, the constraints on the initial condition of total $c\bar{c}$ and more precise measurements of a larger number of quarkonium states should bring progress. For bottomonium, the sensitivity to the heavy-quark potential, the Debye mass, and heavy quark diffusion coefficient have been demonstrated in the literature. However, the classes of models used to describe the data are extensive. There is the need for constraints on initial densities and precise feed-down information to exploit the observed sensitivity beyond the scope of model-dependent statements.

\subsection{Specific Shear Viscosity}
\label{subsec:QGPdensityShear}

\begin{figure}
    \centering
    \includegraphics[width=1\textwidth]{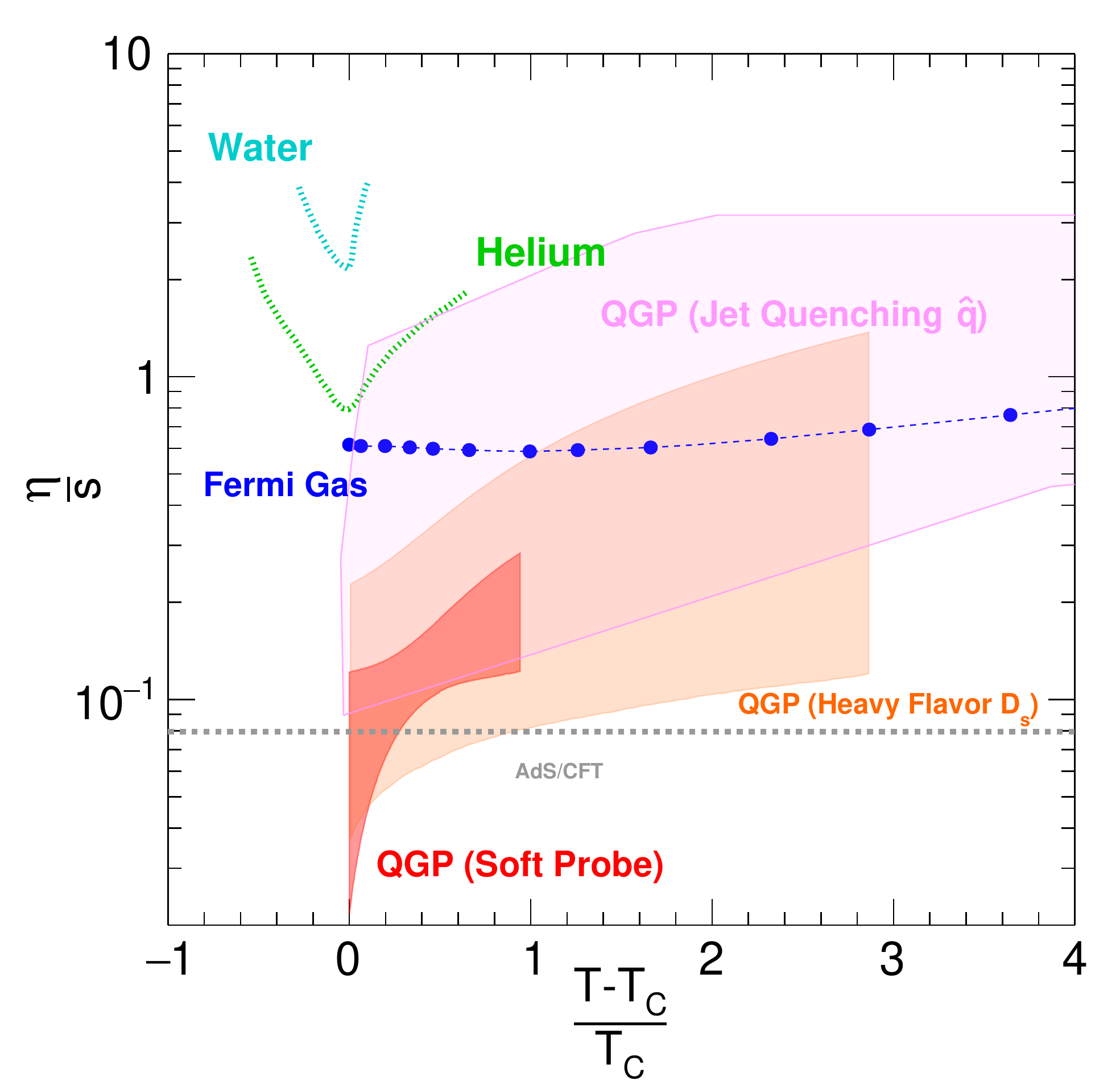} 
\caption{Compilation of the specific shear viscosity as a function of the temperature of the medium. To compare the properties of various media close to phase transition temperature, the temperature is normalized by the (pseudo-~)~critical temperature of the medium. A model-dependent translation of jet quenching parameter $\hat{q}$ and heavy quark diffusion coefficient $D_s$ to specific shear viscosity is performed, and the results are shown as QGP (Jet quenching $\hat{q}$) and QGP (Heavy Flavor $D_{s}$) in the figure. }
    \label{fig:CompilationShearViscosity}
\end{figure}

The ratio of shear viscosity and entropy, often referred to as specific shear viscosity, is a dimensionless quantity and an important transport coefficient of matter. It has been studied extensively in various systems. Figure~\ref{fig:CompilationShearViscosity} summarizes the current status of the extracted specific shear viscosity ($\eta/s$). The $\eta/s$ extracted from a Bayesian analysis (labeled as "QGP (Soft Probe)") on low transverse momentum spectra and hydrodynamics flow is compared to other media such as water, helium, a calculation of fermionic quantum liquids (Fermi gas)~\cite{Enss:2010qh}. A calculation based on a string theory method for a large class of strongly interacting quantum field theories whose dual description involves black holes in anti-de Sitter space (AdS/CFT)~\cite{Kovtun:2004de}. The results from Bayesian analysis are close to calculations from AdS/CFT and smaller than those from other media. Namely, a factor of 10-20 smaller than water and helium and a factor of around five smaller than the calculation of Fermi gases. This shows that QGP is very strongly interacting and the closest medium to ideal fluids.

It is of great interest to check this conclusion using hard probes. In order to compare the transport properties extracted from jet quenching and heavy quark diffusion in the medium, we performed model-dependent translations. As an example, we chose to use eq.(\eqref{eq:sec5_qhat}),
as discussed in Sec.~\ref{subsec:qhatToMediumProperties}. This relation holds under the assumption of multiple soft emissions and for a gas of massless particles. Other equations-of-state would yield a different result, but this allows us to have an overall qualitative picture of the landscape of the $\eta/s$ phase space. 

The results are close to the extracted values from the hydrodynamic flow. However, the spread of the extracted $\hat{q}$ values between various theoretical models is still substantial due to the underlying jet quenching mechanisms, initial state, and hydrodynamic models used in the theoretical calculations. Note that the extended reach of $\eta/s$ shown in the jet quenching result at large temperature comes from a model-dependent parameterization, dominated by the data-driven extractions from theoretical models~\cite{JETSCAPE:2021ehl,Andres:2016iys}.

Nonetheless, we can still check if, within the large uncertainties, there is consistency between heavy quark and the picture seen by the soft probes. The heavy quark spatial diffusion coefficient $D_s$ values extracted from a Bayesian analysis~\cite{Xu:2017obm} is also translated to $\eta/s$ using the relation described in Sec.~\ref{subsec:DsEtaOverS}. The results from soft probes, jet quenching and heavy flavor are consistent with each other, despite the large uncertainties associated with theory and experimental accuracy.

\subsection{Outlook}
\label{subsec:Outlook}

Jets have a well-established theory in vacuum, and its modifications in heavy-ions are well understood. However, a complete jet-hydro coupled description is still limited, blurring a quantitative extraction of QGP transport properties. On the other hand, it is still possible to extract the big picture from the current state-of-art jet quenching models. The extracted transport properties are consistent with that of soft probes.

The heavy quark transport coefficient extracted from heavy flavor hadrons could benefit from the extensive heavy-ion data to be collected in Run 3+4 at the LHC and with sPHENIX at RHIC. Further theoretical developments may provide a better understanding of charm and beauty hadronization. These will be discussed in Chapter~\ref{sec:hadronization}.

The different modeling of the early stages of the heavy-ion collisions, such as the initial effects described in Chapter~\ref{sec:initial}, and the time-dependent evolution of QGP, contributed significantly to the theoretical uncertainties associated with heavy flavor and jet calculations. With hard probes, it is possible to extract useful information in a time-differential way to be sensitive to different parts of the medium evolution. Those ideas and future opportunities will be discussed in Chapters~\ref{sec:timescale} and~\ref{sec:outlook}.

\clearpage

%% file: QGPcontent.tex
\section{Hadronization}
\label{sec:hadronization}

The transition of partonic degrees of freedom produced in high-energy collisions to hadrons detected in the experiment, i.e., hadronization, is subject to modeling since we cannot calculate this process. It is one of the dominating uncertainties for modeling heavy quarks in nucleus-nucleus collisions. Hence, it is an important limitation for quantifying heavy-quark propagation in heavy-ion collisions within the model assumptions considered today~\cite{Rapp:2018qla}. 

The question of the existence of bound states in the QGP and our view of hadronization are intertwined. Among the most notable cases are bound states with more than one heavy quark, see the discussion in Chapter~\ref{sec:acceleration_HFQQ}. Also, for open-heavy flavor hadrons, it has been argued that their existence above the usually considered transition temperature has a crucial impact on the extracted medium properties~\cite{Nahrgang:2013xaa}. In contrast, the definition of jets via infrared and collinear algorithms tries to minimize the impact of different hadronization mechanisms on the observable. However, the investigation of jets in nucleus-nucleus collisions considers physics questions and observables sensitive to small momentum scales, where the description of hadronization matters. The observables sensitive to jet-induced medium response and their modeling are prime examples (discussed in Chapter~\ref{sec:acceleration}).        
In this chapter, we introduce the competing descriptions of hadronization in section~\ref{subsec:hadconcepts}. We present recent observations concerning particle production as a function of the number of charged particles and related modeling in section~\ref{subsec:hadtrans}. We then discuss selected measurements on hadronization of heavy-flavor and jets in section~\ref{subsec:hadHFjets}. We point to opportunities that can help to improve our description of heavy-flavor and jets. Finally, we discuss how measurements in heavy-ion collisions can contribute to the field of hadron spectroscopy usually performed in $e^+e^-$ and pp~collisions. We will discuss the first steps in this new research direction in section~\ref{subsec:hadstruc}.

\subsection{Hadronization in high-energy physics and heavy-ion physics}
\label{subsec:hadconcepts}
\begin{figure}
    \centering
    \includegraphics[width=0.35\textwidth]{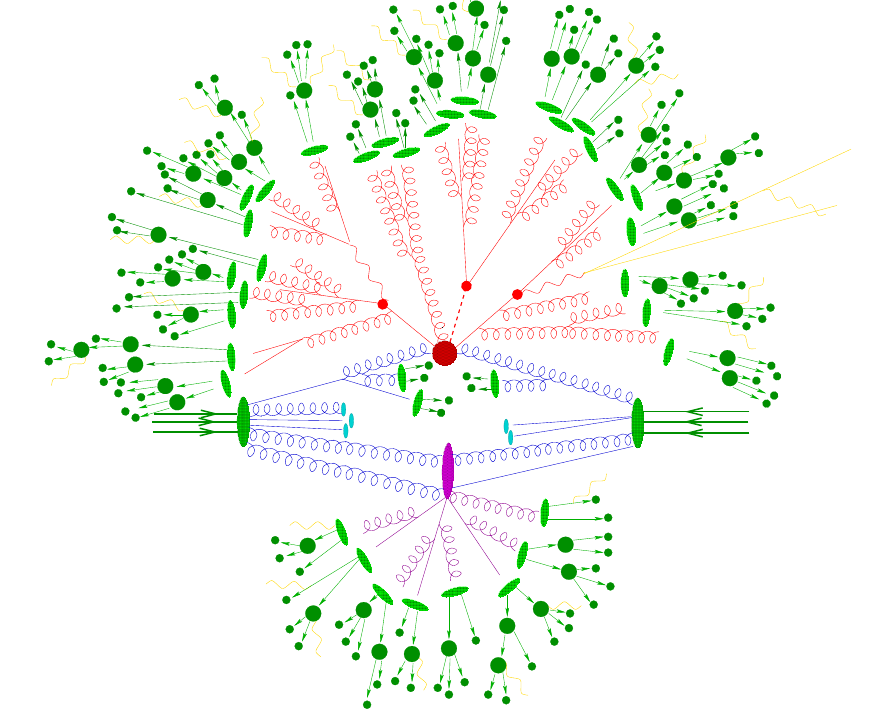} 
    \includegraphics[width=0.6\textwidth]{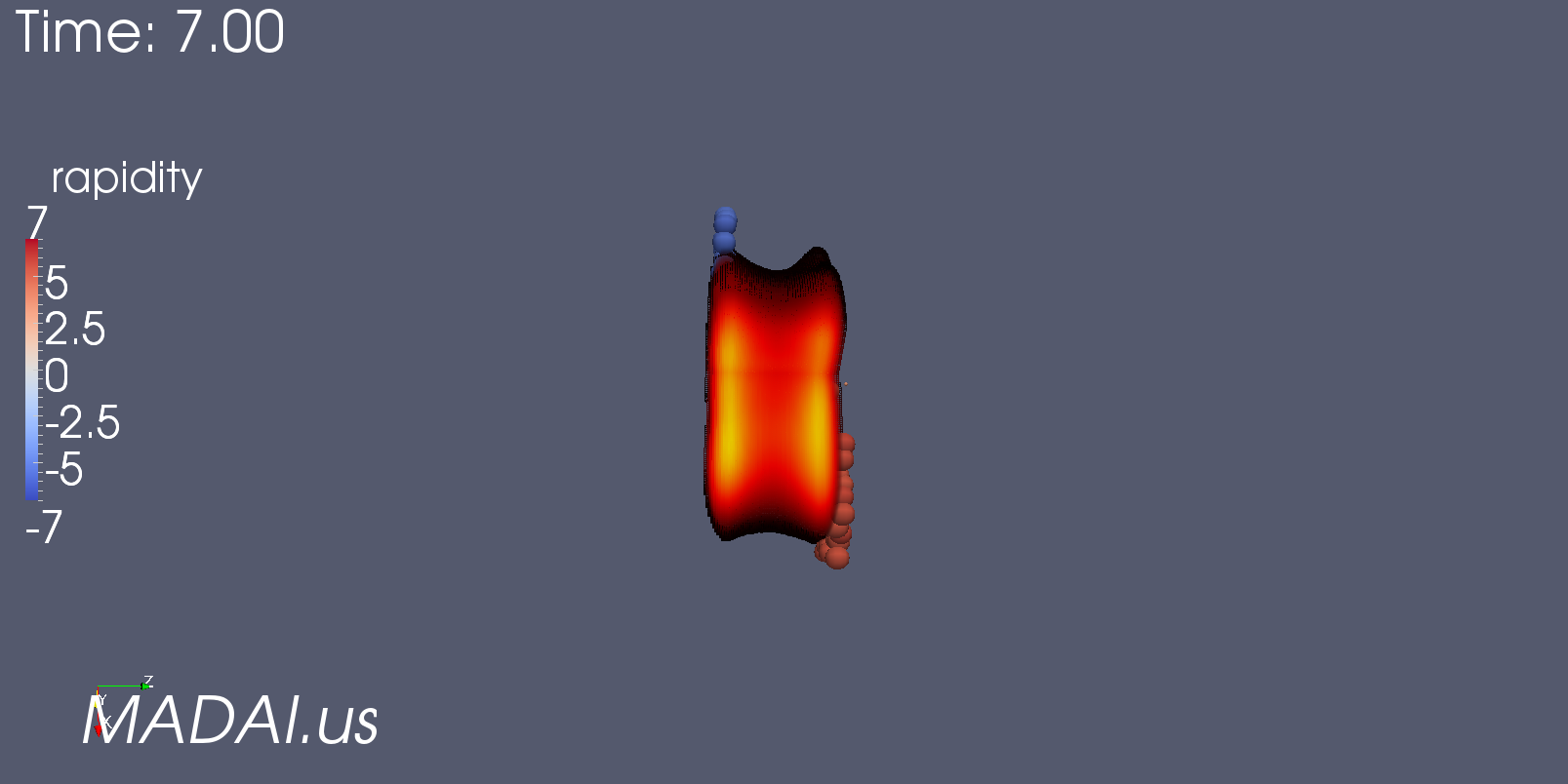}
    \caption{Visualization of a Sherpa $t\bar{t}h$ event  at the LHC taken from \cite{Gleisberg:2008ta} and a visualization of heavy-ion collision at RHIC from the Madai visualization. 
    }
    \label{fig:hadronizationcartoon}
\end{figure}

Let us first consider the description of hadronic final states in non-hadronic resonance production in $e^+e^-$~collisions. A color-neutral quark-antiquark pair is created from a single virtual photon or weak gauge-boson. Subsequently, colored degrees of freedom are created by radiation, see~\ref{sec:acceleration}. In this set-up, hadronization takes place at the end of the parton shower and consists of translating partonic degrees of freedom into hadrons. The particlization of a parton shower in any collision system can be associated with this situation. To describe high-energy proton-proton collisions, the event is seen as the product of multiple (semi)-hard partonic scatterings treated with a parton shower down to a low scale. The produced partons need to be translated into hadrons analog to the situation in $e^+e^-$ as shown pictorially in Fig.~\ref{fig:hadronizationcartoon} (left). This concept is implemented in most of the event generators as Pythia~\cite{Sjostrand:2007gs}, Herwig~\cite{Bahr:2008pv,Bellm:2015jjp} and Sherpa~\cite{Gleisberg:2008ta}, which provide a fully exclusive description of the final state. 

The two most common approaches to model hadronization microscopically in high-energy physics event generators are the Lund-string model~\cite{Andersson:1983ia} employed in Pythia~\cite{Sjostrand:2007gs} and the cluster hadronization model~\cite{Webber:1983if} present in both Herwig~\cite{Bahr:2008pv,Bellm:2015jjp} and Sherpa~\cite{Gleisberg:2008ta}. A pedagogic introduction can be found in~\cite{Campbell:2017hsr}. To gain predictive power in hadronic collisions, hadronization is combined with 'universality' assumptions ('jet universality'): the non-perturbative parameters are, to a large extent, constrained with measurements in $e^+e^-$ collisions with defined partonic kinematics and the same model parameters are then used for the description of hadronic collisions. This approach has to be complemented with the modeling of beam remnants in hadron-hadron collisions and the multiple parton interactions. 

In addition to event generators, perturbative QCD calculations for single-particle production can be carried out. Instead of a full exclusive description of the collision, they provide inclusive rates or cross-sections. In these calculations, non-perturbative but universal fragmentation functions are used to translate the produced (heavy) quark or gluons into a hadron multiplicity with a given momentum, see for a review~\cite{Metz:2016swz}. 

In proton-proton and proton-antiproton collisions, the universality ansatz works well for unidentified hadrons at sufficiently high transverse momenta to describe the fragmentation of a high energetic parton into hadrons with the fragmentation function taken from $e^+e^-$ and deep-inelastic scattering. The universality assumption appears to hold also in nucleus-nucleus collisions at sufficiently large transverse momentum for light flavor hadrons. The production of identified hadrons at RHIC~\cite{STAR:2006uve} and the LHC~\cite{ALICE:2013cdo,ALICE:2014juv} indicate particle species ratios in pp and nucleus-nucleus collisions that are consistent with unity above a transverse momentum of 10~GeV/$c$ for light-flavor hadrons~\footnote{We note that the description of identified hadron spectra within pQCD calculations based on fragmentation functions is already not perfect in pp~collisions, see e.g. in~\cite{ALICE:2020jsh}.}. 

However, in the case of event generators, the required treatment of beam remnants in the hadronic collision implies that the hadronization of partons is not independent of the surrounding environment in specific situations. The beam remnants influence hadronization in hadronic collisions close to beam rapidity. This treatment breaks the universality of fragmentation functions, see, for example, the results by E791 Collaboration close to beam rapidity~\cite{Aitala:1996hf}. This observation shows that factorization breaking is necessary for any complete event description. 

In the presence of QGP, standard descriptions of particle production at low transverse momentum are based on the applicability of thermodynamic concepts. Part of the lifetime of the final-stage evolution of a heavy-ion collision can be treated with hydrodynamics. The thermodynamic system is translated into final state hadrons as the energy density and temperature drop due to the system's expansion. At this point, the statistical hadronization model is usually  to predict inclusive identified particle yields (see for a recent account in~\cite{Andronic:2017pug}). A detailed review can be found in~\cite{Braun-Munzinger:2003pwq}. In this approach, the hadron densities are calculated from a partition function containing hadrons and resonances, the hadron-resonance gas~(HRG). The system is described within a grand-canonical ensemble at the chemical freeze-out temperature $T_{CF}$, and a baryochemical potential $\mu_B$ related to the net-baryon number. The chemical freeze-out is the moment of the system evolution that can already be described by hadronic degrees of freedom. Very close to this moment, the system falls out of chemical equilibrium, i.e., the particle species abundances are fixed. At low baryochemical potential, the HRG describes the equation of state calculated in lattice QCD below the experimentally determined freeze-out temperature. This observation provides a connection between experimental data and ab-initio QCD calculations. We refer to reference~\cite{Andronic:2017pug} and the references therein for details. The extracted chemical freeze-out temperatures are close to the chiral cross-over temperature~$T_c$ seen in lattice QCD calculations~\cite{Borsanyi:2013bia,Bazavov:2014pvz}. 
 
In hydrodynamic simulations~\footnote{An introduction to the underlying concepts is given in Ref.~\cite{Ollitrault:2007du}} aiming at a description of the exclusive final state, the transition of the simulated fluid cells is modeled via a Cooper-Frye formalism~\cite{Cooper:1974mv}. It translates the energy and conserved quantum numbers into single hadron densities according to the HRG, see for an introduction and further references in~\cite{Gale:2013da}. Each fluid cell's transition is modeled at a fixed energy density or temperature. The corresponding switching temperature between fluid and gas description is commonly set to $T_c$, see for instance, for recent global fits to experimental data~\cite{Bernhard:2019bmu,Nijs:2020ors}. In the most common picture, the hadrons 'produced' via this freeze-out, according to single-particle densities in equilibrium, are then put into a hadronic 'afterburner'. This hadronic transport code (see, e.g.,~\cite{Bass:1998ca} for an introduction) treats the hadron scatterings after the chemical freeze-out until the densities of hadrons drop such that interactions cease, e.g.  in~\cite{Bernhard:2019bmu,Nijs:2020ors}. In addition, strong and electromagnetic hadron decays need to be considered. Finally, the hadrons stream freely to the detector. A fluid simulation visualization is shown in Fig.~\ref{fig:hadronizationcartoon}. This prescription provides a method to hadronize thermalized energy deposition. However, it does not connect the microscopic degrees of freedom of non-thermal partons to hadrons in the final state. 

\subsection{Hadronization as a function of charged particle multiplicity in different collision systems}
\label{subsec:hadtrans}

\begin{figure}
    \centering
    \includegraphics[width=0.45\textwidth]{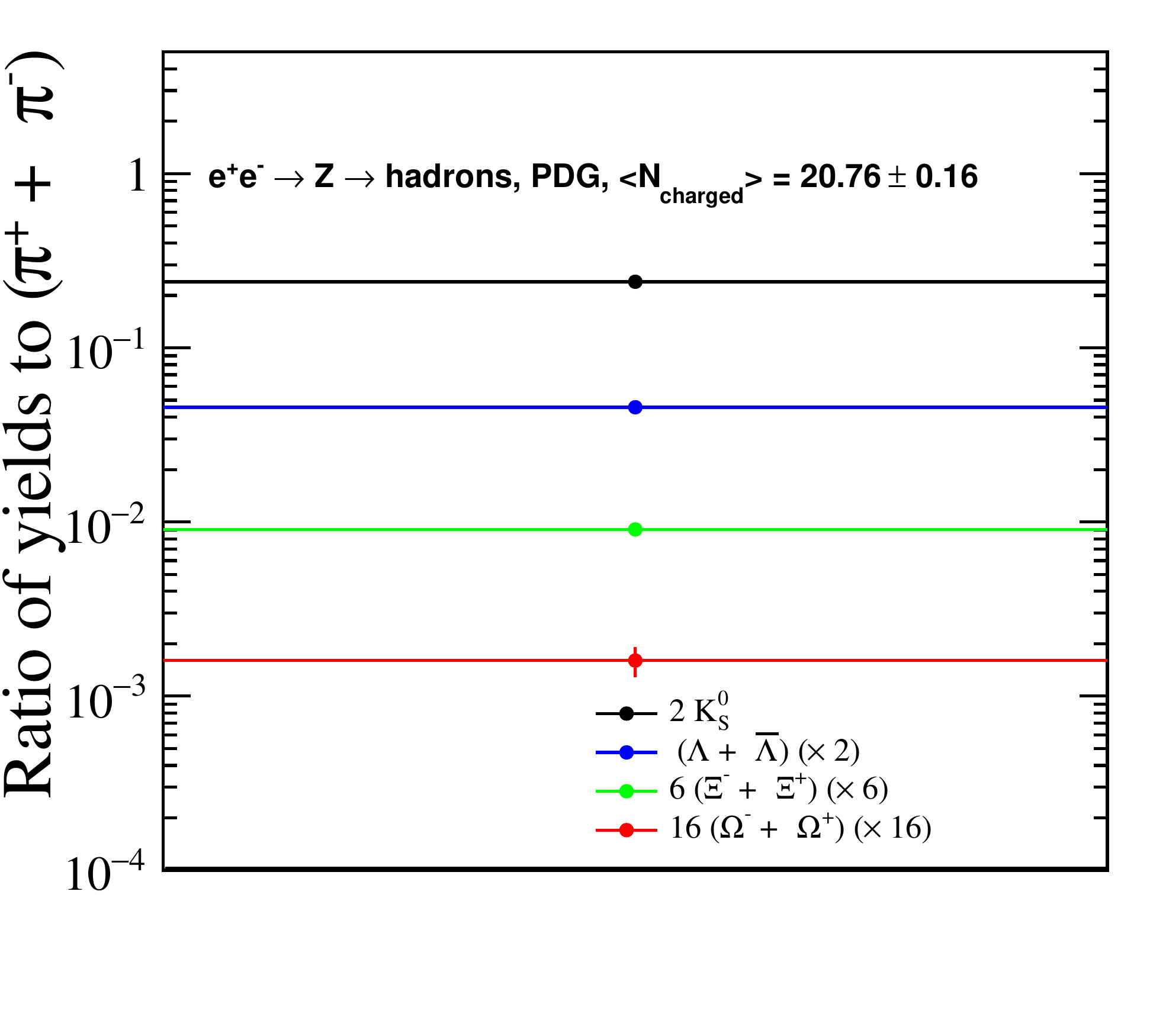}
    \includegraphics[width=0.32\textwidth]{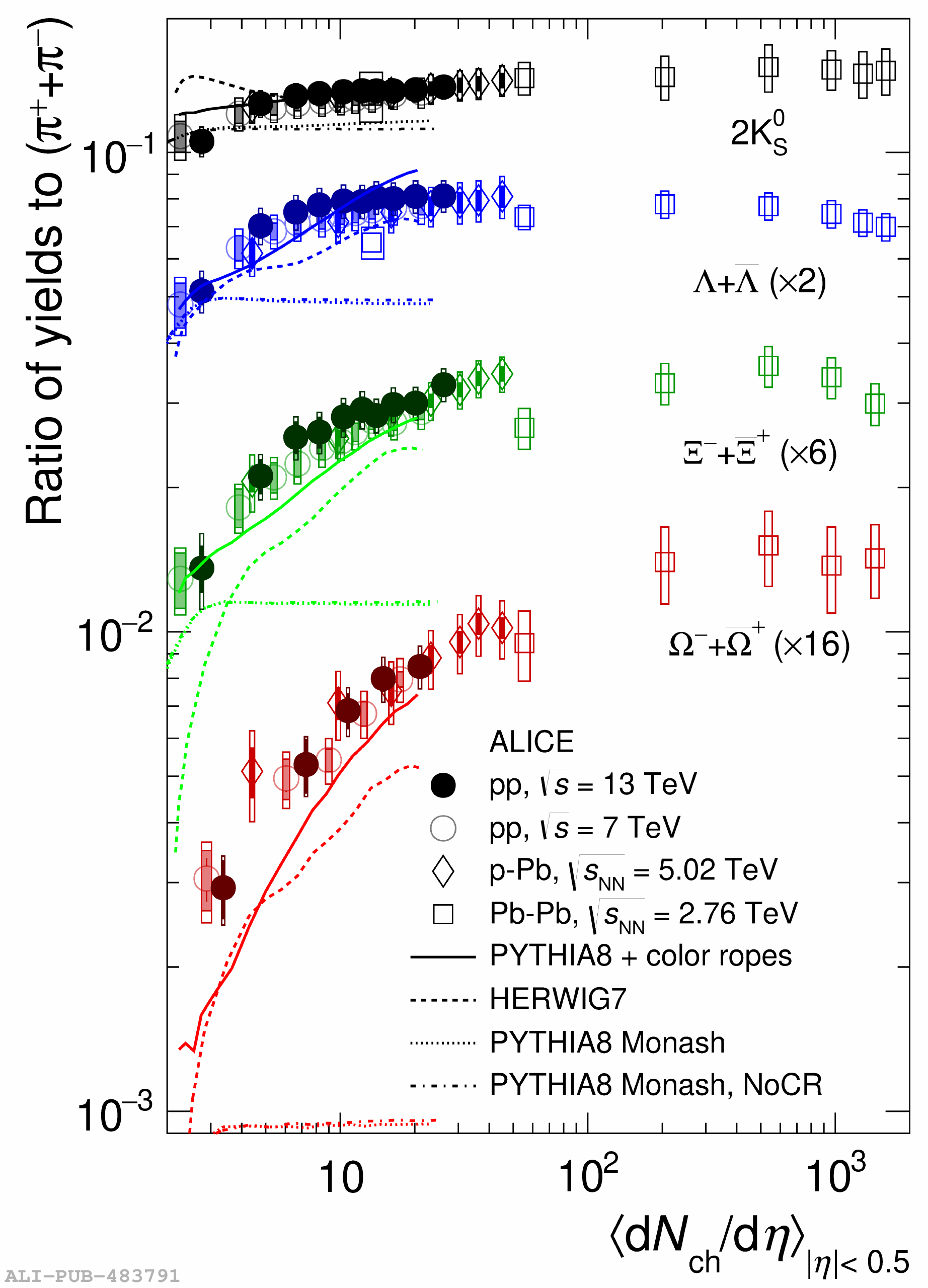} 
    \caption{Left: Strange particle ratios with respect to pions measured at Z-resonance based on PDF compilation~\cite{Zyla:2020zbs}, Right: strange particle yield ratios measured in several hadronic collision systems at the LHC by ALICE taken from~\cite{Acharya:2020zji} including data points from~\cite{Acharya:2019kyh,Acharya:2018orn,Adam:2015vsf,ALICE:2017jyt}. 
    } 
    \label{fig:hadronization}
\end{figure}

In addition to investigating hadronization as a function of transverse momentum, as briefly mentioned in~\ref{subsec:hadconcepts}, observables sensitive to hadronization have been studied extensively in different collision systems: proton-proton, proton(deuteron)-nucleus, and nucleus-nucleus at RHIC and the LHC. 
Results of $p_T$-integrated hadron yield containing a variable degree of strangeness normalized to charged pions show that hadronization of light-flavor particles at low transverse momentum is a function of the charged particle multiplicity independently of the collision system, within experimental uncertainties~\cite{ALICE:2017jyt}. Fig.~\ref{fig:hadronization} shows a compilation of the particle yields ratio (right) with the corresponding measurement in $e^+e^-$~collisions at the Z-peak combined by the Particle Data Group~\cite{Zyla:2020zbs} (left). With standard hadronization modeling from high-energy physics, represented in the figure with the Pythia Monash tune without color reconnection~\cite{Skands:2014pea}, these particle ratios show very little dependence on charged-particle multiplicity and are approaching the measurements in $e^+e^-$~collisions that were used as input constraints. The grand-canonical~\footnote{I.e., grand-canonical with respect to the conserved quantum numbers: baryon, strangeness, and electric charge quantum numbers.} hadron resonance gas model (not shown in the figure) reproduces the values at the largest charged-particle multiplicity in PbPb collisions, above $dN/d\eta \approx $ 100. The charged particle production in a collision system measures the total entropy produced in the collisions. Thus, within the fluid paradigm of a hadronic collision, this entropy can be connected to a volume. It is possible to observe a transition between a situation close to $e^+e^-$~constraints and a description living in the thermodynamic limit of large volume QCD matter as a function of charged-particle multiplicity.

To describe these observations, high-energy physics event generators need to introduce additional mechanisms unimportant for the $e^+e^-$-modeling. In particular, they need to enhance the generation of baryons at high charged particle multiplicities and the generation of strangeness. The considered changes take into account the dense environment of hadronic collisions via color reconnections beyond leading color~\cite{Christiansen:2015yqa} allowing new color topologies for baryon production (a mechanism already studied for precision studies in $e^+e^-$~collisions~\cite{Schael:2013ita}), a detailed space-time picture~\cite{Ferreres-Sole:2018vgo}, hadronic rescattering~\cite{Bierlich:2021poz} or color string shoving~\cite{Bierlich:2017vhg}, the latter being an extension of pythia for describing proton-nucleus and nucleus-nucleus collisions~\cite{Bierlich:2018xfw}.  Fig.~\ref{fig:hadronization} (right) includes as examples 3 types of modified hadronization: the HERWIG 7 curve includes baryonic ropes~\cite{Bellm:2015jjp} as mechanism, the Pythia Monash~\cite{Skands:2014pea} with color reconnections is shown and a Pythia version including color ropes~\cite{Bierlich:2018xfw}. These modifications yield to qualitative agreement of the multiplicity dependence. 

Alternatively, event generators like EPOS~\cite{Werner:2012xh} use two-component hadronization models, like the core-corona picture also in~\cite{Kanakubo:2019ogh}, that are successful for light-flavor hadron production. The decision to hadronize a given piece of matter via a Cooper-Frye or Lund-string hadronization is taken dynamically. The chosen hadronization mechanism for the microscopic object is decided based on the energy lost in the traversed medium. In the case of EPOS, it is a string. If the lost energy is sufficiently large, the string is assumed to be part of the fluid ('core'). At higher string energies, a string fragmentation model is used ('corona'). The hadronization modification as a function of charged particle multiplicity results from different fractions of core and corona hadronization. 

Statistical hadronization models, initially designed for dense systems, have been applied extensively to pp, pPb, and $e^+e^-$ collisions~\footnote{The application of statistical hadronization models to $e^+e^-$~collisions is motivated by the observation of thermal features. This fact has led to interpretations different from the thermal hadron production assumed in heavy-ion collisions. In the latter case, statistical hadronization has been argued to be a signature of equilibrated QCD matter hadronization with a freeze-out close to the cross-over phase boundary~\cite{BraunMunzinger:2003zz}. In contrast to this interpretation, the statistical hadronization has been seen as a generic property of hadronization~\cite{Stock:1999hm,Heinz:2000ba} or from the excited QCD vacuum~\cite{Castorina:2007eb}. The validity of thermal fits to experimental results in $e^+e^-$ has been discussed in~\cite{Becattini:1995if,Andronic:2008ev,Becattini:2010sk}. A grand-canonical treatment fails in $e^+e^-$, pp and $p$Pb~collisions such that a thermodynamic interpretation may only be robust in heavy-ion collisions. The implementation and interpretation of conservation laws and additional parameters required to describe the data in a small collision system are beyond the scope of this review.}. For hadronic collisions at the LHC, by treating either strangeness~\cite{Acharya:2018orn} alone or baryon number and strangeness in a canonical hadronization formalism~\cite{Vovchenko:2019kes,Cleymans:2020fsc}, most of the hadronization patterns can be reproduced with an accuracy of 15$\%$. The $\phi$~meson shows a qualitatively different behavior in~\cite{Vovchenko:2019kes} compared to data since it does not carry net strangeness and, as such, is not suppressed within this model.

Coalescence or recombination approaches are often employed in heavy-ion collision phenomenology. They are based on the idea that a hadron in a dense environment can be formed from the abundant hadron´s constituents  that are sufficiently close in phase space. 
A coalescence or recombination ansatz between partons has been employed early on to describe hadronization in hadronic collisions near beam rapidity~\cite{Das:1977cp}. In this situation, the application of fragmentation functions fails since the presence of a quark reservoir from the beam particles cannot be neglected as in a fragmentation approach. In this case, different algorithms also consider the combination with the remaining partons of the beam particles. The experimental results on baryon production and azimuthal correlations at RHIC suggested the application of this approach to bridge the low-transverse momentum regime described by hydrodynamics and to the fragmentation region at large transverse momentum, see for an early review focused on RHIC findings~\cite{Fries:2008hs}. In the simplest formulation, the hadronization is instantaneous and it is modeled via an overlap integral of the parton constituents and the composite objects characterized by Wigner wave functions in momentum and coordinate space in an instantaneous approximation. Four-momentum conservation is violated in this model in $2\to 1$ processes for meson production and $3 \to 1$ processes for baryon production. To avoid this violation, the idea has been further developed into a transport model, where resonance formation cures the violation of 4-momentum conservation and respects the equilibrium limit~\cite{Ravagli:2007xx,He:2019vgs}. Interestingly, the recent developments of high-energy physics generators, that are employed to describe hadronization in dense environments, and the two-component models do not single out a specific kinematic regime of a couple of GeV in transverse momentum with a different hadronization mechanism. The production of nuclei has been discussed in terms of coalescence approaches, where the coalescing particles are typically the nucleons and not the quarks, see for an early review~\cite{CSERNAI1986223}. An introduction to the concepts and an extended reading list can be found in ref.~\cite{Braun-Munzinger:2018hat}.  

In summary, at the current stage, there is no standard description of hadronization in hadronic collisions that spans from small to large particle number of produced particles and from low to large transverse momenta.

\subsection{Heavy-flavor and jet hadronization }
\label{subsec:hadHFjets}

For the description of heavy-quark and jet hadronization in nucleus-nucleus collisions, the two limiting cases of hadronization, statistical and universal fragmentation, must be recovered based on the experimental observations. However, under the conditions realized in experimental measurements, these limits may not be reached. At low transverse momentum or mass scales and for large medium space-time dimensions, statistical hadronization respecting the constraints of conserved quantum numbers as charm or beauty should be approached. At high transverse momentum and small-medium space-time dimensions, hadronization modeling based on transfer from $e^+e^-$ collisions should provide a suitable description. In the latter case, the hadronization happens outside the QGP medium due to time dilation associated with the Lorentz boost relative to the medium. 

An extensive measurement program of charm and beauty hadrons, including baryons and quarkonium, from proton-proton up to nucleus-nucleus collisions, has been started by the RHIC and LHC collaborations. In proton-proton and proton-nucleus collisions, the transverse momentum spectra and rapidity distributions of heavy-flavor mesons are described within the sizeable theoretical uncertainties by perturbative QCD calculations. The calculations are based on fixed-order NLO perturbation theory and next-to-leading logarithm resummation as in the Fixed-order-Next-to-leading-logarithm(FONLL)~\cite{Cacciari:1998it,Cacciari:2001td,Cacciari:2012ny} or the general-mass variable-flavor-number~\cite{Kniehl:2004fy,Kniehl:2012ti,Benzke:2017yjn,Kramer:2017gct,Helenius:2018uul}~schemes. Both type of calculations  employ fragmentation functions extracted from $e^+e^-$. Selected comparisons with data can be found in~\cite{Andronic:2015wma}. 

\begin{figure}
    \centering
    \includegraphics[width=0.45\textwidth]{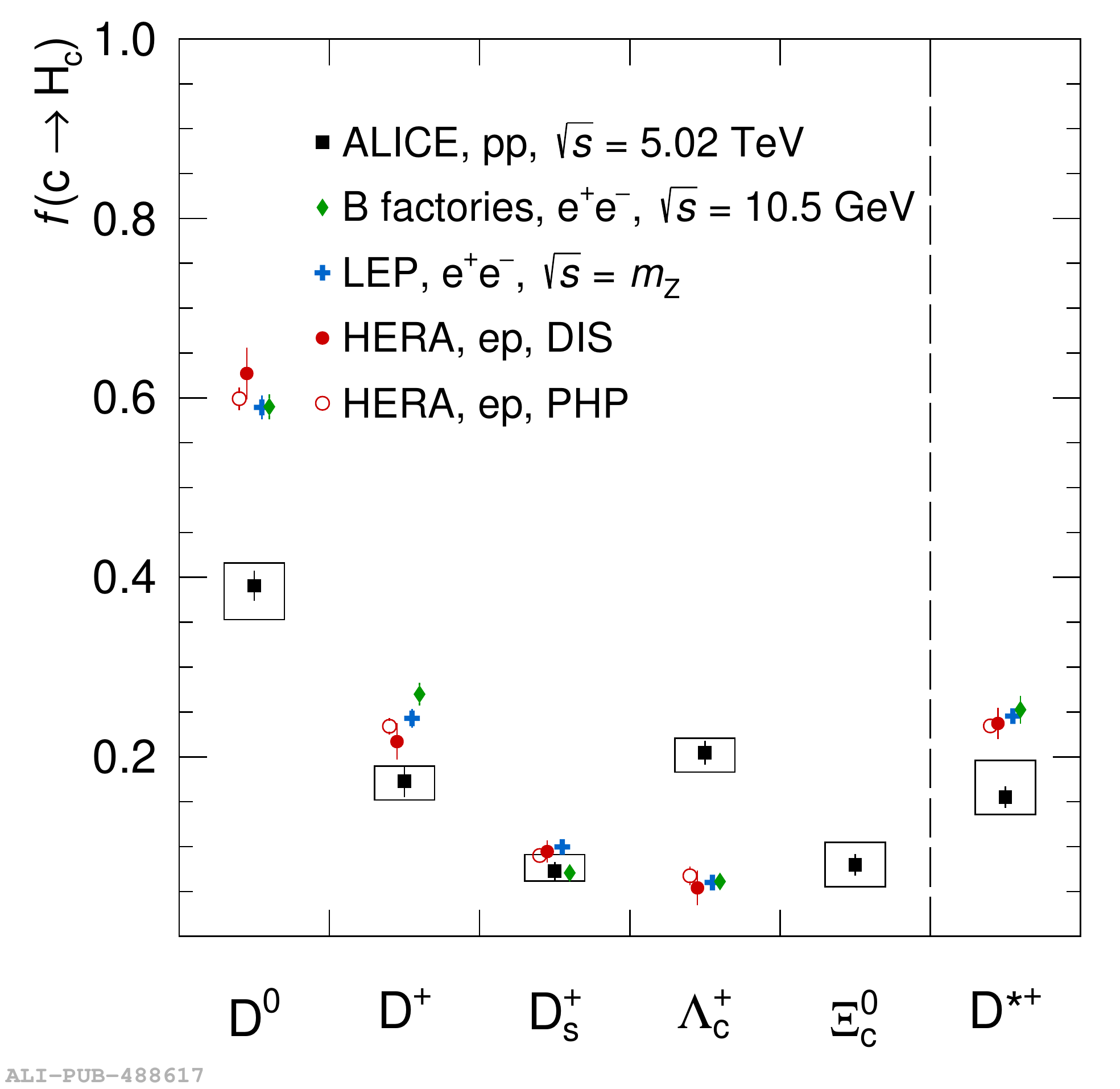}
    \includegraphics[width=0.5\textwidth]{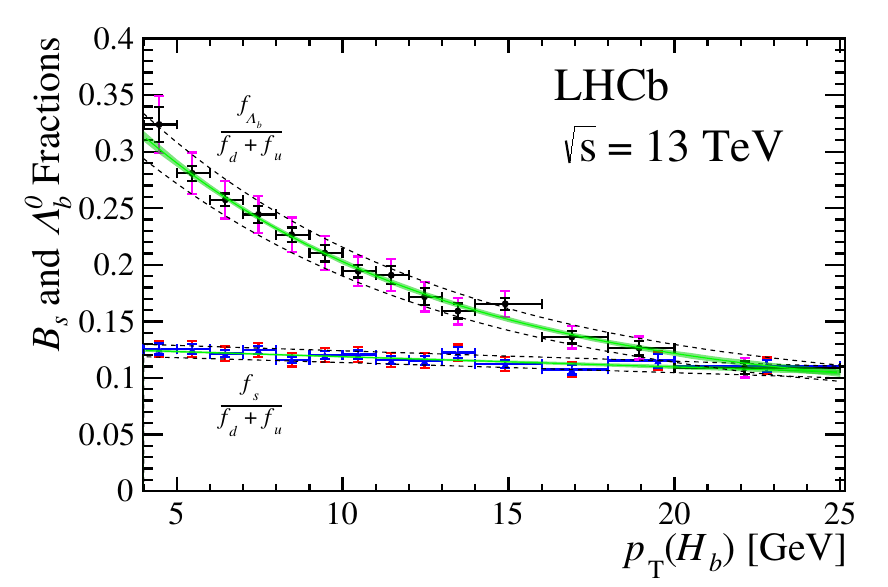}
    \caption{Left: hadronization fractions of charm into various hadrons at the LHC compared to measurements in $e^+e^-$ and $e$p collisions. Right: Fraction of $L_b$ over fraction of $b$ as a function of transverse momentum.}
    \label{fig:hadr_hf_pp}
\end{figure}

While the production of heavy-flavor mesons seems to fairly agree with the naive expectation from $e^+e^-$~collisions, the situation differes for heavy-flavor baryons and for quarkonia. There is a larger fraction of heavy quarks hadronize into baryons  than the naïve expectation from $e^+e^-$ in pp collisions at the LHC both in the case of charm~\cite{Acharya:2017kfy,ALICE:2017dja,ALICE:2020wla,ALICE:2020wfu,ALICE:2021psx,ALICE:2021dhb,ALICE:2021rzj} and beauty~\cite{LHCb:2011leg,LHCb:2014ofc,Aaij:2019pqz} quarks. For instance, the $p_T$-integrated charm hadron measurements by ALICE at $\sqrt{s}=$~5 TeV in pp~collisions~\cite{ALICE:2021dhb} shown in Figure~\ref{fig:hadr_hf_pp} indicate a large difference between the observed fraction of charm hadronizing in baryons compared to $e^+e^-$ and lepton-proton collisions. In addition, the $\Xi^0_{\rm c}$ ground state contributes significantly to the overall charm production. In pp collisions at $\sqrt{s}=$13 TeV, ALICE measured the states $\Sigma^{0,++}$~\cite{ALICE:2021rzj} that decay strongly to about 100~\% into $\Lambda_C$~\cite{Zyla:2020zbs}. This measurement provides further information on the source of the larger $\Lambda_c$-fraction. We note that the LHCb measurement of $\Lambda_c$~baryons at $\sqrt{s}=$7~TeV~\cite{Aaij:2013mga} shows a much smaller baryon over meson ratio than ALICE. Both measurements are not reconcilable within one fragmentation function set~\cite{Kniehl:2020szu}. Further experimental investigations are crucial to clarify the situation and provide a differential distribution of charm baryon production as a function of rapidity. 

The fraction of beauty baryons to the overall beauty production~\cite{LHCb:2011leg,LHCb:2014ofc,Aaij:2019pqz} rises towards low transverse momentum at forward rapidity as measured by LHCb. This indicates the dependence of beauty hadronization on the collision system as observed for charm production (see Figure~\ref{fig:hadr_hf_pp}). The first preliminary measurements of charm baryons as a function of multiplicity in pp~collisions have been shown by ALICE~\cite{Innocenti:2021nsi}. However, the current precision prevents conclusions concerning a dependence of the charm hadronization fractions as a function of multiplicity.  

The consistently observed larger fraction of heavy~quarks hadronizing in baryons compared to $e^+e^-$~collisions may not be surprising given the strong modifications of hadronization patterns observed for light-flavor hadrons. Therefore, it might be argued that the consideration of heavy-flavor hadron production in view of hadronization is not particularly interesting given the fact that the experimental data is scarcer and often less precise. However, in contrast to light-flavor hadrons, the charm and beauty quarks provide an additional handle due to their perturbative production down to low transverse momentum and heavy-flavor quantum number conserved in the strong interaction. Effectively, once they are produced, the heavy quark number charm and beauty are conserved during a time scale relevant to QGP physics. This conservation law is an additional constraint for the description. In addition, the relevance of small collision systems for the modeling might be questioned if the main focus is the investigation of QGP physics. In this respect, it is important to note that the duration of a hydrodynamic description in corresponding simulations, if applicable, becomes shorter in small systems, see, e.g., in~\cite{OrjuelaKoop:2015etn}, and hence, the modeling of heavy-flavor and jet observables become more sensitive to the initial state and the last stage of the collision process. Therefore, the modifications at a relatively small number of charged particles compared to central nucleus-nucleus collision constrain the hadronization modeling.

In order to describe the large baryon production fraction, different mechanisms are invoked within the different model categories: statistical hadronization treatments, event generators starting from $e^+e^-$~modeling, and coalescence models. Within Pythia, the introduction of color reconnections of the heavy quark with more than one other (usually light) quark in the event induces an increase of baryons in environments with larger local color density~\cite{Christiansen:2015yqa} naturally. However, the currently available parameter choices do not describe the $\Xi^0_{\rm c}$ yield~\cite{ALICE:2021psx}. The parameter set also influences the description of light-flavor production that is abundantly available, and they should be considered when aiming for a universal description of hadronization, which is not yet the standard for discussing experimental results.

The statistical hadronization model, successfully used for describing nucleus-nucleus collisions (grand-canonical for light-flavor and thermal weights for charm production constrained by the total charm from hard particle production), yields a $\Lambda_c/D^0$  ratio of 0.22~\cite{ANDRONIC2008149}. In order to match the value of around 0.5 observed by ALICE in pp~collisions, the additional introduction of larger mass excited charm baryons is discussed~\cite{He:2019tik,He:2019vgs}. These baryons, predicted in a relativistic quark model, produce the $\Lambda_C$-baryons decay products. This mechanism can serve as an explanation of the available charmed baryon measurements~\cite{He:2019tik,He:2019vgs}. However, the current model cannot describe the $\Xi^0_{\rm c}$ yield~\cite{ALICE:2021psx}.

A coalescence model approach that uses thermal weights for the relative abundance of the different charm baryons~\cite{Song:2018tpv} is not able to describe the $\Xi^0_{\rm c}$ production in~\cite{ALICE:2021psx}. A coalescence description acting at low transverse momentum combined with fragmentation at higher transverse momentum is also used to describe the baryon production in data~\cite{Minissale:2020bif}.

In summary, there is not yet a unified description of heavy-flavor hadronization patterns in pp~collisions, but there is a growing body of increasingly precise experimental data to test different hypotheses.

 Given the modified hadronization fraction into baryons in pp~collisions compared to $e^+e^-$~collisions, further modifications might be observed for different collision systems. Significant differences between pp to pPb collisions would need to be accounted for by constraints on nuclear parton distributions via nuclear modification factor measurements discussed in Chapter~\ref{sec:initial}. Therefore, heavy-flavor baryon measurements in pPb~collisions are crucial to establish whether we can use pPb nuclear modification factor data on single hadron species to constrain nPDFs. 
 
\begin{figure}
    \centering
    \includegraphics[width=0.48\textwidth]{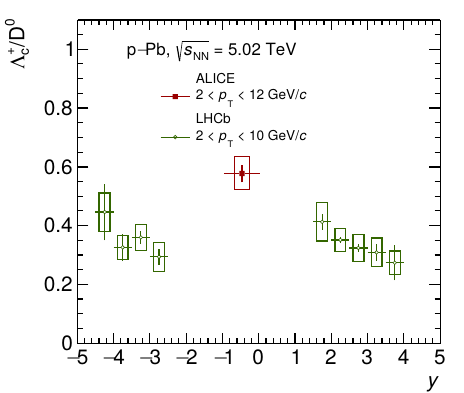}
    \includegraphics[width=0.44\textwidth]{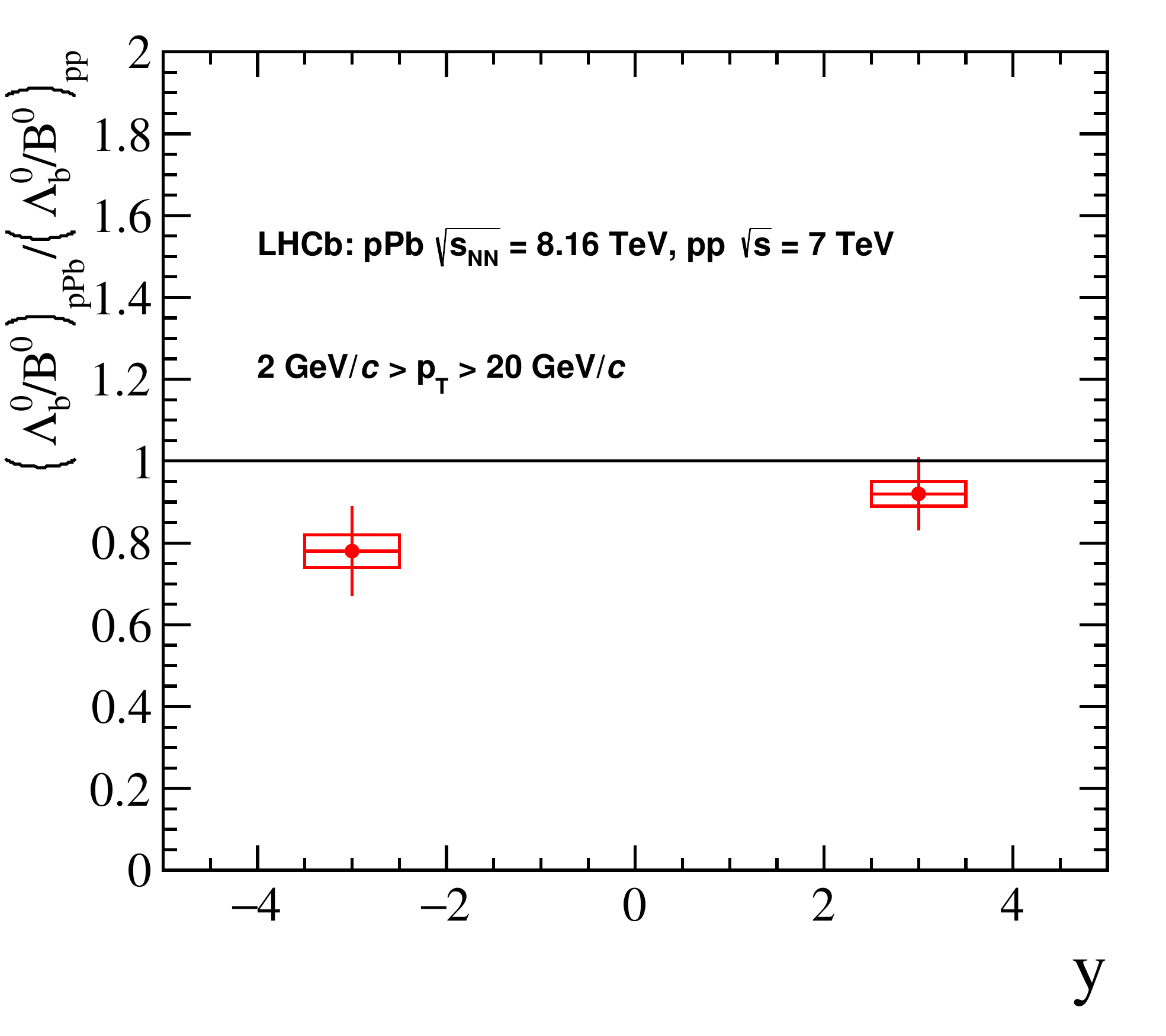}
    \caption{The measurements of the $\Lambda_c/D$~ratio measured by ALICE~\cite{ALICE:2020wfu,ALICE:2020wla} and LHCb~\cite{Aaij:2018iyy} taken from~\cite{ALICE:2020wla} and the double ratio of production cross section $\frac{\sigma_{\Lambda_B}/\sigma_{B}|_{pPb}}{\sigma_{\Lambda_B}/\sigma_{B}|_{pp}}$ measured by LHCb as a function of rapidity taken from~\cite{Aaij:2019lkm}.
    }
    \label{fig:baryons}
\end{figure}

Figure~\ref{fig:baryons} shows the $\Lambda_c^+/D^0$-ratio in pPb measured by ALICE~\cite{ALICE:2020wla,ALICE:2020wfu} and by LHCb~\cite{Aaij:2019lkm}. The measured $\Lambda_c/D$~ratio between the two experiments is closer than in pp~collisions but warrants future studies. The comparison between the baryon-to-meson ratio between pp and pPb by ALICE~\cite{ALICE:2020wla} reveals that the central value for the $p_T$~integrated $\Lambda_c/D$~ratio is slightly larger. However, both values are consistent within about 1~$\sigma$. The $p_T$~dependence appears slightly different between the two collision systems, but the precision does not allow for strong conclusions.
 
LHCb also compared the baryon-to-meson ratio in the beauty sector between pp and pPb collisions~\cite{Aaij:2019lkm}. The ratio of $\Lambda_b$ baryons in pp collisions and pPb is consistent with unity at forward rapidity and deviates from unity by about 20\% corresponding to $1.9 \sigma$ as shown in Fig.~\ref{fig:baryons}. 

Overall, the baryon fractions in pp and in pPb collisions are similar. The current measurements indicate the first hints for differences that will become evident with more detailed investigations.

In nucleus-nucleus collisions, measuring heavy-flavor baryons is challenging due to the sizeable combinatorial background, particularly at low transverse momenta. At RHIC, the STAR collaboration measured, in AuA~\cite{Adam:2019hpq} at $sqrt{s_{NN}}=200$~GeV, a $\Lambda_c/D$ ratio of 1.2 at 2.5~GeV/$c<p_T<3.5$~GeV/$c$ decreasing towards a central value of around 0.5 at larger $p_T$. The first measurements at the LHC in PbPb collisions at $\sqrt{s_{NN}}$=5.02~TeV were targeting larger transverse momenta: ALICE reported a large ratio of $\Lambda_c/D$~\cite{Acharya:2018ckj} of around 1 at $p_T$ 6-12~GeV/c, CMS of around 0.2 for $10$GeV$/c>p_T>20$GeV$/c$ 10 and 20~\cite{Sirunyan:2019fnc}. Recently, ALICE extended the $\Lambda_c$ measurement in PbPb collisions down to a transverse momentum of 1~GeV/$c$~\cite{ALICE:2021bib}. The data shows a sizeable transverse momentum dependence. The $\Lambda_c/D$ exhibits a value of around 0.4 at the lowest transverse momentum and a maximum of 1.1 in the range $4$~GeV$/c<p_T<6$~GeV/$c$ in most central collisions before decreasing again to 0.2 above 10~GeV$c$. An extrapolation down to 0~transverse momentum allows ALICE to provide a $p_T$~integrated $\Lambda_c/D$~ratio of 0.48$^{+0.13}_{-0.12}$. The measurements point to a larger transverse momentum integrated baryon-to-meson ratio compared to $e^+e^-$~fragmentation expectation as observed in pp and pPb. The transverse momentum-dependent measurements down to the lowest transverse momentum provide a rich source for model constraints.

In addition to the ratio of baryon-to-meson production, the investigation of heavy-flavor hadrons with a $s$-valence quark is interesting to check any differences with respect to the observations in pp collisions. It also provides additional test concepts of heavy-quark hadronization from a equilibrated strangeness QGP in central nucleus-nucleus collisions. The ratio of $D_s$ over $D^0$-meson production has been investigated by ALICE~\cite{ALICE:2015dry,ALICE:2018lyv} and STAR~\cite{STAR:2021tte}. The measurements show a larger fraction of heavy-quark mesons appearing as a state with a strange valence quark in nucleus-nucleus collisions than in pp and in $e^+e^-$~collisions as predicted early by statistical model calculations~\cite{Andronic:2003zv,Kuznetsova:2006bh} and the resonance recombination model~\cite{He:2012df}. In the b-hadron sector, CMS released first measurements of $B_s$ compared to $B^+$~\cite{CMS:2018eso,CMS:2021mzx}. The central values of the $B_s$-to-$B^+$-ratio are 2-4 times larger in PbPb than in pp collisions, but the large uncertainties prevent us from excluding the pp~value. At the present stage, different type of models, combined coalescence-fragmentation models~\cite{Plumari:2017ntm}, resonance recombination model with fragmentation~\cite{He:2019vgs}, a 4-momentum conserving coalescence model combined with fragmentation~\cite{Cao:2019iqs} as well as statistical approaches~\cite{Andronic:2021erx}, can account for the $D_s$ observations in central nucleus-nucleus collisions. 

So far, we discussed heavy-flavor hadrons with one heavy quark, open heavy-flavor hadrons. The investigation of quarkonium production in view of hadronization is very interesting.  
In high-energy $e^+e^-$ collisions, quarkonium production is very far from a thermal description in contrast to open heavy flavor hadrons. The prompt J/$\psi$ production at the Z-peak~\footnote{The so-called non-prompt contribution is dominating and stemming from weak decays of beauty-hadrons.}~\cite{L3:1998ndc} is underpredicted in a canonical statistical model by nearly two orders of magnitude~\cite{Andronic:2009sv}. In contrast, the $\Lambda_c/D$~ratio is well described within the same model.

\begin{figure}
    \centering
    \includegraphics[width=0.49\textwidth]{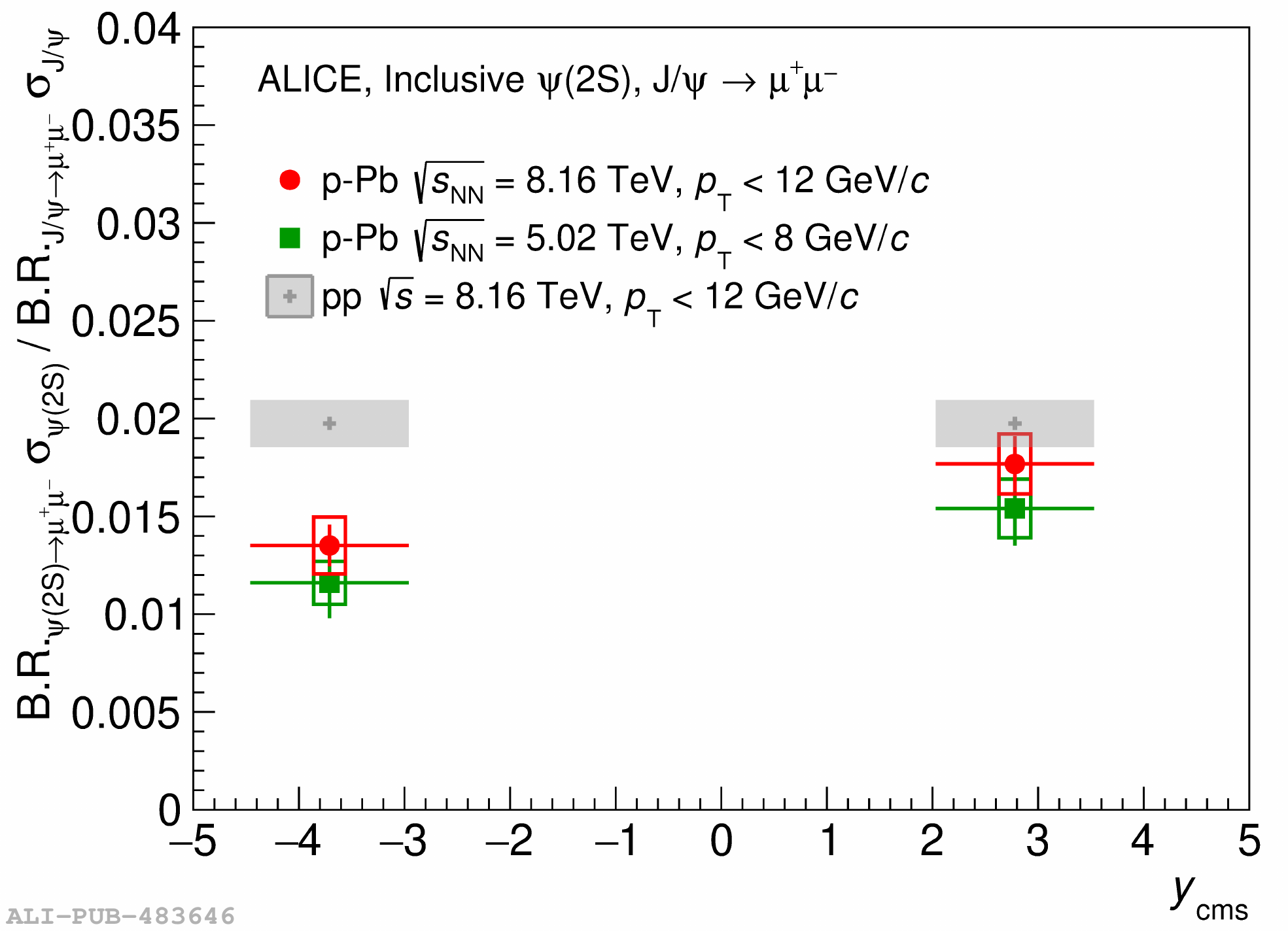}
    \includegraphics[width=0.49\textwidth]{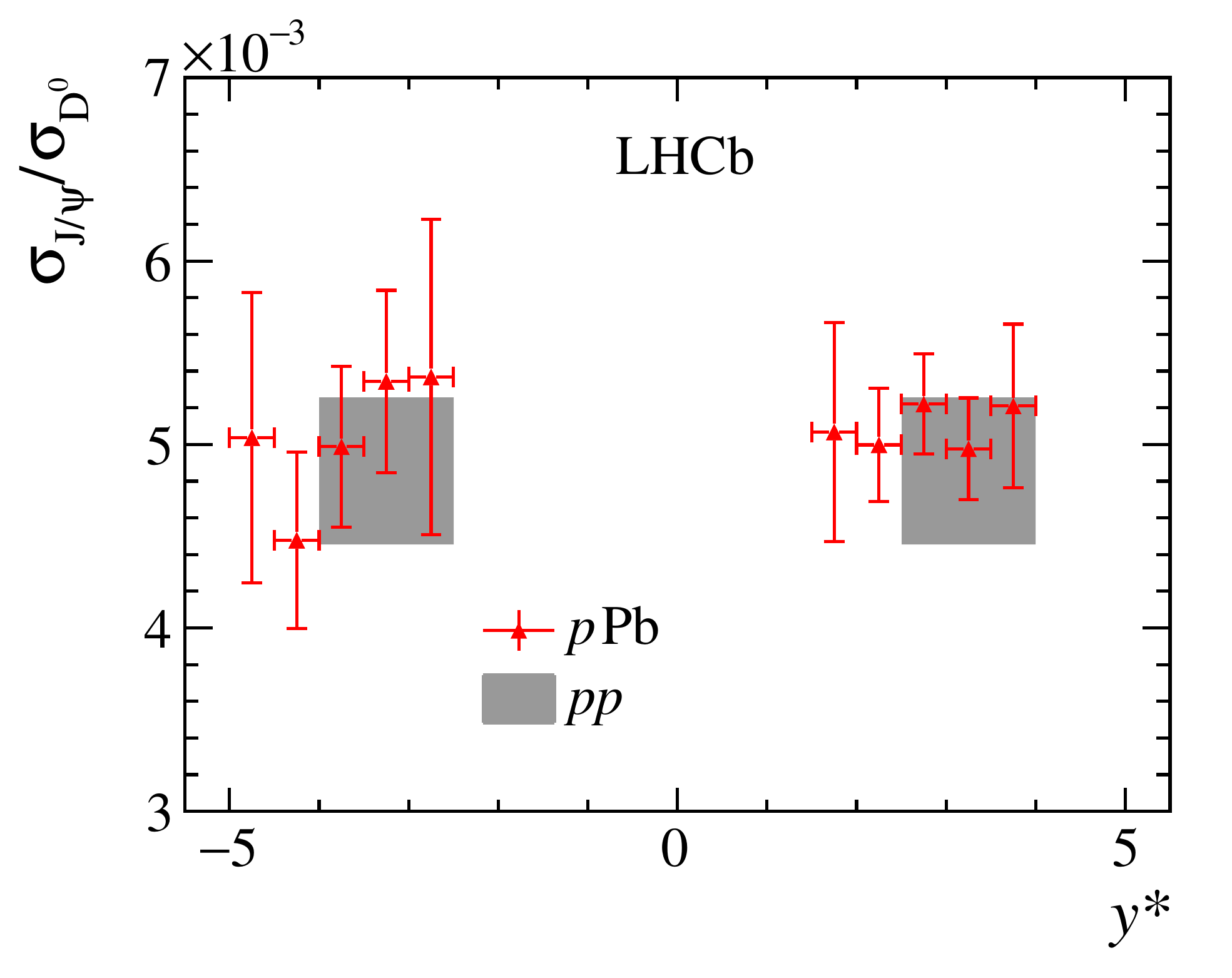}
    \caption{$\psi$(2S)/J/$\psi$~ratio in pPb collisions measured by ALICE~\cite{ALICE:2020vjy}, J/$\psi$/D ratio in pPb compared with the same ratio in pp collisions by LHCb~\cite{LHCb:2017yua}.
    \label{fig:doubleratio}}
\end{figure}

The description of inclusive quarkonium production in pp~collisions has progressed significantly in recent years~\cite{Lansberg:2019adr}. All of the considered approaches assume factorization of the heavy-quark-antiquark pair production with respect to the final state hadron formation. However, they do not consider the characteristics of the hadronic environment. 
A subtle breaking of this factorization with respect to the final state has been found when comparing the $\psi$(2S) to J/$\psi$ production ratio in proton-proton with proton-lead collisions~\cite{PHENIX:2013pmn,Abelev:2014zpa,LHCb:2016vqr,PHENIX:2016vmz,Adam:2016ohd,Aaboud:2017cif,Sirunyan:2018pse,Acharya:2020wwy,Acharya:2020rvc} as well as for $\Upsilon$~\cite{Aaboud:2017cif,Aaij:2018scz,Acharya:2019lqc}, most notably at low transverse momentum and on the nuclear fragmentation side shown in Fig.~\ref{fig:doubleratio}. However, the ratio of J/$\psi$ over D-meson production is consistent, within uncertainties, between the two collision systems. This observation indicates that the excited state nuclear modification does not strongly impact the inclusive J/$\psi$~production, although the latter includes the feed-down from $\psi$(2S) and $\chi_{c1,2}$~\cite{Aaij:2017gcy}. A first measurement of $\chi_{c1}/\chi_{c2}$ ratio has been performed by LHCb in pPb~collisions, where current statistical uncertainty prevents from strong conclusions~\cite{Aaij:2021wfo}. 

The most discussed explanations for the excited state quarkonium production modification with respect to the ground states are rescattering phenomena, either of partonic or hadronic nature~\cite{Ferreiro:2014bia,Du:2015wha,Chen:2016dke,Ma:2017rsu,Ferreiro:2018wbd}, as well as applying transport equation approaches~\cite{Liu:2013via,Du:2018wsj} as employed in nucleus-nucleus collisions. In order to investigate these phenomena, ratios or self-normalized double ratios of different states have been measured in pp~collisions as a function of charged tracks in the event or forward energy deposition~\cite{Chatrchyan:2013nza,Sirunyan:2020zzb}. 
In summary, the investigation of excited quarkonia in proton-nucleus and high-multiplicity pp~collisions shows that already relatively small charged particle multiplicities lead to a violation of a factorized ansatz of the bound state production. However, a precise physical picture of the involved processes can not yet be extracted. 

After the heavy-flavor bound state production, we discuss hadronization in jets. A first measurement of the baryon-to-meson ratio $\Lambda/K^0_S$ in charged jets in pPb and pp~collisions above $p^{ch}_{T,jet}>10$~GeV/$c$~\cite{ALICE:2021cvd} has been performed. It indicates that the particle ratio differs from the one measured in the underlying event and the inclusive production at the same transverse momentum. The observed values are well within the expectation from the Pythia generator. Preliminary data by ALICE suggest a similar behavior in PbPb collisions~\cite{KUCERA2016181}. Hence, the jet provides a hard scale, and its core seems to hadronize independently of its environment according to this first measurement. This observation may open up the possibility of using the 'chemistry', i.e., the observed particle abundance, for tagging. Consequentially, the investigation of the jet-medium response could be supported by studies of the particle species abundances. Furthermore, one may address whether the jet equilibrates synchronous in terms of energy-momentum, kinematically, and its particle species, chemically. The interface between modern jet physics theory and hadronization modeling would need further development to realize these ideas. Given heavy-ion-inspired modeling, a hybrid approach between Lund-string hadronization and coalescence is under development for jet hadronization~\cite{Fries:2019vws}.

The presented measurements are a starting point for better modeling of hadronization. This is not only required to understand the transition between different collision systems. It is also necessary for precision QGP studies involving hadronic final states that are not in the limiting cases of Fig.~\ref{fig:hadronizationcartoon}, i.e., full thermalization or modelization based on $e^+e^-$~collisions. The domain where equilibrium is not yet fully reached is most interesting for extracting the internal structure of QGP beyond the determination of 'integrated' properties. As we argued, these non-thermalized modes provide a window into the QGP. Heavy-flavor hadrons are particularly interesting in this context due to their conservation during the QGP lifetime, tagging effectively well-identified quarks. They are, therefore, a focus of future facilities as discussed in~\ref{sec:outlook}.

\subsection{Heavy-flavor hadronization as a probe of hadron structure}
\label{subsec:hadstruc}

Since the discovery of the $X(3872)$-state in 2003~\cite{PhysRevLett.91.262001}, a large number of mesonic and baryonic 'states' containing heavy charm and beauty quarks have been discovered in $e^+e^-$ and pp collisions that {\it do not} fit into the conventional classification of $q\bar{q}$ mesons and $qqq$ baryons within quark models in terms of mass and quantum numbers, see for a review~\cite{Lebed:2016hpi}.   
For these states, different descriptions have been proposed: compact objects with non-conventional valence quark composition as tetra and pentaquark states, molecules of mesons, baryons, loosely bound states built from colored objects, and effects at kinematic thresholds. Quantum mechanical superposition leads to the superposition of different scenarios. In the case of the most precisely measured state $X(3872)$, the quantum numbers have been experimentally determined by CDF~\cite{CDF:2006ocq} and LHCb~\cite{LHCb:2013kgk}. However,  the discussion of the precise nature of the state is actively ongoing.  

Prompt production of $X(3872)$, i.e., not as the b-hadron decay products, but originating directly from the primary vertex, has been measured at Tevatron~\cite{Bauer:2004bc} and the LHC~\cite{Chatrchyan:2013cld,LHCb:2021ten}. Its prompt production has been discussed as a constraint on the nature of the state~\cite{Suzuki:2005ha,Bignamini:2009sk,Artoisenet:2009wk}. 

The investigation of these states in the experimentally challenging environment of heavy-ion collisions and as a function of charged particle multiplicity in pp and pPb collisions may appear as an additional complication. However, it introduces a new, comparatively large length scale, the size of the created QCD~matter, and the hadronization concepts from QGP physics. In this case, we investigate a thermodynamic system that hadronizes in equilibrium at large particle multiplicity  for light-flavor hadrons. The ¨size¨ of this system that can be dialed in experimentally in a first approximation via the number of final state particles: the 3-volume of the thermodynamic system at the moment of decoupling within a hydrodynamic description scales with the charged particles measured in the final state. In addition, if rescattering plays an important role in the production yield and depends on the object's size, its impact may also be controlled via the number of produced particles in the vicinity of the exotic particle. Therefore, the measurement of the exotic states in heavy-ion collisions and as a function of charged particle multiplicity in nucleus-nucleus collisions can give us access to the nature of the state within a hadronization ansatz that depends on the size of the hadron. Furthermore, suppose the thermal particle production is applicable~\footnote{This has been argued to be the case based on the assumption that a compact multi-quark object is formed at the phase boundary evolving to the final wave function unperturbed to first approximation~\cite{Braun-Munzinger:2018hat}.}. In that case, the existence of new states can be tested~\cite{Andronic:2017pug} in heavy-ion collisions where the particle density is given essentially by the mass and quantum numbers of the particle and the freeze-out temperature, which offers different production probabilities than other production modes. 

\begin{figure}
    \centering
    \includegraphics[width=0.5\textwidth]{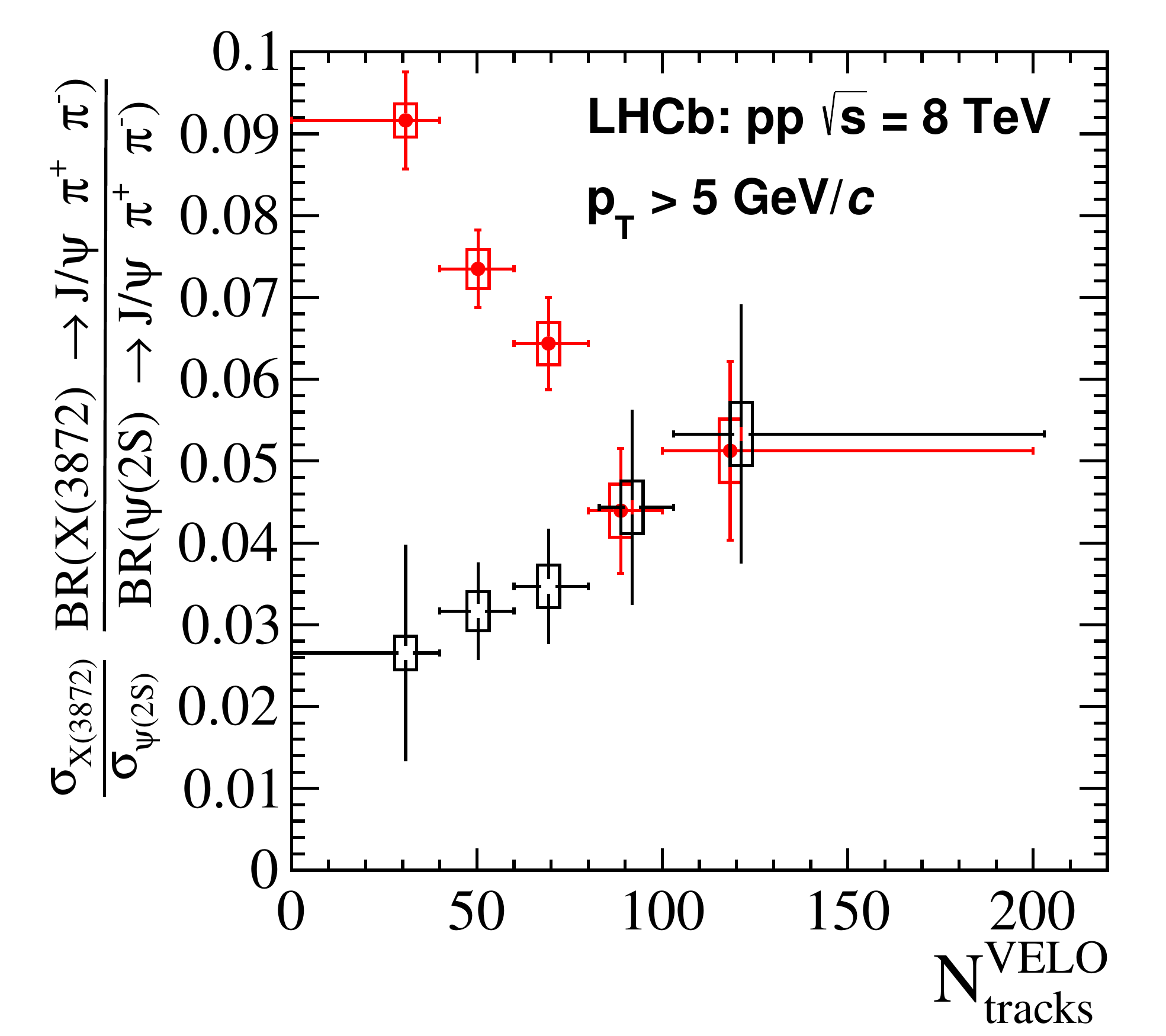}
    \includegraphics[width=0.4\textwidth]{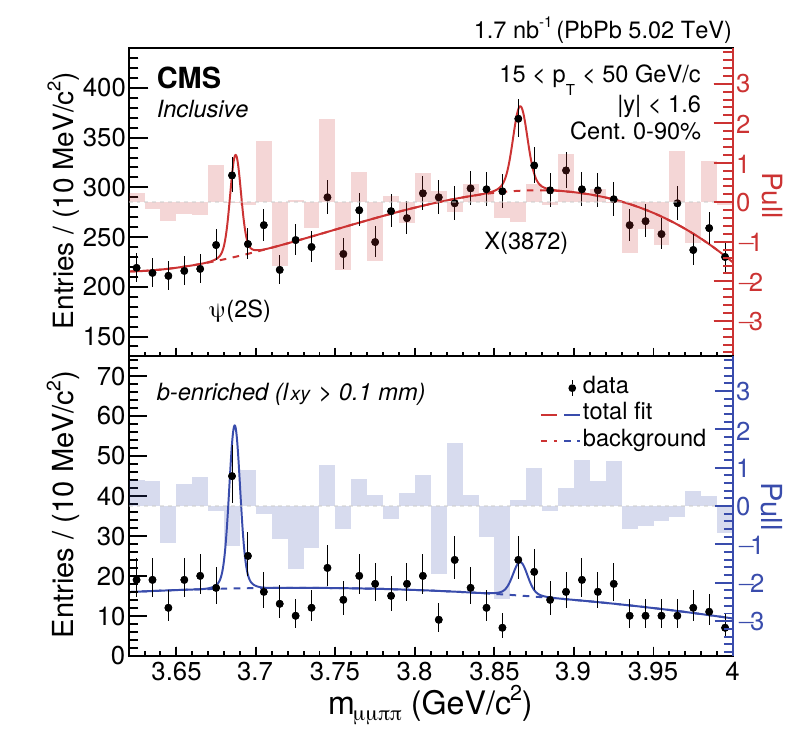}
    \caption{$X(3872)$ measurements: Left: LHCb measurement of prompt and non-prompt component as a function of multiplicity compared to comover model; right: signal extraction of inclusive and b-enriched X(3872) by CMS in PbPb collisions.}
    \label{fig:Xandpsi2s}
\end{figure}

The first measurement of $X(3872)$ production has been performed by the LHCb collaboration as a function of charged particle multiplicity~\cite{Aaij:2020hpf} in pp~collisions. The result is expressed with the ratio $\sigma(X(3872))\times Br(X(3872) \to J/\psi \pi^+ \pi^-)/(\sigma(\psi(2S))\times Br(\psi(2S) \to J/\psi \pi^+ \pi^-)$ and it is shown in Fig.~\ref{fig:Xandpsi2s} as a function of multiplicity. First theoretical calculations treat the LHCb observation in terms of rescattering and deduce a very strong sensitivity of the measurement to the type of hadron favoring a rather compact object~\cite{Esposito:2020ywk}. The model assumptions have been criticized in ~\cite{Braaten:2020iqw} and further theoretical work will be required.

CMS released the first measurement of $X(3872)$ production in nucleus-nucleus collisions~\cite{CMS:2021qpw}, where the signal extraction is shown in Fig.~\ref{fig:Xandpsi2s}. The two peaks show clearly that a significant prompt production can be observed. The central value of the ratio of $N(X(3872))\times BR(X(3872) \to J/\psi \pi^+ \pi^-) $ $/(N(\psi(2S))\times BR(\psi(2S) \to J/\psi \pi^+ \pi^-)$ measured at $15<p_T<50$GeV/$c$ is with $1.08\pm 0.49(stat.) \pm 0.52 (syst.)$ surprisingly large. Although the large uncertainties prevent strong conclusions, the evidence of prompt X(3872) signal is intriguing and hints that an additional mechanism other than comoving particle-induced dissociation is in play. Currently, the theoretical models vary wildly in their predictions. However, they also indicate a strong sensitivity to the type of state within their model assumptions~\cite{Wu:2020zbx,Zhang:2020dwn}. In the case of a molecular interpretation of the X(3872), its production can be put into perspective with respect to weakly bound nuclei as deuteron and hypertriton are also produced abundantly in heavy-ion collisions and measured from pp to nucleus-nucleus collisions at RHIC and the LHC, see for a recent review on LHC measurements~\cite{Braun-Munzinger:2018hat}. Given a large amount of available and upcoming data, a clarification of the appropriate picture of the production mechanism~\cite{Andronic:2017pug,Bellini:2020cbj} is in sight, and it can be then applied to QCD exotica. In the case of the $X(3872)$, three ingredients can be identified: precise measurement of the absolute branching fraction of the X(3872) in the $J/\psi+pipi$ final state, the clarification of the different coalescence model calculations yielding to very different results and an improvement of the experimental precision. Furthermore, the field of femtoscopic correlations, currently mainly exploited in the strange-quark sector~\cite{Acharya:2020asf}, can be applied to the heavy-flavor sector and may teach about the interaction of charm hadrons with other hadrons giving insights into the underlying forces of exotic hadron states. A first femtoscopic measurement involving a charm hadron has been recently presented~\cite{ALICE:2022enj}. 
These first measurements of ALICE, LHCb and CMS are a preview of what will be possible with larger luminosities and future data takings at RHIC and the LHC.

\section{Jets and heavy-flavor in small collision systems, collectivity and energy loss}
\label{sec:smallsystem}

In the previous chapter~\ref{sec:hadronization}, we discussed hadronization measurements in different collision systems. We observe modifications in the hadronization patterns at large charged-particle multiplicity in pp and in pPb~collisions that resemble results in nucleus-nucleus collisions. This experimental finding provokes the question of whether the key QGP signatures, collective expansion~\cite{PhysRevD.27.140}, and the energy loss of high momentum particles~\cite{Bjorken:1982tu}, are observed in collision systems other than nucleus-nucleus collisions.
By searching for these signatures, we may learn the nature of the produced system in high multiplicity pp and pPb collisions and whether QGP physics concepts can be applied in these cases. 
This chapter addresses the contribution of heavy-flavor and jets observables to these conceptual questions.

We discuss measurements of azimuthal anisotropies of heavy-flavor particles in section~\ref{small:azim}. Finally, we review the search status for partonic energy loss in small collision systems in section~\ref{small:loss}.

\subsection{Azimuthal anisotropies of heavy-flavor production}
\label{small:azim}

A long-range azimuthal angle correlation between charged particles in high multiplicity has been observed in pp collisions by CMS with the first LHC data~\cite{CMS:2010ifv}. This observation has been confirmed in various small collision systems at the LHC and RHIC~\cite{CMS:2012qk,ALICE:2012eyl,ALICE:2013snk,PHENIX:2013ktj,ATLAS:2012cix,ATLAS:2014qaj,CMS:2015yux,PHENIX:2015idk,LHCb:2015coe,PHENIX:2018lia,STAR:2019zaf,ALICE:2019zfl}. Based on the comparison of the experimental data with model calculations, 
the paradigm has been established that in pPb collisions, but potentially also in high-multiplicity pp, coordinate-space geometry fluctuations are transformed into final state momentum anisotropies. This transformation requires final state interactions that can be realized either in a strongly coupled system described by hydrodynamics or in a dilute parton transport model, see for a discussion in~\cite{Nagle:2018nvi}. Initial state momentum-space correlations can potentially also play a role, see for a review in~\cite{Schlichting:2016sqo}. 

Beyond the investigation of hadron-hadron collisions, long-range azimuthal correlations have been searched for in $e^+e^-$ or $e$-hadron collisions. No clear signature is found neither in $e^+e^-$ in jet production at the Z~resonance with ALEPH~\cite{Badea:2019vey} or at collision energies around the $\Upsilon(4S)$~resonance with Belle~\cite{Belle:2020mdh,Belle:2022fvl,Belle:2022ars}. Similarly, no signature of collective expansion has been found in deep inelastic $ep$~collisions at $\sqrt{s}=318$~GeV at HERA with ZEUS~\cite{ZEUS:2019jya}. However,  two-particle long-range azimuthal correlations have been seen in photonuclear PbPb collisions at the LHC with ATLAS~\cite{ATLAS:2021jhn} after a template background subtraction.

\begin{figure}
    \centering
    \includegraphics[width=0.5\textwidth]{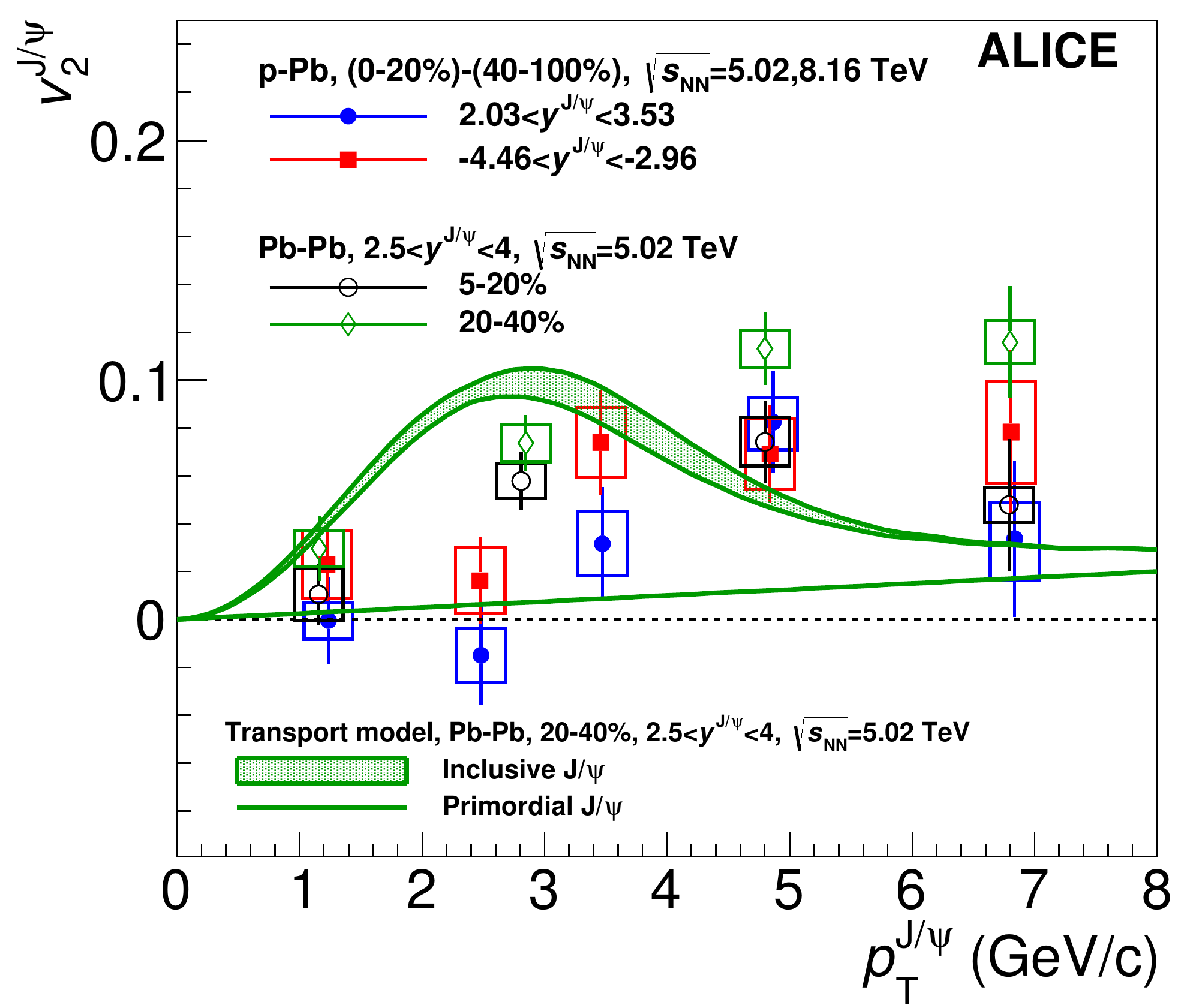}
    \caption{Measurement of J/$\psi$-$v_2$ measured by ALICE in p-Pb collisions and in PbPb collisions compared to a model calculation. The figure is adapted from original publication~\cite{Acharya:2017tfn}.  }
    \label{fig:sec_coll_v2jpsi}
\end{figure}

The investigation of azimuthal correlations for heavy-flavor hadrons in small collision systems is interesting since the relaxation time is comparable to the QGP lifetime already in nucleus-nucleus collisions~\cite{Rapp:2018qla}. This circumstance poses the question to which extent they show signs of final-state interactions in smaller collision systems. Most prominently, heavy-flavor azimuthal correlations, both based on leptons from semileptonic decays~\cite{Acharya:2018dxy,Adam:2015bka} as well as from fully reconstructed hadronic final state, have been experimentally observed~\cite{Sirunyan:2018toe,Sirunyan:2020obi}. 
Furthermore, ALICE collaboration found a sizeable long-range azimuthal correlation between the J/$\psi$ and charged particles in pPb collisions~\cite{Acharya:2017tfn}, after employing a subtraction of the low charged-particle multiplicity correlation. This observation was confirmed by CMS in a slightly different kinematic domain~\cite{Sirunyan:2018kiz}, see Fig.~\ref{fig:sec_coll_v2jpsi}. In pp collisions, muons from charm decays were observed with sizeable azimuthal long-range correlation after a background template subtraction, whereas muons from beauty showed no significant $v_2$ within similar experimental uncertainties~\cite{Aad:2019aol}. 

The large second Fourier coefficient $v_2$ of the $J/\psi$~\cite{Acharya:2017tfn,Sirunyan:2018kiz} cannot be accounted for in models that describe well the nuclear modification and $v_2$ of $J/\psi$ in PbPb measurements, see e.g., in~\cite{Du:2018wsj}. It is difficult to conceive a mechanism to provide such large azimuthal anisotropies in small collision systems~\cite{Du:2018wsj}. So far, no plausible explanation within transport models applied to nucleus-nucleus collisions has been found. Models that assume that the source of the correlation originates from the initial-state wave function in momentum space can account for the large azimuthal asymmetry~\cite{Zhang:2019dth}. However, this type of set-up cannot describe light-flavor hadrons measurements~\cite{Nagle:2018nvi}. 

These heavy-flavor anisotropy puzzles show that the basic mechanisms behind the experimental observations are not yet fully understood. To progress towards a quantitative understanding, the current modeling of exclusive event description must accurately reproduce the experimental observations of non-related background phenomena, commonly called non-flow contributions.

\subsection{Search for energy loss in small systems and peripheral collisions}
\label{small:loss}

In contrast to long-range azimuthal correlations, jet quenching has not been observed in small collision systems. However, all jet quenching models predict a, possibly small, signal of energy loss in case a droplet of Quark-Gluon Plasma is formed. The observation or the setting of clear limits of this signature is necessary to form a coherent view of high-energy collisions with hadronic final states across collision systems. So far, this has not been achieved, and we comment on the encountered problems~\footnote{It is interesting to compare this onset question with the situation at lower beam energies measured by the STAR collaboration, where the high transverse momentum hadron suppression associated with jet quenching is overcome by other effects for collision energies below $\sqrt{s_{NN}}=39$~GeV in AuAu collisions~\cite{STAR:2017ieb}. The low and high collision energy systems are very different, particularly the estimated initial energy density. However, the diagnosis of jet quenching via model calculations in both systems suffers from the disparity between the unclear description of the 'bulk' in these collision systems and the idealized sequential space-time picture with a long-lived hydrodynamic phase in central heavy-ion collisions at large collision energies.}. 

Currently, no sign of jet quenching has been observed in pPb collisions, based on high-transverse momentum particle suppression in single hadron spectra compared to pp collisions~\cite{ALICE:2012mj,ALICE:2014nqx,ALICE:2014xsp,CMS:2015ved,ATLAS:2016xpn} or dijet momentum imbalance~\cite{Chatrchyan:2014hqa}.
Two conceptual problems address parton energy loss in collisions involving light nuclei and protons as initial states. First, the predicted effects are relatively small due to the space-time evolution of the created system, see for a compilation of different collision systems with a simple geometric set-up in~\cite{Huss:2020whe}. The second difficulty is selecting specific collision geometries in small collision systems. The most striking jet quenching modifications in nucleus-nucleus collisions are observed by selecting events based on centrality class (impact parameter). Experimentally, the centrality definition relies on a correlation between the impact parameter with the final state particle multiplicity. However, this correlation is very wide in small collision systems and sensitive to correlations and effects unrelated to the geometry of the collision, making it challenging to experimentally select collisions with small impact parameters, see, for a study in pPb collisions,~\cite{Adam:2014qja}. In peripheral nucleus-nucleus collisions, the nuclear modification factor is also sensitive to biases~\cite{Loizides:2017sqq} that are not accounted for in the observable definition of peripheral collisions. In order to avoid these problems, one could use perturbative QCD calculations as the cross-section reference and concentrate on centrality integrated properties. However, this is not viable given the small size of the expected effects in proton-proton and proton-nucleus collisions, more precisely, of the order of 10\% for impact parameter averaged collisions in~\cite{Huss:2020whe} for pPb collisions. One should also consider the precision limits of perturbative calculations due to the non-perturbative input given by the initial state and hadronization. While uncertainties from the limits of perturbative QCD applicability and deviations from universal hadronization are expected to be only relevant at relatively low momenta, the nuclear modifications of nuclear PDFs are relevant at high momentum and at the 10\% precision level. As we discussed in section~\ref{sec:initial}, the modification of dijet kinematics in pPb collisions is used for constraints on nuclear PDFs. 

In specific kinematic regimes and collision systems, energy loss in 'cold nuclear matter', i.e., the modification of spectra due to radiation when traversing the hadron, can also be phenomenologically relevant. It can be an additional background source for either search for partonic energy loss or constraints for nuclear parton distribution functions. In proton-nucleus collisions, coherent energy loss interference between the initial state and final state radiation from small-angle scattering within the nucleus is suggested to play an important role at forward rapidity for quarkonium~\cite{Arleo:2012rs}, open heavy-flavor~\cite{Arleo:2021bpv} and light-flavor hadrons~\cite{Arleo:2020hat}. The average medium-induced energy loss in this mechanism has a transverse mass dependence given 1/$M_T$, see for a good comparison with large-angle scattering in~\cite{Arleo:2010rb}. The associated transport coefficient $\hat{q}$ amounts to O(0.07)~GeV/$c$ extracted from quarkonium~\cite{Arleo:2020hat,Arleo:2021bpv} and the effect is parametrically at transverse momenta above 10 GeV around midrapidity. Another possibility to circumvent the problem is the usage of coincidence measurements of a high transverse momentum track, and a jet~\cite{ALICE:2015mdb} in pPb collisions, where the observation is consistent with no energy loss and the translation into a quantitative limit is model assumption dependent.

In order to search for partonic energy loss in small collision systems, an OO~run has been proposed that features more significant jet quenching effects compared to pp and proton-lead collisions due to the different nuclear geometry at similar charged particle multiplicity compared to pp and pPb~collisions~\cite{Citron:2018lsq,Brewer:2021kiv}. 

%% file: time.tex
\section{Time scale of QGP evolution}
\label{sec:timescale}

QGP studies using hard probes have brought substantial qualitative breakthroughs. The continuous progress towards the theoretical description of in-medium effects and the experimental control over the high multiplicity background allowed us to identify several novel phenomena related to the different wavelength resolution probing probes. Examples like decoherence effects induced by multiple scatterings or the presence of a jet-induced medium response component are novel findings regarding initial predictions. Despite current theoretical uncertainties, a first approach toward the quantitative characterization of average QGP properties was also possible. Nonetheless, the capabilities of hard probes to provide more insightful information about the inner QGP workings are far from being fully explored. We hope to learn about the QGP timescales, bridging the independent inputs from soft and hard probes. As seen in chapter~\ref{sec:QGPDensity}, the uncertainties related to the onset of quenching effects and QGP evolution dominate our estimates of jet transport coefficients. The following steps for the future are clear, but the path ahead still brings difficulties that must be surpassed. In this chapter, we will try to critically assess the unexplored potentials of using high momentum objects as unique probes to study the time and hence the temperature dependence of QGP properties. We illustrate possible exploration avenues that recent works have put forward. 

Hard probes are initially produced in a hard scattering and thus traverse the full extent of matter created during the heavy-ion collision. In contrast to soft probes, low momentum objects resulting from the fireball expansion, hard probes do not provide only a final picture of the process. They develop within the fast-evolving medium and thus can be used as unique objects to measure time-dependent QGP quantities. These can be separated into the following:
\begin{itemize}
	\item Quarkonia: bound states of two heavy-quarks whose mass state dictates if the wave function is formed at early or later times of the medium expansion.
	\item Open heavy-flavor particles: hadronization of a heavy-quark produced in the hard scattering and that experience the full medium evolution; 
	\item Jets: Result of the fragmentation of a parton produced during the high-momentum transfer process. 
\end{itemize}

The initial idea to use quarkonia states as a hot and dense QCD matter probe was straightforward~\cite{Matsui:1986dk}. In vacuum, heavy bound-states can be described by the non-relativistic Schroedinger equation, whose potential has the typical contributions from the long and short-range distance behavior:
\begin{equation}
V(r) = \sigma r - \frac{\alpha}{r} \, .
\end{equation}
In a static medium, the following ansatz has been used early on~\cite{Karsch:1987pv}:
\begin{equation}
V(r) = - \frac{\alpha}{r} \text{e}^{- \mu (T) r} + \frac{\sigma}{\mu(T)} (1-\text{e}^{- \mu(T) r } ) \approx - \frac{\alpha}{r} \text{e}^{\mu (T) r} \, 
\end{equation}
where the Debye screening radius in a static medium at temperature $T$ is given by $r_D (T)= \mu(T)^{-1}$ and the last equality approximation holds for when the temperature of the medium is much larger than the critical temperature $T > T_C$ (i.e., a QGP phase exists). Because the Debye mass is proportional to the medium temperature, $\mu \simeq g T$, the Debye radius decreases with increasing temperature. Different bound states have different dissociation temperatures. As such, a sequential \textit{melting} pattern would appear. One could then use the resulting yield of quarkonia species as a \textit{thermometer} of the produced medium. This original idea~\cite{Matsui:1986dk} has been strongly modified based on insights in lattice QCD and effective field theory allowing us to address the temperature range in the transition region between hadrons and QGP, see ref.~\cite{Rothkopf:2019ipj}. In addition to this real potential, it has been realized that there is an imaginary component of the quark-antiquark potential~\cite{Laine:2006ns,PhysRevD.78.014017} leading to quarkonium dissociation in the thermal heat bath as discussed in Chapter~\ref{sec:acceleration}. However, there is an immediate inherent difficulty for the direct applicability of the potential ansatz: quarkonia has to reach thermal equilibrium with QGP in order to use the knowledge extracted from lattice QCD calculations that determine the in-medium properties of quarkonia states. Given the fast-evolving medium's time scale, an additional time dependence on the binding potential is expected. If the QGP evolution is adiabatic, the potential evolves slowly, and a system at a given eigenstate would remain in the same eigenstate. However, if the potential evolves rapidly, the system cannot accommodate such changes, resulting in reshuffling a given quarkonium state. An adiabatic condition holds if the timescale related to the energy gap of the bound state, $\tau$, is much smaller than the timescale associated with the medium evolution, $\tau_m$. This is the case, for instance, of bottomonia states. However, other channels lie above this dissociation threshold, and a reshuffle is expected. The rapid evolution of QGP blurs the simple sequential-melting picture, and the role of quarkonia as thermometers is questionable. To overcome this difficulty, progress towards using approaches based on open quantum systems or effective time-dependent potential has been pursued more recently and is reviewed in~\cite{Rothkopf:2019ipj}. An intuitive picture for bottomonium modification in this theoretical ansatz is to see QGP as a sieve acting on the quark-antiquark pair traversing the medium, for which references are given in Sec.~\ref{sec:acceleration_HFQQ}. 

The existence of an imaginary part of the potential and the real-time dynamics of the heavy-quark-pair-heat-bath interaction were not the only factors in the evolution of quarkonia as QGP probes that started from the simple picture of quarkonia as a \textit{thermometer}. It was also realized that (re)generation from the heavy-quarks within QGP could counter the expected suppression~\cite{Thews:2000rj,Braun-Munzinger:2000csl}. This is quite natural once one allows for transitions as indicated by an imaginary potential. Since there is a medium-induced process making the transition between color-singlet to color-octet states, the opposite reaction should also be allowed. The interpretation of the obtained yield of quarkonium is thus more complex than what was initially considered, as it is also sensitive to different physical phenomena. The yield and spectra of regenerated quarkonia relate to the abundance of open heavy-flavor in the system and corresponding (partially) thermalized $p_T$-spectra. It immediately follows that the yield and spectra of regenerated quarkonia are, in principle, sensitive to the abundance and momentum spectra of open-charm states in the system. At RHIC and the LHC energies, the $c\bar{c}$ pair multiplicity becomes large to boost charmonium production. Two approaches are currently employed to describe charmonium regeneration: a statistical approach, where the additional component is mainly produced at the chemical freeze-out, and kinetic recombination, which yields a continuous regeneration over the QGP lifetime. Despite the details of each model, for very high energy densities, both may lead to a $J/\Psi$ enhancement. We refer the reader for chapter \ref{sec:acceleration_HFQQ} for further details on the experimental findings that show this process at work and the main directions to decide between these two directions. The fact that charm observables at low transverse momentum show features in terms of transverse momentum spectrum, particle yield ratios and $v_n$~coefficients close to kinetic and relative chemical equilibrations, indicate at least a partial loss of memory. This observation indicates that charm is a probe sensitive to the late stages of the medium history. The precise relative role of dynamical in-medium interaction and hadronization is still to be clarified. 

While at LHC energies there are evidences of quarkonium regeneration, the Future Circular Collider (FCC), expected to operate at a $\sqrt{s} = 39~\rm{TeV}$ for $PbPb$ collisions, might bring qualitative novel insights. The average temperature of the produced QGP would be slightly larger than at the LHC (about $500~\rm{MeV}$), but the anticipated thermalization time would decrease by a factor 2. As such, the overall increase of the initial QGP temperature at FCC energies will allow having a sizable fraction of the thermal gluons and light quarks. Assuming that we have a thermal-like exponential distribution, there is a large probability of surpassing the threshold to create $c\bar{c}$ pairs (twice the charm quark mass, $2\times 1.5\rm{GeV}$). As such, $c\bar{c}$ pairs can be abundantly produced, and several independent groups reported predictions for the thermal production of charm quarks at the new collider, all of them hinting at a sizeable contribution~\cite{Mangano:2017tke}.
Considering dynamical kinetic equations with an evolving medium, the charmonium yield has been reported to increase by $20\%$ to $80\%$ concerning a baseline without thermal production. The large gap is mainly driven by uncertainties on the QGP thermalization time, as they rely on rough extrapolations from prior LHC knowledge. However, this effect could overcompensate the screening effects even stronger than the pure regeneration from perturbatively produced charm resulting in a $R_{AA} > 1$. Statistical approaches also predict a nuclear modification factor above unity along the same lines. Predictions for higher quarkonia state masses at FCC energies have also been reported. Depending on the interaction potential, the $\upsilon(1S)$ state could be enhanced, while the excited states $2S$ and $3S$ would be partially or totally suppressed in QGP. 

It is thus clear that, despite challenging, the interplay between sequential recombination and regeneration can be used to test the heavy $q\bar{q}$ interaction with the medium and identify the potential interaction dependence on the timescale of the process. Nevertheless, quarkonia states are not the only probes able to probe the QGP time structure. Open heavy-flavor production also depends on the interaction potential, and its spectra also reflects the QGP transport properties (see section \ref{sec:acceleration_transport}). 

Traditionally, when looking at heavy-flavor observables, one assumes an early state production thanks to the large mass of the heavy-quark pair and tries to establish contact to thermal concepts and material properties for low~$p_T$ production. The state-of-the-art perturbative integrated transverse momentum and low-$p_T$ heavy-flavor cross-section are dominated by fixed-order calculations uncertainties with the heavy-quark mass as a hard scale. These quarks are usually plugged into the start of the medium evolution from the beginning in today's modeling. At high transverse momentum, the heavy quark is identified as the leading parton of the jet based on the fragmentation function measurements in $e^+e^-$~collisions. The leading parton assumption is also an excellent approximation in pp~collisions for the selections employed by ALICE to reveal the dead-cone effect~\cite{ALICE:2021aqk}.
In this review, we followed both commonly adopted pictures, early production and leading parton approximation. However, given the time dependence of the parton-medium interaction, it is important to note that heavy-quark pairs can be produced via gluon splitting $g \to Q\bar{Q}$ with the gluon having the virtuality of the $Q\bar{Q}$~mass in the radiation patterns of a jet~\footnote{As a side remark, the production of massless photons at lowest $p_T$ amenable for pQCD, is in fact dominated by fragmentation, see for instance in~\cite{Klasen:2013mga} for a baseline calculation in nucleus-nucleus collisions.}. This process can also occur late in the shower, and a large phase space is available at the LHC. Exploiting this feature via jet substructure measurements to tag the transition from a gluon to a heavy-quark pair with controlled virtuality down to low scales could open a new venue to dial into the time development of a parton shower in the interaction of QGP. The measurement of J/$\psi$~fragmentation in jets by LHCb~\cite{LHCb:2017llq} in pp~collisions as well as the measurement by CMS~\cite{CMS:2021puf} in pp and PbPb collisions ($30~GeV/c>p^{\text{jet}}_T>40~GeV/c$, J/$\psi$~$p_T>6.5$~GeV/$c$) indicate, in the explored kinematics, that the momentum carried by this low~mass $Q\bar{Q}$ is softer than expected in leading order calculations. The natural interpretation is a significant fraction of late production within the parton shower. This is an intriguing first observation that can be exploited in the future to decompose the radiation as a function of time in QGP.     

\begin{figure}
    \centering
    \includegraphics[width=0.45\textwidth]{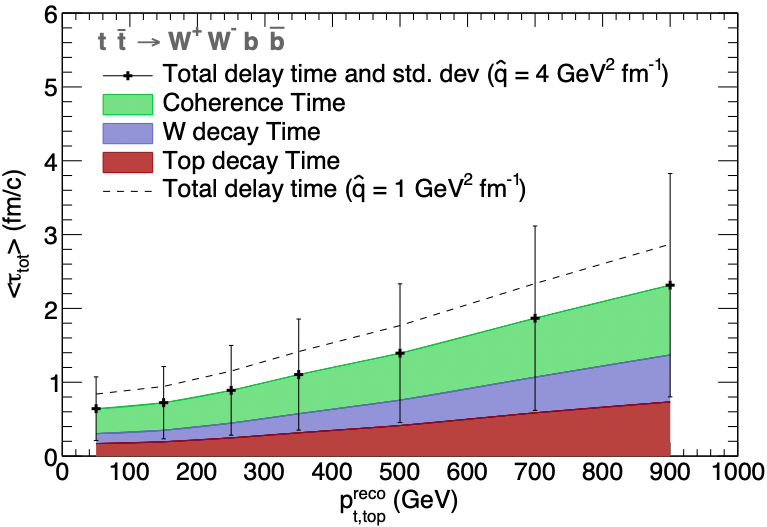}
    \includegraphics[width=0.53\textwidth]{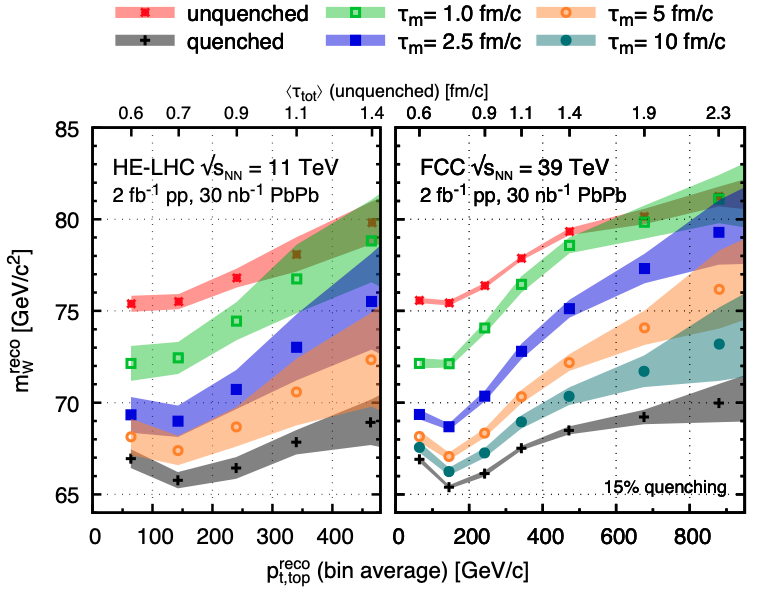}
    \caption{ (Left) Total delay time as given by Eq.~\eqref{eq:sec8_time}, for $\hat{q} = 4~\rm{GeV}^2/\rm{fm}$ (markers) and $\hat{q} = 1~\rm{GeV}^2/\rm{fm}$ (dash line). The standard deviation for the former case is also illustrated as error bars. The average contribution of each component is shown as coloured stacked bands (see legend). (Right) Reconstructed $W$-mass as a function of $p_{t,Top}^{reco}$ for HE-LHC (left) and FCC (right) centre of mass energies. The different colors refer to different finite size media, together with the $pp$ results (at red) and the fully quenched results (at black, see text). Figures taken from~\cite{Apolinario:2017sob}}
    \label{fig:sec8_time}
\end{figure}

However, novel ways of using this decay to create delayed probes inside the medium are being developed. One of these examples is to use top-initiated jets~\cite{Apolinario:2017sob}. The hadronic decay of top quarks (via de $b + W \rightarrow b +q\bar{q}$) provides one excellent time-delayed probe of the medium. After the top decays, the $W$-boson, as a colorless object, will propagate through the medium without medium-induced effects. Once it decays, it will produce a color singlet antenna formed by the $q\bar{q}$ pair that will remain in a color singlet until the medium can resolve the two quarks as independent sources. As mentioned in~\ref{sec:accleration_radiative_energy_loss}, the decoherence times goes as~\cite{Mehtar-Tani:2010ebp}:
\begin{equation}
	t_d = \left( \frac{12}{\hat{q} \theta^2} \right)^{1/3}\, ,
\end{equation} 
where $\theta$ is the antenna opening angle and $\hat{q}$ the jet transport coefficient. Summing the three components, it is possible to create probes that are formed inside the medium after an average time given by~\cite{Apolinario:2017sob}:
\begin{equation}
	\left\langle \tau_{tot} \right\rangle = \tau_{top} \gamma_{T,top} + \tau_{W} \gamma_{T,W} + t_d \, ,
	\label{eq:sec8_time}
\end{equation}
with the transverse boost $\gamma_{T,X} = (p_{T,X}^2/m_X^2 +1 )^{1/2}$ for the particle $X$ with transverse momentum $p_{T,X}$ and mass $m_{X}$. One can control the average decay time by measuring the top-initiated jet. In figure~\ref{fig:sec8_time}, left panel,  the available timescales (top axis) for different reconstructed $p_{T,top}$ and two different values of fixed $\hat{q}$ are shown. The less-dense (smaller $\hat{q}$) the medium is, the more the QCD antenna remains in a color singlet state. The different contributions from this decoherence time and the top and W decay are also shown in different stacked colors. To minimize the background effects to this particular decay channel, semi-leptonic decay of a $t\bar{t}$ event could be used for tagging. The reconstructed $W$-boson mass would provide the observable. If the total delay time surpasses the medium lifetime, the reconstructed $W$-mass would be compatible with a $pp$ system. For shorter delay times, the QCD antenna will experience in-medium energy loss, decreasing its mass. As such, a measure of the $W$-boson mass as a function of the reconstructed top $p_{T,top}$, would allow a tomographic analysis of the created medium in heavy-ion collisions. Given enough statistics, this and similar analyses could be used to probe the different QGP evolution timescales. In Fig.~\ref{fig:sec8_time}, right panel, it is shown the expected reconstructed $W$-boson masses for different finite-size media. For each of them, the total energy loss experienced by all event particles is $15\%$ of their initial energy. The exception is the decay products of the $W$-boson that will lose energy proportionally to the medium extent that they will propagate after $\tau_{tot}$, which is assumed to be linear. The $pp$ expectation baseline is shown in red, while the fully quenched scenario is in black. At FCC energies, $\sqrt{s_{\rm{NN}}} = 39~\rm{TeV}$, it will be possible to extensively use these channels to perform a complete survey of the spacetime structure of the produced QGP. However, at the HE-LHC energies, $\sqrt{s_{\rm{NN}}} = 11~\rm{TeV}$, the available statistics is limited, and one can only constrain limiting scenarios for short or long-lived media.

Single-particle measurements can carry some potential to inspect the time structure of the collision process. In theory, jets, being multi-scale objects that develop through the extended medium, are affected by different wavelength phenomena that occur over time.
The challenge is knowing how to access such information from the collection of final-state particles we denote as a jet. Jet substructure has been intensively explored for heavy-ion studies to oversee the multitude of processes that an in-medium parton shower experiences. The information is always separated in terms of momentum scales. Not only is the available experimental information, but it is also the scale that separates perturbative from non-perturbative effects. Alternatively, one can also use angles or radial distances from pQCD-inspired parton showers. In the presence of a medium, medium-induced radiation is not constrained to the dominant angular ordering behavior (see chapter ~\ref{sec:acceleration_energy_loss}). The medium destroys the interference pattern between the two emitters, and an additional anti-angular ordered contribution, proportional to the medium characteristics, appears.
In addition, QGP and its induced effects evolve as a function of time. The timescale of the collision process will then be highly correlated to the magnitude of medium-induced modifications that we want to identify concerning a vacuum baseline. Although these considerations are still in their infancy, they are currently being pursued to unlock the full exploitation of jets as probes of QGP. For instance, one of the most recent attempts was to use the concept of formation time~\cite{Apolinario:2020uvt}, 
\begin{equation}
\tau_{form} \approx \frac{E}{Q^2} \approx \frac{1}{2 E z (1-z) (1 - \cos \theta ) } \, ,
\label{eq:sec4_formation_time}
\end{equation}
where $E$ is the energy of the incoming particle with virtuality $Q$ that splits into a pair with opening angle $\theta$, each carrying a fraction $z$ and $1-z$ of the parent parton. By identifying jet populations with different $\tau_{form}$, classified as "early-" ($\tau_{form} < 1~\rm{fm/c}$) or "late-" ($\tau_{form} > 3~\rm{fm/c}$) initiated jets, it was found a strongly and weakly-modified jets respectively. This was done by reclustering a jet with the generalized $k_T$-family with $p = 0.5$ ($\tau$ algorithm), where:
\begin{equation}
d_{ij} \approx p_{T,i} \theta_{i,j}^2 \sim \tau_{form}^{-1}
\end{equation}
and identifying the first soft-drop emission (with $z_{cut}$ = 0.1). The connection of these results to the concept of a \textit{time} is still blurred. It is not clear whether this selection is just a consequence of kinematics, where jets of small opening angles, with typically longer formation times, are known to survive more in the medium (the jet selection bias discussed in ~\ref{sec:acceleration_model_independent}). In this sense, using this type of novel reclustering tool may not add much more value to a possible jet selection based on angle and/or transverse momentum.
Nonetheless, this and other preliminary studies~\cite{Apolinario:2020uvt,Caucal:2021bae,Mehtar-Tani:2019rrk} seem to indicate that, while the connection to a \textit{time} is not yet understood, these novel tools seem to perform slightly better with current state-of-art jet quenching Monte Carlo. The possibility to use those simply as a jet quenching classifier at LHC energies may already boost current heavy-ion studies. Future efforts will establish how robust these tools are in a heavy-ion background and if they can build a connection to the QGP lifetime.

%% file: outlook.tex
\section{Conclusions and outlook}
\label{sec:outlook}

A large experimental, phenomenological and theoretical effort is dedicated to studying heavy quarks and jets in heavy-ion collisions to characterize QGP. In this review dedicated to these observables, we started with three guiding questions:
\begin{itemize}
    \item How does the strongly coupled QGP emerge from an asymptotic free gauge theory?
    \item What is the substructure of QGP when probed at various resolution lengths? 
    \item What are the QGP transport properties?
\end{itemize}
Let us give some tentative answers. 

The last question is already a tall order but presumably the most concrete and modest one to attack. Despite the large variety of models and uncertainties behind the underlying assumptions, all approaches yield large stopping power of the medium and small spatial diffusion coefficients for heavy quarks. For jet transport, we compiled different estimates on $\hat{q}/T^3$ and discussed the remaining sizeable differences in the approaches. We related the transport coefficients from model calculations within approximate relations to the entropy to shear viscosity ratio. The derived parameters from hard probes with theoretical models are consistent with Bayesian fits to soft particle production and correlations. Although the associated theoretical uncertainties on the extraction of jet quenching, charm diffusion transport coefficients and subsequent translation to specific shear viscosity are still considerable, it is a starting point for quantitative extraction of QGP properties. Increasing experimental precision and improving theory input will allow excluding specific model scenarios in the future. Simultaneously, it will become increasingly important to fix the initial conditions, unify the hydrodynamics description of the system, and quantify better hadronization in theoretical models to achieve this goal. 

Detecting the QGP substructure across different length scales remains an open quest. In this respect, jets and heavy quarks give access to information that is not yet available via soft particle production. We explained that modern jet substructure measurements could provide information on quasi-particle scattering centers. Quarkonium measurements can address the fate of heavy-quark bound states in the QGP. Experimental results indicated strong modifications of quarkonium observables commonly associated with mechanisms related to deconfinement and its connection to open heavy-flavor. So far, we did not manage to resolve quasi-particles, and the question of the fate of heavy-quark bound states in the QGP is answered differently by different classes of models. These questions are longstanding in heavy-ion physics, but new perspectives based on future experimental precision and the theory progress give us hope that we can progress significantly in the coming years with the help of hard probes. 

Understanding the emergence of QGP from the underlying theory can be addressed in different ways. Pragmatically, we proceed via two more specific data-driven questions: how do the transitions occur from the initial state to the hydrodynamic description, and how does the system transition to hadrons? How far does the strongly coupled QGP description hold in less idealized situations than in central nucleus-nucleus collisions~\footnote{We know from studies in $e^+e^-$~collisions and $e$p that collective patterns in azimuthal correlations are not present within the experimental precision, see the discussion in Section~\ref{small:azim}. It seems that collectivity can be indeed switched off in collisions involving point-like particles. }? We pointed out how to explore the time dimension of different hard processes in the first question. Furthermore, the same interactions that control the approach from the initial state to a hydrodynamic description also control the parton-QGP interaction in jet physics. The non-observation of jet quenching in proton-nucleus and proton-proton collisions is still a vital feature to be understood, given the success of a hydrodynamical description in proton-nucleus collisions at the LHC. The planned oxygen-oxygen collisions at the LHC with similar densities, but a more controlled geometry than proton-lead collisions, should allow us to address this question in the next years~\cite{Brewer:2021kiv}. 

Hadronization, a genuinely non-perturbative process, is associated with the transition from the sQGP to hadrons. The thermal limit of particle production can be tested. Interesting patterns in heavy-flavor hadron production, particularly concerning baryon production, arise in pp and pPb collisions indicating the break-down of factorization with respect to the final state. This observation needs to be addressed both in theoretical approaches and with additional experimental measurements. A better conceptual consensus based on the available and upcoming data and precise modeling of these hadronization phenomena will also help us quantify QGP in large collision systems.

Beyond the current status, we look into a future with large new samples and data from new collision systems, but also qualitative advances with the detector upgrades at RHIC and the LHC. At the LHC, the upgrade of ALICE being currently commissioned targets the analysis of the full available data stream. The phase 2 upgrades of CMS will considerably improve the tracking performance in terms of resolutions and efficiencies and enlarge its acceptance to 8 units of rapidity~\cite{CMS:2017lum}. Furthermore, the CMS MIP Timing Detector will have particle identification capabilities over a wide pseudo-rapidity range~\cite{Butler:2019rpu}, improving the heavy-flavor hadron reconstruction significantly. LHCb will enhance its capabilities in nucleus-nucleus collisions~\cite{Citron:2018lsq} and will provide precise heavy-flavor measurements at the forward rapidity in proton-nucleus collisions~\cite{LHCb-CONF-2018-005}. In addition, the installation of a storage cell allows up to 100 times larger statistics in the fixed-target mode of LHCb starting with Run 3~\cite{Bursche:2018orf,LHCbCollaboration:2673690}. Studies for the possibility of a fixed-target mode of ALICE are investigated for Run 4. The investigation of small collision systems is complemented with the planned OO~run~\cite{Brewer:2021kiv}. 
At RHIC, the brand new sPHENIX detector~\cite{PHENIX:2015siv} will start to take data in 2023, promising to perform jet measurements and $\Upsilon$~measurements so far only available at the LHC and collect a large fraction of the data stream. 

On the horizon of 2030, we will see a new generation of heavy-ion experiments and the arrival of an electron-ion collider in the United States. At the LHC, the heavy-ion community is interested in a new detector set-up with the main emphasis on heavy-flavor observables~\cite{Adamova:2019vkf}. LHCb is undergoing a complete upgrade stretching its capabilities towards most central collisions in collider mode~\cite{LHCb:2018roe}. The experiments at the LHC will profit significantly from luminosity increases foreseen with the help of light-ion data taking indicated in the HL-LHC Yellow Report~\cite{Citron:2018lsq}. The initial state will be more strongly constrained by studies at the electron-ion collider~\cite{Accardi:2012qut,AbdulKhalek:2021gbh} and further improvements based on measurements at the LHC in ultra-peripheral collisions as well as inclusive measurements at the low-$x$ frontier as the ones presented in this review. 

Beyond currently available hadron beams, the future circular hadron-hadron collider (FCC-hh) has been proposed as a future energy frontier machine in the 21st century~\cite{Mangano:2017tke}. This machine with a PbPb collision energy of 39.3~TeV will offer unique opportunities to study hard probes in new kinematics regimes. The thermodynamic system arises from a very dense system of gluons in the initial state that can be tested by high-energy particle production. Charm quarks may become active degrees of freedom that can be thermally produced at higher collision energy. In particular, the time dependence of the parton interaction with QGP can be investigated with boosted objects such as the top production at large transverse momentum introduced in~\ref{sec:timescale}. The path ahead will establish more independence between the hard and soft sectors. The degeneracy provided by both probes will boost our sensitivity to time differential measurements and deepen our understanding of the theory of strong interactions.  

In summary, heavy-quark and jets are versatile tools for investigating QGP since they experience the whole system evolution. Different from soft probes, they could constrain the medium properties at various moments of the system's lifetime. The community has just become quantitative in extracting average medium properties with hard probes and developing handles to dive into the system's time evolution. This information can confirm soft-sector findings and enlarge our knowledge of the medium at higher energy scales and temperatures. Most importantly, they can potentially probe the full-time QGP evolution in heavy-ion collisions. The recent advances and upcoming programs give hope to progress on the field's open conceptual questions and contribute to our understanding of QCD as a whole.

%% file: acknowledgements.tex
\section{Acknowledgements}

The authors would like to thank Carlota Andrés, Valerio Bertone, Xabier Feal, Rabah Abdul Khalek, Aleksi Kurkela, Guilherme Milhano, Daniel Pablos, Sarah Porteboeuf and Carlos Salgado for useful discussions. The authors also thank the comments and feedback provided by Anton Andronic, Nestor Armesto, Guilherme Milhano, Francesco Prino when reading the draft.

LA acknowledges the financial support OE - Portugal, Funda\c{c}\~{a}o para a Ci\^{e}ncia e Tecnologia (FCT) under contract 2021.03209.CEECIND and projects EXPL/FIS-PAR/0905/2021 and CERN/FIS- PAR/0032/2021.
MW acknowldeges support by the GLUODYNAMICS project funded
by the "P2IO LabEx (ANR-10-LABX-0038)" in the frame-
work "Investissements d’Avenir" (ANR-11-IDEX-0003-01)
managed by the Agence Nationale de la Recherche (ANR),
France. YL acknowledges support by Department of Energy, Office of Science (US), under
Grant No. DE-SC0011088.